\documentclass{aa}  
\usepackage[normalem]{ulem}
\usepackage{siunitx}
\usepackage{graphicx}
\usepackage{txfonts}
\usepackage{xcolor}
\usepackage[T1]{fontenc}
\usepackage{ae,aecompl}
\usepackage{longtable}
\usepackage{booktabs}
\usepackage{comment}
\usepackage{lscape}
\usepackage{subcaption}
\usepackage{supertabular}
\usepackage{lipsum}
\usepackage{bm}         
\usepackage{natbib,twoopt}
\newcolumntype{T}{>{\tiny}l}
\usepackage[colorlinks=true,linkcolor=red,citecolor    = blue,breaklinks=true]{hyperref} 
\bibpunct{(}{)}{;}{a}{}{,}             
\makeatletter
  \newcommandtwoopt{\citeads}[3][][]{\href{http://adsabs.harvard.edu/abs/#3}%
    {\def\hyper@linkstart##1##2{}%
     \let\hyper@linkend\@empty\citealp[#1][#2]{#3}}}
  \newcommandtwoopt{\citepads}[3][][]{\href{http://adsabs.harvard.edu/abs/#3}%
    {\def\hyper@linkstart##1##2{}%
     \let\hyper@linkend\@empty\citep[#1][#2]{#3}}}
  \newcommandtwoopt{\citetads}[3][][]{\href{http://adsabs.harvard.edu/abs/#3}%
    {\def\hyper@linkstart##1##2{}%
     \let\hyper@linkend\@empty\citet[#1][#2]{#3}}}
  \newcommandtwoopt{\citeyearads}[3][][]%
    {\href{http://adsabs.harvard.edu/abs/#3}
    {\def\hyper@linkstart##1##2{}%
     \let\hyper@linkend\@empty\citeyear[#1][#2]{#3}}}
\makeatother

\newcommand{\kms}{\,km\,s$^{-1}$} 
\newcommand{\porb}{$P_{\rm{orb}}$} 
 
\newcommand{\Msol}{M$_\odot$}

\newcommand{\vgamma}{$v_\gamma$}
\newcommand{\vgammares}{$\epsilon v_\gamma$} 
\newcommand{\SSS}[1]{\textcolor{black}{#1}} 

\newcommand{\gc}{\texttt{VELOCE}}
\newcommand{\veloce}{\texttt{VELOCE}} 
\newcommand{\gaia}{\textit{Gaia}} 
\newcommand{\bonaCep}{258} 

\newcommand{\nbincep}{76} 
\newcommand{\nbincepnew}{32} 
\newcommand{\nbincepcand}{14} 
\newcommand{\nnonbincep}{24} 
\newcommand{\norbits}{33} 
\newcommand{\norbitsNew}{18} 
\newcommand{\norbitsKnown}{15} 
\newcommand{\norbitsTent}{3} 
\newcommand{\npolydeg}{30} 
\newcommand{\npolyNew}{16} 
\newcommand{\npolyKnown}{14} 
\newcommand{\nrvtfY}{12} 
\newcommand{\nrvtfYNew}{4} 
\newcommand{\nrvtfYKnown}{8} 
\newcommand{\nrvtfNO}{24} 

\newcommand{\hermes}{\texttt{Hermes}} 
\newcommand{\coralie}{\texttt{Coralie}}

\newcommand{\sscomm}[1]{\texttt{\textbf{\color{black}#1}}}
\newcommand{\referee}[1]{\color{black}#1}
\newcommand{\refereetwo}[1]{\color{black}#1}

\begin{document} 

   \titlerunning{\veloce~II -- Galactic spectroscopic binary Cepheids}
    \authorrunning{Shetye, et al.}
    
   \title{VELOcities of CEpheids (\veloce) \\ 
   II.~Systematic Search for Spectroscopic Binary Cepheids}

    \author{Shreeya S. Shetye\inst{1,3}\thanks{\email{shreeya.shetye@kuleuven.be}}
    \and Giordano Viviani \inst{1}\thanks{\email{giordano.viviani@epfl.ch}}
    \and Richard I. Anderson\inst{1}\thanks{\email{richard.anderson@epfl.ch}}
   \and Nami Mowlavi \inst{2}
   \and Laurent Eyer \inst{2}
   \and Nancy R. Evans \inst{4}
\and L\'aszl\'o Szabados \inst{5}
          }

    \institute{Institute of Physics, \'Ecole Polytechnique F\'ed\'erale de Lausanne (EPFL), Observatoire de Sauverny, 1290 Versoix, Switzerland
    \and D\'epartement d'Astronomie, Universit\'e de Gen\`eve, Chemin Pegasi 51, 1290 Versoix, Switzerland
    \and
    Instituut voor Sterrenkunde, KU Leuven, Celestijnenlaan 200D bus 2401, Leuven, 3001, Belgium
    \and Smithsonian Astrophysical Observatory, MS 4, 60 Garden St., Cambridge, MA 02138, USA
    \and Konkoly Observatory, HUN-REN Research Centre for Astronomy and Earth Sciences, MTA Centre of Excellence, Konkoly Thege Mikl\'os \'ut 15-17, H-1121 Budapest, Hungary}

   \date{Received \today; accepted date here}

\abstract{
Classical Cepheids provide valuable insights into the evolution of stellar multiplicity among intermediate-mass stars. Here, we present a systematic investigation of single-lined spectroscopic binaries (SB1) based on high-precision velocities measured by the VELOcities of CEpheids (\gc) project. 
We detected \nbincep\ ($29\%$) SB1 systems among the \bonaCep~Milky Way Cepheids in the first \gc\ data release, \nbincepnew~($43\%$) of which were not previously known to be SB1 systems. We determined 30~precise and \norbitsTent~tentative orbital solutions, \norbitsNew~($53\%$) of which are reported for the first time. This large set of Cepheid orbits provides a detailed view of the eccentricity $e$ and orbital period \porb\ distribution among evolved intermediate-mass stars, ranging from $e \in [0.0, 0.8]$ and \porb $\in [240, 9\,000]$\,d. Orbital motion on timescales exceeding the 11~yr \gc\ baseline was investigated using a template fitting technique applied to literature data. Particularly interesting objects include a) R~Cru, the Cepheid with the shortest orbital period in the Milky Way (240\,d), b) ASAS~J103158-5814.7, a short-period overtone Cepheid exhibiting time-dependent pulsation amplitudes as well as orbital motion, c) 17 triple systems with outer visual companions, among other interesting objects.
Most \gc\ Cepheids (21/23) that exhibit evidence for a companion based on \gaia\ proper motion anomaly are also spectroscopic binaries, whereas the remaining do not exhibit significant ($> 3\sigma$) orbital RV variations. \gaia~quality flags, notably the Renormalized Unit Weight Error (RUWE), do not allow to reliably identify Cepheid binaries, although statistically the average RUWE of SB1 Cepheids is slightly higher than that of non-SB1 Cepheids. Comparison with \gaia\ photometric amplitudes in $G-$, $Bp$, and $Rp$ also does not allow to identify spectroscopic binaries among the full \gc\ sample, indicating that the photometric amplitudes in this wavelength range are not sufficiently informative of companion stars.
}


   \maketitle
   
\section{Introduction}


Classical Cepheids (henceforth: Cepheids) are evolved, luminous, intermediate-mass ($\sim$~3 - 11~\Msol), radially pulsating stars. For more than a century, Cepheids have been crucial calibrators of the extragalactic distance scale \citep{Hertzsprung1913} thanks to the famous period-luminosity relation (PLR or Leavitt Law, \citealt{Leavitt1912}). The unceasing interest in Cepheids is not only due to their cosmological relevance, but also because these are crucial populations to understand stellar physics  and are important tracers in galactic studies \citep[and references therein]{Luck2011, Genovali2014, Poggio2021, Kovtyukh2022}. 

Given the importance of Cepheids as distance tracers, many astrophysical effects have been explored as potential biases of the Leavitt law, including multiplicity \citep[e.g.][]{Mochejska2000, Kiss2005}. For certain individual Cepheids, light contributed by companions can indeed affect distance estimates, the most extreme cases of this kind include V1334~Cyg in the Milky Way \citep{Gallenne2018} and a sample of Cepheids with double-lined spectroscopic companions in the Large Magellanic Cloud \citep[which are indeed selected using their Leavitt law outlier nature]{Pilecki2021}. However, \citet{2018AndersonbinH0} showed that statistically speaking, Cepheid companions cannot bias the extragalactic distance scale or the Hubble constant measured therefrom. A similar result was recently obtained using population synthesis \citep{Karczmarek2022}. However, orbital motion on timescales of up to several years may affect the accuracy of parallax measurements of Cepheids \citep{Anderson2015,Anderson2016rv,Benedict2022}. 

Parallaxes of Cepheids provided by the third data release (DR3) of the ESA mission \gaia\ do not yet account for orbital motion due to companions. However, this is expected to be available in future data releases, and DR3 already provided a large number of non-variable stars with non-single star solutions \citep{GaiaDR3Bin2022}. Combining astrometric and radial velocity data to determine orbits is particularly useful because it allows to measure the total mass of a system \citep[e.g.][]{Gallenne2019} and thus provides key information for constraining stellar models in the context of the mass discrepancy problem \citep[e.g.][]{Caputo2005,Anderson2016rot}.

In the past few decades, numerous complementary techniques are used for the detection of Cepheid binarity/multiplicity. 
Direct spectroscopic identification of low-mass companions at any separation is possible using  X-ray and ultraviolet (UV) observations \citep{Evans1992IUE, Evans2013, Evans2022}. 
The detection of orbital motion through radial velocity (RV) variations is effective to a wide range of companion masses and spectral types, however, it necessitates long temporal baselines \citep{Evans2015, Anderson2016rv}. 
Conversely, photometric binary detection methods, such as those using photometric amplitudes \citep{Coulsen1985b, Klagyvik2009} or \referee{photometric} colors and ratios of amplitudes of different colors \citep{Gieren1985},
require confirmation of the binary detection through subsequent spectroscopic investigations.
Cepheid companions can also be detected using optical long-baseline interferometry, if the contrast and angular separation between the Cepheid and its companion are sufficiently small \citep{Gallenne2014IAU, Gallenne2015, Gallenne2016, Gallenne2019}.

\textit{Gaia} proper motions have been introduced as indicators of multiplicity
\citep{Kervella2019b, Kervella2019a, Kervella2022}. While proper motion analysis is a powerful tool to detect resolved companions, it cannot be applied to the brightest stars that saturate the \textit{Gaia} detectors. 
Furthermore, \gaia~ astrometric quality flags like the reduced $\chi^2$ statistic are used as indicators of binarity assuming that the orbital motion leads to a photocentre wobble \citep{Vasily2020, Stassun2021}.  
In fact, several cuts on \gaia~quality flags like the RUWE and astrometric excess noise have been defined to filter the potentially multiple systems \citep{GaiaDR3Bin2022}. However, these cuts are established using stars where astrometric processing is simpler than in high-amplitude chromatically variable stars, such as classical Cepheids. It remains unclear if the same cuts can be applied to Cepheids, which are brighter than most stars used for calibrating \gaia~ systematics (cf. \cite{Khan2023} and references therein), and whose variability can cause excess noise in astrometric measurements.

The traditional approach of RV variations provides an unambiguous detection of a binary companion and helps constrain the orbital parameters. Dedicated RV monitoring campaigns have unraveled the orbital elements distribution of Cepheids (e.g., \citealt{Evans2015, BinCepDB}). Low-mass companions such as the one of $\delta\,$Cep \citep{Anderson2015} or companions of fainter Cepheids ($m_V > 8$\,mag) require significant observational effort and precision to be detected, and this can only be achieved by long-term RV monitoring. 

The VELOcities of CEpheids (\gc) project provides unprecedented RV time series of classical Cepheids (\cite{paperI}; Paper I henceforth). Taking advantage of the 11-year baseline of \gc, here we report the results of a systematic study of
single-lined spectroscopic binaries (SB1). Where the data allow it, we provide estimates of orbital periods, or describe the orbital motion using different methods.
The paper is structured as follows. The observations and data analysis are presented in Section\,\ref{sec:datanal}, in Section~\ref{sec:methods} we describe the different methods we used to analyze the SB1 Cepheids, followed by results in
Section\,\ref{sec:res}, which is divided into sub-sections for results from each method used to describe the binary. 
In Section~\ref{sec:sample}, we present the correlation between derived orbital parameters and compare our results with the \gaia~binarity indicators.
We briefly discuss and summarize this work in Section\,\ref{sec:disc}. 
The RV data as well as any fitted models, including orbital solutions and polynomial trends, used here are made publicly available  as part of Paper~I. 
In the Appendix, we have presented supporting figures (Sections~\ref{sec:guidingimage}, \ref{sec:orbitfits}, \ref{Appendix:Vgamma_Orbit_figures}, \ref{app:RVTF-noFigs})
and tables~(Sections~\ref{app:Gaiaastrometry}, \ref{app:SUCygRVTF}).

\section{Sample selection and available data\label{sec:data}\label{sec:datanal}}
\gc~comprises of precise Cepheid RV time-series data collected from both hemispheres using the high-resolution spectrographs \coralie~\citep{Coralie2001} and \hermes~\citep{Hermes2011}. We used a minimum of 30 RV measurements per star, extending to up to 300 RV measurements for some cases.
These observations cover a temporal baseline of nearly eleven years. 
Most of the \gc~observations targeted signal-to-noise ratios (SNRs) of at least 20, more commonly between 25 to 30, per pixel near 5500~\AA. With such SNR we have typical RV uncertainties in the order of 50 ms$^{-1}$. A detailed description of the observations and the RV data can be found in Paper~I. 

We identify SB1 candidates from \bonaCep~classical Cepheids observed by the \gc\ project (cf. Paper~I for details) via time-variable pulsation-averaged velocities, $v_\gamma$. Orbital motion in Cepheids introduces Doppler shifts on timescales of typically one year or longer that are mostly well separated from the pulsational timescales (of order days to months). Using \gc\ data, orbital motion can be detected in several different ways, depending on the availability and quality of the measurements (Section~\ref{sec:methods}). 
The cleanest examples in our sample are Cepheids where clear, and possibly repeating, variations of $v_\gamma$ are discernible by eye in the pulsation (Fourier series) fit residuals. Such cases generally have orbital semi-amplitudes of several \kms\ and  orbital periods of $\lesssim 3$\,yr.

While the selection function of Cepheids in \gc\ is rather complex, cf. Paper~I, we consider \gc\ neither particularly biased towards the detection of SB1 systems, nor against it. For example, \gc\ initially specifically targeted Cepheids with no prior RV information and Cepheids residing near star clusters \citep{Anderson2013}. Later, Cepheids bright and nearby enough to measure interferometric radii were added \citep[e.g.][]{Anderson2016vlti,Breitfelder2016}, and long-period Cepheids were targeted specifically for their use in distance ladder calibration \citep{Anderson2016rv}. \gc\ also followed Cepheids with the goal of detecting modulated variability \citep[e.g.][]{Anderson2014,Anderson2016c2c,Anderson2019polaris,Anderson2020}.  Of course, stars found to exhibit orbital motion were followed up with the goal of characterizing orbits. However, only few stars were added to specifically search for suspected orbital signals. 

\begin{figure*}
\begin{center}
\includegraphics[scale=0.55]{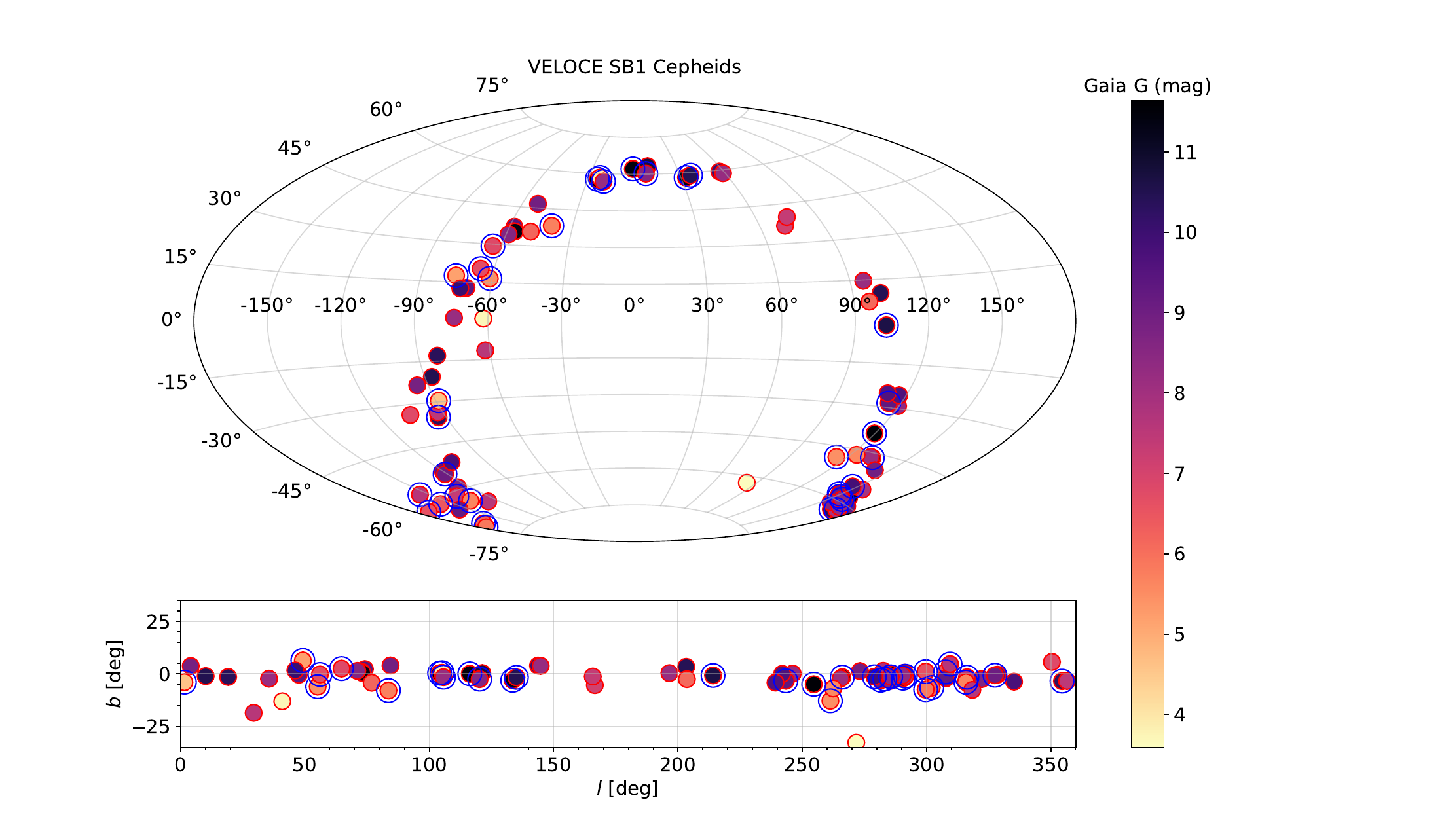}
\caption{\label{skymap}Location of the sample SB1 Cepheids on the sky in equatorial (Top panel) and in galactic (Bottom panel) coordinate system. The symbols are color coded in the \textit{Gaia G} magnitude. The symbols surrounded with blue circles are systems for which we provide an orbital solution in the current work.  } 
\end{center}
\end{figure*}

Table\,\ref{tab:BasicInfo} contains the list of stars considered in the present paper, ordered by whether their nature as spectroscopic binary was first reported here, was previously discussed in the literature. 
In Figure~\ref{skymap}, we present the on-sky distribution of the sample SB1 Cepheids and the range in \textit{Gaia G}~mag that they span. 

\onecolumn
\begin{tiny}
\sisetup{round-mode=places}
\begin{longtable}{@{}lTcS[round-precision=2]S[round-precision=4]cccc@{}}
\caption{Sample of binary Cepheids and candidates from the literature considered here.} \label{tab:BasicInfo}\\  

\toprule
Cepheid    &   \textit{Gaia} DR3 source ID & $\langle{m_{\rm{V}}}\rangle$ & \textit{Gaia}~ $\langle{m_{\rm{G}}}\rangle$      & { $P_{\rm puls}$}             & Status  & References & Evidence & Triple? \\
& & {(mag)} & {(mag)} & {(d)}  & & & & \\
\midrule
\endfirsthead
\multicolumn{4}{l}{\tablename\ \thetable\ -- \textit{Continued from previous page}} \\
\toprule
Cepheid                                          &  \textit{Gaia} DR3 source ID & $\langle{m_{\rm{V}}}\rangle$ & \textit{Gaia}~ $\langle{m_{\rm{G}}}\rangle$      & { $P_{\rm puls}$}             & Status  & References & Evidence & Triple? \\
& & {(mag)} & {(mag)} & {(d)} & & & & \\
\midrule
\endhead
\hline \multicolumn{9}{r}{\textit{Continued on next page}} \\
\endfoot
\bottomrule
\endlastfoot

\toprule
\multicolumn{9}{c}{Newly discovered spectroscopic binary Cepheids} \\
\midrule
AQ Pup                                        & 5597379741549105280              & 8.54  & 8.315241 & 30.19651884653673 & Susp  &     \hyperlink{R1}{R1}       & Trend      & Y        \\
ASAS~J064540+0330.4                           &                                  &       & & 3.014697133  & New   &            & Orbit   & Y           \\
ASAS~J064553+1003.8                           & 3350719221309022720              & 10.78 & 10.4372 & 2.68005122 & New   &            & Trend     &         \\
ASAS~J084951$-$4627.2                           & 5329675052782690944              & 11.03 & 10.8466835        & 3.788888725 & New   &            & Orbit    &          \\
ASAS~J100814$-$5856.6                           & 5258280842214814336              & 11.48 & 11.400031         & 3.766487195 & New   &            & Orbit       &       \\
ASAS~J103158$-$5814.7                           & 5351436724362450304              & 11.06 & 11.084759         & 1.119204370277835 & New   &            & Trend       &       \\
ASAS~J155847$-$5341.8                           & 5980814424369504256              & 9.99  & 9.450114          & 2.807504772141829 & New   &            & Trend       &       \\
ASAS~J174108$-$2328.5                           & 4116471792510901248              & 9.55  & 8.86463           & 3.779663972147002 & New   &            & Trend       &       \\
ASAS~J174603$-$3528.1                           & 4041107790775862272              & 10.94 & 10.312113         & 2.572441087 & New   &            & Orbit       &       \\
$\beta$ Dor                                       & 4757601523650165120              & 3.76  & 3.5934615         &  9.842731815006351  & New   &            & RVTF-Yes    &       \\
DR Vel                                        & 5313887130948758016              & 9.25  & 8.937748          & 11.19953465405213 & New &            & Trend     &         \\
FO Car                                        & 5241780677399802624              & 10.96 & 10.349283         & 10.35689201 & New   &            & Orbit       & Y       \\
GX Car                                        & 5257664497238811776              & 9.42  & 9.092877          & 7.1969302207 & New   &            & Orbit      &        \\
IT Car                                        & 5337191279937200256              & 7.9   & 7.8237414         & 7.532987155 & New   &            & Orbit       &       \\
MY Pup                                        & 5506374096132016512              & 5.65  & 5.465354          & 5.694105544 & New   &            & Orbit     &         \\
NT Pup                                        & 5537860123428416640              &       & 11.558196         & 15.56031755 & New   &            & Orbit        &      \\
OX Cam                                        & 473293889810320128               & 10.81 & 10.056244         & 5.06549147733 & New   &            & Trend     &         \\
RY Sco                                        & 4041690364529590144              & 7.51  & 7.4897914         & 20.3267495455 & New  &          & Trend      & Y        \\ 
RY Vel                                        & 5355057622307185280              & 7.86  & 7.8796782         & 28.17420905 & New   &            & Trend     &         \\
SX Vel                                        & 5329838158460391296              & 8.33  & 8.120997          & 9.550730552  & New   &            & Trend      &        \\
SZ Aql                                        & 4267549637851481344              & 7.92  & 8.201151          & 17.14086139 & New   &            & RVTF-Yes     &      \\
V0391 Nor                                     & 5881995546318024704              & 8.99  & 8.663814          & 4.373318918 & New   &            & Trend      &        \\
V0402 Cyg                                     & 2060021625508894592              & 9.87  & 9.520483          & 4.364924688 & New   &            & RVTF-Yes    & Y       \\
V0407 Cas                                     & 2012787293154800896              &       & 11.468034         & 4.5661016185 & New   &            & Orbit      & Y        \\
V0492 Cyg                                     & 2058874388200159872              &       & 11.564838         & 7.5780101633 & New   &            & Trend      &        \\
V0659 Cen                                     & 5868451040512196224              & 6.49  & 6.474747          & 5.6239867361 & Susp  &   \hyperlink{R2}{R2}         & Orbit   &Y           \\
V0827 Cas                                     & 430814876547869440               & 11.57 & 10.743507         & 3.1986031771 & New   &            & Trend       &Y       \\
V1162 Aql                                     & 4190143160245024256              & 7.81  & 7.5933595         & 5.3762072917 & New   &            & Trend     &         \\
V1803 Aql                                     & 4319018739208020096              & 10.35 & 9.390986          & 8.6281300519 & New   &            & Trend      &        \\
V2475 Cyg                                     & 2055881689337758336              &       & 11.413767         & 11.555709049 & New   &            & Trend       &       \\
VZ Pup                                        & 5600052040150252800              & 10.15 & 9.341325          & 23.17433227 & New   &            & Orbit       &       \\
X Cyg                                         & 1870258975238302208              & 6.47  & 6.199011          & 16.38656086  & New   &            & RVTF-Yes     &      \\
\midrule
\multicolumn{9}{c}{First orbital solutions for literature SB1 Cepheids}                \\
\midrule
BP Cir                                        & 5877460679352962048              & 7.52  & 7.326591          & 2.398125386 & Known &   \hyperlink{R4}{R4}           & Orbit    &          \\
FN Vel                                        & 5307761545524640256              & 10.25 & 9.854069          & 5.324151611 & Known & \gc,~\hyperlink{R5}{R5}, \hyperlink{R6}{R6}           & Orbit  &            \\
MU Cep                                        & 2200018111723748224              & 12.3  & 11.635483         & 3.767849084 & Known &  \gc,~\hyperlink{R6}{R6}           & Orbit      &        \\
R Cru                                         & 6054935874795049216              & 6.42  & 6.5845823         & 5.825671671 & Known  &    \hyperlink{R7}{R7}         & Orbit    &          \\
R Mus                                         & 5855468247702904704              & 6.33  & 6.2306046         & 7.510323654 & Known &   \hyperlink{R3}{R3}         & Orbit     &         \\
VY Per                                        & 459035766618689280               & 11.32 & 10.47096          & 5.531907552  & Known &  \hyperlink{R24}{R24}          & Orbit    &          \\
\midrule
\multicolumn{9}{c}{Literature SB1 confirmed by \gc}     \\
\midrule
$\alpha$ UMi & & 2.02 & &3.97201276471 & Known& \gc,~\hyperlink{R32}{R32} & Orbit \footnote{ We refer the reader to \cite{Anderson2019polaris} for the detailed orbital solution of $\alpha$ UMi using \gc~data.}  & \\
AD Pup                                        & 5614312705966204288              & 9.99  & 9.634962          & 13.59759898 & Susp  &   \hyperlink{R9}{R9}          & Trend      &        \\ 
AH Vel                                        & 5519380081746387328              & 5.76  & 5.587531          & 4.227158226 & Known  &    \hyperlink{R3}{R3}          & Trend     &        \\
AQ Car                                        & 5254662177677566464              & 8.84  & 8.642382          & 9.769574135 & Susp  &  \gc,~\hyperlink{R10}{R10}          & Trend     &         \\
AW Per                                        & 174489098011145216               & 7.51  & 7.085648          & 6.463825837 & Known &  \hyperlink{R11}{R11}            & Trend      &Y        \\
AX Cir                                        & 5873984023533350400              & 5.96  & 5.6583986         & 5.273442402  & Known &   \hyperlink{R4}{R4}         & Orbit     &         \\
CD Cyg                                        & 2058374144759464064              & 8.35  & 8.587173          & 17.07617397 & Known  &   \gc,~\hyperlink{R10}{R10}         & RVTF-Yes    &       \\
$\delta$ Cep                                  & 2200153454733285248              & 3.75  & 3.8505998         & 5.366267257 & Known &   \gc,~\hyperlink{R12}{R12}          & Orbit      &        \\
DL Cas                                        & 428620663657823232               & 8.63  & 8.57736           & 8.000933725 & Known &  \hyperlink{R13}{R13}, \hyperlink{R14}{R14}            & Orbit    &          \\
$\eta$ Aql                                       & 4240272953377646592              & 3.8   & 3.7476714         & 7.176834062 & Known &   \hyperlink{R15}{R15}         & RVTF-Yes    &       \\
FF Aql                                        & 4514145288240593408              & 5.38  & 5.170657          & 4.470984457 & Known &    \hyperlink{R16}{R16}         & Orbit       &Y       \\
FR Car                                        & 5337764640889249536              & 9.64  & 9.326704          & 10.7168931674  & Known &  \hyperlink{R17}{R17}           & Trend     &         \\
KN Cen                                        & 5864135319959353600              & 9.86  & 9.240802          & 34.027314893 & Known &    \gc, \hyperlink{R10}{R10}       & Trend    &          \\
LR TrA                                        & 5824226655600824704              & 7.8   & 7.588107          & 2.428288147 & Known &     \hyperlink{R7}{R7}       & Trend     &         \\
RS Ori                                        & 3368698813404804352              & 8.42  & 8.193092          & 7.5669250799  & Known &  \hyperlink{R18}{R18}           & RVTF-Yes     &      \\
RV Sco                                        & 6026412893938675712              & 6.61  & 6.7674227         & 6.061321383  & Susp  &  \hyperlink{R19}{R19}           & Trend     & Y         \\
RW Cam                                        & 473043922712140928               & 8.72  & 8.229256          & 16.41565439 & Known &    \hyperlink{R2}{R2}        & Trend     &         \\
RX Aur                                        & 200708636406382720               & 7.62  & 7.36267           & 11.62479025 & Known &    \hyperlink{R20}{R20}         & Trend    &          \\
S Mus                                         & 5855852527008107008              & 8.33  & 5.8812647         & 9.660008318  & Known &   \hyperlink{R16}{R16}          & Orbit     &         \\
S Sge                                         & 1820309639468685824              & 5.36  & 5.461868          & 8.382083647 & Known &    \hyperlink{R20}{R20}         & Orbit     & Y         \\
SS CMa                                        & 5616601820448126336              & 9.84  & 9.5626545         & 12.35247015 & Known &  \gc,~\hyperlink{R10}{R10}           & RVTF-Yes    &       \\
SU Cyg                                        & 2031776202613700480              & 6.44  & 6.798264          & 3.845750744484619 & Known &    \hyperlink{R21}{R21}         & Orbit     &         \\
SY Nor                                        & 5884729035255064064              & 9.77  & 9.040833          & 12.64521312 & Known &  \hyperlink{R2}{R2}, \hyperlink{R14}{R14}          & Orbit     &         \\
SZ Cyg                                        & 2071433765909167232              & 9.37  & 8.952725          & 15.109624588 & Known & \gc,~\hyperlink{R10}{R10}           & RVTF-Yes   &        \\
T Mon                                         & 3324535073449061504              & 5.98  & 6.124317          & 27.03550151 & Known &   \hyperlink{R13}{R13}, \hyperlink{R20}{R20}      & Trend    &          \\
TX Mon                                        & 3106349738382214144              & 10.67 & 10.601333         & 8.702117829  & Known &   \hyperlink{R8}{R8}         & Orbit     &         \\
U Vul                                         & 1825621002188696448              & 7.16  & 6.700444          & 7.990663373 & Known &  \hyperlink{R17}{R17}          & Orbit     &         \\
UX Per                                        & 506699870563323264               & 11.26 & 11.410275         & 4.565708977 & Known &   \hyperlink{R8}{R8}          & Trend    &Y          \\
UZ Sct                                        & 4104869264724284544              & 10.91 & 10.364139         & 14.74752861 & Susp  &  \hyperlink{R22}{R22}          & Trend     &Y      \\
V0340 Ara                                     & 5937099633141128448              & 10.03 & 9.741439          & 20.81548748 & Known &   \hyperlink{R7}{R7}         & RVTF-Yes    &       \\
V0916 Aql         & 4313179507891570304 & 10.86  & 10.064854 & 13.44360207  & Known & \hyperlink{R33}{R33}  & RVTF-Yes & \\
V1334 Cyg                                     & 1964855904803120640              & 5.88  & 5.724996          & 3.332489725 & Known &  \hyperlink{R23}{R23}          & Orbit    &Y          \\
VY Sgr                                        & 4094007532908816896              & 11.36 & 10.495356         & 13.55852787 & Known &    \hyperlink{R22}{R22}        & RVTF-Yes   &        \\
W Sgr                                         & 4050309195613114624              & 4.69  & 4.585245          & 7.595078119 & Known &   \hyperlink{R16}{R16}         & Orbit     &Y         \\
XX Cen                                        & 5871922507947292032              & 7.3   & 7.621872          & 10.95195648  & Known &   \hyperlink{R13}{R13}         & Orbit      &        \\
XZ Car                                        & 5338036117182452096              & 8.67  & 8.311601          & 16.65206949 & Susp  &     \gc,~\hyperlink{R10}{R10}      & Trend    &       \\
YZ Car                                        & 5255254711361371520              & 8.24  & 8.387514          & 18.16955271  & Known &   \hyperlink{R10}{R10}, \hyperlink{R16}{R16}         & Orbit &              \\
Z Lac                                         & 2007201567928631296              & 8.57  & 8.185639          & 10.88588420 & Known &   \hyperlink{R13}{R13}         & Orbit     &Y         \\
\midrule
\multicolumn{9}{c}{Literature SB1 not confirmed by \gc}  \\
\midrule
AA Gem                                        & 3430067092837622272              & 9.91  & 9.393121          & 11.30370465 & Known &   \hyperlink{R8}{R8}         & RVTF-No  &          \\
AN Aur                                        & 201574982848108416               & 10.21 & 10.013501         & 10.28851242 & Known &  \hyperlink{R25}{R25}          & RVTF-No   &         \\
AV Sgr                                        & 4069645924308096512              & 11.49 & 10.239907         & 15.32936111 & Known &  \hyperlink{R17}{R17}          & RVTF-No    &        \\
BB Sgr                                        & 4085919765884068736              & 6.69  & 6.5842776         & 6.637132461 & Susp  &     \hyperlink{R26}{R26}       & RVTF-No   &         \\
BG Cru                                        & 6058439910929477120              & 5.53  & 5.2722187         & 3.34253966  & Known &    \hyperlink{R7}{R7}        & RVTF-No   &         \\
BG Vel                                        & 5324034867356093056              & 7.69  & 7.220882          & 6.923857628 & Susp  &   \gc, \hyperlink{R31}{R31}        &  RVTF-No   &       \\
CS Ori                                        & 3330259852538068608              & 11.29 & 11.259904         & 3.889114374 & Known & \hyperlink{R8}{R8}           & RVTF-No   &         \\
FN Aql                                        & 4269036830424588800              & 7.96  & 8.005132          & 9.481369081 & Known &  \hyperlink{R27}{R27}          & RVTF-No    &        \\
RT Aur                                        & 3435571660360952704              & 5.55  & 5.3362513         & 3.728430532 & Susp &   \hyperlink{R28}{R28}         & RVTF-No    &        \\
RU Sct                                        & 4258436301367796480              & 8.82  & 8.713349          & 19.70764299 & Known &   \hyperlink{R17}{R17}         & RVTF-No   &         \\
RZ Vel                                        & 5523162573544337408              & 7.26  & 6.850103          & 20.40200461 & Known &  \gc, \hyperlink{R31}{R31}          & RVTF-No    &        \\
S Nor                                         & 5835124087174043136              & 6.49  & 6.175861          & 9.754719024 & Known &   \hyperlink{R1}{R1}          & RVTF-No     &         \\
SV Per                                        & 203496585576324224               & 8.86  & 8.6964445         & 11.12935697 & Known &  \hyperlink{R2}{R2}          & RVTF-No    &        \\
TY Sct                                        & 4256744187345732224              & 11    & 9.899057          & 11.05382753 & Known &   \hyperlink{R22}{R22}         & RVTF-No   &         \\
TZ Mon                                        & 3112475495616980992              &       & 10.510648         & 7.428215310 & Known &     \hyperlink{R17}{R17}       & RVTF-No    &        \\
U Sgr                                         & 4092905375639902464              & 6.68  & 6.42923           & 6.745297856 & Known &    \hyperlink{R14}{R14}        & RVTF-No   &         \\
VW Cen                                        & 5864955727424819200              & 10.36 & 9.853758          & 15.03680318 & Known &    \hyperlink{R17}{R17}        & RVTF-No   &         \\
VY Car                                        & 5351161399793209984              & 6.87  & 7.337772          & 18.88235666 & Susp  &   \gc,~\hyperlink{R10}{R10}        & RVTF-No    &        \\
WZ Sgr                                        & 4094784475310672128              & 7.45  & 7.682094          & 21.84941843 & Known &     \hyperlink{R30}{R30}       & RVTF-No   &         \\
X Lac                                         & 2004036486267748352              & 8.42  & 8.115799          & 5.445378789 & Known &  \hyperlink{R20}{R20}          & RVTF-No   &         \\
X Pup                                         & 5620098679741674496              & 8.46  & 8.3476515         & 25.96970451 & Susp  &  \gc,~\hyperlink{R10}{R10}          & RVTF-No    &        \\
XX Car                                        & 5238808628736339584              & 9.42  & 9.077229          & 15.70520883 & Known &     \hyperlink{R30}{R30}       & RVTF-No   &         \\
YZ Aur                                        & 188724234539584256               & 10.24 & 9.882999          & 18.19430908 & Known &   \hyperlink{R8}{R8}         & RVTF-No    &        \\
YZ Sgr                                        & 4099189015819292800              & 7.02  & 7.0141516         & 9.553891677 & Known &   \hyperlink{R17}{R17}         & RVTF-No  &  \\

\bottomrule
\end{longtable}
\tablefoot{
Identifiers and magnitudes were compiled from SIMBAD, \gaia\ mean G-band magnitudes from the \gaia\ archive, and pulsation periods from Paper~I. Column \sscomm{`Status'} indicates whether a star's SB1 nature was newly discovered (`New'), or previously known or suspected (`Known' or `Susp'). Stars listed as `Susp' among newly discovered SB1 systems were considered tentative previously.
References to previous studies are abbreviated using alphanumeric codes as follows.
R1: \hypertarget{R1}{\cite{Evans1994a}}, 
R2: \hypertarget{R2}{\cite{Evans1994b}}, 
R3: \hypertarget{R3}{\cite{Evans1982}}, 
R4: \hypertarget{R4}{\cite{Petterson2004Orbits}}, 
R5: \hypertarget{R5}{\cite{Kovtyukh2015}}, 
R6: \hypertarget{R6}{\cite{PhDthesis}}, 
R7: \hypertarget{R7}{\cite{BinCepDB}}, 
R8: \hypertarget{R8}{\cite{Szabados1998}}, 
R9: \hypertarget{R9}{\cite{Szabados2013b}}, 
R10: \hypertarget{R10}{\cite{Anderson2016rv}}, 
R11: \hypertarget{R11}{\cite{Griffin2016}}, 
R12: \hypertarget{R12}{\cite{Anderson2015}},
R13: \hypertarget{R13}{\cite{Evans1995Masses}},
R14: \hypertarget{R14}{\cite{Bersier1994}},
R15: \hypertarget{R15}{\cite{Evans1991}},
R16: \hypertarget{R16}{\cite{Gallenne2019}},
R17: \hypertarget{R17}{\cite{Szabados1996}},
R18: \hypertarget{R18}{\cite{Evans1990}},
R19: \hypertarget{R19}{\cite{Evans1992IUE}},
R20: \hypertarget{R20}{\cite{Gorynya1996}}, 
R21: \hypertarget{R21}{\cite{Wahlgren1998}},
R22: \hypertarget{R22}{\cite{Pont1994}},
R23: \hypertarget{R23}{\cite{Gallenne2013b}},
R24: \hypertarget{R24}{\cite{Szabados1992}},
R25: \hypertarget{R25}{\cite{Madore1977}},
R26: \hypertarget{R26}{\cite{Gieren1982}},
R27: \hypertarget{R27}{\cite{Szabados2014}},
R28: \hypertarget{R28}{\cite{Turner2007}},
R29: \hypertarget{R29}{\cite{Russo1981}},
R30: \hypertarget{R30}{\cite{Bersier2002}},
R31: \hypertarget{R31}{\cite{Szabados2013}},
R32: \hypertarget{R32}{\cite{Anderson2019polaris}}, 
R33: \hypertarget{R33}{\cite{Gorynya1996b}}.
The stars with `\gc' in the `References' column are cases where \gc~data was previously used to establish the binarity or the orbit. 
Column `Evidence' states whether we determine a combined model for orbit and pulsation (`Orbit'), determine a trend in the pulsation residuals (`Trend'), or whether template fitting involving literature RV data indicates a time-variable $v_\gamma$ (`RVTF-Yes') or not (`RVTF-No'). The last column `Triple?' indicates whether the star is likely part of a triple system, usually involving a visually resolved companion.}
\end{tiny}
\twocolumn

\section{Data treatment/Methods} \label{sec:methods}

\subsection{RV curve models adopted}\label{sec:model}

In most Cepheids, the RV variability due to pulsation dominates the RV time series with pulsational peak-to-peak amplitudes typically
ranging from $20 - 70$\kms. It is customary to model the pulsational variability using a Fourier series with a number of harmonics adapted to match the complexity of the RV variability. Paper~I describes this process for all Cepheid RV curves in \gc. In short, the RV signal due to pulsation is modeled as:
\begin{equation}
v_r(t) = v_{\gamma} + \sum_{n}{a_{n} \cos{(2 n \pi \phi)} + b_{n}
  \sin{(2 n \pi \phi)}}\,,
  \label{eq:Fseries}
\end{equation}
where $n$ represents the order of the harmonic, $t$ is time (in heliocentric
Julian Date) and $\phi = (t - E) / P_{\rm{puls}}$ is the pulsation phase with epoch\footnote{Defined for all \gc~stars such that $\phi=0$ coincides with $v_\gamma$ on the steeper descending branch of the RV curve close to the mean observing date.} $E$ and the pulsation period, $P_{\rm{puls}}$. 
As described in Paper I, the optimal model is determined by fitting each RV time series using up to 20 Fourier harmonics and then performing model comparisons using the F-test and the Bayesian Information Criterion (BIC). 

The key evidence for the spectroscopic binary nature presented here is a long-term modulation of $v_\gamma = v_\gamma(t)$, typically on timescales $> 1$\,yr and much longer than the pulsation period. We thus identify binaries by the correlated fit residuals they produce (Eq.\,\ref{eq:Fseries}). Orbital motion is then either modeled using a Keplerian orbit or approximated via linear, quadratic, or even higher order polynomials if the orbit is not sufficiently sampled to determine the full orbital solution.

As noted by \citet{Anderson2014}, Cepheids can exhibit modulated RV curves that can be misinterpreted as showing evidence of orbital motion. We note that all targets discussed here are bona fide spectroscopic binaries as identified by the absence of modulated line shape variability \citep{Anderson2016c2c}, although a few of the binaries presented here also show additionally modulated spectral variability. A detailed description of modulated RV variability of \gc~Cepheids will be presented in future work.

\subsection{Keplerian fitting}\label{Sect:Kepfits}

In the case of spectroscopic binaries, $v_\gamma = v_\gamma(t)$ in Eq.\,\ref{eq:Fseries} varies with time. The degree to which such variation can be used to infer the orbital elements depends largely on the sampling, baseline, and precision. Depending on these factors, we employed two methods for fitting Keplerian orbits to the data.

When fitting pulsational variability and orbital motion simultaneously, we fitted the Fourier series (Eq.\,\ref{eq:Fseries}) and a Keplerian model simultaneously using the non-linear least squares routine \texttt{leastsq} available in \texttt{scipy.optimize} \citep{2020SciPy-NMeth}. Following e.g.~\cite{Hilditch}, the orbital motion is represented as:
\begin{equation}
v_\gamma(t) = v_{\gamma,0} + K \left[ \cos{\left(\omega + \theta\right)} + e \cos{\omega}
\right] \,,
\label{eq:Keplerian}
\end{equation}
where $\omega$ denotes argument of periastron, $K$ semi-amplitude (in \kms), $e$ eccentricity,
$\theta$ true anomaly, and $v_{\gamma,0}$ the systemic RV relative to the solar system barycenter. We note that $v_\gamma$ (without subscript 0) refers to the pulsation averaged velocity (cf. Paper~I), which coincides with $v_{\gamma,0}$ if an orbital solution is determined or if the star does not have a companion. The uncertainties on the parameters are determined using the covariance matrix returned by the fitting routine, which incorporates statistical information on the reliability of the parameters estimated. Two examples of combined Fourier series plus Keplerian orbits for classical Cepheids are illustrated in Figures \ref{fig:Example_kepler1} and \ref{fig:Example_kepler2}.

\label{sec:mcmc}
To understand and mitigate any dependence of our results on starting values, as well as to avoid convergence to a local minimum, we also implemented a Markov Chain Monte Carlo (MCMC) method for determining Keplerian orbits. 
The MCMC fits further allow to explore the posterior distribution of the orbital parameters to identify possible correlations and to rigorously quantify uncertainties \citep[e.g.][]{Ford_2005, Ford_2006}. However, we have not yet implemented an MCMC fit to  simultaneously model pulsational and orbital variations; this is foreseen for future work. 
Thus, all MCMC orbital solutions presented here use template fitting residuals of \gc~(c.f. Section 5.1 of Paper I for more details) as well as literature radial velocity data, where pulsational variability has been removed (cf. Section \ref{sec:RVTF}). 
The MCMC orbits presented in this work utilize an extended baseline, achieved through the combination of \gc~with literature datasets (V+L).

After testing different orbital parameterizations, we found that the following implementation converged the fastest \citep[cf. also][]{Fulton_2018}:
\begin{equation}
    \mathbf{x} = (v_{\gamma, 0},\ \ln P_{\rm{orb}},\  \ln K,\ e,\ \omega,\ \phi_0) 
\end{equation}
where $\phi_0$ is the orbital phase to the next pericenter passage at $t=E$.
We opted for a uniform prior for all parameters, and set sharp boundaries far away from possible solutions so that they would not alter the posterior distribution.
Initial guesses were obtained by computing the Maximum A Posteriori (MAP)\footnote{In the case of uniform priors, the MAP method coincides with the more known Maximum Likelihood Estimate (MLE) method.} probability estimates from a grid of starting values of the parameters.
The resulting orbits allowed us to visualize the possible different sets of parameters that could explain our data and have a sense of the stability of each solution by counting how many grid points converged to the same orbital parameters. We then chose the best parameters as the initial guess for our MCMC (an example of MCMC orbital fitting is provided in Figure \ref{fig:vgammaOrbitUVul}).

We used the MCMC implementation provided by the python package \texttt{emcee} \citep{emcee} and, as suggested by the authors, used the autocorrelation time, $\tau_{ACT}$, to determine whether the chain was sufficiently long. In particular, the convergence criterion consisted of the following inequalities: 
\begin{equation}
    \langle \tau_{ACT} \rangle < N_{\text{iteration}}/ 60
\end{equation}
\begin{equation}
    \frac{\Delta\tau_{ACT}}{\tau_{ACT}} < 0.01
\end{equation}
After running the MCMC, we visually inspected the resulting posterior distributions to check the absence of significantly non-gaussian distributions (Figure~\ref{fig:SUCyg_corner}). We also verified that the mean and standard deviation were compatible with the 16th, 50th, and 84th percentiles.
We use the mean and standard deviation as the final results for the orbital estimates and its uncertainties. 
Finally, we also investigated circular orbits in a few cases where $e$ was either very small or not significantly detected. In this case:
\begin{equation}
  \mathbf{x} = (v_{\gamma, 0}, \ln P_{\rm{orb}}, \ln K, \phi_0) \ ,
\end{equation}
whereby $\phi_0$ represents the orbital phase to the next outgoing node at $t=E$.

Using orbital elements, we calculate the projected semimajor axis as:
\begin{equation}
    a \sin{i}\ \mathrm{[au]} = \frac{1}{2\pi G} K~ P_{orb}~ \sqrt{1-e^2}
\end{equation}
and the mass function as
\begin{equation}
    f_{\mathrm{mass}}\ \mathrm{[M_\odot]}= \frac{1}{2\pi G} K^3~ P_{orb}~ {(1-e^2)}^\frac{3}{2} \ ,
\end{equation}
where $G$ denotes the gravitational constant and the mass of the Sun was assumed to be $M_\odot = 1.98840987 \cdot 10^{30}$\,kg (as used in \texttt{astropy.constants}).

\subsection{Orbital trends represented using polynomials} \label{Sect:Polyfits}
Cepheids whose pulsation residuals computed using Eq.\,\ref{eq:Fseries} indicated the presence of incompletely sampled orbital motion were modeled using the sum of Eq.\,\ref{eq:Fseries}  and a low-order polynomial of the form:
\begin{equation}
    v_{\mathrm{trend}}(t) = \sum_{k=1,...,i}{ p_k \cdot (t - E)^k } \ .
\end{equation}
The polynomial degree $i$ adopted depends on the structure observed in the pulsation residuals (e.g., linear or non-linear) and the overall v$_{\gamma(t)}$ variation over the \gc~temporal baseline. We prioritized low values of $i$, typically $< 3$, and sought to obtain a reasonably close fit to the data. 
Additional data are required to determine full orbital solutions in these cases. 
Please note that in the present study, Fourier+polynomial fitting was exclusively employed for incompletely sampled orbits within \gc. This differs from the approach in Paper~1, where polynomials were utilized to represent modulated variability of any origin.

We note that $v_{\mathrm{trend}}(t)$ represents the change of the pulsation averaged velocity $v_\gamma(t)$ over time, so that $v_\gamma$ in Eq.\,\ref{eq:Fseries} becomes the pulsation-averaged velocity near the midpoint of the RV time-series. Introducing such polynomials allows to faithfully recover the pulsational variability and to partially quantify the orbital motion already sampled by the available data. However, it should be noted that in this case $v_\gamma$ is specific to the time-series and not generally valid for the star. Specifically, $v_\gamma$ then does not refer to the center of mass line-of-sight velocity of the system with respect to the Sun.

\subsection{RV template fitting (RVTF)}\label{sec:RVTF}
We investigated variations of $v_\gamma$ over timescales exceeding the \gc\ baseline using RV datasets from the literature. To this end, we adopted the literature RV zero-point differences from Paper~I and the RV template fitting (RVTF) approach developed in \citet{Anderson2016rv,Anderson2019polaris}. In this approach, the pulsational RV curve is assumed to be fully characterized by the high-quality pulsation fits obtained in Paper~I using \gc\ data of each star. The template fitting technique then fits the \gc\ pulsation models to the available literature data while solving for two parameters, an offset in $v_\gamma(t)$ with respect to \gc, and a phase shift $\Delta \phi$. While the former is required to investigate binarity, the latter allows to efficiently deal with pulsation periods changing over the course of several decades without having to assume a functional form for such period changes (cf. Section~5.2 of Paper~1).

In Paper~I, we developed a semi-automated approach for clustering the available data sets using a Kernel Density Estimation algorithm whose bandwidth was varied according to data availability and quality. For binaries, we double checked the most appropriate bandwidth to use to avoid fitting data exhibiting significant variations of $v_\gamma$ as one epoch. 
Furthermore, we visually inspected all template fits to make sure they covered a sufficiently broad range in phase to be informative. Template fits restricted to an insufficiently narrow range in phase were discarded. 
For each clustered template fit and per literature reference, we obtained the mean date, range of dates, fit parameters $\Delta v_{\gamma}$, $\Delta \phi$, their uncertainties, as well as the time-series epoch residuals (denoted as \vgammares~henceforth). 

To investigate the significance of orbital signatures, we calculated the SNR between velocity offset and square-summed uncertainties as follows:  
\begin{equation}
    \mathrm{SNR_{SB1}} = \frac{|\Delta v_{\gamma}|}{\sqrt{(\sigma_v)^2 + (\sigma_{v_{\gamma}})^2 }} \ ,
    \label{eq:RVTFSNR}
\end{equation}
where $\Delta v_{\gamma}$ is the difference between the zero-point-corrected \footnote{These zero-points were established based on stars with stable \vgamma, more details can be found in Section~5.2 of Paper~1.} $v_{\gamma}$ obtained by template fitting and the pulsation averaged velocity from Paper~I. $\sigma_v$ denotes the uncertainty on $v_{\gamma}$, and $\sigma_{v_{\gamma}}$ is the uncertainty on the instrumental zero-point correction $v_{\gamma}$.
Using Eq.\,\ref{eq:RVTFSNR}, we searched for stars exhibiting deviations of more than $\mathrm{SNR_{SB1}} > 3$ and rejected any spurious $v_{\gamma}$ variations among closely neighboring cluster sets following visual inspection, cf. Section~\ref{sect:RVTFresults}.

Where feasible, we determined full orbital solutions using our MCMC algorighm (cf. Section~\ref{sec:mcmc}) using the \vgammares~from template fitting as input. In practice, this often worked better than trying to fit a combined Fourier series and Keplerian model, since period changes could be effectively removed using the template fitting algorithm.
The \vgamma~and \vgammares~of literature datasets used in the orbital fits with MCMC algorithm and in the RVTF analysis are included in the \gc~data files published as part of Paper~1.

\referee{Please note that a thorough analysis comparing \gaia~DR3 RVs with \gc~RVs was performed in Paper I. The main arguments against including \gaia~data in the current work were as follows: Firstly, their temporal baselines mostly overlapped with existing data, and no significant improvements in orbital coverage were identified. Secondly, we observed that certain stars displayed temporal trends in Gaia radial velocities (RVs) over timescales where \gc~RVs showed no variations in pulsation-averaged velocity. However, a clear pattern regarding when and for which stars this occurred could not be discerned. In light of these disparities, \gaia~RVs were omitted as a literature source for the spectroscopic binary search employing the method described above. With extended baselines, increased observation frequency, and improved data processing, future data releases with \gaia~DR4 and DR5 are poised to excel in detecting SB1 systems and refining orbital solutions.}

\section{Results}\label{sec:res}

\begin{table*}
\begin{center}
\caption{An overview of the results from the current work. }\label{tab:ResultsOverview}
    \begin{tabular}{l|l}
    \hline 
    \bottomrule
        Total SB1 detections & \nbincep \\
        No. of new SB1 detections & \nbincepnew \\
        No. of orbits & \norbits~(15~(New) + \norbitsKnown~(Known) + \norbitsTent~(Tentative)) \\
        No. of stars fitted with polynomial & \npolydeg~(\npolyNew~(New) + \npolyKnown~(Known)) \\
        No. of stars with RVTF-Yes & \nrvtfY~(\nrvtfYNew~(New) + \nrvtfYKnown~(Known))\\
        No. of stars with RVTF-No  & \nrvtfNO\\
        \hline
        \bottomrule
    \end{tabular}
    \end{center}
\end{table*}

This section describes the results of our systematic search for spectroscopic binary Cepheids using \gc\ data. An overview of the results is presented in Table~\ref{tab:ResultsOverview}.
Tables ~\ref{tab:orbits} and \ref{tab:orbits2} list orbital parameters for \norbitsNew~
Cepheids whose orbital parameters are being reported for the first time and \norbitsKnown~Cepheids whose orbits had previously been reported in the literature, respectively. Overall, we identified 
\nbincep\ bona fide SB1 Cepheids and \nbincepcand\ additional SB1 candidates in the \gc\ sample. \nbincepnew~of these are reported as SB1 systems for the first time. 
Unless otherwise specified, we here discuss only bona fide SB1 Cepheids.

We first discuss fully determined orbits in Section \ref{sec:fullorbits}. Sections~\ref{sec:trends} and \ref{sect:RVTFresults} present Cepheids with incompletely sampled orbits represented by polynomials and investigated using template fitting, respectively. We discuss the spectroscopic binary fraction in Section~\ref{Sec:bf} and provide brief notes on individual Cepheids in Section~\ref{sec:individual}. 

\subsection{Full orbital solutions\label{sec:fullorbits}}

This section presents the results obtained by fitting full orbital solutions to the RV time-series. We determined complete orbital solutions for \norbits~stars (including \norbitsTent~tentative orbital fits). Section~\ref{Sect:velOrbits} begins with results based on \gc\ data alone, followed by results based on combined \gc\ and (zero-point offset-corrected) literature RVs in Section~\ref{Sec:VplusLOrbits}. Tables~\ref{tab:orbits} and \ref{tab:orbits2} distinguish these solution types using the tags `V' and `V+L', respectively.

\begin{table*}
\caption{Orbital elements of the new spectroscopic binary Cepheids' orbits presented in this work. Table notes are provided at the end of Table~\ref{tab:orbits2}.}
\label{tab:orbits}

\sisetup{round-mode=places}
\renewcommand{\arraystretch}{1.1}
\setlength{\tabcolsep}{2pt}
\begin{tabular}{lS[round-precision=4] S[round-precision=2]|lS[round-precision=2]S[round-precision=3]S[round-precision=3]S[round-precision=4,scientific-notation=false]S[round-precision=1,scientific-notation=true]S[round-precision=2]lS[round-precision=2]}
\hline
Cepheid & {P$_{puls}$}      &  {T$_0$-2.4M} & data  & {P$_{orb}$ (d)} & {$e$} & {K (km s$^{-1}$)}       & {$a\sin i$ (au)}      & {fmass}      & {$\omega$ (deg)} & {M$_1$}   & {rms} \\
& {(d)} & & & {$\sigma_{P_{orb}}$} & {$\sigma_{e}$} & {$\sigma_K$} & {$\sigma_{asini}$} & {$\sigma_{fmass}$} & {$\sigma_{argperi}$} & {M$_\odot$} & {(km s$^{-1}$)}   \\
\hline
\multicolumn{12}{c}{New orbits from current work} \\
\hline
ASAS  & 3.014697133408803 & 59013.58575344974 & V & 901.1760531590489 & 0.443593746450277 & 5.809711445759326 & \num{0.4313116250892591} & \num{0.01317714054652676} & 203.142591304 &5.0 & 0.133\\
J064540$+$0330.4 & &2.667587775430524 &\ref{fig:ASAS0330Orbit} & 2.806494032185666 & 0.01050618837511512 & 0.05922255146770738& \num{0.005234284581294802}& \num{0.000465490075482364}& 1.394968813413105& & \\
ASAS  & 3.788888724510723 & 57854.63875269861 & V & 1578.270370403218 & 0.4317409114470686 & 14.75959639476132 & \num{1.93138902404954} & \num{0.3857566686373262} & 145.4440616594043 & 4.9 & 0.262\\
J084951$-$4627.2 & &6.602481796197955 & \ref{fig:ASAS4627Orbit} & 3.414114938275642 & 0.005441155009462919 & 0.15804536707794886& \num{0.021823591580016777}& \num{0.012861720949208835}& 1.373889550063201 & & \\
ASAS  & 3.766487195371095 & 58067.70688788355 & V & 2458.573090091723 & 0.4541113884548805 & 13.77734906716542 & \num{2.774010508344804} & \num{0.4710044717326998} & 20.85343234564256 &4.2 & 0.229\\
J100814$-$5856.6  & &1.794347925389624 &  \ref{fig:ASAS5856Orbit} & 1.223559263204456 & 0.004081272623992712 & 0.08279734709242276& \num{0.01793806741298391}& \num{0.00911312947038303}& 0.2892597075843363 & & \\
ASAS  & 2.572441087108965 & 58980.80126826584 & V & 2302.725208306967 & 0.3578096449198623 & 4.144754560614341 & \num{0.8192203436258875} & \num{0.01382881374673608} & 344.02 & 4.9 & 0.221\\
J174603$-$3528.1 & &97.27173501408603 & \ref{fig:ASAS3528Orbit} & 41.73888051354215 & 0.1527422059840751 & 1.4762701580705688& \num{0.29664339141326707}& \num{0.015005689014259178}& 13.49201774524856 && \\
FN Vel & 5.324151610552629 & 56407.86628361183 & V & 471.7948006837752 & 0.2177680787462827 & 21.90765138450724 & \num{0.9272727416071631} & \num{0.4777307003082529} & 42.0693661046049 &6.4 & 0.157\\
 & &0.3573467862223459 &\ref{fig:FNVelOrbit} & 0.061948776626779 & 0.000880084997224785 & 0.02517814615827142& \num{0.001088737360366946}& \num{0.001673372250775628}& 0.2096668128744358&& \\
FO Car & 10.35689199698813 & 58244.26824002632 & V & 1667.573959415696 & 0.3776307286940477 & 14.93244550173746 & \num{2.119410032514236} & \num{0.4566061118364683} & 85.2554258350761 &6.1 & 0.155\\
 & &2.987994264883472 &\ref{fig:FOCarOrbit} & 2.818161989677259 & 0.002008642205319132 & 0.030433639831150775& \num{0.00591634164523468}& \num{0.003139786079346267}& 0.5859750012241262 && \\
&  &  58244.8872 &V+L   &   1662.2515   &    0.3775     &    14.9105    &    \num{2.1096}    &\num{0.453350509072} &    85.7245   &&   0.5770  \\   
&&    1.8067  & \ref{fig:VgammaOrbitFOCar}    &    1.5697     &    0.0018     &    0.0253     &   \num{0.0044}     & \num{0.002573410638} &    0.2656     & & \\    
GX Car & 7.196930220776081 & 56836.7814623098 & V & 2630.341289263578 & 0.03763108729038361 & 5.034360757838175 & \num{1.216346128294652} & \num{0.03469079870813331} & 6.686556729790391 &5.6 & 0.118\\
 & &32.46333981197005 &\ref{fig:GXCarOrbit} & 5.815660311948826 & 0.003477106283726837 & 0.017037941205533765& \num{0.004919716897497693}& \num{0.000360728012406864}& 4.443133364617084 && \\
IT Car & 7.53298715457763 & 59460.47494369336 & V & 1249.33376107457 & 0.5842820043609697 & 8.934432374975387 & \num{0.8326635625564061} & \num{0.04933086159790254} & 116.6953644476931 &6.0 & 0.091\\
 & &0.6629198720467369 &\ref{fig:ITCarOrbit}  & 0.7850895375993879 & 0.002023028251786716 & 0.024613659597038786& \num{0.002787303713119983}& \num{0.000487577435848597}& 0.1967484488984583 && \\
MU Cep & 3.767849084007058 & 56071.15670422189 & V & 2028.137878058533 & 0.4344790946379841 & 6.87979099960882 & \num{1.155187886257922} & \num{0.04998387228797467} & 10.29107526015332 &5.8 & 0.186\\
 & &3.022715587775972 &\ref{fig:MUCepOrbit} & 2.702546186858977 & 0.00381485646213739 & 0.03538823493190611& \num{0.006576336667704069}& \num{0.000832609064380584}& 0.6791036664229223&&\\
MY Pup & 5.694105543732406 & 57936.81382049262 & V & 2082.436584712036 & 0.1031200092720386 & 0.6705502906646842 & \num{0.1276703796731255} & \num{6.40020451858561e-05} & 80.71489181192572 &5.0 & 0.115\\
 & &431.7551979499551 &\ref{fig:MYPupOrbit} & 54.53837872827295 & 0.05169345711219556 & 0.023265117408481382& \num{0.005592358562735549}& \num{6.94686233924046e-06}& 84.81824087722065&&\\
 && 57974.9036 &V+L& 2186.1417 & 0.1411 & 0.6537 &  \num{0.1300} & \num{0.000061391556} &   91.5921 & &  0.3079   \\ 
&&   39.9302     & \ref{fig:VgammaOrbitMYPup}&    10.4375&    0.0099 &   0.0083 &    \num{0.0018} & \num{0.000002384780}   &7.9511 & & \\
NT Pup & 15.56031755483616 & 56679.12539259872 & V & 3653.645160013692 & 0.4655219061988115 & 8.837602926861937 & \num{2.626820669003807} & \num{0.1810946054772082} & 205.92 & 9.7 & 0.413\\
 & &6.427632919985515 &\ref{fig:Example_kepler2}  & 17.42241101877485 & 0.005158687626253751 & 0.051537354879252734& \num{0.021363959069819957}& \num{0.003682076239066864}& 0.6401783074523222& &\\
R Cru & 5.825671671294418 & 57292.64719009478 & V & 237.4230877692552 & 0.01733229320151919 & 1.028651393602178 & \num{0.02244576368237823} & \num{2.67561876498787e-05} & 205.1416441647547 & 5.5 & 0.162\\
 & &26.80000807374719 &\ref{fig:RCruOrbit} & 0.1371146624138122 & 0.01305797463160153 & 0.01220001924533285& \num{0.000266575269320327}& \num{9.52300627545895e-07}& 34.00185174174634 &&\\
&  &  57158.3761   & V+L &  237.6183    &        &    0.9985     &    \num{0.0218 }    & \num{0.000024506706} &         &    & 0.2530     \\
& &    0.1686  & \ref{fig:VgammaOrbitRCru}   &    0.0316     &  &    0.0034     &    \num{0.0001}     & \num{0.000000251771} &    & &  \\ 
V0407 Cas & 4.566101618693201 & 57445.27530997008 & V & 498.3567918527564 & 0.3073428705267426 & 1.863750520285949 & \num{0.0812438556980101} & \num{0.000287976989805758} & 102.4637656930134 &5.3 & 0.089\\
 & &11.76215397502153 &\ref{fig:V0407CasOrbit} & 0.826847390488594 & 0.04091435141186017 & 0.15324517108236904& \num{0.006776139544322989}& \num{7.20434434396489e-05}& 12.37238195106434 & &\\
VY Per & 5.531907552207723 & 57861.59809766626 & V & 806.175370318165 & 0.3823105275838298 & 3.835037899394561 & \num{0.2626005036516619} & \num{0.003716161544189495} & 108.7149089351549 & 5.0 & 0.083\\
 & &10.44863929979464 &\ref{fig:VYPerOrbit} & 0.7803681768816528 & 0.01532057871397825 & 0.08306462537432331& \num{0.005971632620801227}& \num{0.000253316421008961}& 2.319160598450992& & \\ 
 &&57858.1115 & V+L &   805.0835    &   0.3687     &  3.9279   &  \num{0.2702}    & \num{0.004060195869} &   109.3387  & &   0.4541     \\
 &&  15.5833    &\ref{fig:VgammaOrbitVYPer} &    0.6285    &   0.0198   &   0.1274  &   \num{0.0091} & \num{0.000408098156} &  4.1675   &  &  \\ 
VZ Pup & 23.17433226605626 & 58234.07958569905 & V & 3674.134935397733 & 0.1506311993774707 & 2.036214439160547 & \num{0.6798346578164179} & \num{0.003104321346446353} & 128.7384937246364 & 8.3 & 0.375\\
 & &241.003158493706 &\ref{fig:VZPupOrbit} & 222.8626096798067 & 0.0358165382625608 & 0.054107312720699394& \num{0.045176349404805825}& \num{0.000315183265567539}& 9.431472644636052& &\\
  & &  58184.4737  &V+L&   4026.3311&    0.2507    &  2.1117  &  \num{0.7566} & \num{0.003564224929} & 126.9455 & &    0.5796     \\
 &&  21.2455  & \ref{fig:VgammaOrbitVZPup} &   17.8197 &  0.0066 & 0.0133 & \num{0.0060} & \num{0.000071426619} & 1.3452 & &  \\   
\hline
\multicolumn{12}{c}{New `Tentative' orbits from current work}  \\
\hline 
  BP Cir     & 2.3981  &  55730.4207 &V+L  &   4339.4764   &    0.7792     &    2.1848     &    \num{0.5463}     & \num{0.001154780674} &   228.6850    &  &  0.3546   \\  
&  &    11.6028  &\ref{fig:VgammaOrbitBPCir}  &    8.4883     &    0.0060     &    0.0551     &    \num{0.0153}     & \num{0.000096579817} &    0.8075     & &\\  
 R Mus   &  7.5103   &  58352.0685 &V+L   &   4446.7222   &    0.5508     &    1.6941     &    \num{0.5779}     &\num{0.001302274961} &   220.2264    &   & 1.7469  \\   
 &&    19.2841 & \ref{fig:VgammaOrbitRMus}    &    19.0390    &    0.0046     &    0.0092     & \num{0.0045}     & \num{0.000026213890} &    1.1529     &  & \\   
 V0659 Cen   & 5.6240 &  52812.1094 &V+L  &   9276.0565   &    0.3231     &    8.2599     & \num{6.6651}     &\num{0.459089102283} &   182.5242    &    & 0.8553   \\  
&&    81.5196 & \ref{fig:VgammaOrbitV0659}    &   130.3835    &    0.0522     &    1.5191     & \num{1.2358}     &\num{0.254706347198} &    0.8974  & &   \\    

\hline

\end{tabular}
\end{table*}

\begin{table*}[h!]
\caption{
Orbital elements of \gc~SB1 Cepheids with literature-known orbits. } 
\label{tab:orbits2}

\sisetup{round-mode=places}
\renewcommand{\arraystretch}{1.1}
\setlength{\tabcolsep}{2pt}
\begin{tabular}{lS[round-precision=4] S[round-precision=2]|lS[round-precision=2]S[round-precision=3]S[round-precision=3]S[round-precision=4,scientific-notation=false]S[round-precision=1,scientific-notation=true]S[round-precision=2]lS[round-precision=2]}
\hline
Cepheid & {P$_{puls}$}      &  {T$_0$-2.4M} & data  & {P$_{orb}$ (d)} & {$e$} & {K (km s$^{-1}$)}       & {$a\sin i$ (au)}      & {fmass}      & {$\omega$ (deg)} & {M$_1$}  & rms \\
& {(d)} & & & {$\sigma_{P_{orb}}$} & {$\sigma_{e}$} & {$\sigma_K$} & {$\sigma_{asini}$} & {$\sigma_{fmass}$} & {$\sigma_{argperi}$} & {$M_\odot$} & {(km s$^{-1}$)}   \\
\hline
\multicolumn{12}{c}{Update to literature-known SB1s orbits}\\
\hline
 AX Cir     & 5.2734   &  54483.5890 & V+L  &   6284.5104   &    0.1868     &    10.9544    &    \num{6.2167}     & \num{0.811567806783} &   211.3387   & 5.0&    1.7559     \\
& &    33.0390   & \ref{fig:VgammaOrbitAXCir} &    12.4068    &    0.0090     &    0.2292     &    \num{0.1311}     & \num{0.051133530835} &    1.2120     & &\\
$\delta$ Cep & 5.366267257089509 & 58696.03234986413 & V & 3450.957035562507 & 0.7454743850888806 & 3.011549689276684 & \num{0.6367376561090067} & \num{0.002891142319520845} & 242.75 & 5.0& 0.093\\
 & &13.61372517348621 & \ref{fig:Example_kepler1} & 19.57125542804286 & 0.005706986138448262 & 0.018144569383598517& 
\num{0.008058388937584987}& \num{9.94908298833544e-05}& 0.5617219184939692 & & \\
DL Cas & 8.000933735036956 & 58113.84512060594 & V & 684.8891234680277 & 0.3440391513061821 & 16.65601836715227 & \num{0.9845660154935494} & \num{0.2713694513997766} & 25.92173105585034 &7.1 & 0.3\\
 & &1.001059654523932 & \ref{fig:DLCasOrbit} & 0.2107166542092301 & 0.004704356925474212 & 0.09841095232985729& \num{0.006099088772422727}& \num{0.005037623191381817}& 0.5985269811096878 & & \\
 & &  57434.6741   & V+L & 684.6749    &    0.3412     &    16.3428    &    \num{0.9668}     & \num{0.257176168641} &    29.2450    & &  1.4267   \\ 
 & &    0.9895 & \ref{fig:VgammaOrbitDLCas}    &    0.0550     &    0.0041     &    0.1020     &    \num{0.0062}     & \num{0.004968173692} &    0.6071   & &  \\  
FF Aql & 4.470984457346529 & 58253.28050232952 & V & 1433.200855329908 & 0.04993207939167316 & 4.787798666159667 & \num{0.6299547123978971} & \num{0.01623232920643414} & 304.86663924071 & 5.8 & 0.137\\
 & &35.40727381406652 &\ref{fig:FFAqlOrbit} & 5.658316961177006 & 0.0110332470424485 & 0.028314033449573818& \num{0.004492811863782082}& \num{0.000296251480493672}& 6.545512072819402 & & \\
&  &  58302.2342 &V+L  &   1431.7001   &    0.0682     &    4.8766     &   \num{0.6403}     &\num{0.017083474488} &   317.4064    & &   3.6782    \\ 
& &    7.8632 & \ref{fig:VgammaOrbitFFAql}     &    0.5506     &    0.0044     &    0.0144     &    \num{0.0019}     &\num{0.000152669433} &    2.0172     & & \\    
S Mus & 9.660008317542447 & 56663.88923652747 & V & 505.2732125520661 & 0.08656044976780858 & 14.78607318592295 & \num{0.684154896298129} & \num{0.1672929894006781} & 197.9168450319443 & 6.0 & 0.092\\
 & &1.39017882131802 &\ref{fig:SMusOrbit} & 0.08587125786883504 & 0.001763025895347488 & 0.018935259195755248& \num{0.000890058626667136}& \num{0.000647953716999878}& 0.9597442229013265 & & \\
 &  &  56152.5697 &V+L  &   505.1579    &    0.0808     &    14.7771    & \num{0.6839}     &\num{0.167246202252} &   194.1058    & &  2.1842  \\   
& &    2.4262 & \ref{fig:VgammaOrbitSMus}    &    0.0943     &    0.0039     &    0.0196     &  \num{0.0009}     &\num{0.000684489788} &    1.7517 & &    \\    
S Sge & 8.38208364744103 & 58141.05913046924 & V & 675.8354607892479 & 0.2362868963752267 & 15.52505506727696 & \num{0.9371461814522907} & \num{0.2403294770628674} & 200.4685454215376 &6.0 & 0.251\\
 & &2.157402957868327 &\ref{fig:SSgeOrbit} & 0.2738711716891759 & 0.003107921963042392 & 0.07100518865048161& \num{0.004364210661689279}& \num{0.003346263716409553}& 0.9779117452222771 & &\\
& &  57468.2034   &  V+L   &   676.0146    &    0.2477     &    15.6260    &    \num{0.9407}     & \num{0.243035290021} &   201.5275    &&   1.9845     \\
 & &    1.1866  & \ref{fig:VgammaOrbitSSge}   &    0.0645     &    0.0033     &    0.0504     &   \num{0.0031}     & \num{0.002438729460} &    0.5406     &     & \\    
SU Cyg   & 3.8458      &  56395.1941 &V+L &   548.2609    &    0.3419     &    29.4436    &    \num{1.3944}    &\num{1.203381072413} &   220.4813    &5.2  &  3.2758   \\  
 &&    1.7432& \ref{fig:VgammaOrbitSUCyg}     &    0.1049     &    0.0051     &    0.2176     &    \num{0.0107}     & \num{0.027626923924} &    1.5715     & &  \\ 
 SY Nor & 12.64521312231367 & 57784.69753610548 & V & 551.9245436080665 & 0.001078934794884134 & 15.3285918072429 & \num{0.7776607412187174} & \num{0.2059103053823602} & 255.1615506980371 &6.8 & 0.24\\
 & &321.3458048812978 &\ref{fig:SYNorOrbit} & 0.7681767781684938 & 0.003211402618513705 & 0.05516545863306972& \num{0.003000697981260003}& \num{0.002241524564164997}& 196.3367361997652 & & \\
 & & 57945.6719 & V+L &  551.9623    &&    15.3354  &  \num{0.7781}  & \num{0.206262134063} && &   0.4863     \\
& &  0.1620  &\ref{fig:VgammaOrbitSYNor} &  0.0561  & &   0.0132  & \num{0.0007} & \num{0.000532422283} & & &  \\  
 TX Mon & 8.702117829778473 & 59204.55388742112 & V & 1602.203642354045 & 0.5832039099737136 & 11.98483434771373 & \num{1.433799605460985} & \num{0.1531429160817492} & 295.48908039684227 &7.1 & 0.179\\
 & &4.49963557097468 & \ref{fig:TXMonOrbit}& 2.465452383902561 & 0.003770241810022369 & 0.1238675390012225& \num{0.01572551564261065}& \num{0.004994608703862224}& 0.4480496672137408 & & \\
 &&  57599.4070 &V+L  &   1606.9799   &    0.5845     &    11.9798    & \num{1.4359}     &\num{0.152939514539} &   295.1232  & &   0.7276    \\ 
&&    1.9134 & \ref{fig:VgammaOrbitTXMon}     &    0.9132     &    0.0015     &    0.0480     & \num{0.0061}     &\num{0.001943354569} &    0.2582  &&    \\ 
U Vul & 7.990663373892701 & 59935.90143035789 & V & 2528.773049067898 & 0.5507450108503994 & 3.154151183132403 & \num{0.6119521600069149} & \num{0.004779689446520979} & 1.368937809204958 &7.1 &  0.115\\
 & &5.28215497701398 & \ref{fig:UVulOrbit} & 8.570434192624152 & 0.005285193696731184 & 0.04548910130066881& \num{0.009419619423098552}& \num{0.000215909523643451}& 0.6860796195597175 & & \\
&&  57404.9228 &V+L  &   2521.2277   &    0.5552     &    3.1332     & \num{0.6039}     & \num{0.004622385636} &   359.3633    & &    1.0242     \\
& &    2.4673   & \ref{fig:vgammaOrbitUVul}   &    0.9388     &    0.0028     &    0.0169     & \num{0.0036}     &\num{0.000081372919} &    0.3287 & &       \\  
V1334 Cyg & 3.332489724704534 & 57172.485139377 & V & 1947.791776560139 & 0.2288643588499294 & 14.20814821371665 & \num{2.476314669174745} & \num{0.5338256129154219} & 227.109564436883 &5.2 & 0.051\\
 & &1.743218178343061 & \ref{fig:V1334CygOrbit} & 2.082868120313132 & 0.0014164422042397 & 0.03689010332313571& \num{0.007004885497712641}& \num{0.004232688558989027}& 0.3726783635046596 & & \\
 &&57186.8041  & V+L & 1931.1655 &  0.2358  & 14.4919 & \num{2.4999}  & \num{0.558904847064} & 231.0933 & &  0.8889     \\
   &&   2.6158   & \ref{fig:VgammaOrbitV1334} &   1.1184  & 0.0028  & 0.0488 & \num{0.0087} & \num{0.005776463179} &  0.5590   &&  \\   
W Sgr & 7.595078119317044 & 58121.3986863512 & V & 2006.304073873486 & 0.2722159383468885 & 1.507356405008145 & \num{0.2674871066699145} & \num{0.000634135434621928} & 317.91714659286095 & 5.5 & 0.195\\
 & &26.28361504841031 &\ref{fig:WSgrOrbit} & 17.7785914198005 & 0.03863734599027836 & 0.01321302241135146& \num{0.004510933043917224}& \num{2.78687540741288e-05}& 4.432164925503169 & & \\
XX Cen & 10.9519565013132 & 58211.32582117668 & V & 722.4397800252201 & 0.2950516194619363 & 5.39627472893656 & \num{0.3423941713868294} & \num{0.01025752910450199} & 283.11430753867876 &6.4 & 0.121\\
 & &1.662733643490934 &\ref{fig:XXCenOrbit} & 0.4220664727635537 & 0.004266949548127247 & 0.026175633160291843& \num{0.001738208483646338}& \num{0.000155298787401806}& 0.9206456007225363 &&  \\
YZ Car & 18.1695527147354 & 57936.96363336287 & V & 827.2048874452315 & 0.02890674506347962 & 10.38019235780192 & \num{0.7889426426473563} & \num{0.09571435569224941} & 352.67178522584834 & 7.4& 0.242\\
 & &38.6815687466984 & \ref{fig:YZCarOrbit}& 1.209967441679768 & 0.004198257117699893 & 0.06157274894960601& \num{0.004820949771850162}& \num{0.001709362138432453}& 16.68301194015072 & & \\
  && 57120.7259  &V+L&  830.0247  &&  10.19146 & \num{0.7776}  &\num{0.091037900048} &&   & 1.8727\\
    && 0.1534 & \ref{fig:VgammaOrbitYZCar} &0.1120 &&  0.0085 & \num{0.0007} &\num{0.000227823567} & & & \\
Z Lac & 10.88588420006621 & 58534.56031476192 & V & 382.7656776383766 & 0.01379688911769722 & 10.7735733033636 & \num{0.3790180398544885} & \num{0.04956563109291869} & 65.00793404113135& 8.0 & 0.095\\
 & &54.20982824729158 &\ref{fig:ZLacOrbit} & 0.1408141188326439 & 0.01998167061052529 & 0.3216924606507447& \num{0.01131859360485355}& \num{0.004440227124693783}& 43.45464221194877 & & \\
   \hline
\end{tabular}
\tablefoot{\scriptsize The pulsation period and epoch are retrieved after the RV fitting following the method described in Paper~1. In column~4, `V' denotes when the orbits were strictly fitted with \gc~data alone, while `V+L' denotes when the orbit was fitted using the combination of $v_{\gamma}$ residuals from \gc\ and literature datasets (See Section~\ref{sec:mcmc} for more details). 
The second row in the `data' column contains links to the orbital fit figures presented in the paper.
The next columns list the different orbital elements, namely, orbital period (P$_{orb}$), eccentricity ($e$), semi-amplitude (K), projected semi-major axis (asini), mass function (fmass), argument of periastron ($\omega$). 
We estimated the individual Cepheid masses ($M_1$) using period-mass relations based on Geneva stellar evolution models \citep{Ekstrom2012}, our method was similar to the one described in \cite{Anderson2016rv}.
The last column presents the `rms' (root mean square) of the fit between the Fourier+Keplerian model and the RV data for `V' orbits and between the keplerian model and epoch residuals for `V+L' orbits.}
\end{table*}

\subsubsection{Orbits from \gc\ data alone (V)}\label{Sect:velOrbits}

\begin{figure*}[h!]
    \centering
    \includegraphics[scale=0.5]{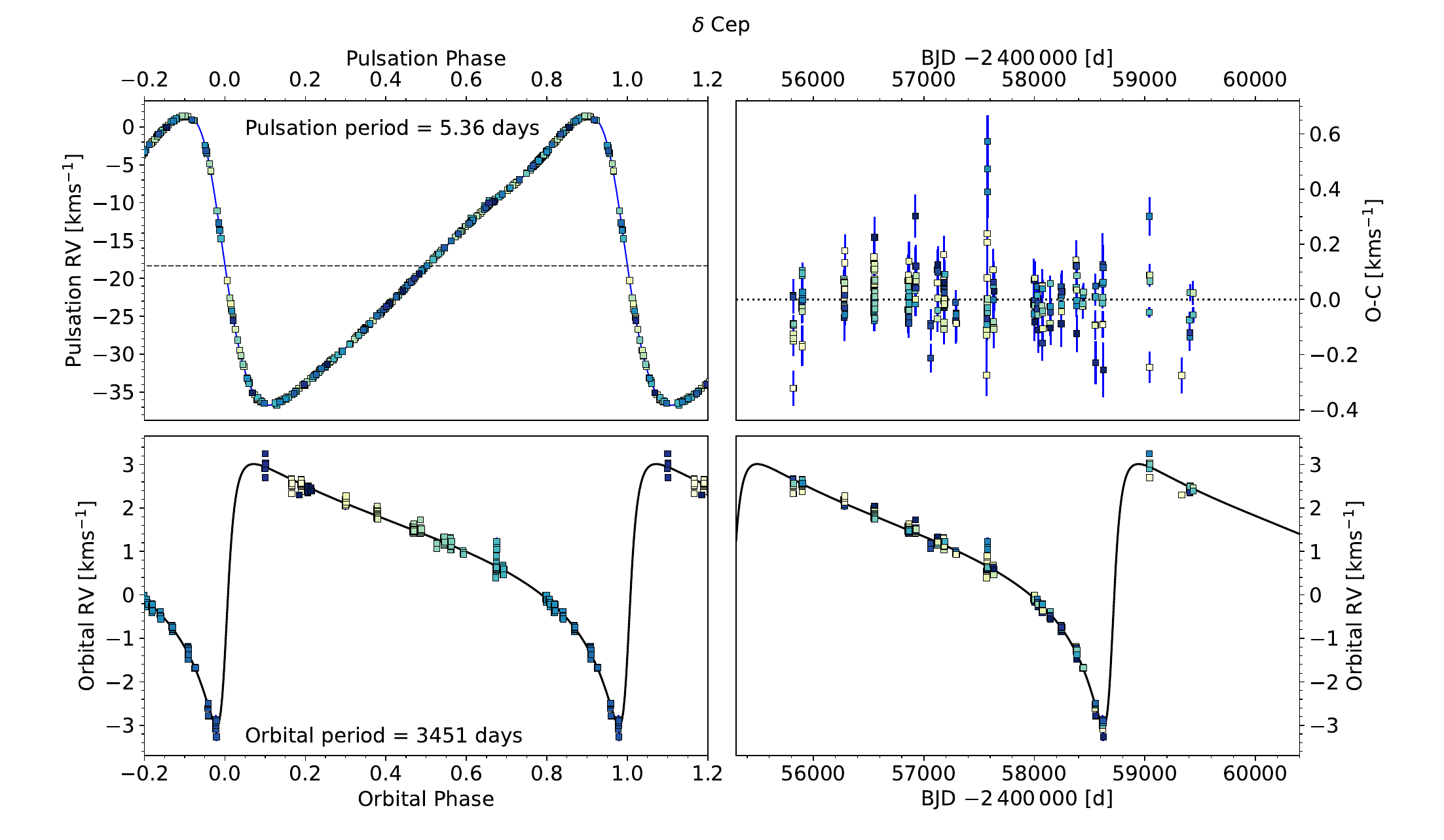}
    \caption{Pulsation and orbital fit of the naked-eye Cepheid prototype $\delta$~Cephei. Left top: pulsation RV variability where the systemic velocity, $v_{\gamma,0}$ is indicated by a dashed line. Right top: residuals after fitting the Fourier series and Keplerian orbit against observation date. Left bottom: phase-folded orbital RV variation against orbital phase. Right bottom: orbital RV variation versus the observation date.  }
    \label{fig:Example_kepler1}
\end{figure*}

\begin{figure*}[h!]
    \centering
    \includegraphics[scale=0.5]{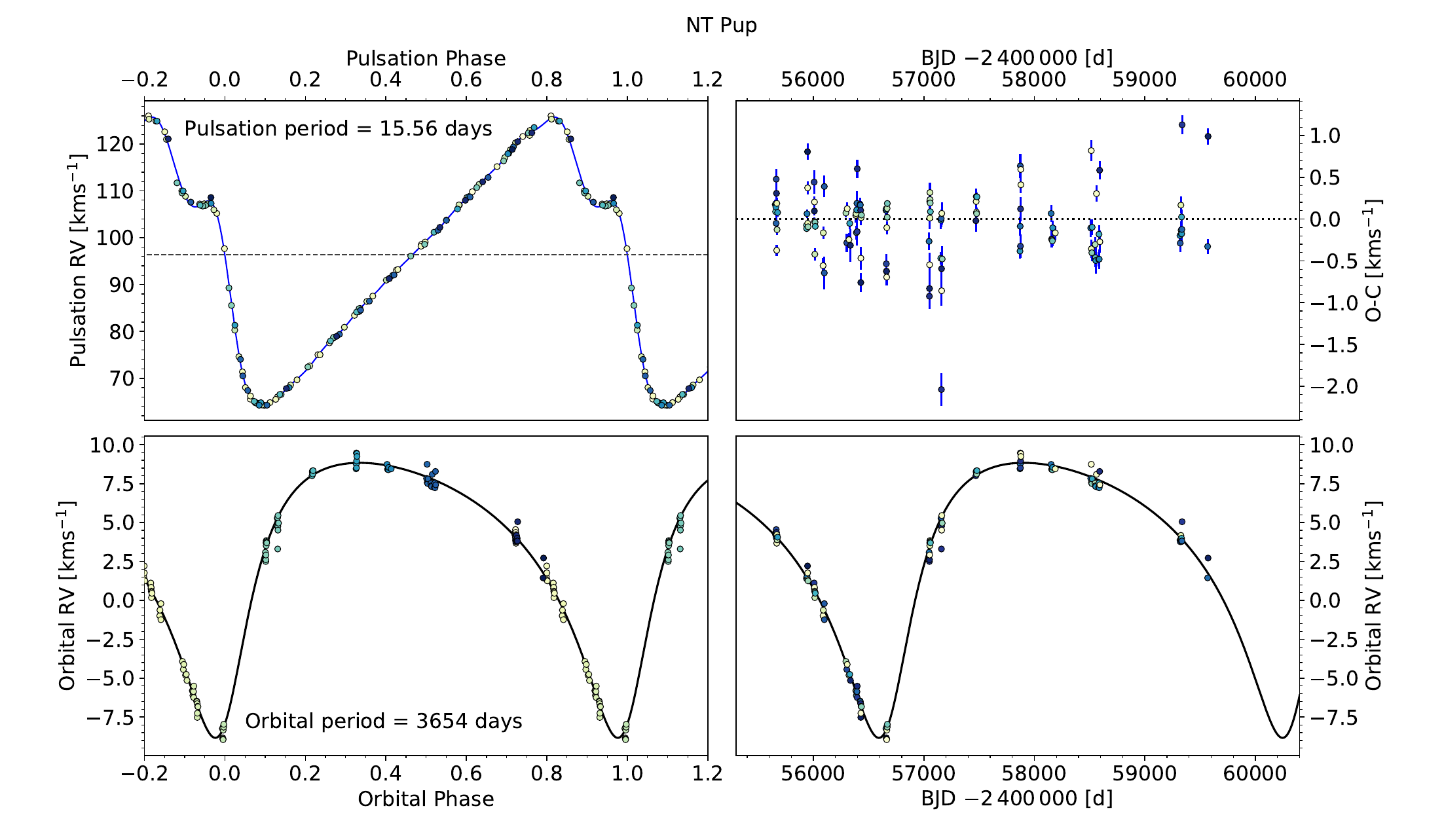}
    \caption{Pulsation and orbital fit of one of the faintest binaries in our sample, NT~Puppis. Figure description is same as Figure~\ref{fig:Example_kepler1}.  }
    \label{fig:Example_kepler2}
\end{figure*}

Thanks to the high RV precision and data homogeneity, orbits derived using \gc\ data alone are typically very accurate.  However, the available 11~yr baseline is not always sufficient to sample orbits fully, and occasional gaps in the time series may unfortunately coincide with particularly important orbital phases, preventing a definitive determination of all orbital parameters.  All systems presented here were fitted using the combined Fourier series plus Keplerian model, which simultaneously solves for pulsational and orbital variability. The resulting orbital periods range from 237 to approximately 4000 days, that is, the baseline available from \gc\ DR1. 

Figures~\ref{fig:Example_kepler1} and \ref{fig:Example_kepler2} illustrate two examples of combined Fourier series plus Keplerian orbits for the naked-eye prototype of classical Cepheids, $\delta$~Cephei, and the faint southern target NT~Puppis, cf. Section\,\ref{sec:individual} for additional details. All other combined fits are illustrated in Appendix\,\ref{sec:orbitfits}.

The choice of starting values for model fit parameters, notably \porb\ and $e$, can impact the determination of orbital solution. We therefore adopted reasonable starting values estimated by inspecting the Fourier series fit residuals. Where starting values could influence the result, we cross-checked the results using the MCMC template fitting algorithm, cf. Section~\ref{sec:mcmc}. For most stars, however, the data set sampled the orbital cycles sufficiently well to allow an initial guess close to the correct orbital period.

The best fit orbital parameters obtained here generally agree with previous orbital solutions published in the literature despite improved uncertainties. However, we obtain significantly different orbital periods for six Cepheids $\delta$~Cep (cf. Figure\,\ref{fig:Example_kepler1}), AX~Cir, MU~Cep, V1334~Cyg, W~Sgr, and XX~Cen. These stars, among others, are discussed in detail in Section~\ref{sec:individual}. 

\subsubsection{Orbits from \gc\ and literature data (V+L)}\label{Sec:VplusLOrbits}

\label{sec:vgammafitting}
\begin{figure*}
    \centering 
    \includegraphics[width=\textwidth]{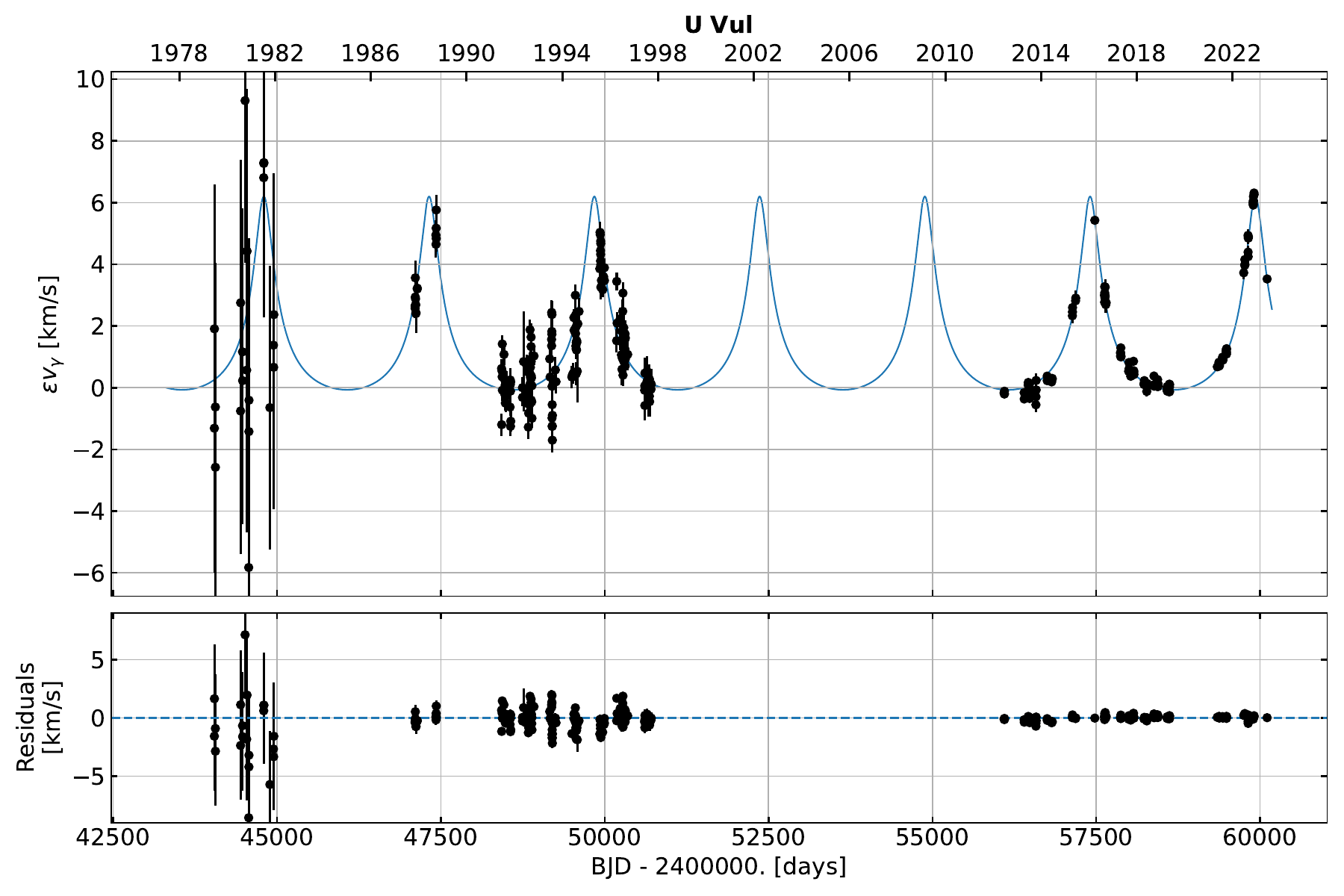}
    \caption{ \label{fig:vgammaOrbitUVul} The orbit of U~Vul determined by fitting $v_{\gamma}$ epoch residuals of literature and \gc~data.}
\end{figure*}

\begin{figure*}
    \centering
    \includegraphics[width=\textwidth]{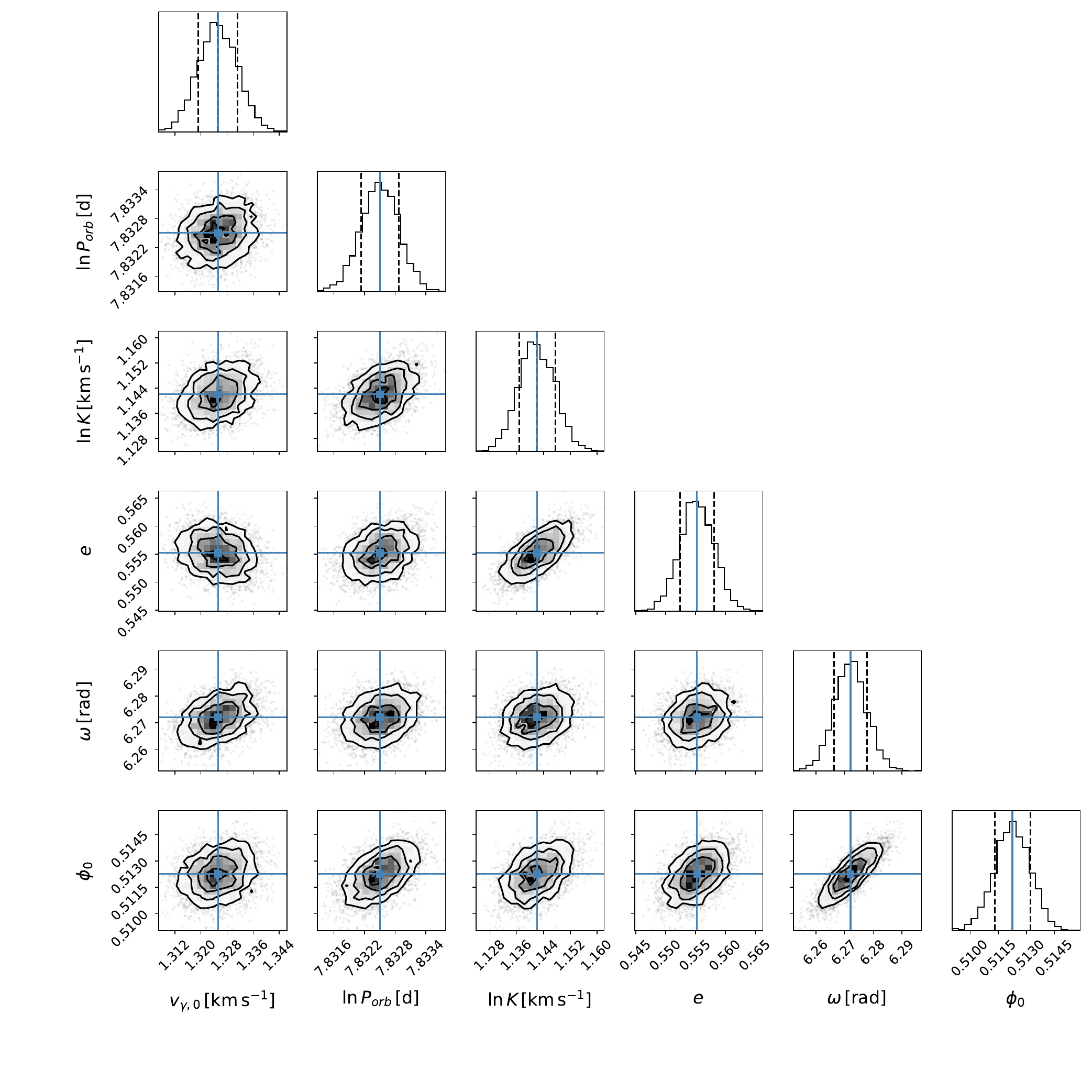}
    \caption{Joint posterior distributions derived from the MCMC sampling for U~Vul. The histograms show marginalized posterior distributions of the parameters.}
    \label{fig:SUCyg_corner}
\end{figure*}

We determined orbital solutions using combined \gc\ and literature data using our MCMC algorithm applied to the template fitting epoch residuals; such fits are labeled as `V+L' in Tables~\ref{tab:orbits} and \ref{tab:orbits2}. Orbits of five newly discovered binaries (FO~Car, MY~Pup, R~Cru, VY~Per, and VZ~Pup) and nine previously known binaries (DL~Cas, FF~Aql, S~Mus, S~Sge, SY~Nor,  TX~Mon, U~Vul, V1334~Cyg, and YZ~Car) 
were obtained using \gc\ data alone as well as in combination with literature RVs.
The `V' and `V+L' orbits for 12 of these 14 agree within the uncertainties, although `V+L' orbits tend to have lower uncertainties, mostly due to better orbital sampling. The exceptions are MY~Pup and VZ~Pup. 
\SSS{The \gc\ baseline only marginally covers the orbital period of MY~Pup, hence, incorporating literature data results in a modification of the period. }
The \porb\ of VZ~Pup differs by $2.4\sigma$ between `V' and `V+L' orbits due to the \porb\ being on the order of the available \gc\ baseline. The `V+L' result is thus clearly preferred, hence, whenever available we use `V+L' orbital estimates in any sample descriptions presented here. 
Lastly, for two stars, AX~Cir and SU~Cyg, we could only obtain a combined fit using \gc\ and literature data. In the case of AX~Cir (Figure~\ref{fig:VgammaOrbitAXCir}), due to its very long orbital period and for SU~Cyg (Figure~\ref{fig:VgammaOrbitSUCyg}) due to the incomplete sampling of orbital phase with \gc~data alone. 

\subsubsection{`Tentative' orbits using \gc\ and literature data\label{sec:tentative} }
We determined first tentative orbital solutions for three stars: BP~Cir, R~Mus, and V0659~Cen. We consider these orbits `tentative' because the literature data were not sufficiently constraining, while the \gc\ data alone were not sufficient to sample a full orbit. Additional observations are required to ascertain the true orbital solutions for these stars. 
Please note that the uncertainties associated with the orbital parameters of the `tentative orbits' listed in Table~\ref{tab:orbits} may not accurately reflect the limitations arising from the insufficient sampling of these long orbits.
\referee{The stated uncertainties are formal uncertainties on the model given the data. However, in cases of poor or incomplete orbital sampling, it is possible that the true orbital period is outside the range of these formal errors. This is why we present these cases separately as `tentative'.}

\subsection{Incompletely sampled orbits modeled as polynomial trends\label{sec:trends}}

\begin{figure*}[ht!]
    \centering
    \includegraphics[scale=0.5]{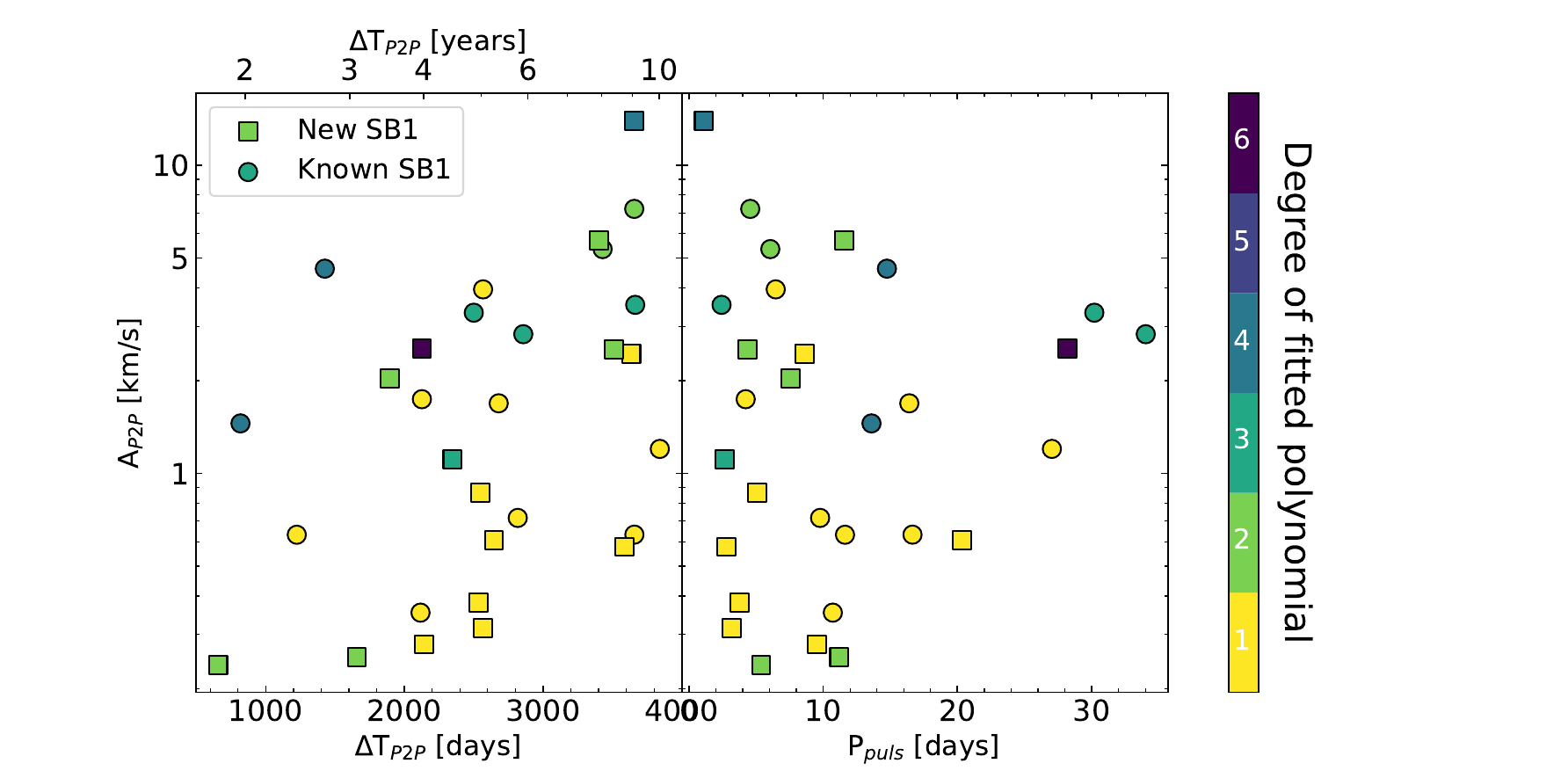} 
    \caption{
    The distribution of A$_{p2p}$, maximum amplitude variation in $v_{\gamma}$, for the stars in our sample where we could fit a polynomial to the $v_{\gamma}$ trend (see Section~\ref{Sect:Polyfits} for details). In the left panel, we have the A$_{p2p}$ vs the baseline in time over which it was calculated and in the right panel, the same against the pulsation period. 
    The newly discovered SB1s from our sample are plotted in filled squares, while the literature-known SB1s are plotted in filled circles. The color bar represents the degree of polynomial fitted in the Fourier+Polynomial model. }
    \label{fig:polydeg_results}
\end{figure*}

\begin{table}[]
\centering
\setlength{\tabcolsep}{2pt}
\caption{List of SB1 from our sample where we found some evidence of orbital motion, and a polynomial was fitted as the orbit could not be sampled/covered adequately.}\label{tab:polydegs} 
\sisetup{round-mode=places}
\begin{tabular}{lcS[round-precision=2]S[round-precision=0]c}
\hline
Cepheid & deg & A$_{p2p}$ &$\Delta$T$_{p2p}$& Trend \\
 & &{(km s$^{-1}$)}&{(d)}& \\
\hline 
\multicolumn{5}{c}{SB1s discovered in the current work}\\
\hline
AQ Pup&3&3.320225667098065&2500.105493999996&up\\
ASAS J064553+1003.8&3&1.108292219298164&2347.0&down\\
ASAS J103158-5814.7&4&13.95310965693483&3658.0&down\\
ASAS J155847-5341.8&1&0.5773923585320375&3587.120128999995&down\\
ASAS J174108-2328.5&1&0.3810367466164744&2532.895360000002&up\\
DR Vel&2&0.2527701514130634&1657.0&down\\
OX Cam&1&0.8663559796850429&2549.063739499994&down\\
RY Sco&1&0.6065881867750207&2645.759737&up\\
RY Vel&6&2.540767774298601&2126.0&up\\
SX Vel&1&0.2784987306025535&2140.056317000002&up\\
V0391 Nor&2&2.524852816530128&3511.0&down\\
V0492 Cyg&2&2.029192361644205&1895.0&up\\
V0827 Cas&1&0.3149317078970242&2568.0395479&up\\
V1162 Aql&2&0.2384042954511223&656.0&up\\
V1803 Aql&1&2.441596674885739&3634.833878499972&down\\
V2475 Cyg&2&5.708763506054583&3404.0&down\\
\hline
\multicolumn{5}{c}{Literature-known SB1s}\\
\hline
AD Pup&4&1.451948182076336&818.0&down\\
AH Vel&1&1.739904163355519&2125.994319999998&up\\
AQ Car&1&0.7165187052266945&2817.144399999997&down\\
AW Per&1&3.952692379094831&2567.101016299996&down\\
FR Car&1&0.3530250875904333&2115.970938999999&up\\
KN Cen&3&2.827676764804899&2857.0&down\\
LR TrA&3&3.521263149100633&3664.0&up\\
RV Sco&2&5.340853915195297&3429.0&up\\
RW Cam&1&1.68723132277184&2679.798328799945&up\\
RX Aur&1&0.6318497871845814&1223.7187595&up\\
T Mon&1&1.199972857501646&3841.348309000001&up\\
UX Per&2&7.212160593830838&3657.0&down\\
UZ Sct&4&4.614123160024221&1426.0&up\\
XZ Car&1&0.6323370049702612&3658.007333999994&up\\
\hline
\end{tabular}

\tablefoot{Column~2 represents the degree of polynomial used in our analysis, Column~3 lists the peak-to-peak amplitude of the (linear or non-linear) $v_{\gamma}$ variation and Column~4 lists the timeline (in days) over which this amplitude was computed. The last column indicates the direction of the trend, up for increasing velocity, down for decreasing.}
\end{table}

We represented \npolyNew~newly detected SB1 systems and \npolyKnown~previously reported SB1 systems using a combined Fourier series plus polynomial trend. Table~\ref{tab:polydegs} quantifies the measured range of the orbital RV variation, A$_{p2p}$, over the temporal baseline \SSS{(denoted by $\Delta$T$_{p2p}$), which is the baseline of the observations over which the A$_{p2p}$ was calculated.}
Among the \npolydeg~systems examined with Fourier plus polynomial fit, 15 Cepheids exhibit non-linear trends. 
Among the 15 stars with linear trends, 5 demonstrate a decreasing trend in \vgamma, whereas the remaining 10 feature increasing \vgamma. Additional observations are required to determine these orbits.  

Figure~\ref{fig:polydeg_results}  presents the distribution of A$_{p2p}$ versus $\Delta$T$_{p2p}$ and the pulsation period. A clear trend of increasing A$_{p2p}$ as we extend to longer temporal baselines is visible. For a fixed value of $K$, this trend arises because longer baselines increasingly sample the full (albeit incomplete) peak-to-peak variation of the orbit. However, there is significant dispersion in the value of $K$ at fixed \porb\ due to different inclinations and mass ratio. 
The smallest trends of between 0.2 - 0.35 \kms\ are found for 
V1162~Aql, DR~Vel, FR~Car and SX~Vel (over $\sim$ 1.8, 4.5, 5.8 and 5.9~yrs respectively).
By comparison, the smallest $K$ among completely solved orbits is $0.67$\,\kms\ (MY Pup, $5.7$~yr). At the long timescale limit, we find $A_{p2p} = 1.2$\,\kms\ variation for T~Mon over the course of nearly $10.5$~yr. By comparison, the lowest $K=2.1$~\kms\ for a completed orbit and a similar \porb\ of 11~yr is found for VZ Pup.

\subsection{Template fitting for stars without orbital signatures in \veloce\ data}\label{sect:RVTFresults}
Paper~I presented a list of SB1 Cepheids exhibiting time-variable average velocities detected using \gc\ data alone. SB1 Cepheids with very long \porb\ could, however, remain undetected based on \gc\ data alone. Hence, we investigated the temporal stability of $v_\gamma$ for all Cepheids in \gc\ using the well-defined RV templates from Section~\ref{sec:RVTF}. To this end, we computed $\mathrm{SNR_{SB1}}$ (Eq.~\ref{eq:RVTFSNR}) and considered a star a likely binary if any template fit resulted in a 3-$\sigma$ deviation ($\mathrm{SNR_{SB1}}$ $>$3) from a constant value of \vgamma. 

\begin{figure}
\centering
    \begin{subfigure}{.5\textwidth}
    \includegraphics[scale=0.28,trim={0 0 0 1.5cm}]{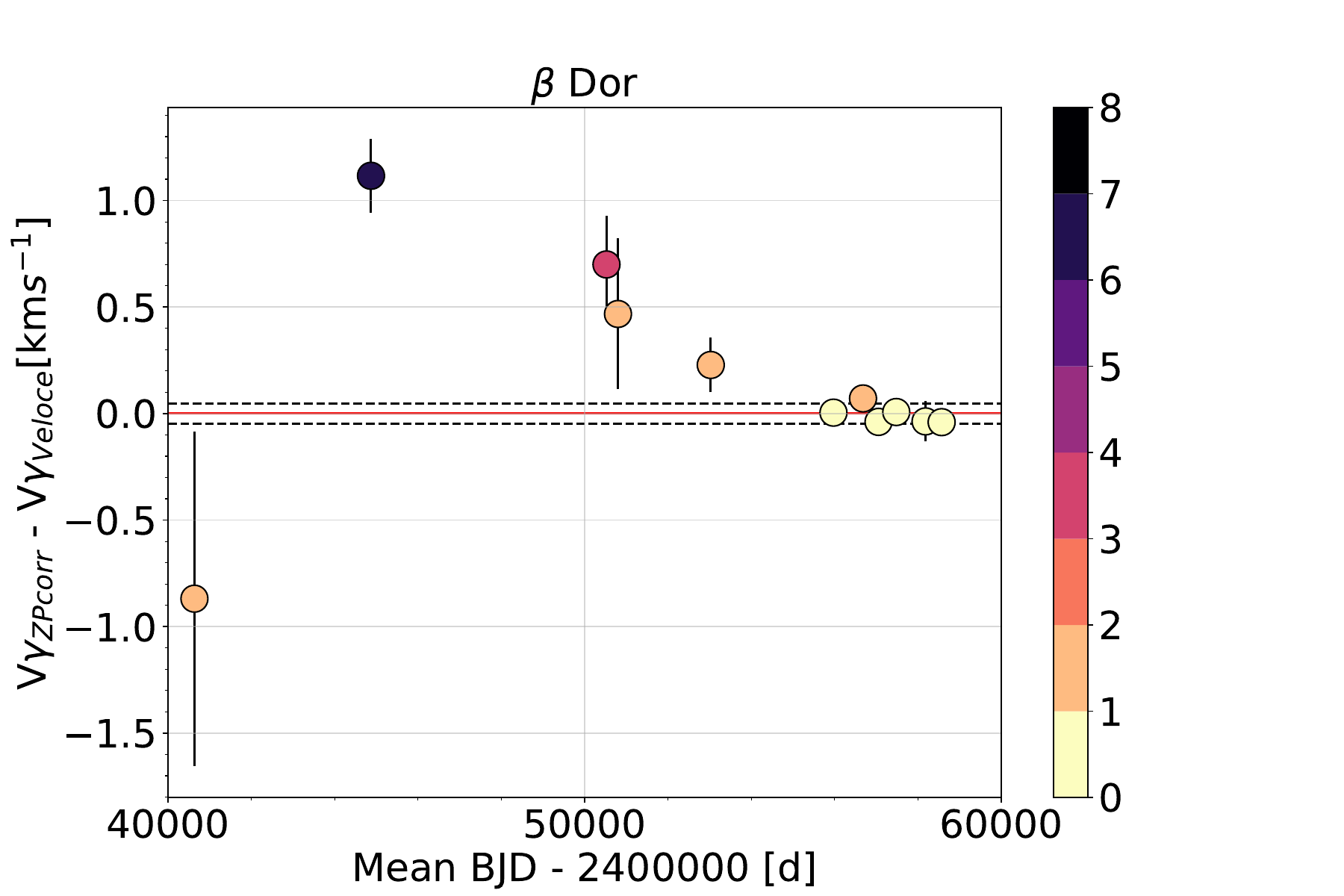}
    \end{subfigure}
    \begin{subfigure}{.5\textwidth}
    \includegraphics[scale=0.28,trim={0 0 0 1.1cm}]{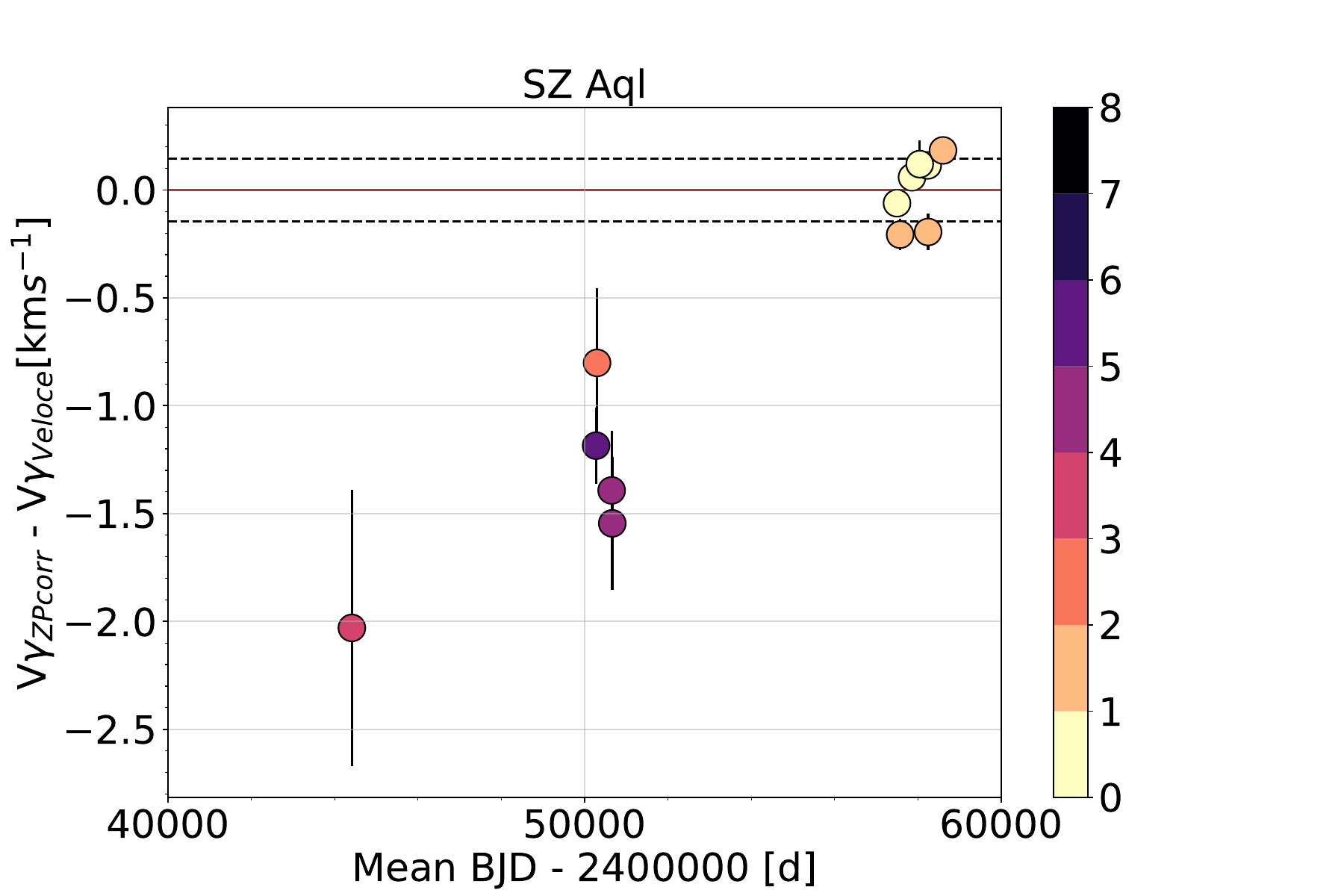}
    \end{subfigure} 
    \begin{subfigure}{.5\textwidth}
    \includegraphics[scale=0.28,trim={0 0 0 1.1cm}]{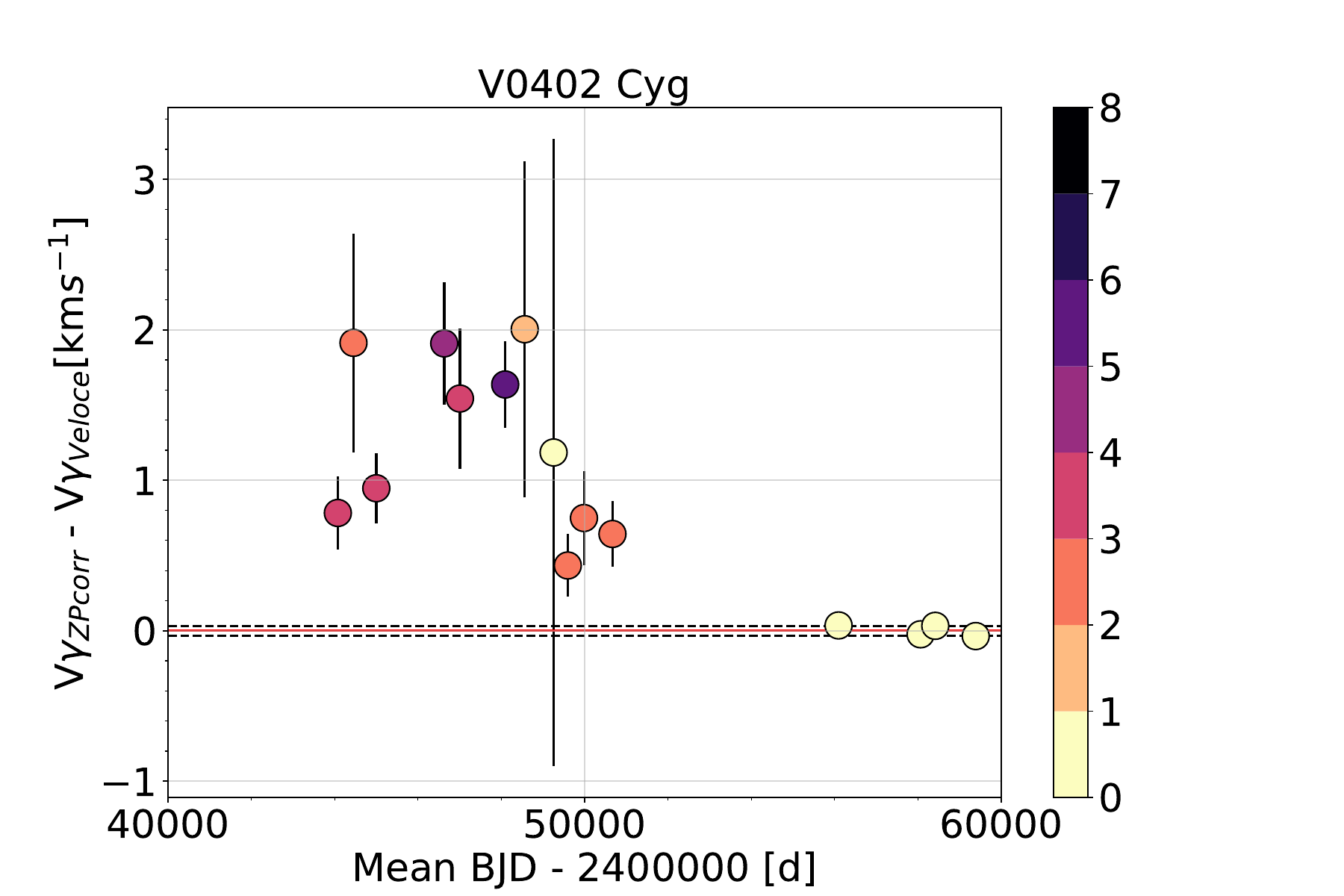}
    \end{subfigure}
    \begin{subfigure}{.5\textwidth}
    \includegraphics[scale=0.28,trim={0 0 0 1.1cm}]{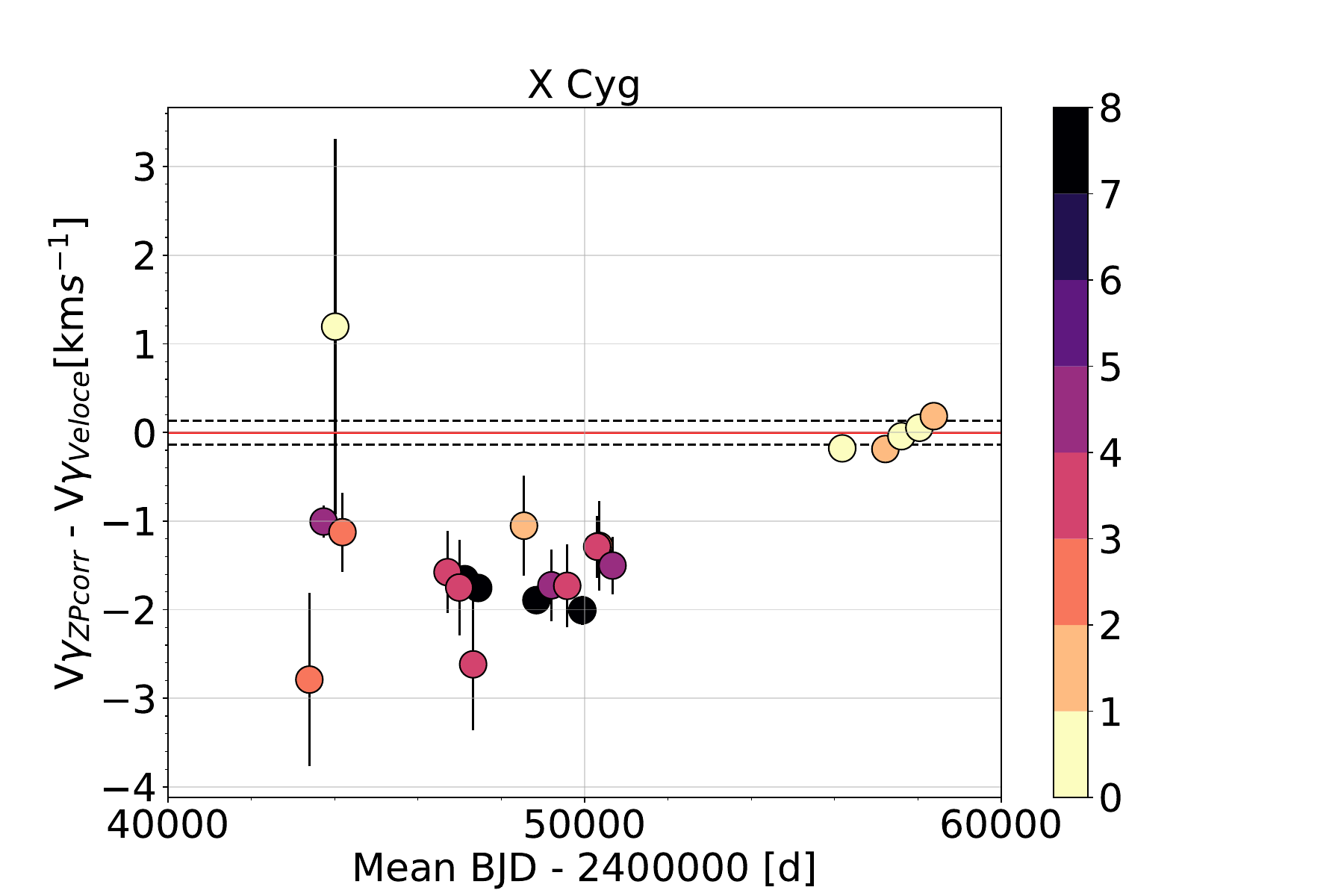}
    \end{subfigure}
\caption{Zero-point corrected $\Delta v_{\gamma}$ for new SB1s detected through RVTF. The dotted lines mark the standard deviation of the \vgamma$_{\gc}$. This figure only includes stars where we have a 3-$\sigma$ detection through our analysis. The color bar represents the $\mathrm{SNR_{SB1}}$ of the various $v_{\gamma}$ measurements where the $\mathrm{SNR_{SB1}}$ was calculated as described in the text.
}\label{fig:vgammaNewDetections}.
\end{figure}

Applying this method to 174 bona fide \gc\ Cepheids with available literature data resulted in the discovery of four additional SB1 candidates: $\beta$~Dor, SZ~Aql, V0402~Cyg, and X~Cyg, cf.  Figure~\ref{fig:vgammaNewDetections}. All four exhibit visually convincing trends in \vgamma\ on the order of $1-3\,$\kms\ on timescales exceeding $15\,000-20\,000$~d.  

\begin{figure*}[ht!]
    \begin{subfigure}{.35\textwidth}
    \includegraphics[scale=0.25, trim={5cm 0 5cm 0}]{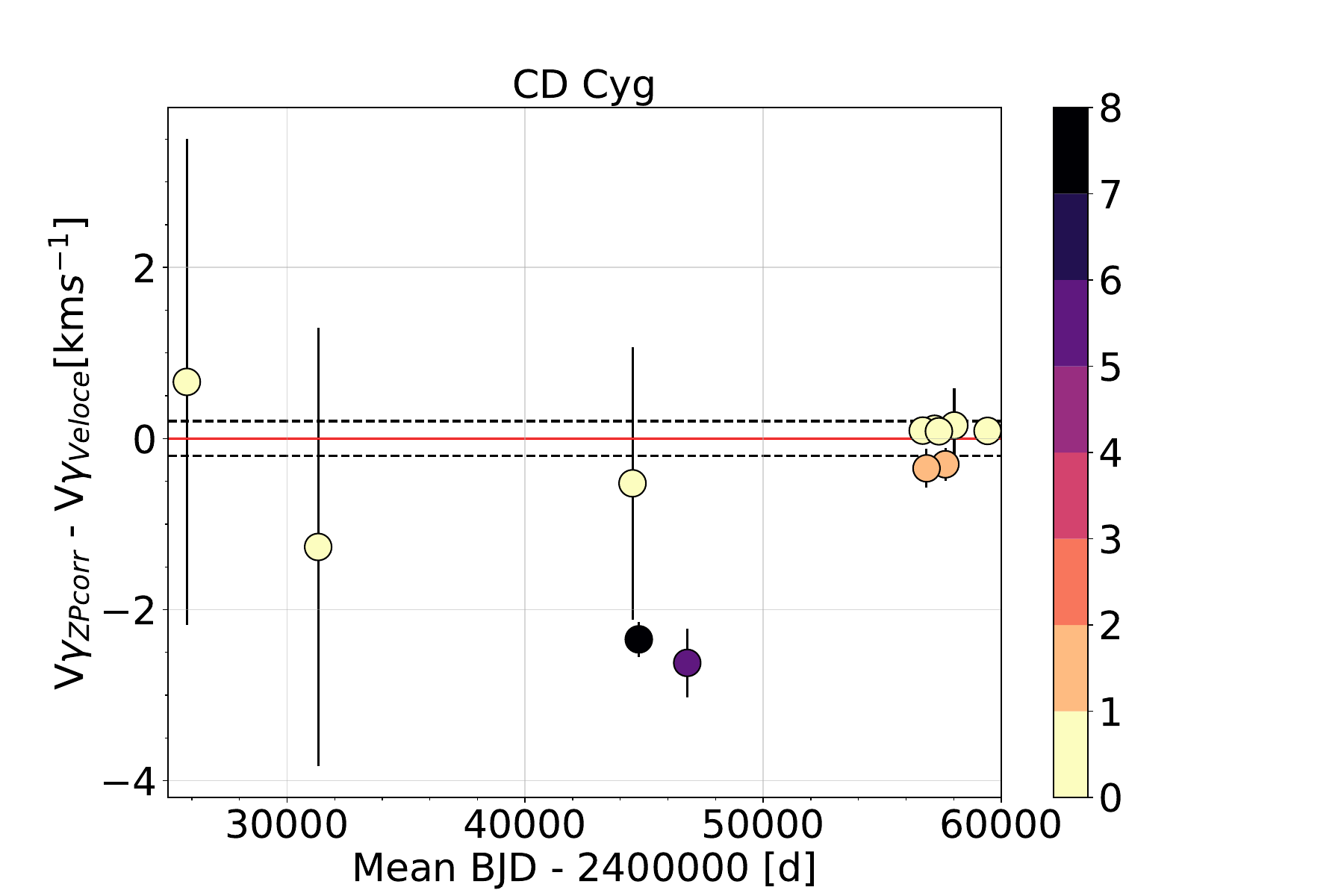}
    \end{subfigure}
    \begin{subfigure}{.35\textwidth}
    \includegraphics[scale=0.25, trim={5cm 0 5cm 0}]{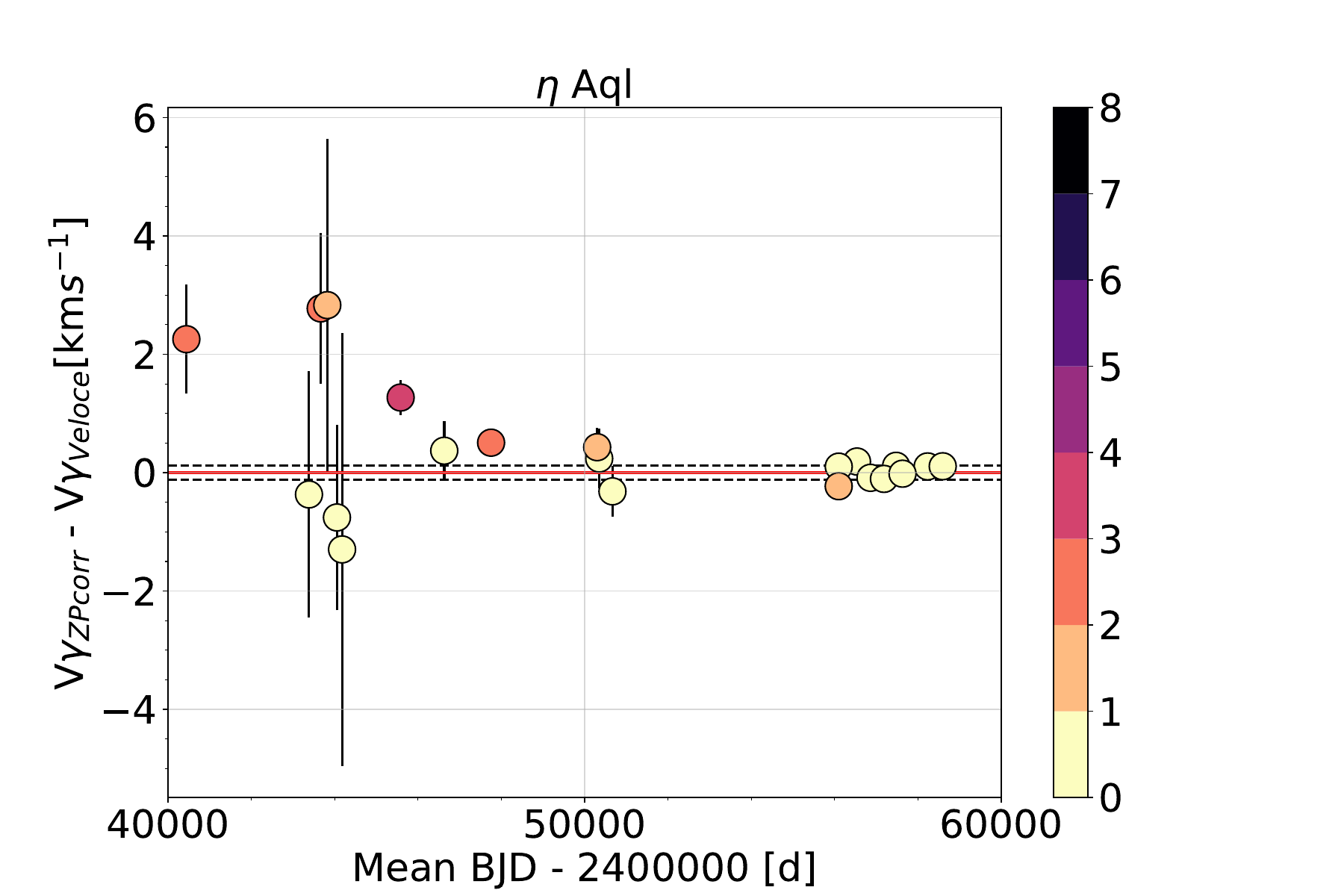}
    \end{subfigure}
    \begin{subfigure}{.35\textwidth}
    \includegraphics[scale=0.25, trim={5cm 0 5cm 0}]{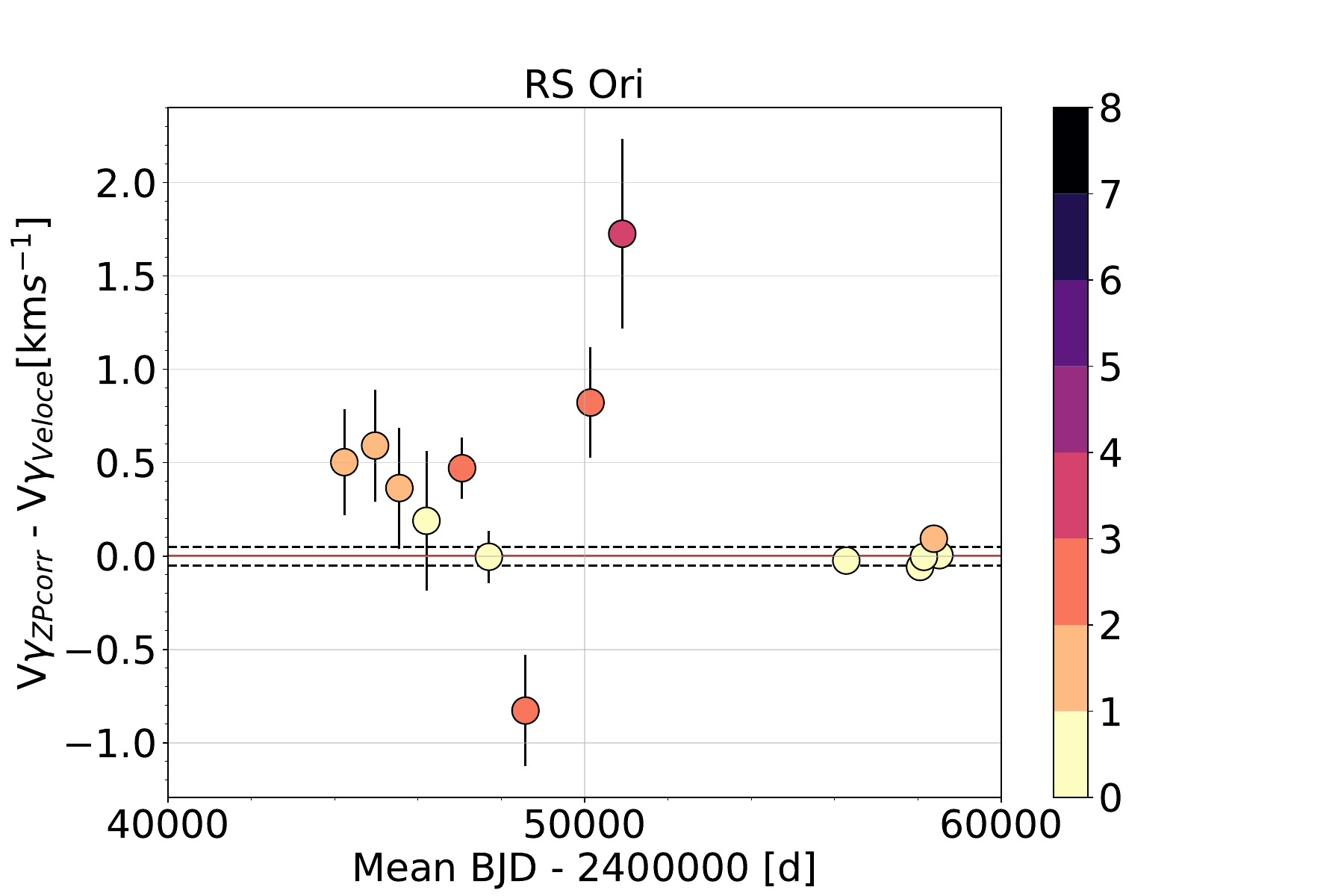}
    \end{subfigure}
    \newline
    
    
    \begin{subfigure}{.35\textwidth}
    \includegraphics[scale=0.25, trim={5cm 0 5cm 0}]{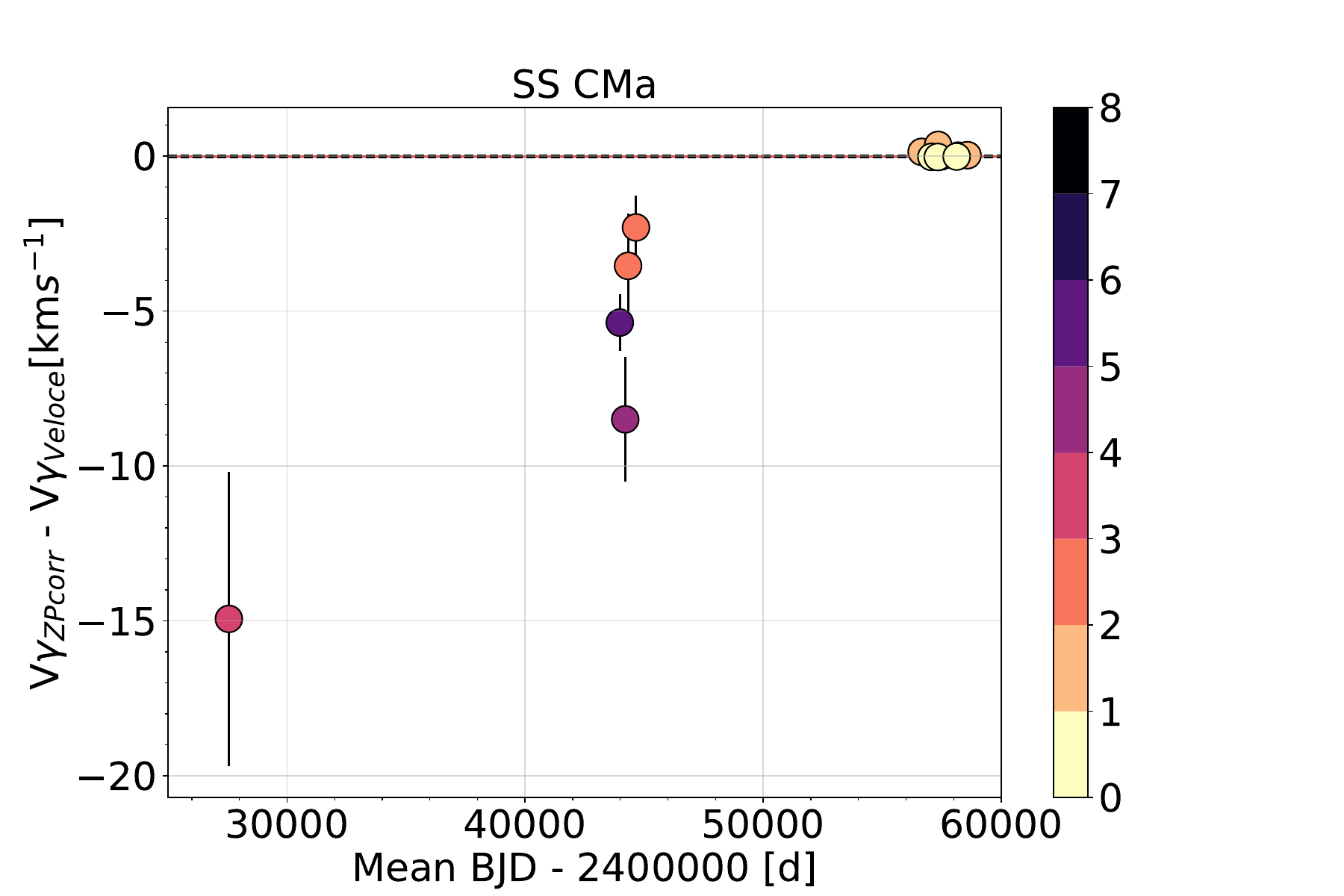}
    \end{subfigure}
    \begin{subfigure}{.35\textwidth}
    \includegraphics[scale=0.25, trim={5cm 0 5cm 0}]{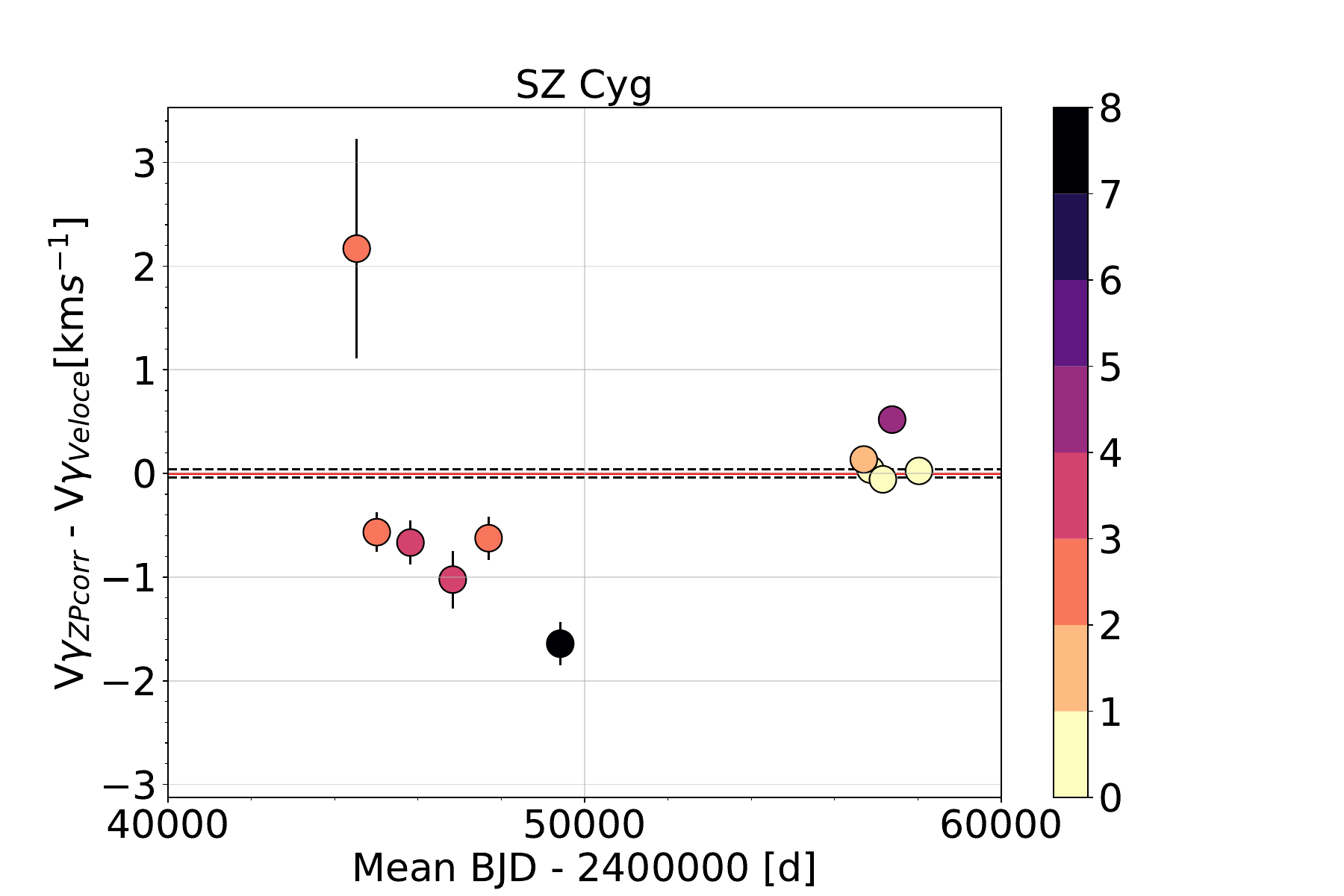}
    \end{subfigure}
     \begin{subfigure}{.35\textwidth}
    \includegraphics[scale=0.25, trim={5cm 0 5cm 0}]{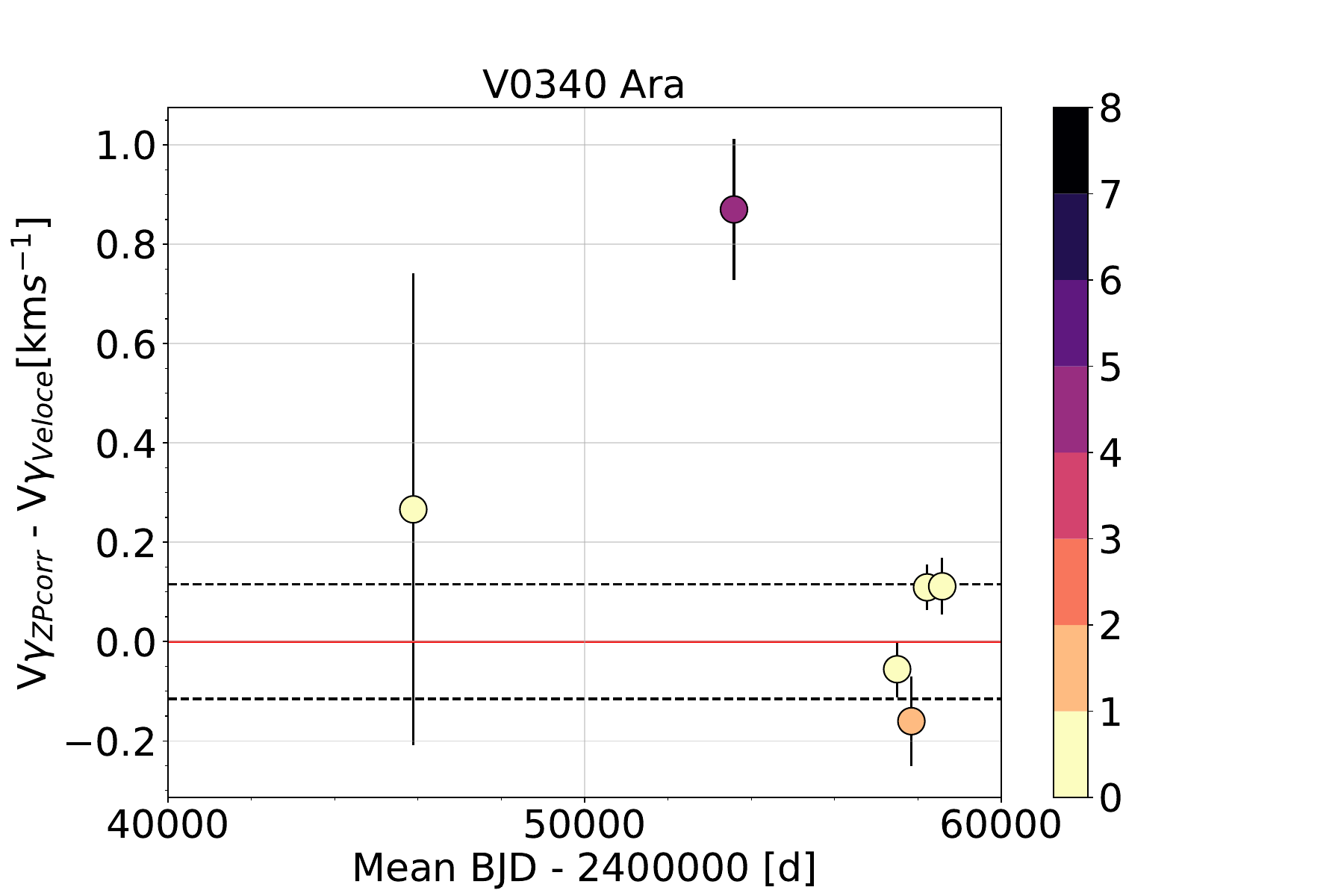}
    \end{subfigure}
    \newline 
   
    \begin{subfigure}{.35\textwidth}
    \includegraphics[scale=0.25, trim={5cm 0 5cm 0}]{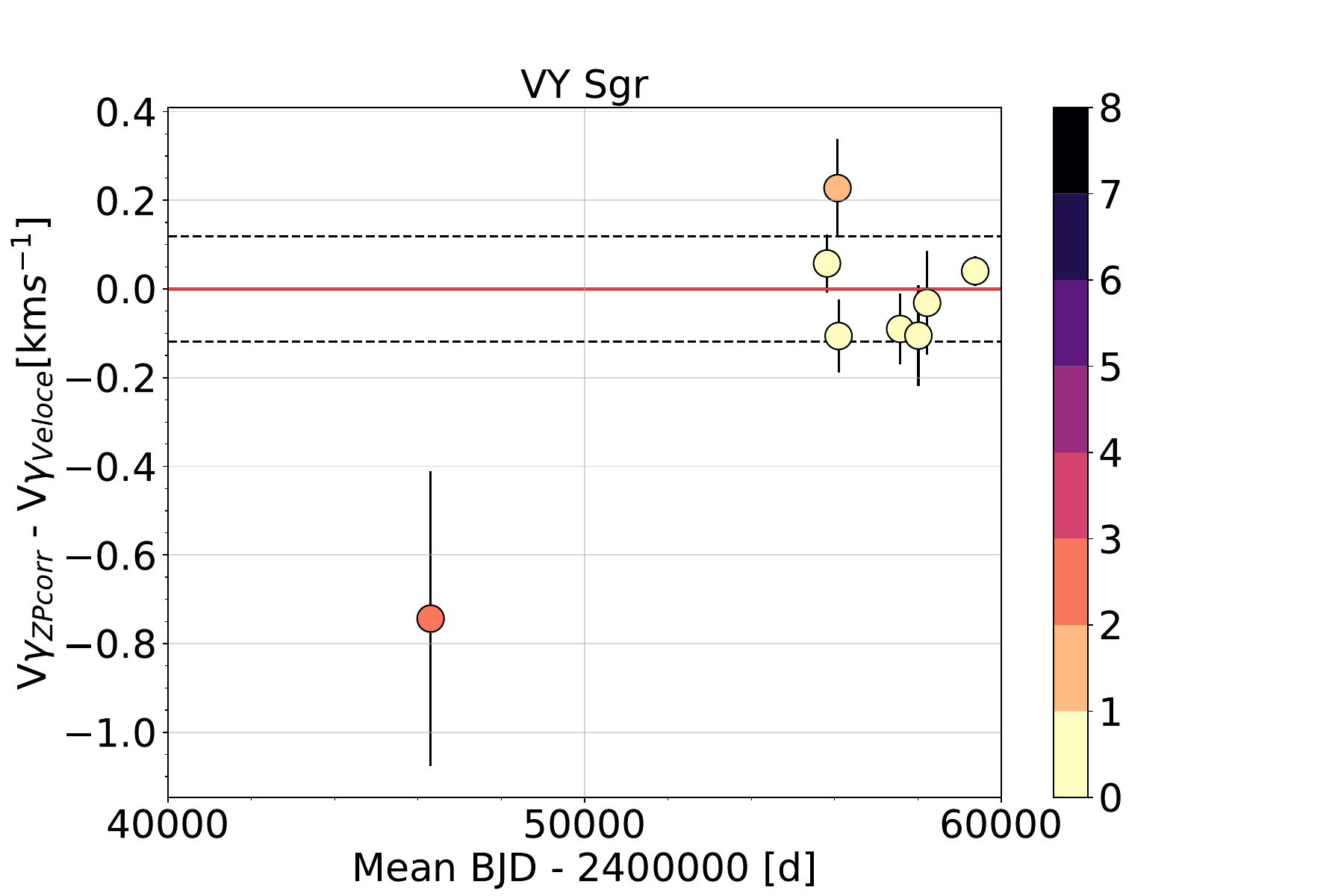}
    \end{subfigure}
    \begin{subfigure}{.35\textwidth}
    \includegraphics[scale=0.25, trim={5cm 0 5cm 0}]{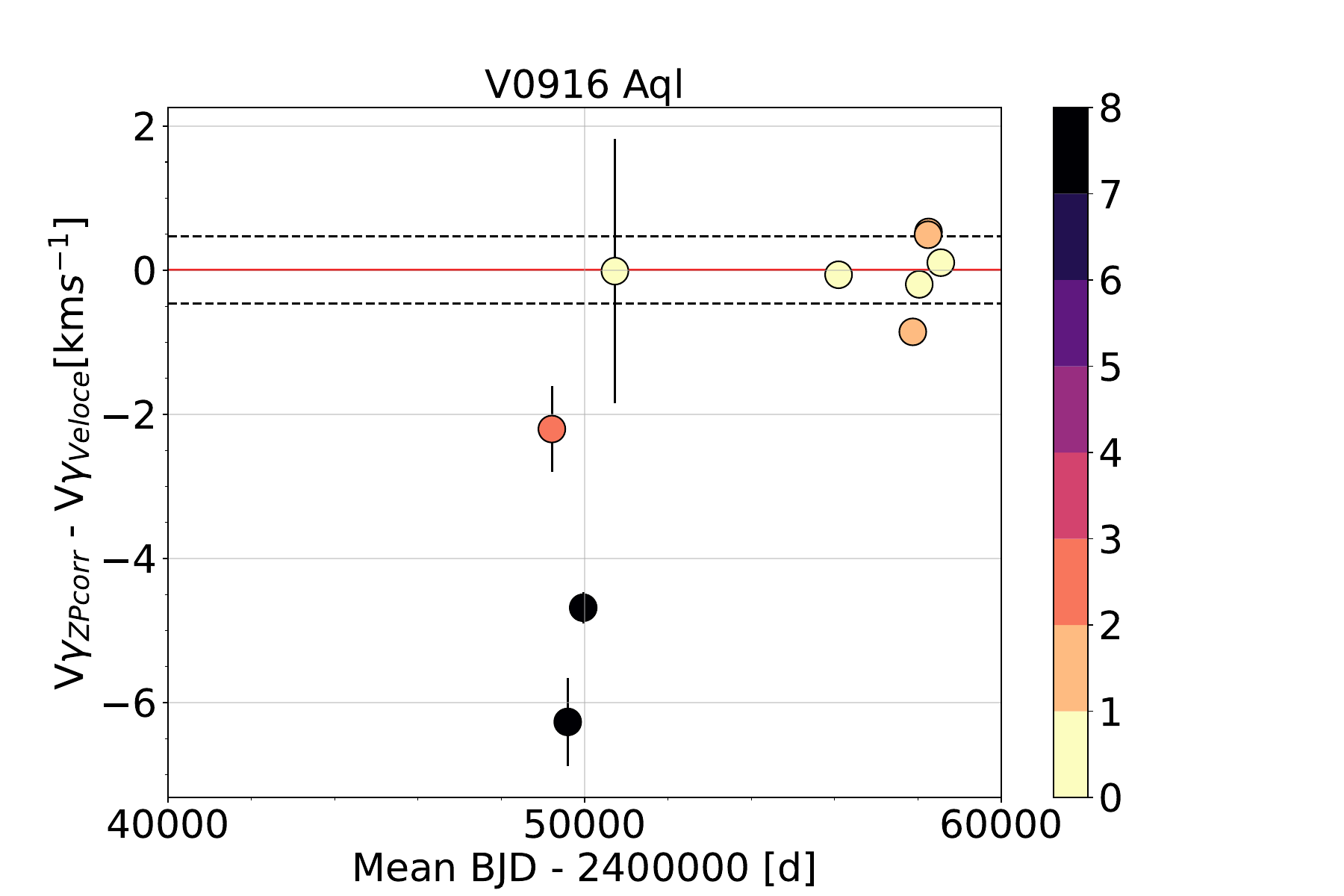}
    \end{subfigure}
\caption{Literature-known SB1 candidates detected through RVTF. The figure description is same as Figure~\ref{fig:vgammaNewDetections}.} 
\label{fig:VgammaforSB1signs}
\end{figure*}

Additionally, we found \nrvtfYKnown~bona fide SB1 Cepheids (CD~Cyg, $\eta$~Aql, RS~Ori, SS~CMa, SZ~Cyg, V0340~Ara, V0916~Aql and VY~Sgr) among a sample of 32 SB1 candidates previously reported in the literature \citep{BinCepDB} where the available baseline of \gc~data alone was insufficient to detect orbital motion. Figure~\ref{fig:VgammaforSB1signs} illustrates these cases. CD~Cyg and SS~CMa were already discussed as potential SB1s from the RVTF analysis of \cite{Anderson2016rv} using \gc~data. Here, we confirm that the deviations in $v_{\gamma}$ indicate of binarity. 

For the remaining \nnonbincep~Cepheids among the 32 candidates, we could not confirm unambiguous orbital motion despite large overlaps in the data used. Ten of these \nnonbincep~Cepheids do not exhibit a $\mathrm{SNR_{SB1}} > 3$, over a temporal baseline of 10~000 - 20~000 days. For the remaining 14 cases, individual \vgamma\ variations are found, although they should be considered spurious variations because closely neighboring \vgamma\ estimates clearly show the absence of a trend at these epochs. 
In Appendix 
Figure ~\ref{fig:VgammafornoSB1signs}, we present all these `non-SB1' cases 
and we have tagged these stars with `RVTF-No' in the `Binarity' column of Table~\ref{tab:BasicInfo}.

Given the high accuracy of the \gc~template fitting procedure, which effectively removes any pulsational variability and period variations, and since instrumental zero-point differences have been corrected using non-binary Cepheids (cf. Paper~I), we conclude that these \nnonbincep~Cepheids should not be considered spectroscopic binaries. 

\subsection{SB1 fraction among \gc~Cepheids}\label{Sec:bf}
\begin{figure*}
    \centering
    \includegraphics[scale=0.5]{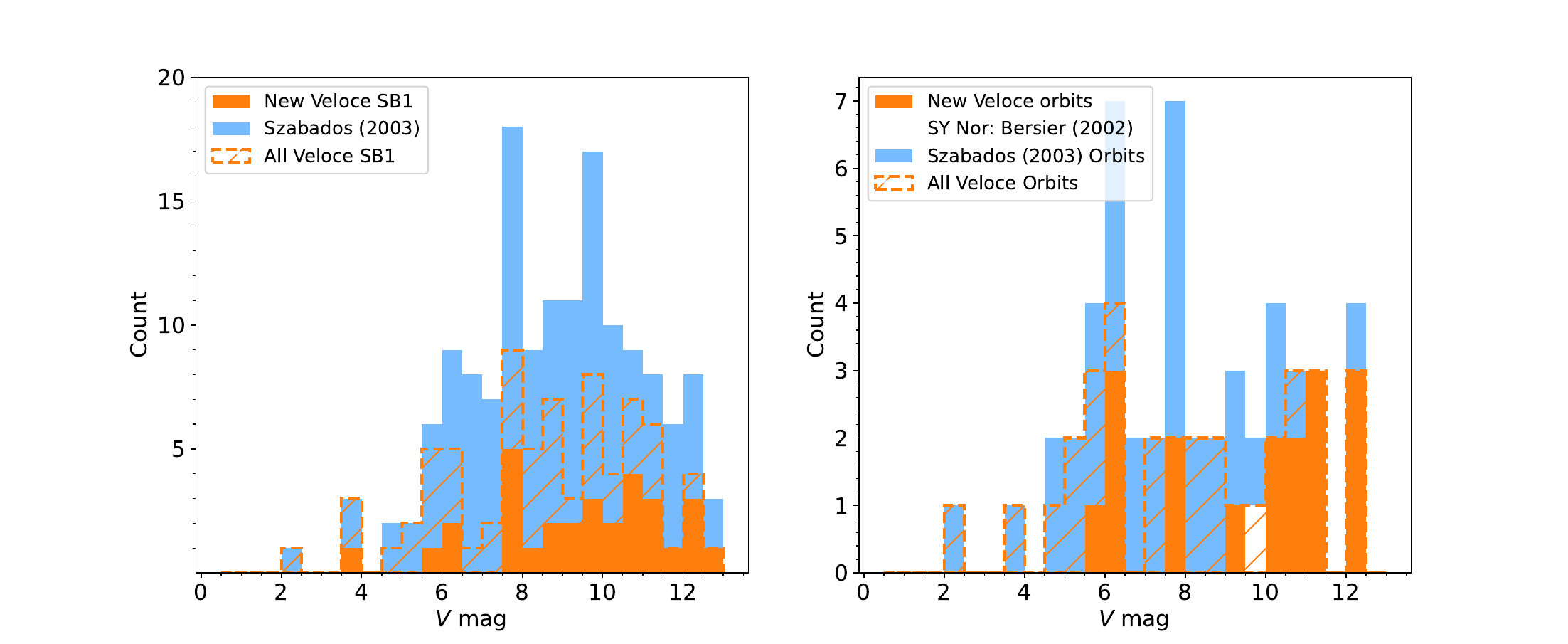}
    \caption{Stacked histogram presenting the new \gc~SB1 detections (left panel) and the new \gc~orbits (right panel) along with the same from \cite{BinCepDB}. SY Nor's orbit is provided by us as well as \cite{Bersier2002}; however, it is missing in the \cite{BinCepDB} database. 
    }
    \label{fig:Histogram_withSzabados}
\end{figure*}
The \nbincep\ bona-fide SB1 Cepheids detected as part of \gc\ yield a lower limit to the spectroscopic binary fraction of $29.6 \pm 3.4\%$ that is fully consistent with the result by \cite{Evans2015} of $29\pm8$\% based on the 40 brightest Cepheids. If all \nbincepcand\ SB1 candidates from Paper~I should be confirmed to be bona-fide SB1 systems, then the \gc~binary fraction would increase to $\sim 35\%$.

Using the subset of Cepheids with fully determined orbital solutions from Tables.\,\ref{tab:orbits} and \ref{tab:orbits2} and assuming that all non-determined orbits have very long orbital periods, we find a lower limit to the fraction of SB1 Cepheids with orbital periods of less than $10$~yr of $15.2 \pm 2.4 \%$, slightly lower albeit consistent with the reported $20 \pm 6\%$ from \citet{Evans2015}. Although our stringent template-based investigation (Section~\ref{sect:RVTFresults}) questioned the veracity of the SB1 nature of \nnonbincep~Cepheids for which such evidence had been previously discussed \citep[cf. references in][]{BinCepDB}, the fraction of SB1 systems among Cepheids we determined remains similar because we also identified new SB1 systems. However, the more accurate determination of time-variable $v_\gamma$ alongside detailed zero-point corrections affords much greater confidence in the determination of the SB1 nature of individual Cepheids. Of course, the SB1 fraction of Cepheids is merely a lower limit of the total fraction of Cepheids with companions due to the possibility of very large separations, unfavourable inclinations, and extreme mass ratios.

Figure~\ref{fig:Histogram_withSzabados} illustrates the magnitude distribution of newly discovered and detected SB1 Cepheids alongside the distribution of  binary systems reported in the online compilation by \cite{BinCepDB}, as well as the distribution of magnitudes for which orbits were determined here. 
\SSS{To ensure consistency in the comparison, we have exclusively included the bona-fide SB1 binaries from \cite{BinCepDB} in Figure~\ref{fig:Histogram_withSzabados}. Therefore, other types of binaries (such as photometric, visual, etc.) from their catalog are not accounted for in this analysis.}
\SSS{Please note that Figure~\ref{fig:Histogram_withSzabados} does not include \nnonbincep\ systems that are not detected as SB1 in our study.}

As the right panel of Figure\,\ref{fig:Histogram_withSzabados} shows, we here double the number of orbital solutions for Cepheids fainter than $\sim 8$-th magnitude, and there are only 14 Cepheids with known orbital solutions that are not included in this work (one Cepheid among these 14, namely T~Mon has incompletely sampled orbit in this study). 
Conversely, we determine the first orbital solutions for \norbitsNew\ Cepheids, most of which fainter than $\gtrsim 8$\,th magnitude. 

\subsection{Notes on individual binaries\label{sec:individual}}

\begin{figure}
    \centering
    \includegraphics[width=0.48\textwidth, trim={0 0 3cm 0}]{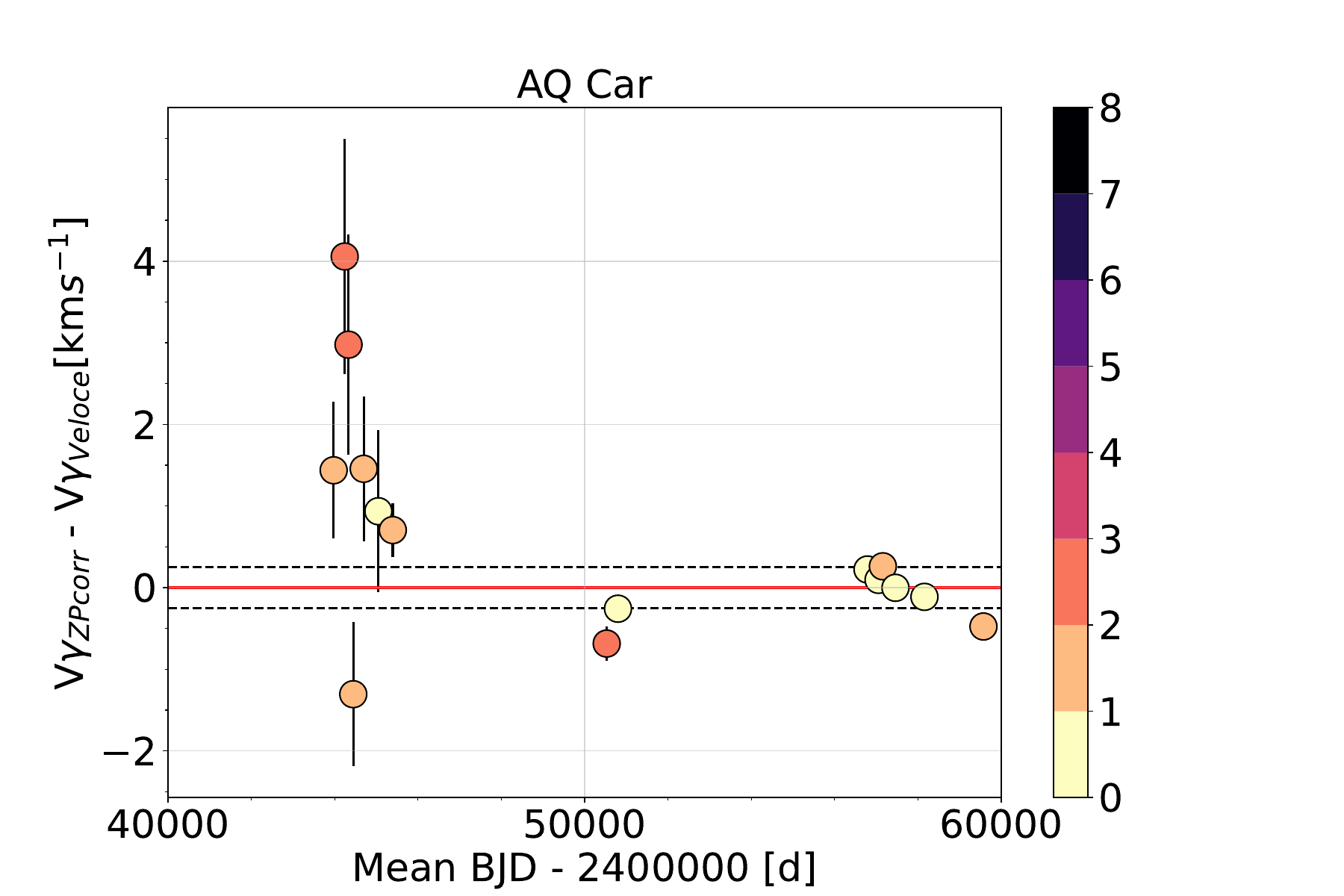}
    \caption{Temporal variation of \vgamma\ for AQ~Car based on RVTF. The data points after BJD $>56000$ are  from \veloce. }
    \label{fig:AQCarRVTF}
\end{figure}

We have presented evidence of the SB1 nature of \nbincep\ Milky Way Cepheids on timescales spanning up to ~40 years, including 
\nbincepnew~stars whose SB1 nature was not previously known. The following provides additional background for specific stars of interest, notably if our results disagree with previous studies. We note that several Cepheids were first reported to be SB1 systems based on early \gc\ results, such as FN~Vel and MU~Cep \citep{PhDthesis}, BG~Vel and RZ~Vel \citep{Szabados2013}, $\delta$~Cep \citep{Anderson2015}, and AQ~Car, CD~Cyg, KN~Cen, SS~CMa, SZ~Cyg, XZ~Car, VY~Car and X~Pup  \citep{Anderson2016rv}.
\SSS{\cite{Anderson2016rv} noticed tentative evidence of \vgamma~variations in VY~Car, AQ~Car, SZ~Cyg and X~Pup. 
However, they ultimately deemed these findings inconclusive due to imprecise data in the literature and the occasional substantial fluctuations in pulsational variability. 
In the current work we successfully detect these \vgamma~variations for SZ~Cyg (when \gc~data is combined with zero-point corrected literature datasets, see Figure~\ref{fig:VgammaforSB1signs}) and AQ~Car (cf. Table~\ref{tab:polydegs}, Figure~\ref{fig:AQCarRVTF}). 
Nevertheless, in the case of VY Car and X Pup, the \vgamma~values, when adjusted for zero-point, do not reveal any consistent variation indicative of binarity (see Figure~\ref{fig:VgammafornoSB1signs}). }

\paragraph{$\alpha$~UMi (Polaris)} was one of the first Cepheids whose nature as an SB1 system was realized \citep{Moore1929}, and much literature exists on this star. Polaris was extensively studied using pre-2019 \gc\ data by \citet{Anderson2019polaris}, who determined a full orbital solution and studied the stability of RV amplitudes as well as the presence of additional periodic signals. Given the very long $\sim 30$-year orbital period, we did not update the orbital solution here and instead refer to the aforementioned publication as well as \citet{Torres2023} for orbital parameters.

\paragraph{$\delta$ Cephei} is the archetype of classical Cepheids and was discovered to be an SB1 system using an earlier subset of \veloce\ data. \citet{Anderson2015} estimated an orbit with \porb=$2201^{+5.73}_{-6.31}$\,d, $K=1.5^{+0.239}_{-0.080}$~\kms, and $e=0.647^{+0.038}_{-0.021}$ by combining new RV data with RV measurements from the literature. A main difficulty of this approach was the correction of zero-point differences for datasets with short ($\sim 1$~yr) baselines. Here, we report a correction to the orbital solution, with \porb=$3450.9 \pm 19.6$\,d, $K=3.01 \pm 0.02$~\kms, and $e=0.745\pm 0.006$, based purely on \gc\ data and hence not subject to the uncertainties associated with literature zero-point offsets. 
\referee{This updated orbit of $\delta$ Cep~was already presented at the 2022 RR~Lyr-Cep conference in La Palma \citep{Shetye2022}.}
\refereetwo{ It agrees reasonably well with the results recently presented by \cite{Nardetto2024} based on a combination of High Accuracy Radial velocity Planet Searcher for the Northern hemisphere (HARPS-N) data and existing literature. Their estimated orbital period of 3401.8 d and eccentricity of 0.71 differ by 2$-\sigma$~ and 1.7$-\sigma$~from ours. These small differences may result from their use of a heterogeneous dataset, or the better coverage of the periastron passage by the HARPS-N data.}
As Figure\,\ref{fig:Example_kepler1} shows, the orbit is now completely sampled and contained within the \gc\ baseline ($3620$\,d). Importantly, the extreme ends of the orbital velocity have been observed and tightly constrain $e$ and $K$. 
Unfortunately, including literature data for a `V+L' orbit does not improve the orbital solution due to the same issues that affected the prior analysis. Extending the baseline with future \hermes\ observations will further improve the orbital period. We note that the mass function and the high brightness contrast imply the companion to be a low-mass star, possibly a white dwarf \citep{Gallenne2016}. 
\SSS{A similar deduction was made in the recent X-ray study of Cepheids by \cite{Evans2022}.
A companion of G spectral type or later would be expected to be detected in X-rays, which puts a lower limit on the mass of the companion.}
Finally, we note that the single-star astrometric solution provided in \gaia\ DR3 clearly indicates poor fit quality\footnote{Specifically, the \gaia\ DR3 astrometric parameters \texttt{astrometric\_gof\_al}$=31$, \texttt{astrometric\_chi2\_al}$=41802$, \texttt{astrometric\_excess\_noise}$=1.23$, \texttt{astrometric\_gof\_noise\_sig}$=1777$, and \texttt{RUWE}$=2.7$ all indicate the presence of unmodeled signal.}. It thus appears likely that \gaia's next data release could deliver an astrometric orbital solution, as predicted by \citet{Anderson2015}.

\paragraph{$\eta$~Aql} was recently studied by \cite{Benedict2022} using spectroscopic, photometric and astrometric (from the Hubble Space Telescope's Fine Guidance Sensor) observations, including unpublished \gc\ data. No significant variations in the RV residuals were detected after removing the Cepheid pulsational RV signature, leading \citet{Benedict2022} to conclude that either $\eta$~Aql has a very long orbital period (longer than their baseline of $\sim$38 years) or that it has a face-on orbit. We considered an extended  temporal baseline by including RV data from \cite{1980SAAOC...1..257L}, \cite{1987ApJS...65..307B}, and \cite{1989ApJS...69..951W} and detect a 3-$\sigma$ variation in the $\Delta$~\vgamma  
(Figure~\ref{fig:VgammaforSB1signs}). If the older datasets are to be believed despite large RV uncertainties of $\sim$1\kms, then $\eta$~Aql is indeed an SB1 Cepheid with a very long orbit of more than 50 years.
\SSS{Verifying whether this spectroscopic companion corresponds to the one identified in the UV imaging and spectra of $\eta$~Aql by \cite{Evans2013}, at a separation of 180 au, necessitates a follow-up of the long orbital period.}

\paragraph{AQ Car} was reported to show tentative signs of time-variable \vgamma\ using an earlier set of \veloce\ and literature data. In Paper~I, we confirmed the SB1 nature of this star. Although weak, $\sim 0.7$ \kms, the signal is clearly present in the \veloce\ data thanks to the baseline of $\sim 4$~yr. Although literature datasets appear imprecise to usefully inform the \vgamma~variations (cf. Figure~\ref{fig:AQCarRVTF}), in actuality, the fluctuations observed in the literature data surpass a 3-$\sigma$ threshold. As $\sigma$ is calculated using \gc~data under the assumption of no orbital motion, consequently, the deviations from a stable \vgamma~are consistently underestimated in this scenario.

\paragraph{AQ Pup} was reported to have a photometric companion \citep{Evans1994a} that was not previously confirmed spectroscopically \citep{Vinko1991}. Here, we report low-amplitude $v_{\gamma}$ variations ($\sim$ 3.3 \kms) on a timescale of 2500 days. Whether this signature corresponds to the photometric companion remains unclear.

\paragraph{ASAS J064540+0330.4}
\SSS{- We present the first orbital fit for ASAS~J064540+0330.4 with a \porb~of 901.18~$\pm$~2.81 days and an eccentricity of 0.44~$\pm$~0.01. As shown in Figure~\ref{fig:ASAS0330Orbit}, the pulsational residuals indicate additional signals beyond the orbital ones. It is important to note that further observational follow-up is necessary to sample the orbital phase better and establish this orbit more firmly.}

\begin{figure}
    \centering
    \includegraphics[scale=0.5]{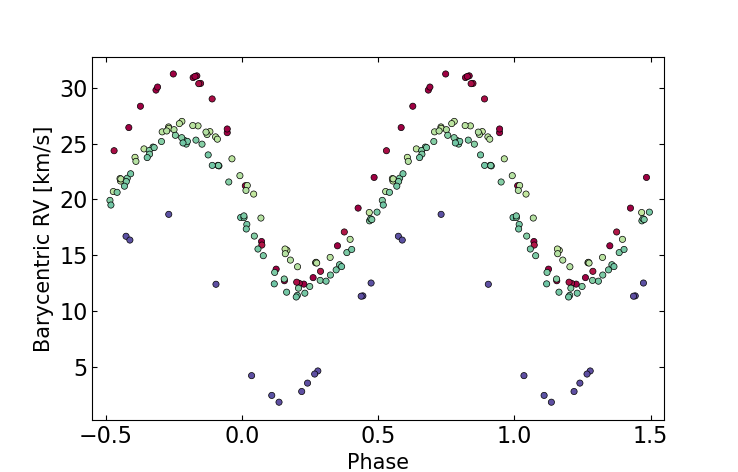}
    \caption{Phase-folded \coralie\ RV of ASASJ103158-5814.7, same star that exhibits orbital motion, variable amplitude and a variable pulsation period. }
    \label{fig:ASASJ103158-5814.7_example}
\end{figure}
\paragraph{ASAS~J103158-5814.7} (GDR3~5351436724362450304) is a particularly interesting star, cf. Figure~\ref{fig:ASASJ103158-5814.7_example} and Paper~I. It exhibits three signals at once: very short-period overtone pulsation (classified as first overtone by Gaia DR3 SOS, $P_{\rm{puls}} = 1.11936$\,d), high-amplitude orbital motion exceeding 13\,\kms, cf. Table\,\ref{tab:polydegs}, and strongly modulated pulsation amplitudes reminiscent of the ``Blazhko'' Cepheid V0473~Lyrae \citep{Burki1982, Molnar2017}. 
At present, the orbital variations do not allow the estimation of an orbital period and we will thus continue monitoring this intriguing star. Unfortunately, this star was not included in the list of Cepheids with time-series RV data in \gaia~DR3.

\SSS{\paragraph{AX Cir} is a known SB1 and its orbit was first derived by \cite{Petterson2004Orbits}. Our \porb\ estimate of $\sim$6285 days differs by 5\% (10-$\sigma$) from that of \cite{Petterson2004Orbits}. While our combined dataset, comprising \gc~and literature data, incompletely samples orbital phases, we obtain an eccentricity that is in total agreement with the previously reported orbital eccentricity for AX Cir. }

\paragraph{FN Vel} was discovered to be an SB1 by \cite{PhDthesis} who provided a first orbital solution. \citet{Kovtyukh2015} showed evidence for the companion using the Ca H \& K lines. Here, we provide an updated orbital solution.

\paragraph{BP Cir} is known to host a hot main-sequence companion \citep{Arellano1985,Evans1992bpcir} based on IUE spectroscopy. \citet{Petterson2004Orbits} reported $v_\gamma$ variations exceeding 5 \kms~and that the star could possibly have a long orbital period. 
Using \gc~alone we have $v_\gamma$ variations of $\sim$2 \kms, however, when we combined \gc~with literature datasets from \cite{Petterson2004Orbits} and \cite{Borgniet2019} we found $v_\gamma$ variations of $\sim$ 4 \kms~at a temporal baseline of 9000 days. \SSS{We estimated BP~Cir's orbital period using \gc~and literature datasets, and found a tentative solution with \porb$\sim$ 4300 days and $e \sim 0.8$, cf. Figure~\ref{fig:VgammaOrbitBPCir}. Nevertheless, due to limitations in the available data, we are unable to definitively determine whether a period of 4300 days or twice that duration provides a more accurate representation.}

\paragraph{FO Car's} observations began in March 2016, and we find a Keplerian  orbit of \porb~$\sim 1660$\,d. To the best of our knowledge, no previous publication\footnote{\citealt{pulsatingbinaries} credited \citealt{Szabados1996} with detecting such evidence. Although FO~Car is not listed in this publication, FR~Car is, and we surmise a typo in \citealt{pulsatingbinaries}.} has presented the case of FO\,Car. However, the online database of binary Cepheids \citep{BinCepDB} lists it as a spectroscopic binary, crediting Szabados \& Berdnikov (2003 in preparation). \citet{Russo1981} found photometric evidence for a companion using a phase shift technique involving optical and UV data. Based on the visual companion identified during the spectroscopic observations (cf. Figure~\ref{fig:FOCar_guiding}) we consider this as the most likely source of the phase shift. Given the angular separation, the visual companion is clearly not the cause of the 4.5~yr orbit, and hence, FO~Car is a triple system composed of the Cepheid with an inner spectroscopic and outer visual companion.

\paragraph{MU Cep} - The spectroscopic binarity of MU\,Cep (not to be confused with the much brighter red supergiant $\mu$\,Cep) was discovered by \cite{PhDthesis}. Here we present an updated orbital solution, cf. Table\,\ref{tab:orbits}, illustrated in Figure\,\ref{fig:MUCepOrbit}. The previously reported  $P_{orb} \sim 1500 \pm 350$\,d exceeded the observational baseline. Thanks to continued follow-up of this faint star ($\langle\ m_V \rangle$ = 12.3~mag), we have achieved nearly complete coverage of MU~Cep's orbital phase and obtain tight orbital constraints, with \porb$=2028\pm2.7$\,d. 

\paragraph{R Cru} is listed as an SB1 in the database by \citet{BinCepDB}, which credits unpublished work by
L. Szabados and D. Bersier with this discovery. More recently, \cite{Evans2020resolved} reported a visual companion of R~Cru at a separation of 1.9 arcseconds based on Hubble Space Telescope (HST) imaging. We have acquired nearly 7~yr worth of very high-precision RV data for this bright Cepheid. The data clearly reveal low-amplitude orbital motion with \porb$\sim$ 238\,d (cf. Figure~\ref{fig:RCruOrbit}), rendering R~Cru the Cepheid with the shortest orbital period in the Milky Way. The template fitting routine further finds evidence for non-linear period changes (see Figure 15 of PaperI), cf. also \cite{Csornyei2022}.  
The relative semimajor axis of $a_{\rm rel} > 274 R_\odot$ implies that the Cepheid and its companion are not in contact, assuming a Cepheid radius of $47 R_\odot$ computed using a P-R relation \citep{Anderson2016rot} and a nearly circular orbit. For a Cepheid mass of around $5 M_\odot$ and a $2 M_\odot$ A2 companion (upper limit from the IUE spectrum, \citealt{Evans1992IUE}), the mass function for a circular orbit yields a minimum inclination of $> 3.3\,$deg, improving slightly over the limit from the mass function alone ($1.6\,$deg). For the range of inclinations $3.3 - 90\,$deg, and assuming a $5 M_\odot$ Cepheid, we thus find the allowed mass range for the companion to be between $0.09 - 2.0\,M_\odot$.

\paragraph{RV Sco} - \citet{POP81} list RV~Sco as a visual multiple system, with a non-variable companion of $m_V \sim$ 13~mag at a separation of 6.0 arcseconds. \cite{Evans1992IUE} did not detect any photometric companion for RV~Sco within their IUE spectral analysis. \cite{Szabados1989} suspected its binarity through its variable $\gamma$-velocity. Here, we confirm the finding as we detected significant (A$_{p2p}\sim$5.3 \kms) amplitude variation in RV resulting into a 4-$\sigma$ detection of binarity. RV~Sco is therefore a triple (SB1 + visual companion) system.

\paragraph{RY Sco} is listed as a visual binary in \cite{BinCepDB}, with references to \citet{POP81, Evans1994a}. We here detect the stars's SB1 nature for the first time, with very low-amplitude orbital RV variations of $\sim$0.6 \kms. 
Combining \gc~with literature datasets and extending to longer baseline, we notice a $v_\gamma$ variation of $\sim$2.5 \kms on a timescale of 41 years. RY~Sco is thus most likely a triple system with both a visual and a spectroscopic companions.

\paragraph{RZ Vel} was reported to be an SB1 Cepheid by \citet{Szabados2013}. 
\SSS{\citet{Szabados2013} used a subset of \gc~data (between BJD 2455650 - 2456400). We extended this \gc~baseline to BJD 2459600 and combined it with literature datasets from \cite{1980SAAOC...1..257L, Coulsen1985b, Bersier2002, Borgniet2019}. Although, data from \cite{Bersier2002} was also included in \cite{Szabados2013}, we do not detect 3-$\sigma$ \vgamma~variations for this or any other recent dataset after the year 2000.}

\paragraph{S Mus} was first identified as a binary using photometric observations \citep{Evans1968}. A first orbital solution was presented by \citet{Evans1982}, with several others since. In particular, \citet{Petterson2004Orbits} provided a precise update to this solution, and \citet{Binnenfeld2022} determined the periodicity using partial periodograms directly computed from the spectra rather than RV measurements. We gathered 78 high-precision \coralie\ RVs over a total baseline of 6.3~yr. These data alone yield an orbital solution in agreement (to within 1 day) with  \citet{Petterson2004Orbits} and \citet{Gallenne2019}, cf. Table\,\ref{tab:orbits2}. As shown by \citet{Evans2015,Evans2020resolved}, the companion is an X-ray active B3V star. 

\paragraph{S Nor} was detected as an SB1 by \cite{Szabados1989}, where they estimated tentative orbital periods of  3300 or 6500 days. \cite{Groenewegen2008} modeled S~Nor's RV data using a circular orbit with a fixed \porb~of 3600 days and concluded that more RV data are needed to better determine this orbit. Here, we do not find this orbital estimate to be consistent with \gc~data. We do not detect any \vgamma~variations with \gc~data alone, however, when \gc~is combined with literature datasets we do obtain a 3-$\sigma$ detection for the oldest RV datasets (Figure~\ref{fig:VgammafornoSB1signs1}). 
\referee{However, the only cluster that exceeds 3-$\sigma$ is based on only three observations and the uncertainty of this particular cluster is likely underestimated. 
Consequently, we doubt the validity of the only $>$3-$\sigma$ outlier and cannot consider the star a bonafide binary. Nevertheless, if the RVTF results are taken at face value, then the orbit would have to be significantly eccentric.} 
Lastly, Figure~\ref{fig:VgammafornoSB1signs1} indicates that a potential orbit clearly has high eccentricity.

\paragraph{SU Cyg} is a solitary case wherein the pulsation model was derived using a combination of \gc~data and zero-point corrected RV data from \cite{Borgniet2019}. While \gc~data provided strong constraints on the orbit, they were insufficient on their own to create a satisfactory Fourier series model. To address this, we included the \cite{Borgniet2019} measurements to formulate the RV template. This template was then used to compute \vgammares, as listed in the appendix Table~\ref{tab:SUCygVgamma}. Finally, employing the MCMC method, we obtained the `V+L' orbit of SU~Cyg with \porb~of $\sim$548~$\pm$~0.1 d and eccentricity of $\sim$0.34~$\pm$~0.005 (Figure~\ref{fig:VgammaOrbitSUCyg}). These orbital estimates are in agreement with $\Delta$\porb~$<$ 1 d (all orbital elements within 3-$\sigma$ difference) the ones in the literature from \cite{Evans1988SUCyg, 
Imbert1984SUCyg, Groenewegen2008}.

\paragraph{TX Mon} was first reported as an SB1 by \citet{Szabados1998}, who estimated a short \porb$\approx 470 \pm 30$\,d. We here confirm the SB1 nature of the star and determine the orbit, albeit at much longer \porb=$1607\pm1$\,d, with $K=11.9\pm0.04$\,\kms. 

\paragraph{V0659 Cen} is host to a hot B6.0V companion detected directly using IUE UV spectroscopy \citep{Evans1994b}. V0659~Cen also has a high X-ray flux potentially originating from its companion \citep{Evans2022}.
Here we present the first RV evidence of the Cepheid's orbital motion with a peak-to-peak RV variation of $\gtrsim 5\,$\kms. Combining \gc~and literature data, we estimate a preliminary orbital period \porb~$\sim 9300 \pm 130$\,d with mild eccentricity ($e = 0.32$). 
\SSS{We designate this orbit as `tentative' due to the ambiguity regarding whether the orbital period is approximately 6000 or 9000 days. It should be emphasized that the current uncertainties, which are around 100 days, do not account for this ambiguity.}
Additional {\it HST} UV spectroscopy will be particularly valuable to measure the mass of the Cepheid (Evans et al. in prep.).

\paragraph{V1334 Cyg} is among the few sample cases where there is a disagreement between `V' and `V+L' estimates, specifically of 0.8\% (3-$\sigma$) in \porb~for V1334~Cyg. Even after combining the \gc~and literature datasets, constraints on the ascending part and extremities of the orbital phase remain limited. As the literature dataset from \cite{Gorynya1992} and \cite{Borgniet2019} add extra constraints on the descending branch of the orbital phase, we prefer the `V+L' orbital estimate for V1334~Cyg. Furthermore, both `V' and `V+L' orbital estimates are in 0.5\% (3-$\sigma$) disagreement with the ones from \citet{Gallenne2018} and 3-$\sigma$ and 2-$\sigma$, respectively, with the ones from \cite{Evans2000v1334}. This maybe attributed to the aforementioned poor sampling at some orbital phases. Future observations, with overall improved sampling will help to elucidate the discrepancy. \referee{ Lastly, \citet{Gallenne2018} demonstrated that \gaia~DR2 parallax is off by 3.6-$\sigma$ compared to the orbital parallax determined from the visual orbit + RVs of both components. This could be another reason for the discrepancy between our orbital parameters and the ones from \cite{Gallenne2018}.}

\paragraph{VY Car} was considered a tentative SB1 candidate by \citet{Anderson2016rv}. Despite improved treatment of instrumental zero-point offsets, the orbital signal has neither disappeared nor clarified. We note that Figure\,\ref{fig:VgammafornoSB1signs} suggests a trend with time. However, none of the epochs deviate from the average by more than 3-$\sigma$. Hence, VY~Car remains an unconfirmed SB1 candidate with a potentially very long orbital period.

\paragraph{VY Per} was reported to exhibit \vgamma\ variations exceeding 5\kms\ by \cite{Szabados1992} over a baseline of two years. We confirm the SB1 nature of this star and obtain a tentative orbital solution, with a \porb~$\sim$806~d and $e \sim 0.4$ (cf. Figure~\ref{fig:VYPerOrbit}). Additionally, the MCMC (V+L) orbital estimate is presented in Table~\ref{tab:orbits} and Figure~\ref{fig:VgammaOrbitVYPer}.
This MCMC indicates that the true orbital period could be a multiple of $805$\,d. Unfortunately, the result remains ambiguous due to the sampling of the data, which also does not include the orbit's descending branch. 

\SSS{\paragraph{W Sgr}'s \porb~using \gc~data differs by 6-$\sigma$, 20-$\sigma$ and 18-$\sigma$ from that of \cite{1989A&A...216..125B}, \cite{Petterson2004Orbits}, and \cite{Gallenne2019} respectively. We would like to note that in our work the eccentricity is not sufficiently constrained (Figure~\ref{fig:WSgrOrbit}), hence, the orbit is probably more eccentric and shorter than our model as well as the \porb~is potentially more uncertain than the fit suggests. Unfortunately, the current dataset does not provide sufficient information for a more accurate fit, and additional observations will be necessary. In light of these considerations, we emphasize that \cite{Petterson2004Orbits} currently offer a more reliable orbital solution.}

\paragraph{X Cyg} was reported to exhibit an orbital signal via the light time travel effect by  \citet{Karoly1990}, who estimated \porb$\approx 55$\,yr.  \citet{Evans2015} and \citet{Hintz2021} did not detect long-term \vgamma\ variations based on $\sim$~11-year baselines. We here identify X~Cyg's $v_\gamma$ variations using much longer temporal baselines of $\sim$ 40 yrs by combining \gc\ data with the literature \citep{1987ApJS...65..307B, 1989ApJS...69..951W, 1998MNRAS.297..825K, Barnes2005, Gorynya1992, Storm2004, Bersier1994, Borgniet2019}. We thus find a total $v_\gamma$ difference of $\sim$ 3 \kms~and a lower limit of 40~yr on \porb, consistent with \cite{Karoly1990}.

\paragraph{X Pup} was mentioned by \citet{Anderson2016rv} to exhibit tentative signs of time-variable \vgamma\ based on a difference between \veloce\ data and the literature. However, the improved corrections of RV zero-points from \veloce-I together with the clustered RVTF analysis do not reveal a significant signature of orbital motion, cf. Figure\,\ref{fig:VgammafornoSB1signs}. 

\paragraph{XX Cen} - \cite{Groenewegen2008} determined an orbital solution with \porb$=924 \pm 1.1$\,d and $K=4.47\pm0.28$\,\kms, and $e$ fixed to $0$. We find a significantly different orbit for XX~Cen based on \gc\ data alone and determine an orbital solution with \porb$=722.44 \pm 0.42$\,d, $K=5.4\pm0.03$\,\kms, that is significantly eccentric with $e = 0.295 \pm 0.004$. 

\section{Sample analysis\label{sec:sample}}

\subsection{Correlations among orbital elements\label{sec:orbits}}

\begin{figure*}[ht!]
    \centering
    \includegraphics[scale=0.7]{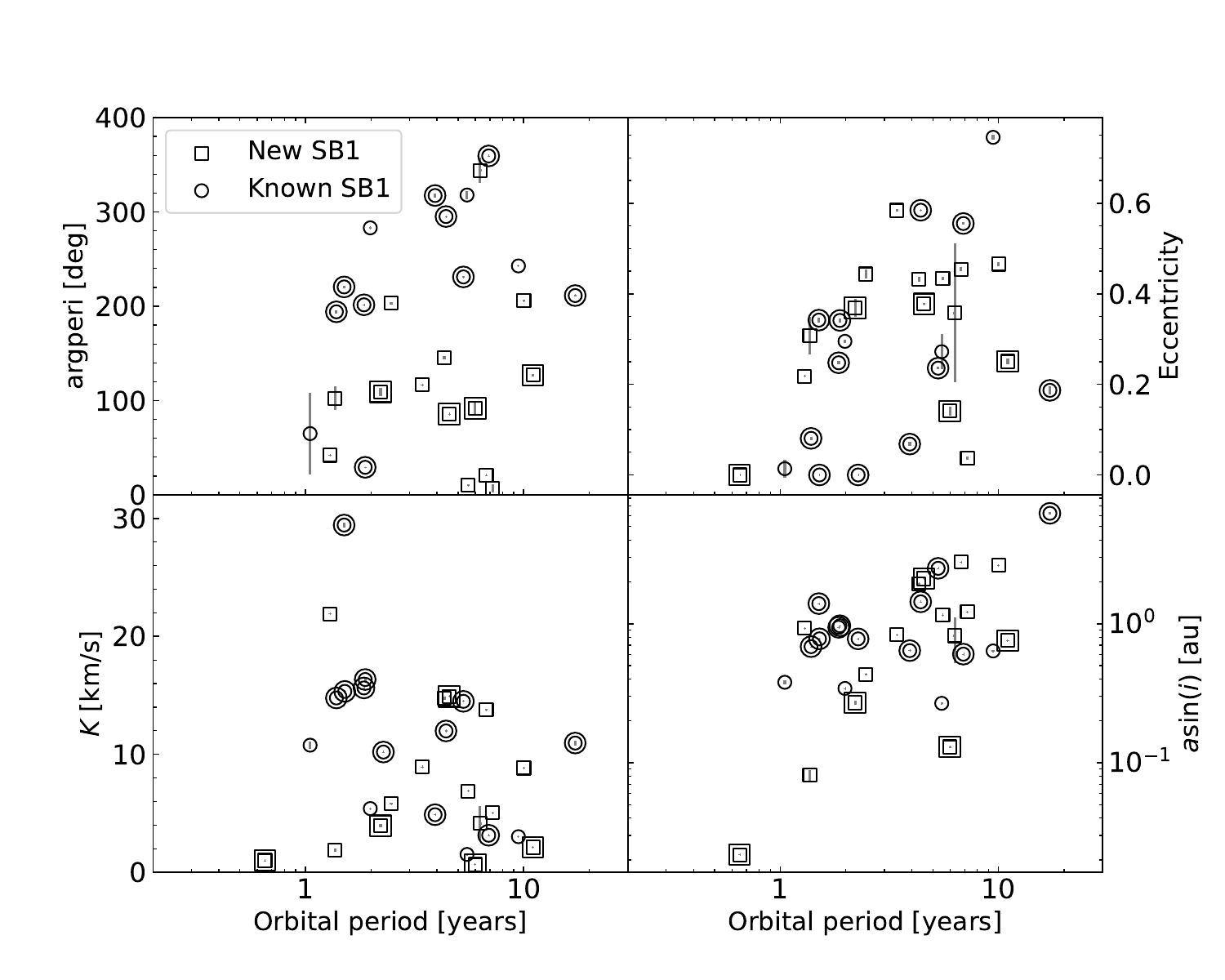}
    \caption{The distribution of different orbital elements (Upper left: argument of periastron $\omega$; upper right: eccentricity $e$; lower left: semi-amplitude $K$ of the orbital RV variations; lower right: projected semi-major axis $a \sin{i}$) versus orbital period \porb\ for Cepheids with orbits determined here.  SB1s with newly reported orbits in this study are plotted using open squares, while \gc~orbital estimates of SB1s with
    literature-known orbits are plotted in open circles. Symbols with double boundaries highlight `V+L' orbits determined using \gc~and zero-point corrected literature RVs. R~Cru, SY~Nor and YZ~Car are absent in the top left panel due to insignificant eccentricity.
    \label{fig:Correlation_Orbitalparams} }
\end{figure*}

Figure~\ref{fig:Correlation_Orbitalparams} shows the correlation between different orbital elements ($\omega$, $K$, $e$, and $a\sin{i}$) with \porb\ for all orbital solutions reported here. 
\SSS{This includes first-time orbital estimates of \norbitsNew~stars and \norbitsKnown~\gc\ orbital estimates for Cepheids with literature-known orbits.}
Newly discovered SB1 systems populate the full range of orbital parameters, indicating that our sample selection was not biased towards the literature-known orbital properties of SB1 Cepheids.
Semi-amplitudes tend to decrease as \porb\ increases. Thanks to the precision of \gc~RV measurements and long-term monitoring, we were able to discover many new SB1 systems with very small semi-amplitudes and long orbital periods. 
\referee{A noticeable correlation exists between $a\sin{i}$ and \porb, and the dispersion along this correlation might stem, at least partially, from fluctuations in the combined masses of the Cepheid and its companion.}

The eccentricity versus orbital period distribution (upper right panel in Figure~\ref{fig:Correlation_Orbitalparams}) is particularly interesting. We find that the maximum eccentricity increases with orbital period. Despite a large range of eccentricities at most \porb\ values, an absence of very high $e < 0.6$ is clearly noticeable at \porb$\lesssim 3$~yr. The two Cepheids with the shortest \porb\ have very low or no noticeable eccentricity. We also notice an absence of circular orbits at \porb$\gtrsim 3$~yr, which appears to become an exclusion region at longer \porb. Since such distant companions have most likely evolved in isolation, this could point to a primordial feature of intermediate-mass (B-star) binaries. However, NBODY simulations show that cluster dynamics increase $e$ for long orbital periods over time (Dinnbier et al., in prep.). Additional long-period orbits are required to investigate this effect.

Most of these trends qualitatively agree with the empirical results by \citet{Evans2015}, as well as with predictions from the Cepheid binary population synthesis study by \cite{Neilson2015} and \citet{Karczmarek2022}, as well as with the dynamical NBODY6 simulations by \cite{Dinnbier24}.
However, we do find several Cepheids at high $5 < P_{orb} < 10$~yr and relatively low eccentricities ($e < 0.2$) where Cepheid binaries have been predicted to have a low probability of occurring \citep{Neilson2015}. Additionally, R~Cru is very close to the minimum \porb\ consistent with the synthetic populations by \cite{Neilson2015} and \citet{Karczmarek2022}.

Incompletely sampled orbits described in Section~\ref{sec:trends} and illustrated in Figure~\ref{fig:polydeg_results} provide partial information on $K$ and \porb. Comparing $A_{\rm p2p}$ with $K$ (in Figure~\ref{fig:Correlation_Orbitalparams}) as a function of \porb, we notice the inverse trends. Shorter orbital periods tend to have larger $K$, whereas we find larger $A_{\rm p2p}$ over longer baselines. This suggests that most short-\porb\ orbits with $K \gtrsim 0.5$\,\kms\ of Cepheids observed by \gc\ have been determined. Very low $A_{\rm p2p}$ values are likely to continue growing over longer baselines. 

\subsection{Cepheids exhibiting proper motion anomaly \label{sec:PMa}}
\citet{Kervella2022} used \textit{Gaia} to detect astrometric binaries using the proper motion anomaly (PMa). This method compares the proper motion vector determined by the ESA \textit{Hipparcos} mission \citep{hipcatalog} and various \gaia\ data releases \citep{GaiaMissionDR1,GaiaSummaryDR1,GaiaDR1TGAS,GaiaDR2summary,GaiaDR2astrometry,GaiaEDR3summary,GaiaEDR3astrometry}.
PMa quantifies the significance of a long-term rotation of the proper motion vector due to orbital motion. Stars with PMa signals exceeding $3\sigma$ were thus labeled as binaries by \citet{Kervella2022}. Proper motion probes the on-sky projection of orbital motion, orthogonal to the line-of-sight variations measured using RVs. 
Therefore, the PMa and RV methods function as complementary tools for binary detection \referee{(cf. \cite{Kervella2019a} for details on the sensitivity function of PMa)}. Consequently, it's typical for some stars identified as SB1 to lack discernible deviations in their proper motions which would categorize them as PMa-True candidates. 
\referee{This could either be if the Cepheids are located at large distances or possess orbital periods too short for detection.
On one hand, companions with orbital periods shorter than $\sim$ 3 years are unlikely to be identified via PMa due to the time window smearing inherent in \gaia~DR3.
On the other hand, extremely long-period companions, spanning hundreds of years, are similarly challenging to detect, representing a limitation shared by VELOCE as well.}
Importantly, a PMa-False flag does not raise concerns about the SB1 detection. 
Among the \nbincep\ 
binary Cepheids in our sample, 19 have been reported to exhibit PMa signals (with binary flag `BinH2EG3b' in \cite{Kervella2022} as 1). 
Evidence for astrometric orbital signals in form of the PMa has been reported for 10 of \norbits~SB1 ($30\%$) Cepheids with orbital solutions presented here (including tentative ones) as well as for 8 of \npolydeg~($25\%$) SB1 systems with long-term \vgamma\ trends exhibit PMa \citep{Kervella2022}. This also includes 2 of the newly discovered SB1 Cepheids, GX~Car and V0659~Cen. Table~\ref{tab:GaiaAstrometry} lists the PMa binary flags from \citet{Kervella2022} for all stars in common with our sample. 

Two systems with reported PMa, RZ~Vel and SV~Per, deserve a specific mention. For these two stars, we found no evidence of time-variable \vgamma. However, SV~Per is known to have a very nearby (0.2") companion that has been resolved using \textit{HST/WFC3} spatial scans \citep{Riess2018} and which is most likely unresolved by \gaia. Time-variable contrast differences due to the Cepheid's pulsation would affect the measurement of the photocenter for SV~Per, which may lead to a spurious PMa signal.  

Finally, we note that none of the Cepheids considered to be non-SB1s in \gc\hbox{---}stars without evidence for time-variable \vgamma\hbox{---}have been reported to exhibit a significant PMa. Specifically, this is true for all stars without evidence for time-variable \vgamma\ based on \gc\ data alone as well as our RVTF analysis (Sect.\,\ref{sec:RVTF}), including stars for which our results contradict previous claims of time-variable \vgamma. 

As this comparison shows, RVs from \gc\ and literature data standardized to the \gc\ zero-points provide the most complete evidence for orbital motion for Cepheids on  timescales of up to a few decades. Nonetheless, we caution that signals on the order of $1\,$\kms\ and lower can be introduced also by modulated variability \citep{Anderson2014}. CCF shape parameters, such as the bisector inverse span and full width at half maximum can be informative to this end \citep{Anderson2016c2c} and will be considered in future work.

\subsection{\textit{Gaia} DR3 astrometric quality flags of SB1 Cepheids}

\begin{figure*}
    \centering
    \includegraphics[scale=0.55]{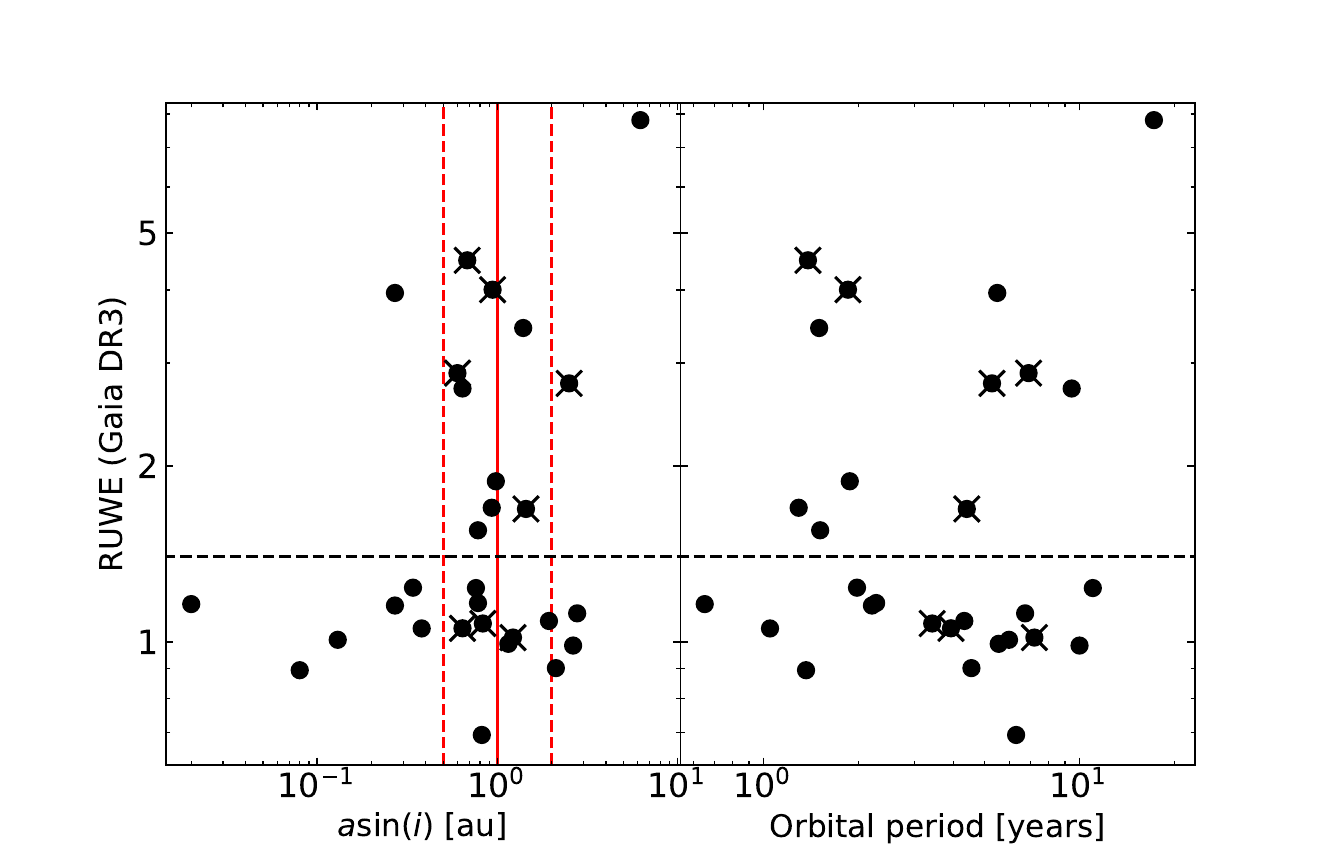}
    
    \caption{Distribution of \textit{Gaia} DR3 RUWE with the semi-major axis (left) and orbital period (right) of our sample systems. The stars tagged as binaries from PMa \citep{Kervella2022} are marked with crosses. The red solid line indicates $a\sin{i}$ of 1~au, and the dashed red lines indicates the same at 0.5 and 2~au.}
    \label{fig:RUWE}
\end{figure*}

\begin{figure}
    \centering
    \includegraphics[scale=0.59]{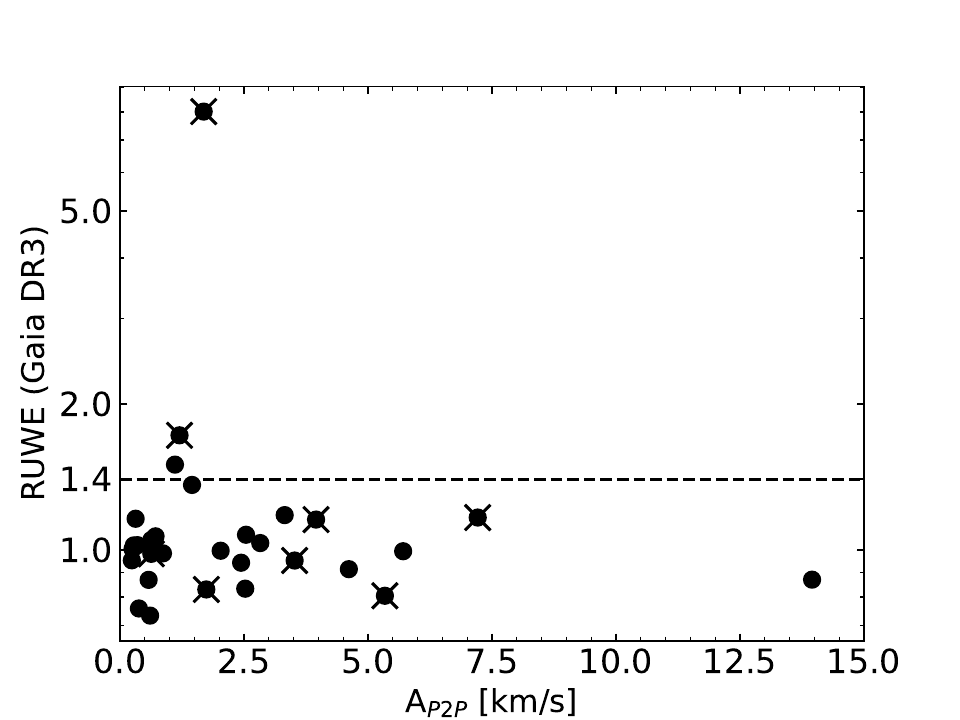}
    \caption{Distribution of \textit{Gaia} DR3 RUWE with the maximum amplitude of the (linear or non-linear) $v_\gamma$ variation for SB1 systems with `Trend'. The stars tagged as binaries from PMa \citep{Kervella2022} are marked with crosses. }
    \label{fig:RUWE_Trends}
\end{figure}

\begin{figure}
    \centering
    \includegraphics[scale=0.61]{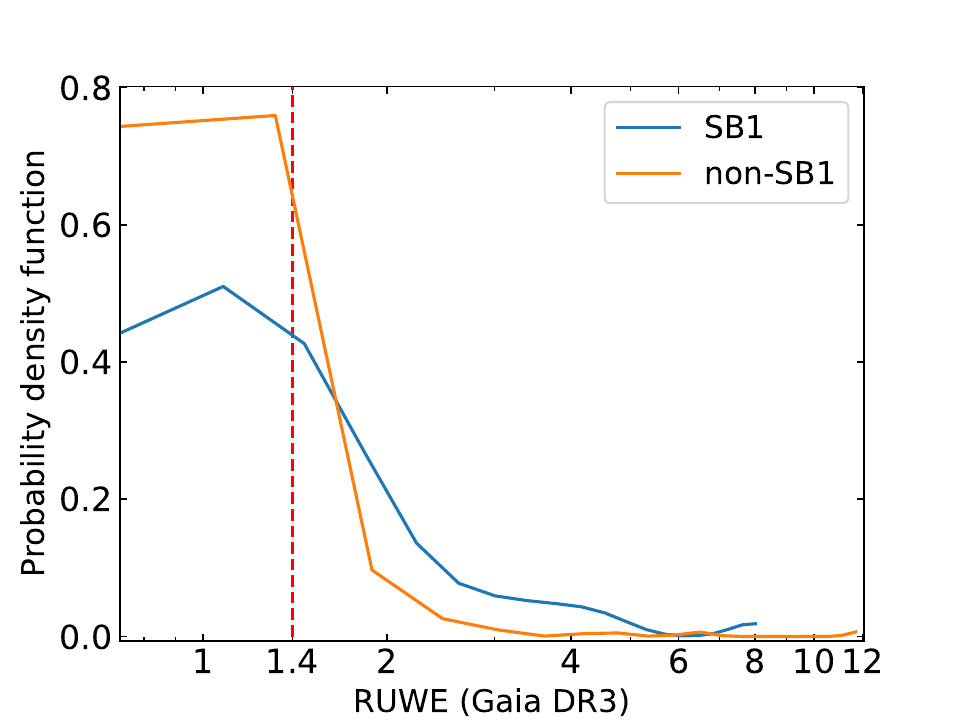}
    \caption{Probability density function of \textit{Gaia} DR3 RUWE for \gc~SB1 and non-SB1 systems.} 
    \label{fig:RUWE_Pdf}
\end{figure}

\gaia\ DR3 reports several astrometric quality flags and parameters to aid the interpretation of the reported astrometric measurements. The Renormalized Unit Weight Error (RUWE) has received particular attention and is often considered an indicator of stellar multiplicity. RUWE is calculated as the square root of the normalized $\chi$-square of the astrometric fit to the along-scan observations \citep{RuweDescription}, and the threshold of RUWE $< 1.4$ is frequently used to indicate well-behaved astrometric solutions \citep{GaiaDR2astrometry, GaiaEDR3astrometry}.
However, several factors can result in elevated RUWE, which does not identify the origin of the excess residuals and includes all possible sources of error in the fit to the astrometric model. The latter include any signals affecting photocenter stability, notably involving marginally resolved visual binaries and unmodeled astrometric orbital signals. 

In Cepheids, the high-amplitude chromatic variability of Cepheids could also contribute to RUWE. For example, \gaia\ may automatically select different gating schemes and window classes depending on the momentaneous brightness of Cepheids that vary by $\sim 1$\,mag in optical bands. Additionally, chromatic variability may lead to noise due to time-variable chromatic diffraction properties. Given these added difficulties, it is likely for RUWE to be elevated for Cepheids compared to non-variable stars. We therefore investigated to what degree excess RUWE and other astrometric quality indicators reported by \gaia\ may serve as an indicator of multiplicity for Cepheids. 

Figure~\ref{fig:RUWE} shows the comparison of \gaia\ DR3 RUWE against the projected semimajor axis and \porb.
The figure shows a significant excess RUWE near $a\sin{i}$ values close to 1\,au as 5 out of 8 stars with the highest RUWE (RUWE > 2) fall into this category (stars located within the dashed red lines in the left panel of Figure~\ref{fig:RUWE}).
Interestingly, a corresponding feature is not so clearly seen near \porb$=1$~yr, suggesting that the size of the orbital ellipse on the sky more directly affects the astrometric solution than the periodicity of the signal. 
The $6300$\,d binary with the highest RUWE of $7.8$ is AX~Cir, whose hot main sequence (likely B6V) companion has been resolved interferometrically \citep{Gallenne2014AXCir}. A photometric contamination is the potential origin of the elevated \gaia~RUWE.

Figure~\ref{fig:RUWE_Trends} shows RUWE in \gaia\ DR3 vs the amplitude of trends, $A_{\rm p2p}$, determined using \gc~data only in Sect.\,\ref{sec:trends}. Given that the majority of these orbital signals occur on timescales longer than the \gc\ baseline, it is not surprising that RUWE~$< 1.4$ for the vast majority of these stars. The most significant outlier at RUWE $\sim 8$ is RW~Cam, which was shown to feature a significant UV excess indicative of a hot main sequence companion \citep{Stepien1968}. The star with the second highest RUWE of $1.7$ is T~Mon, host to a B9.8V companion detected directly from UV spectroscopy \citep{Evans1994TMon} but remained undetected interferometrically in $H-$band \citep{Gallenne2019}. These stars collectively indicate that elevated \gaia~RUWE values in Cepheids may be attributed to unresolved companions rather than orbital motion.

Figure~\ref{fig:RUWE_Pdf} shows the probability density function of RUWE for SB1 and non-SB1 Cepheids in \gc. Approximately 25\% of \gc\ SB1s have a RUWE~$>$~1.4 (see Table~\ref{tab:GaiaAstrometry}). While there is a slightly higher probability for an SB1 Cepheid to have excess RUWE compared to non-SB1 Cepheids, the difference is not very pronounced. Qualitatively, among the newly identified SB1 systems (RVTF), Figure~\ref{fig:RUWE_Pdf} suggests similar probabilities near RUWE $\approx 1.7-1.8$. However, among these, $\beta$~Dor's extreme brightness ($m_V=3.8$\,mag) likely contributes substantially to the high RUWE of $4.5$. 

Finally, we also investigated the \gaia\ astrometric quality indicators  astrometric\_gof\_al and astrometric\_excess\_noise as indicators of multiplicity of Cepheids. However, no significant trends appeared in this comparison aside from a mild increase in astrometric\_excess\_noise near $a\sin{i}\sim 1$\,au.

In summary, we find excess RUWE mainly for Cepheid binaries with projected semimajor axis close to 1au. The parameter astrometric\_excess\_noise is also increased in this range, though the correlation is less clear. In addition to specific orbital configurations, elevated RUWE can arise from photocenter variations in unresolved binaries and in very bright stars, for example. Additionally, the majority of binary Cepheids are not identified by RUWE~$>1.4$. 

\subsection{Amplitude ratios as indicators of companion stars\label{sec:ampratios}}

\begin{figure*}
    \centering
    \includegraphics[width=1\textwidth]{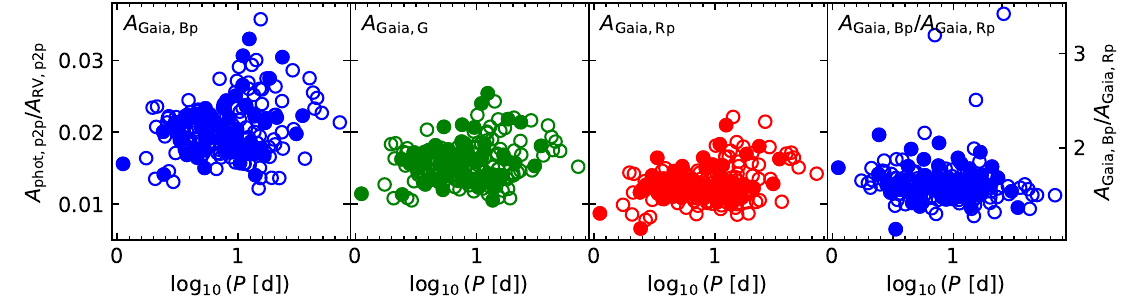}
    \caption{\textit{First three panels from left to right:} photometric-to-RV amplitude ratios from \gaia\ DR3 and \gc\ against the logarithmic pulsation period, with wavelength of the photometric band increasing from left to right. \textit{Right panel:} ratio of \gaia\ photometric amplitudes in $G_{Bp}$ to $G_{Rp}$. Filled circles show SB1 Cepheids reported in \veloce. Other single-mode Cepheids observed by \veloce\ are shown as open circles. There is no clear difference between SB1 Cepheids and all others in any of the amplitudes considered, nor a dependence of such a difference on $\log{P_{\mathrm{puls}}}$.}
    \label{fig:amp_rv_phot}
\end{figure*}

Photometric amplitude ratios can be used to identify Cepheid companion stars in case the photometric contrast is sufficiently low \citep[e.g.][]{Madore1977,1980PASP...92..315M,Evans1994a}. Since most Cepheid companions are hot early-type stars, shorter-wavelength data is generally more sensitive to companions than longer-wavelength photometry. A noticeable exception are double-lined spectroscopic binaries (SB2) in the Large Magellanic Cloud that consist of a Cepheid and a giant companion \citep{Pilecki2021}.

In the Milky Way, \citet{Klagyvik2009} considered low ratios between photometric and RV amplitudes as an indicator for the presence of companion stars. Thanks to \gaia's multi-band photometry, such a comparison is possible here using a large sample of stars. Figure\,\ref{fig:amp_rv_phot} thus shows the ratio between \gaia\ photometric and \veloce\ RV amplitudes in three bands ($G_{Bp}$, $G$, and $G_{Rp}$) as a function of the logarithmic pulsation period. The higher ratios at shorter wavelength are readily explained by the larger photometric amplitudes. SB1 Cepheids identified in \gc\ are shown as filled circles, and stars which are not SB1 systems are shown as open circles. As a consequence of the Cepheid period-color relation, the contrast between Cepheid and companion should increase toward longer periods, rendering photometric detection more difficult. Additionally, any signal in the amplitude ratio should be enhanced at shorter wavelengths. However, the comparison between SB1 Cepheids and all others observed in all three bands reveals no indication that bona fide SB1 Cepheids have reduced amplitudes compared to other Cepheids (Figure\,\ref{fig:amp_rv_phot}). The only potential signature is a slightly reduced range of amplitude ratios for SB1 Cepheids, which, however, clusters in the center of the distribution. 
The comparison between photometric amplitudes in $G_{Bp}$ vs $G_{Rp}$ (not shown) yields the same impression. Hence, we find that \gaia's photometric amplitudes are not good indicators of Cepheid companions.

\section{Discussion and Summary}\label{sec:disc}
We have presented the largest homogeneous investigation of \nbincep\ SB1 Cepheids to date based on the unprecedented RV data set provided by the \gc~project (Paper~I). An additional \nbincepcand\ SB1 candidates have been identified in Paper~I, but are not discussed here due to their nature as candidate binaries. 
\gc\ data allow us to extend the search for binary Cepheids to  fainter magnitudes, adding \nbincepnew\ new SB1 systems, and \norbitsNew\ first orbital determinations of Cepheids, 11 of which are fainter than $m_V > 8$\,mag (cf. Figure~\ref{fig:Histogram_withSzabados}). 
Thanks to these improvements, we sharpened the uncertainty on the SB1 fraction of Cepheids by a factor of $\sim 2.3$. 

We present 30 definitive and \norbitsTent\ tentative orbital solutions, which range from the discovery of the shortest \porb\ in the Milky Way for R Cru (238\,d) to $\sim 17$~yr (AX~Cir), and also include a correction to the orbital solution of $\delta$~Cep. For \norbitsNew~Cepheids, we estimate orbital solutions for the first time. \gc\ observations are particularly sensitive to \porb$\lesssim 5$\,yr, and the available temporal baselines differ from star to star. The ensemble of orbital solutions suggests two peculiar features in the $e-\log(P)$ diagram, which are a possible exclusion zone of low-eccentricity systems (0 $\leq e <$ 0.2) at long orbital periods (\porb~$\gtrsim10$\,yr) as well as a dearth of stars with orbital period in the range of $2.5-3.5$\,yr. Further observations and a more detailed analysis of detection limits will be needed to confirm these features.

Based on our search for SB1 Cepheids and a detailed investigation of evidence for orbital signals using literature, we find a SB1 fraction of $29.6 \pm 3.4\%$. This is a lower limit since 14 SB1 candidates reported in Paper~I were not considered in this fraction. Furthermore, we estimate the fraction of SB1 Cepheids with \porb$< 10$\,yr of $15.2 \pm 2.4 \%$, half of which have \porb\ of less than 2.5 years. 
Our binary fraction estimates agree with previous reports, albeit with $2-3\times$ lower uncertainty. 
Of course, SB1 Cepheids are detectable mainly for orbital periods of $< 100$\,yr, and other methods must be considered to estimate the total binary fraction of Cepheids. \referee{Considerable effort is currently underway to comprehensively determine the binary fraction of Cepheids, employing a diverse array of techniques such as astrometry, spectroscopy, interferometry, and observations across various wavelengths.}
For example, \cite{Evans2022} estimated a total binary fraction of $57 \pm 12\%$ based on SB1 Cepheids, and Cepheids with evidence for companions from X-ray and UV observations.
\referee{It is worth noting that future Gaia data releases, particularly DR4 and DR5, will contribute significantly with an abundance of astrometric orbits, enriching our understanding of these systems even further.
}

Several of the SB1 Cepheids studied here are part of triple systems, including AW~Per \citep{Evans2000awper}, FF~Aql \citep{Udalski1993}, S~Sge \citep{Evans1993SSge},  V1334~Cyg (\citealt{Abt1970, Evans1994b}), V0659~Cen (\citealt{Evans2013, Evans2022}), and W~Sgr \citep{Evans2009}. Here, we further identify FO~Car, RV~Sco, RY~Sco, and UX~Per as triple systems with an inner SB1 and outer visual companion. Comparing the binary fraction and incidence of triples among Cepheids with dynamical NBODY simulations that assume an initial binary fraction of $100\%$ \citep{Dinnbier24} suggests that a significant fraction of Cepheid progenitors (B-type stars) must be formed as triples or even higher-order systems. 

Last, but not least, we find that the \gaia\ DR3 parallax quality indicator RUWE is elevated for Cepheid binaries whose projected semimajor axis, $a \sin{i}$, is close to $1\,$au.
However, upon assessing RUWE for both SB1 and non-SB1 from \gc, we ascertain that \gaia~RUWE is not a reliable indicator for identifying Cepheid spectroscopic binaries and can be noticeably affected by photometric systematics, such as marginally resolved binaries or high apparent brightness. Additionally, RUWE tends to be slightly higher for Cepheids than for non-variable stars.

An examination of PMa in relation to various orbital properties of our stars does not reveal any correlation between the orbital orientation and PMa detection. \referee{Furthermore, all \gc~Cepheids with previously reported PMa detections are also identified to be SB1 systems.}

Precision radial velocities provide the most sensitive diagnostic of multiplicity of Cepheids to date. Future \gaia~data releases will allow to determine systemic masses of Cepheids using the combined astrometric and spectroscopic orbital signals. In \gaia\ DR3, only one Cepheid was reported with a non-single-star astrometric solution. Suspiciously, the orbital period of RX~Cam was given as $33\,$months, matching exactly the observational baseline of \gaia\ DR3. However, RX Cam was unfortunately not observed as part of the \veloce\ project. The upcoming fourth \gaia\ data release in combination with \veloce\ data will be a treasure trove for determining accurate Cepheid masses and to establish a well-sampled mass-luminosity relation required to elucidate the physics of Cepheids, notably with respect to mixing processes, such as convection and rotation, required to explain the mass discrepancy problem \citep[e.g][]{PradaMoroni2012,Anderson2014}. 

\section*{Acknowledgments}
RIA, SS, and GV acknowledge support from the European Research Council (ERC) under the European Union's Horizon 2020 research and innovation programme (Grant Agreement No. 947660). RIA, and SS further acknowledge support through a Swiss National Science Foundation Eccellenza Professorial Fellowship (award PCEFP2\_194638). 
SS would further like to extend their acknowledgement to the Research Foundation-Flanders (grant number: 1239522N). 
This work has made use of data from the European Space Agency (ESA) mission {\it Gaia} (\url{https://www.cosmos.esa.int/gaia}), processed by the {\it Gaia}
Data Processing and Analysis Consortium (DPAC, \url{https://www.cosmos.esa.int/web/gaia/dpac/consortium}). Funding for the DPAC has been provided by national institutions, in particular the institutions participating in the {\it Gaia} Multilateral Agreement.
The lead authors would like to acknowledge the very useful compilation of information on binary Cepheids provided by Laszlo Szabados at Konkoly Observatory (\cite{BinCepDB}, \url{https://cep.konkoly.hu/intro.html}).

\bibliographystyle{aa}
\bibliography{RVCatBib}

\begin{appendix} 
\section{Guiding camera Image\label{sec:guidingimage}}
\begin{figure}
    \centering
    \includegraphics[scale=0.6]{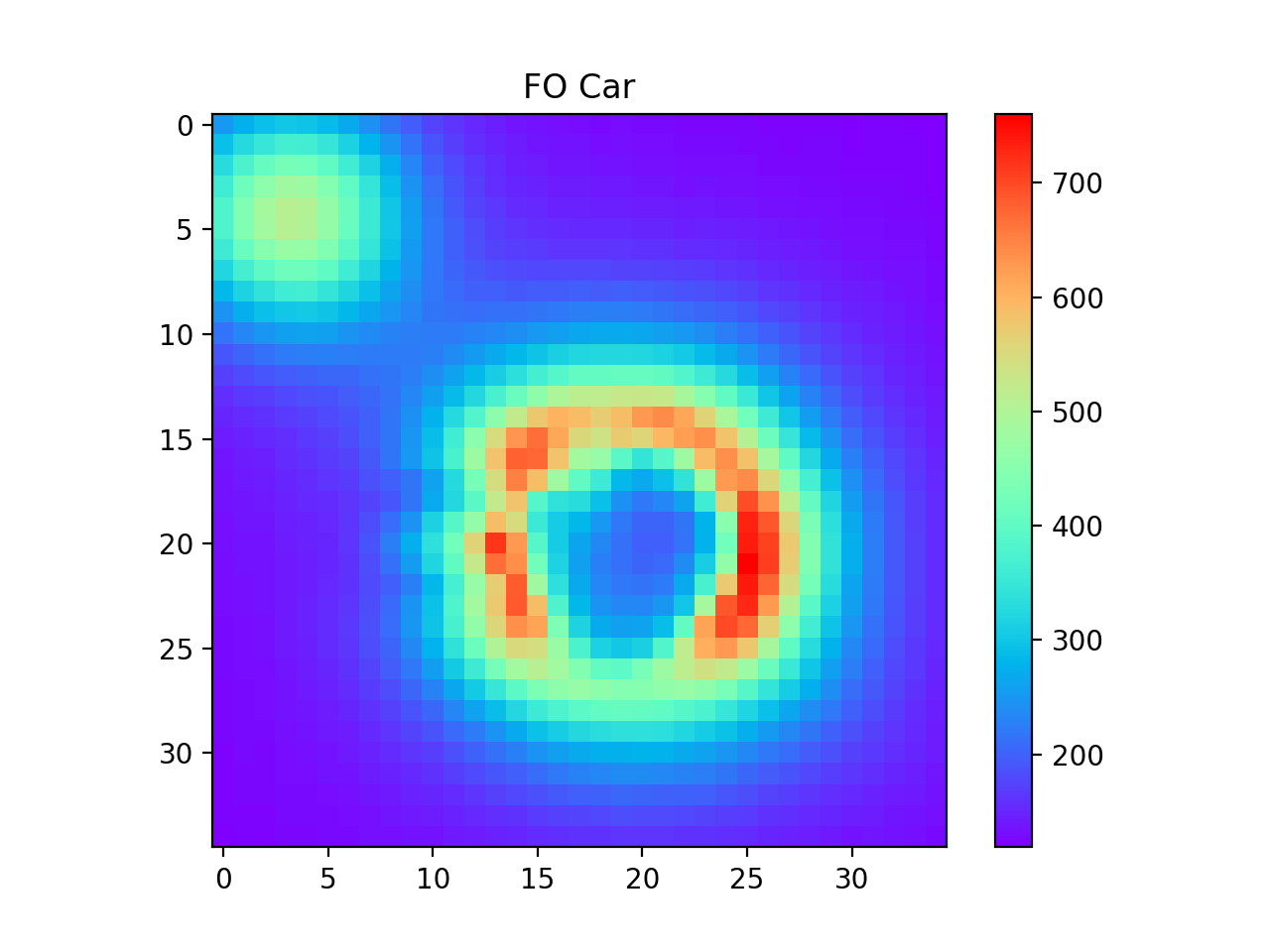}
    \caption{\coralie\ guiding camera image of FO Car. The center of the Cepheid is dark because of the fiber placement during the integration. The object at the top left is an outer (visual) companion, which renders the SB1 Cepheid FO Car a triple system.}
    \label{fig:FOCar_guiding}
\end{figure}

When available, the \coralie~and \hermes~guiding camera images were utilized to examine for triple systems. In Figure~\ref{fig:FOCar_guiding}, we present a guiding camera image of FO~Car obtained from \coralie. 

\section{Orbital fitting\label{sec:orbitfits}}
\begin{figure*}
    \centering
    \includegraphics[scale=0.5]{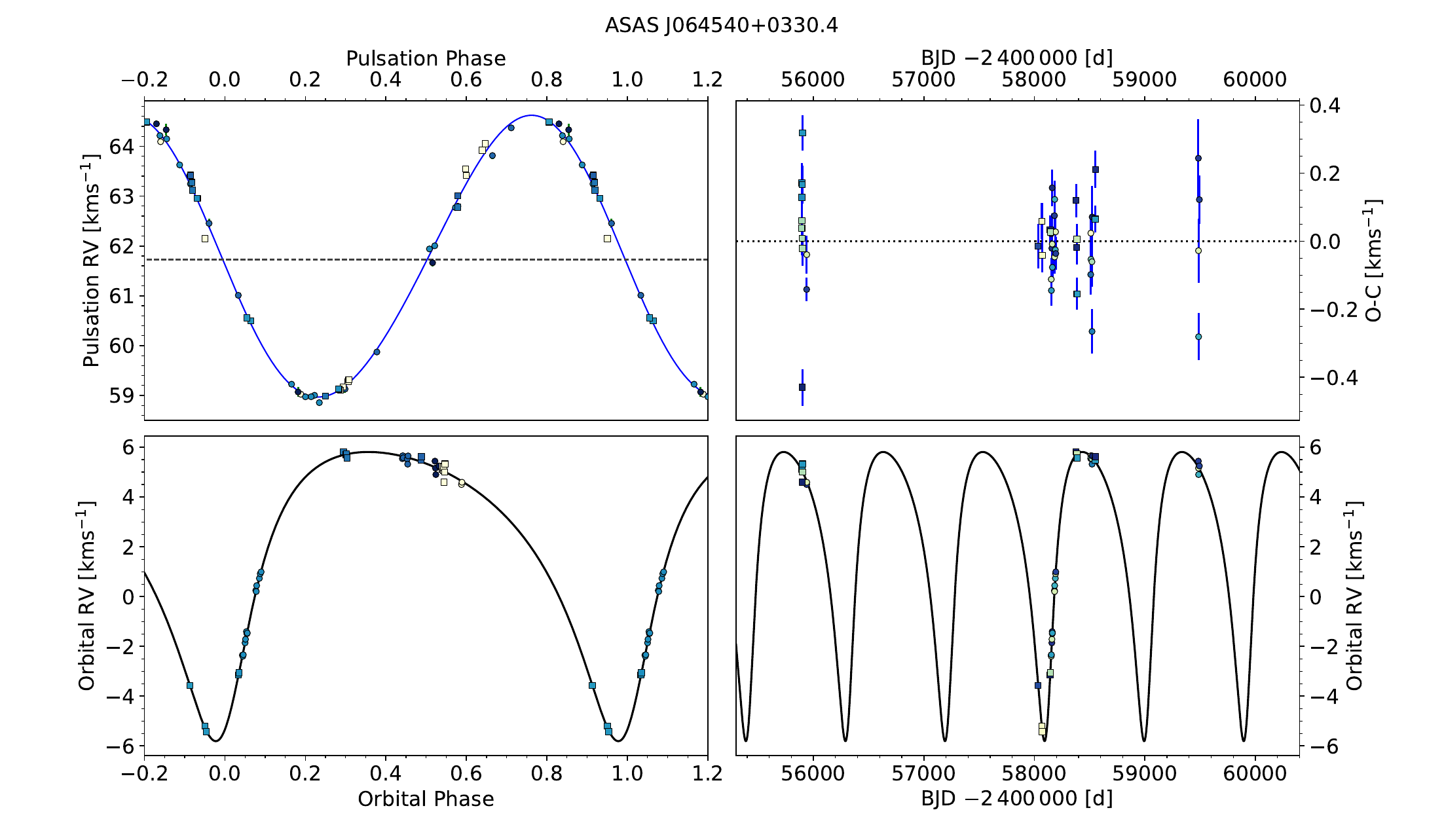}
    \caption{Pulsational and orbital fit of ASAS J064540+0330.4. Figure description is same as Figure~\ref{fig:Example_kepler1}.}
    \label{fig:ASAS0330Orbit}
\end{figure*}

\begin{figure*}
    \centering
    \includegraphics[scale=0.5]{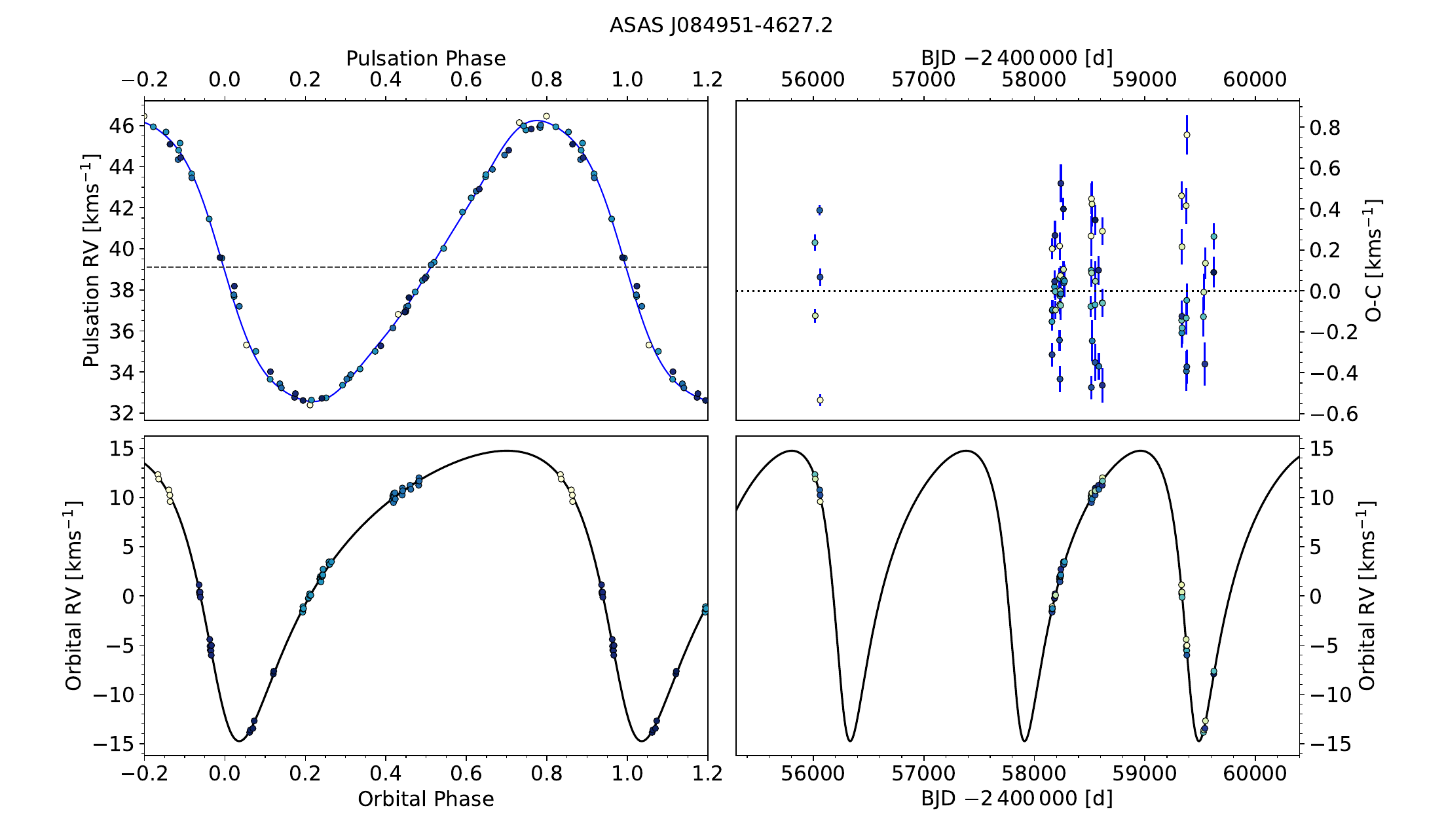}
    \caption{Pulsational and orbital fit of ASAS J084951-4627.2. Figure description is same as Figure~\ref{fig:Example_kepler1}. The pulsational RV curve amplitude modulation is apparent in the top left panel.}
    \label{fig:ASAS4627Orbit}
\end{figure*}

\begin{figure*}
    \centering
    \includegraphics[scale=0.5]{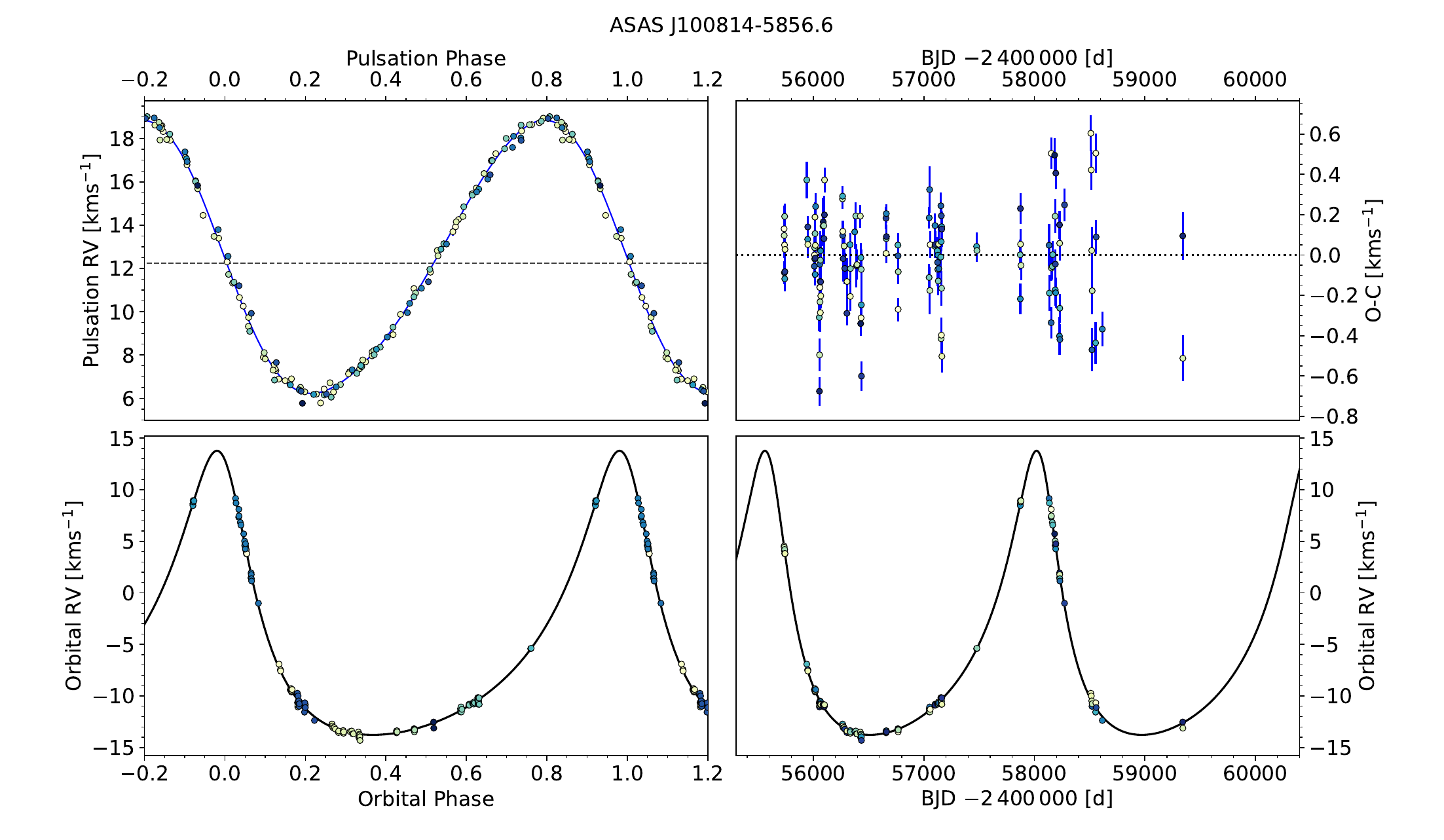}
    \caption{Pulsational and orbital fit of ASAS J100814-5856.6. Figure description is same as Figure~\ref{fig:Example_kepler1}. }
    \label{fig:ASAS5856Orbit}
\end{figure*}

\begin{figure*}
    \centering
    \includegraphics[scale=0.5]{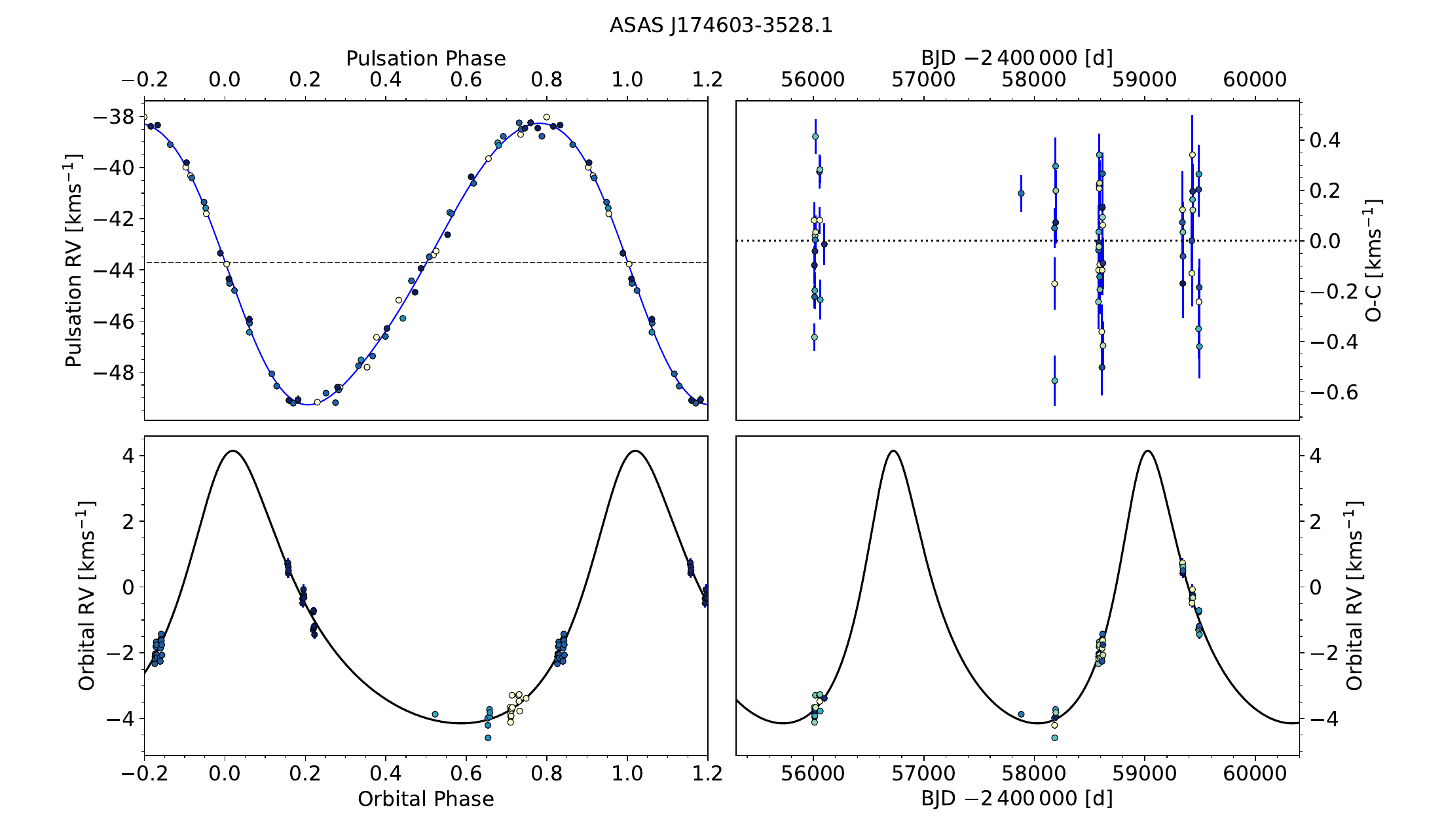}
    \caption{Pulsational and orbital fit of ASAS J174603-3528.1. Figure description is same as Figure~\ref{fig:Example_kepler1}.}
    \label{fig:ASAS3528Orbit}
\end{figure*}

\begin{figure*}
    \centering
    \includegraphics[scale=0.5]{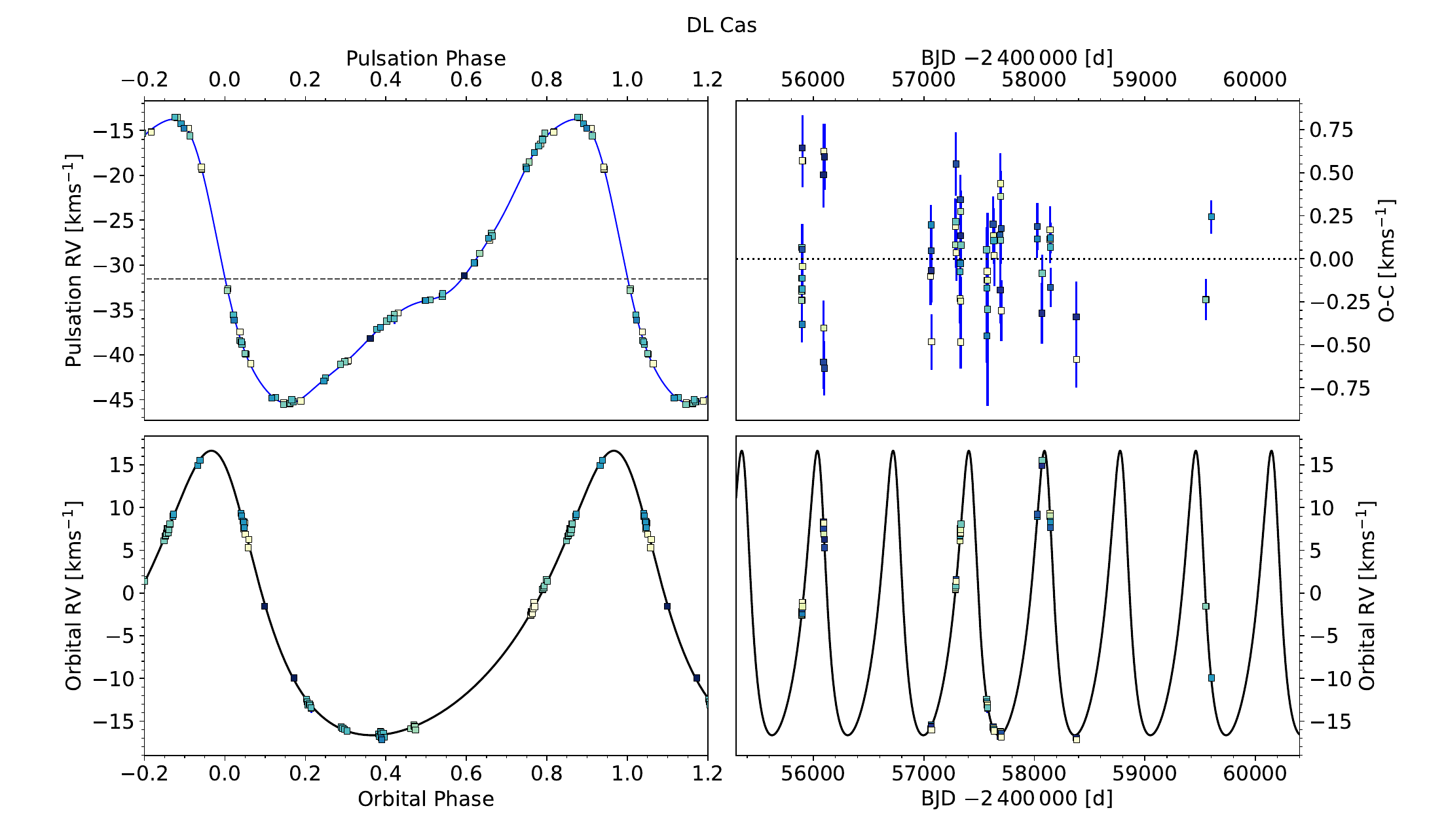}
    \caption{Pulsational and orbital fit of DL Cas. Figure description is same as Figure~\ref{fig:Example_kepler1}.}
    \label{fig:DLCasOrbit}
\end{figure*}

\begin{figure*}
    \centering
    \includegraphics[scale=0.5]{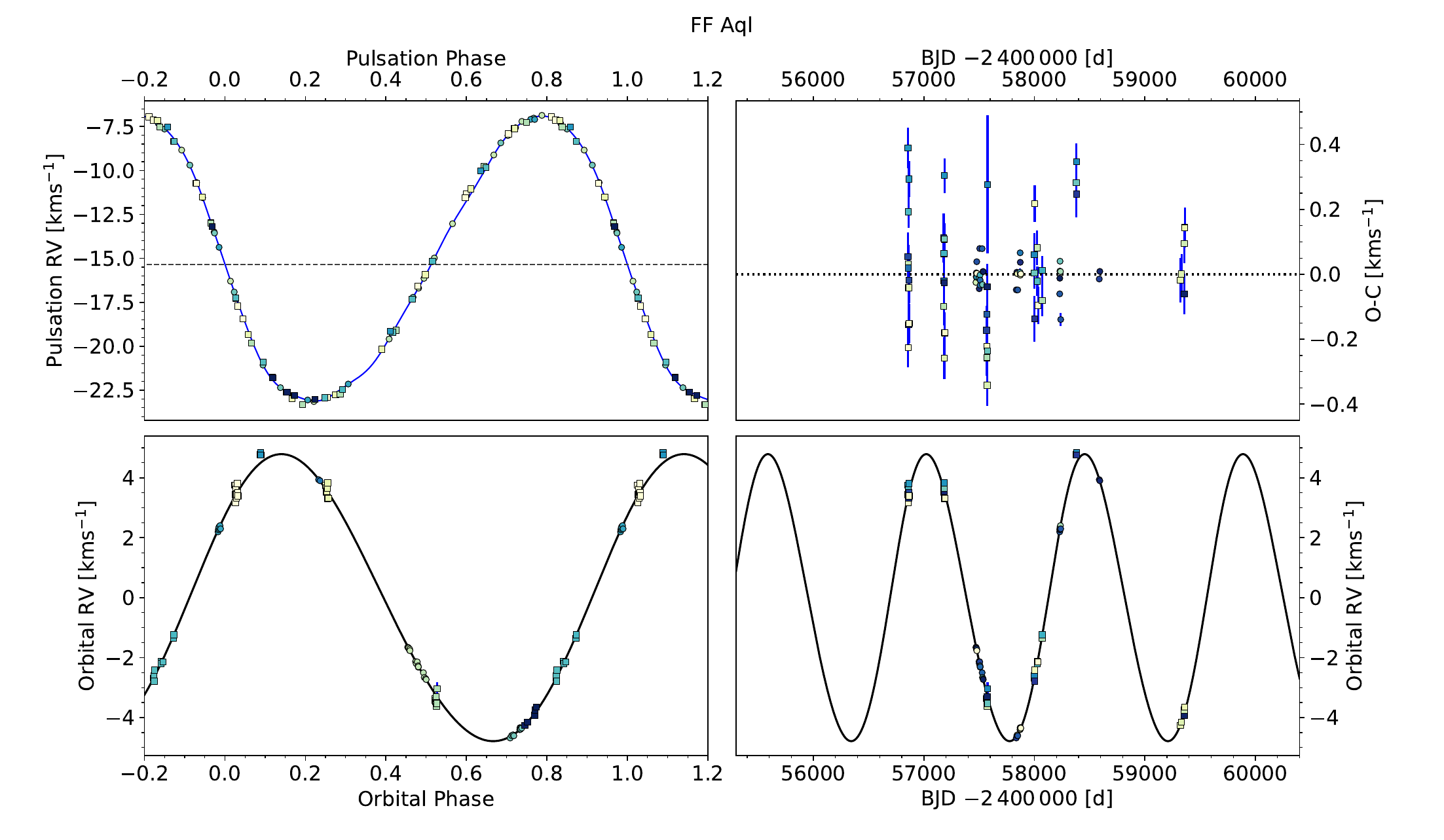}
    \caption{Pulsational and orbital fit of FF Aql. Figure description is same as Figure~\ref{fig:Example_kepler1}.}
    \label{fig:FFAqlOrbit}
\end{figure*}

\begin{figure*}
    \centering
    \includegraphics[scale=0.5]{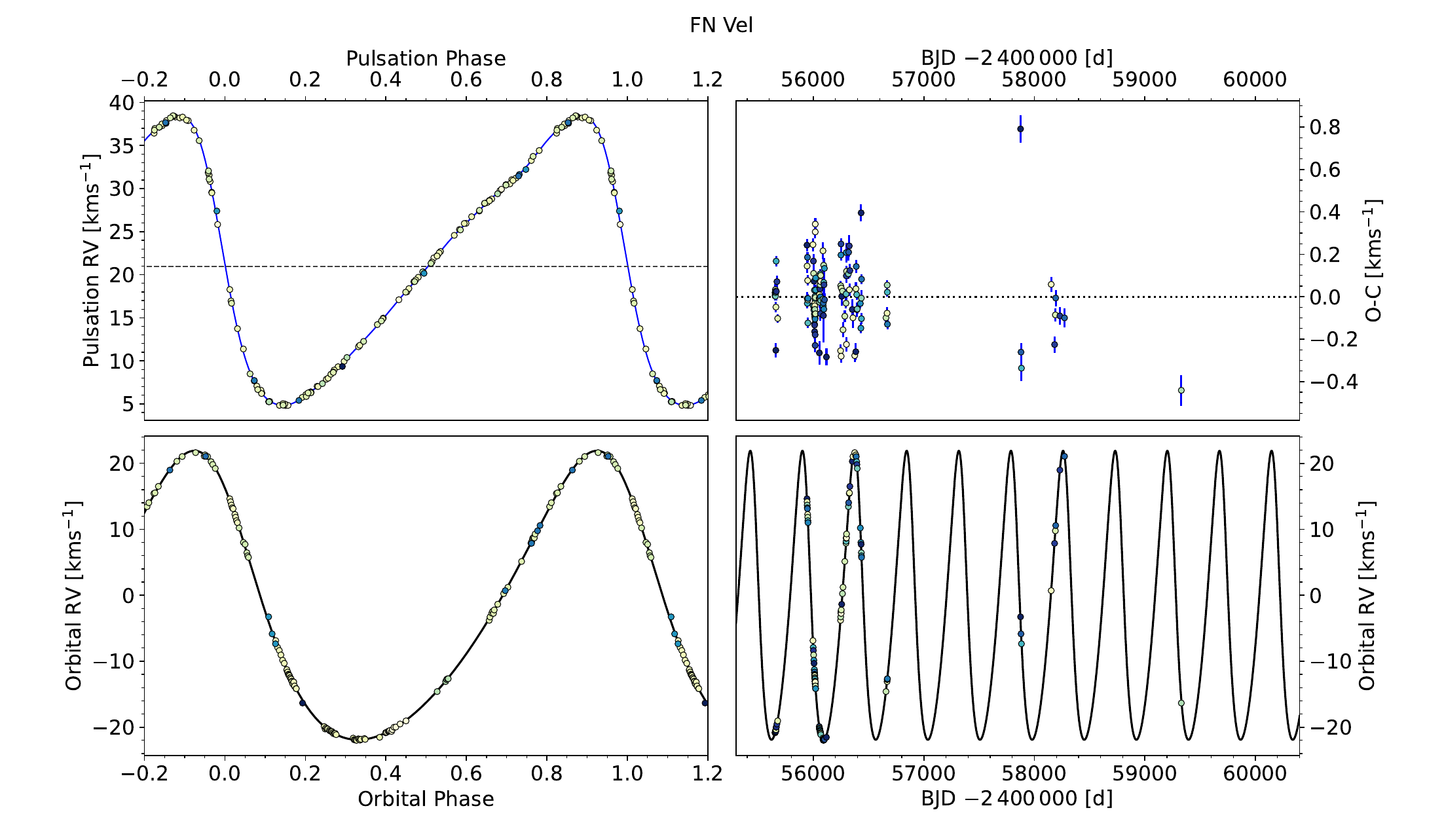}
    \caption{Pulsational and orbital fit of FN Vel. Figure description is same as Figure~\ref{fig:Example_kepler1}.}
    \label{fig:FNVelOrbit}
\end{figure*}

\begin{figure*}
    \centering
    \includegraphics[scale=0.5]{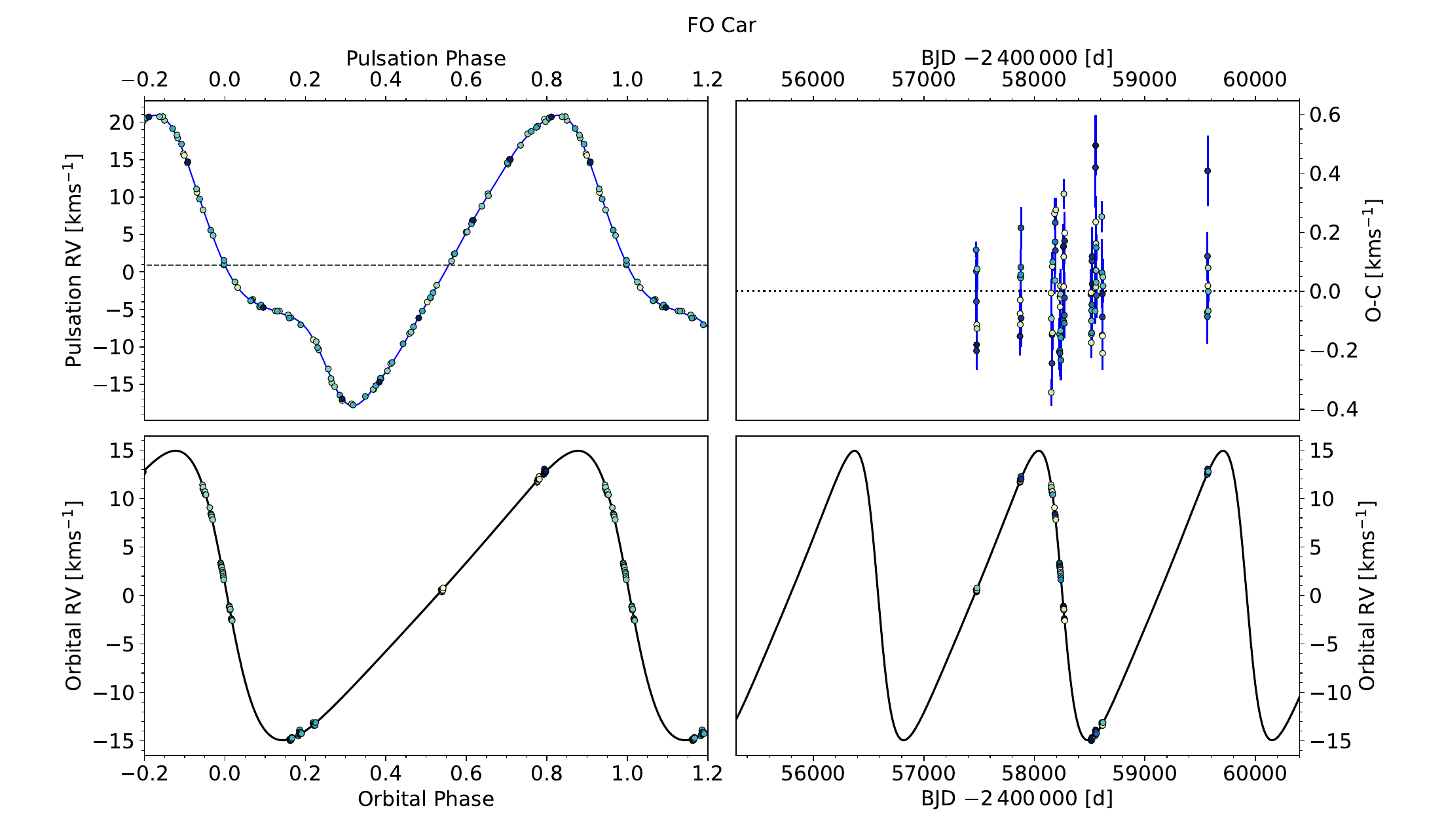}
    \caption{Pulsational and orbital fit of FO Car. Figure description is same as Figure~\ref{fig:Example_kepler1}.}
    \label{fig:FOCarOrbit}
\end{figure*}

\begin{figure*}
    \centering
    \includegraphics[scale=0.5]{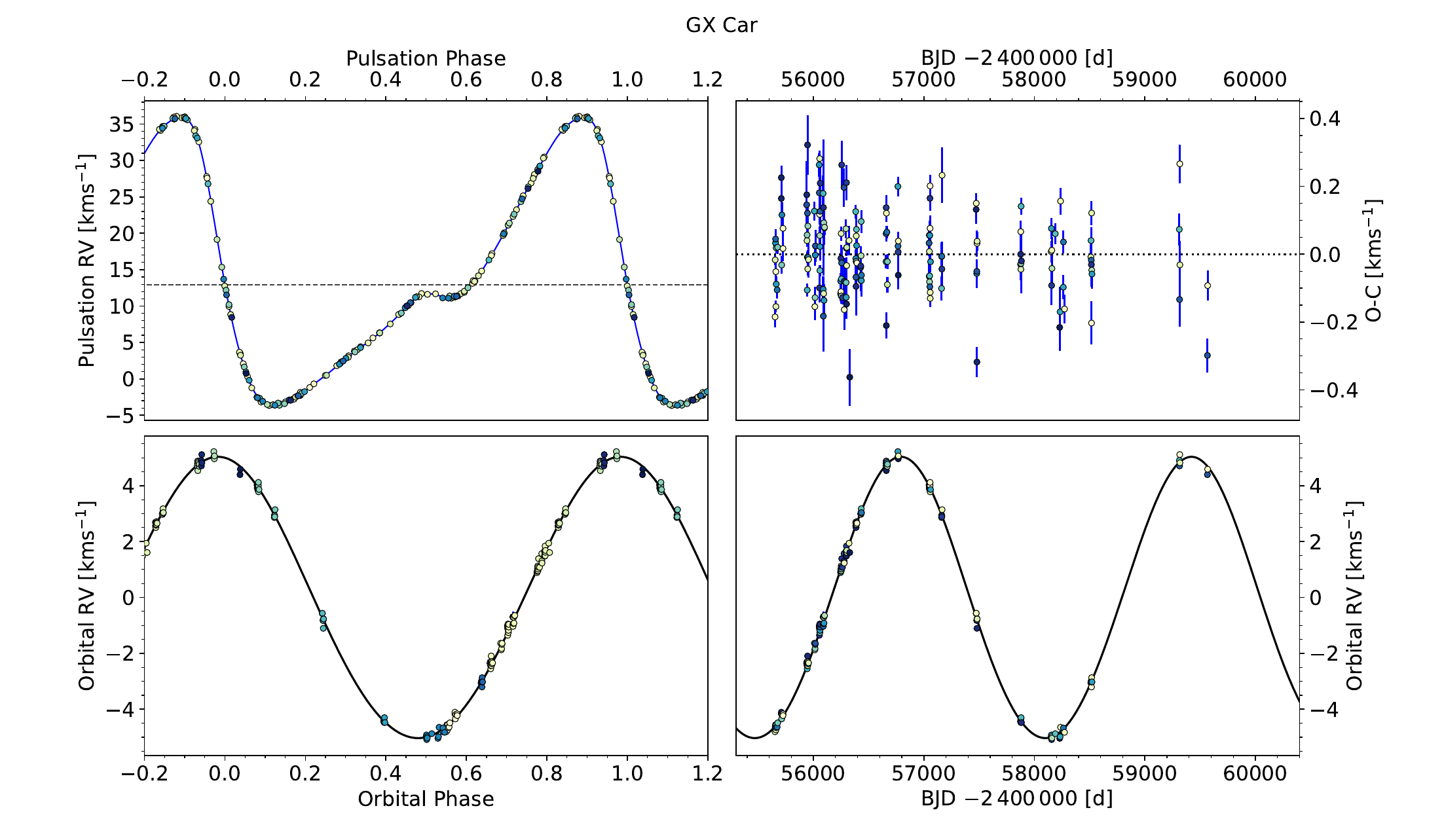}
    \caption{Pulsational and orbital fit of GX Car. Figure description is same as Figure~\ref{fig:Example_kepler1}.}
    \label{fig:GXCarOrbit}
\end{figure*}

\begin{figure*}
    \centering
    \includegraphics[scale=0.5]{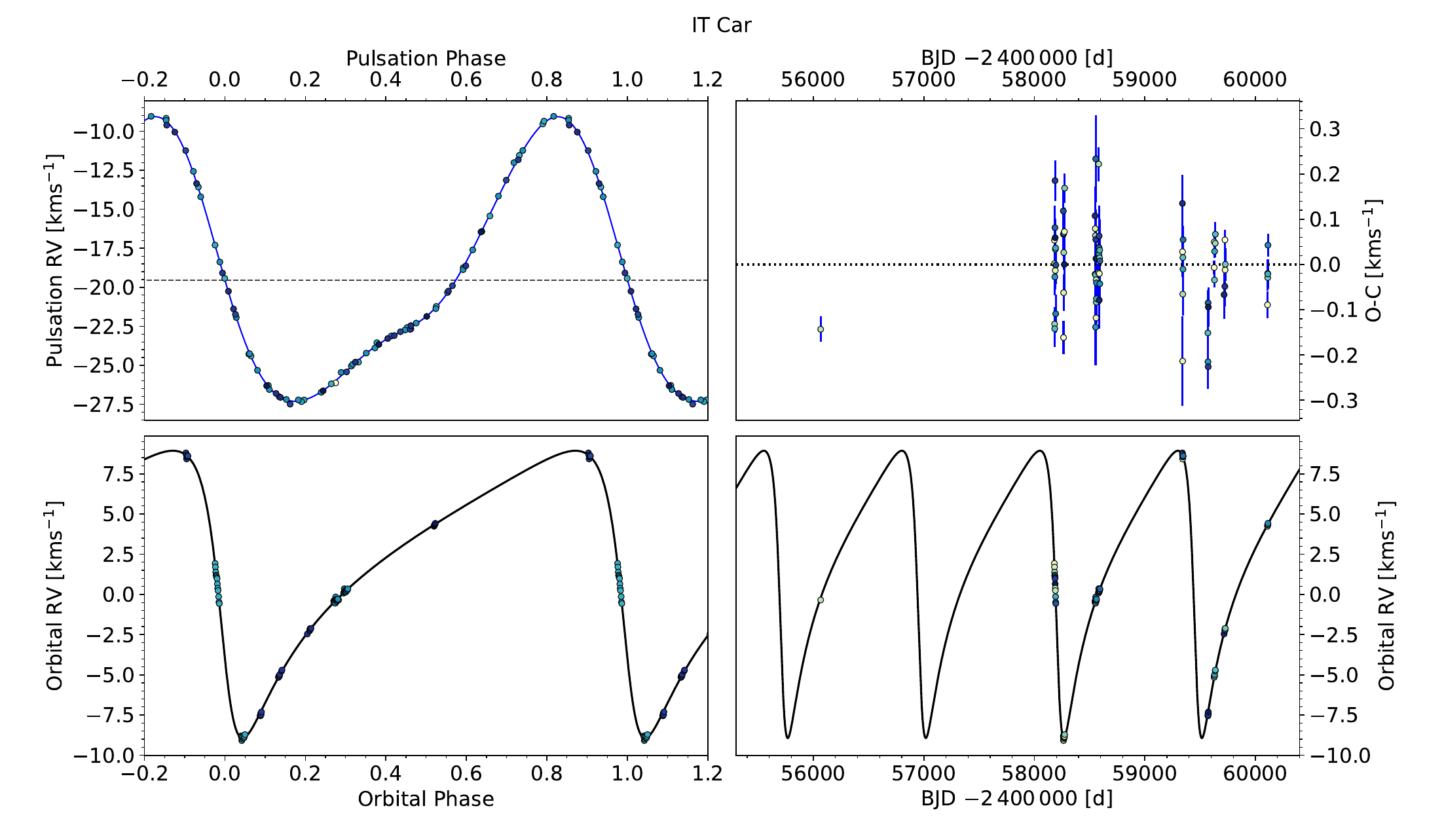}
    \caption{Pulsational and orbital fit of IT Car. Figure description is same as Figure~\ref{fig:Example_kepler1}.}
    \label{fig:ITCarOrbit}
\end{figure*}

\begin{figure*}
    \centering
    \includegraphics[scale=0.5]{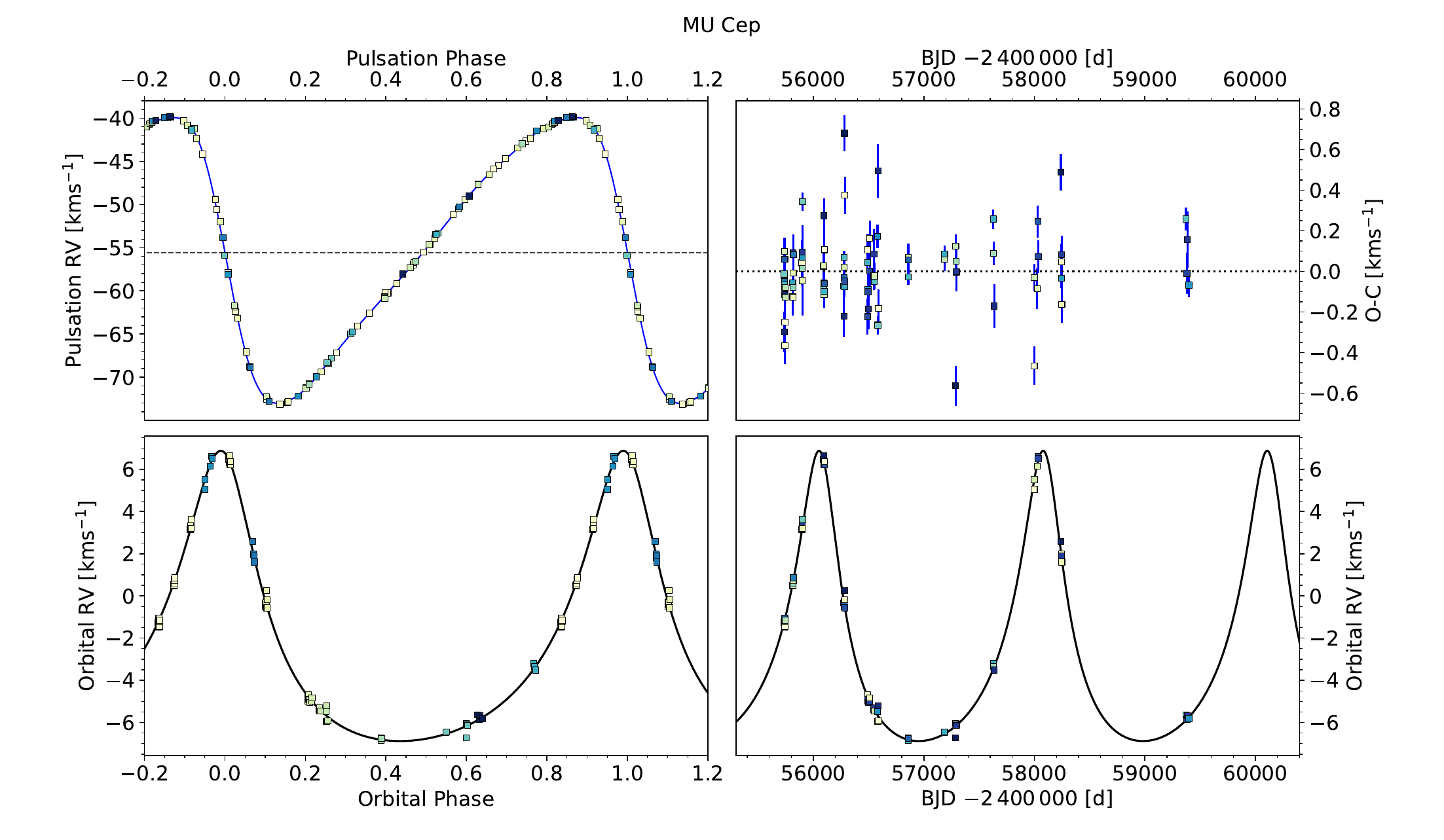}
    \caption{Pulsational and orbital fit of MU Cep. Figure description is same as Figure~\ref{fig:Example_kepler1}.}
    \label{fig:MUCepOrbit}
\end{figure*}

\begin{figure*}
    \centering
    \includegraphics[scale=0.5]{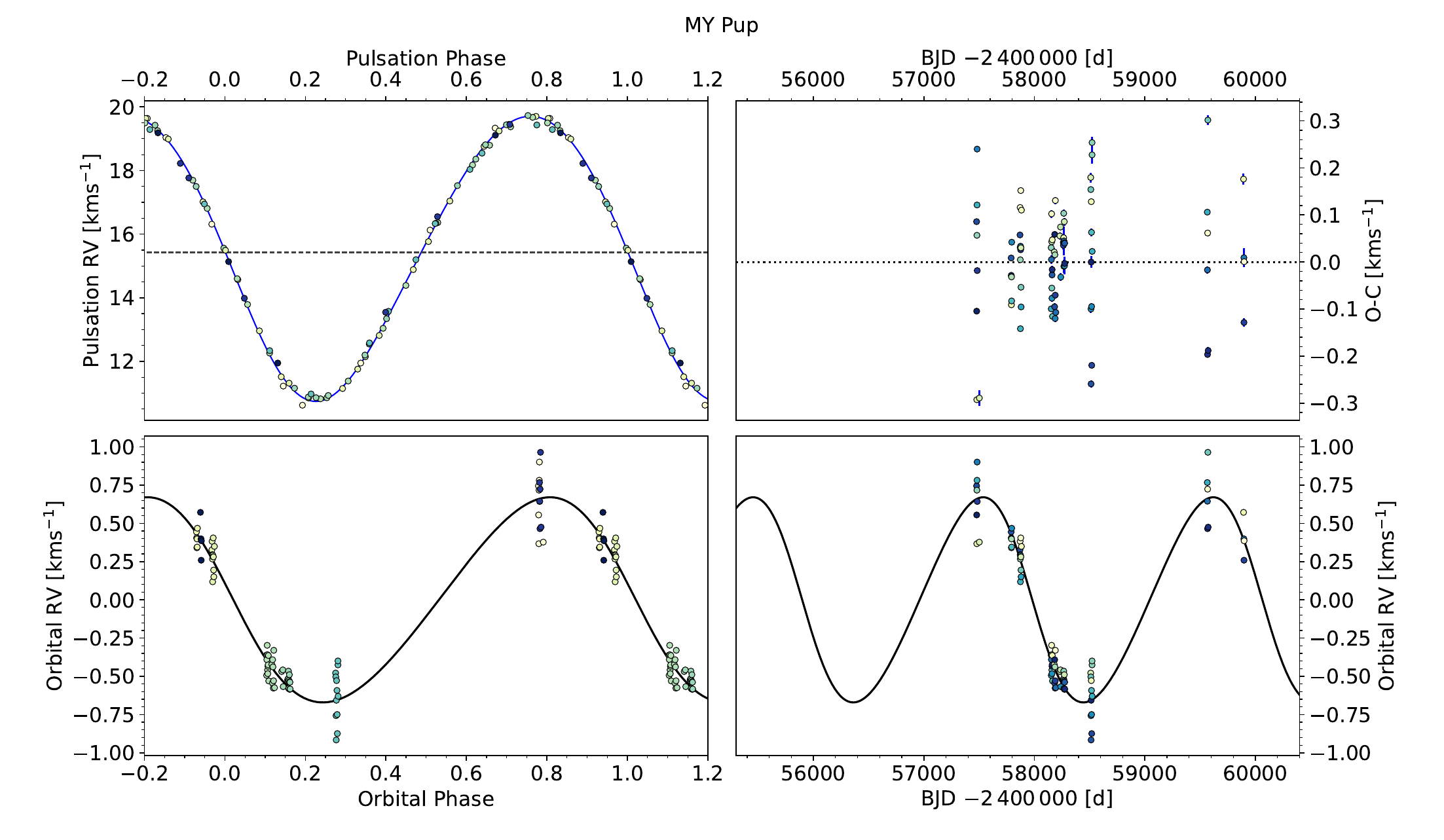}
    \caption{Pulsational and orbital fit of MY Pup. Figure description is same as Figure~\ref{fig:Example_kepler1}.}
    \label{fig:MYPupOrbit}
\end{figure*}

\begin{figure*}
    \centering
    \includegraphics[scale=0.5]{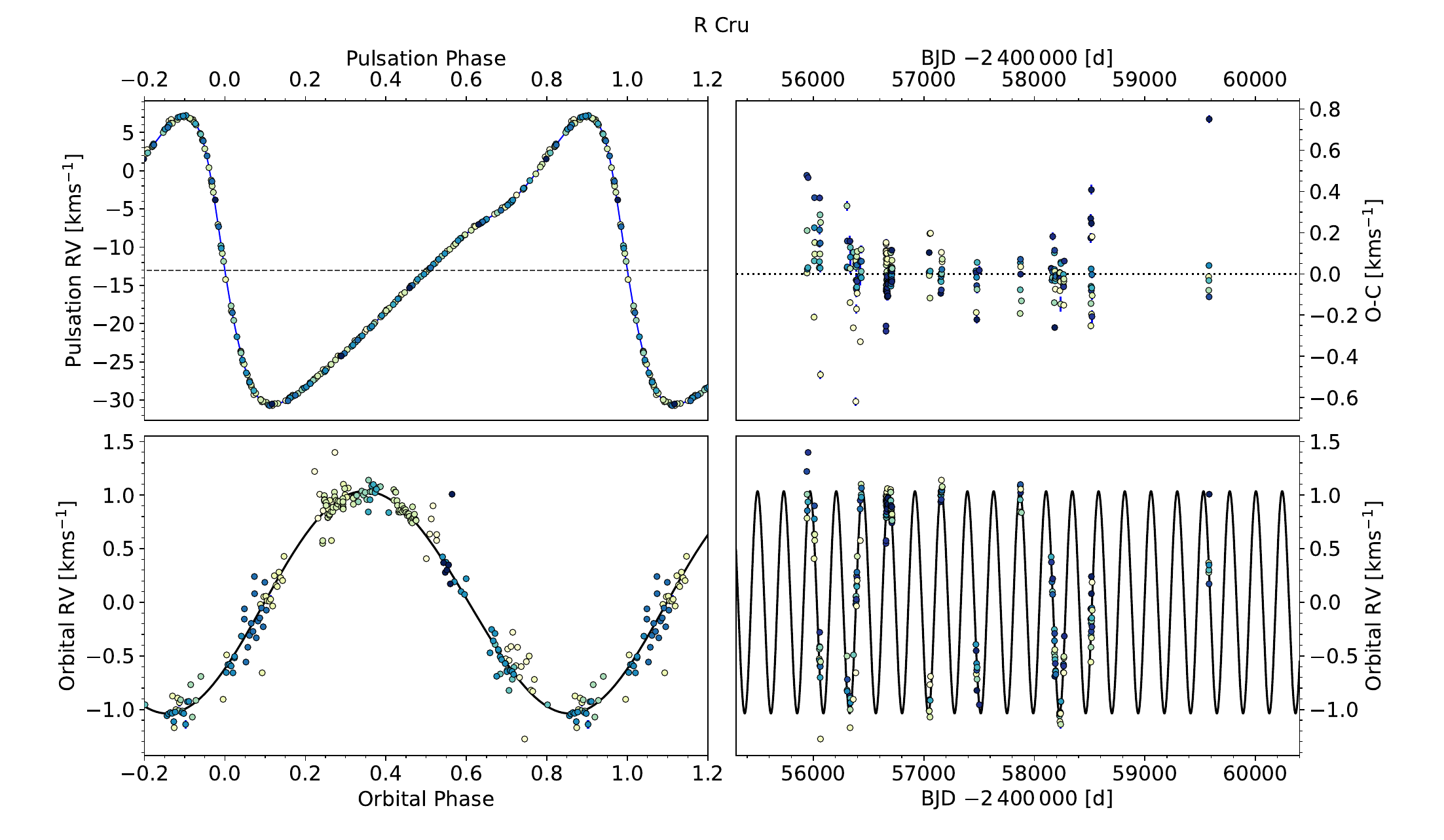}
    \caption{Pulsational and orbital fit of R Cru. Figure description is same as Figure~\ref{fig:Example_kepler1}.}
    \label{fig:RCruOrbit}
\end{figure*}

\begin{figure*}
    \centering
    \includegraphics[scale=0.5]{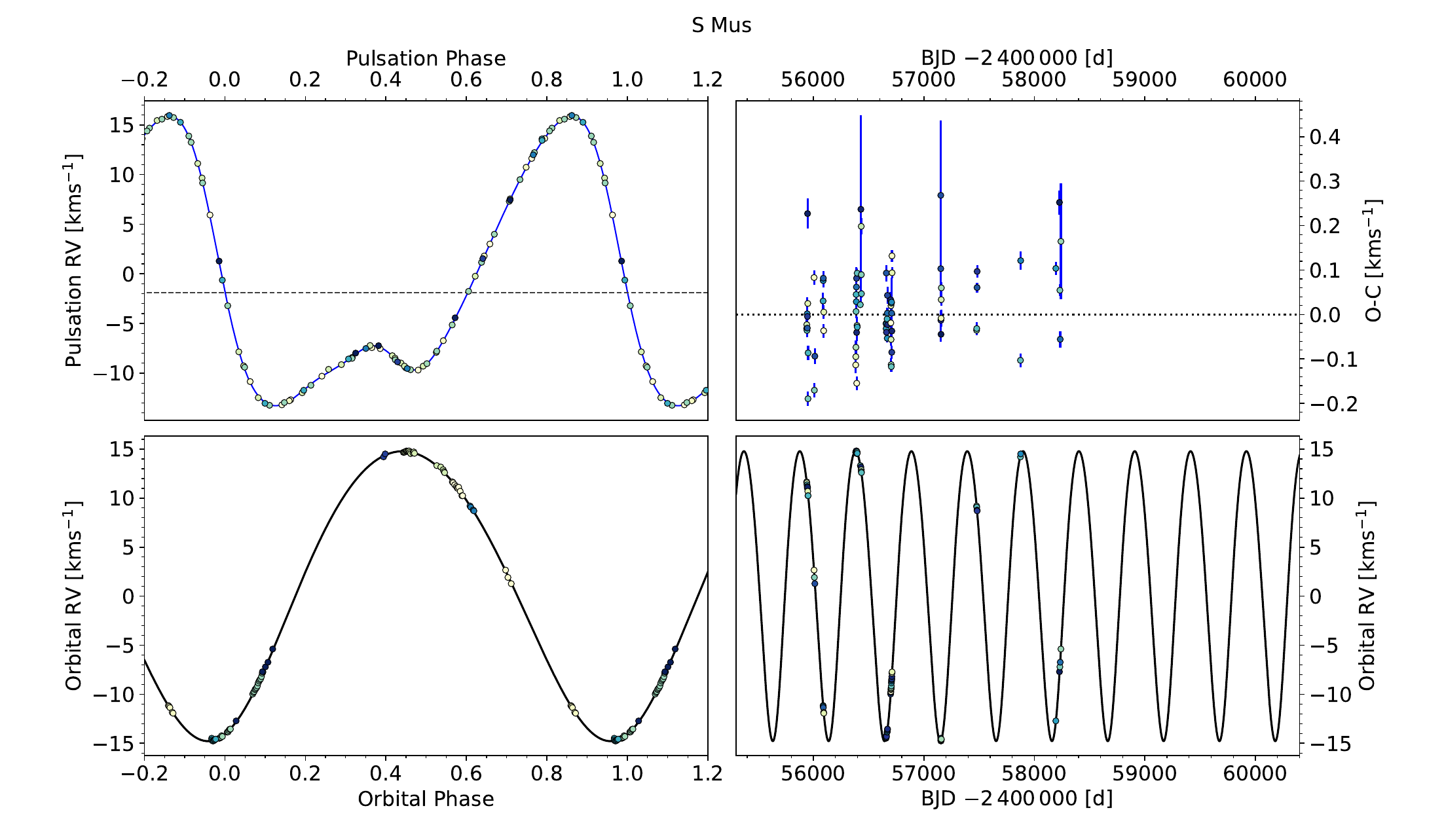}
    \caption{Pulsational and orbital fit of S Mus. Figure description is same as Figure~\ref{fig:Example_kepler1}.}
    \label{fig:SMusOrbit}
\end{figure*}

\begin{figure*}
    \centering
    \includegraphics[scale=0.5]{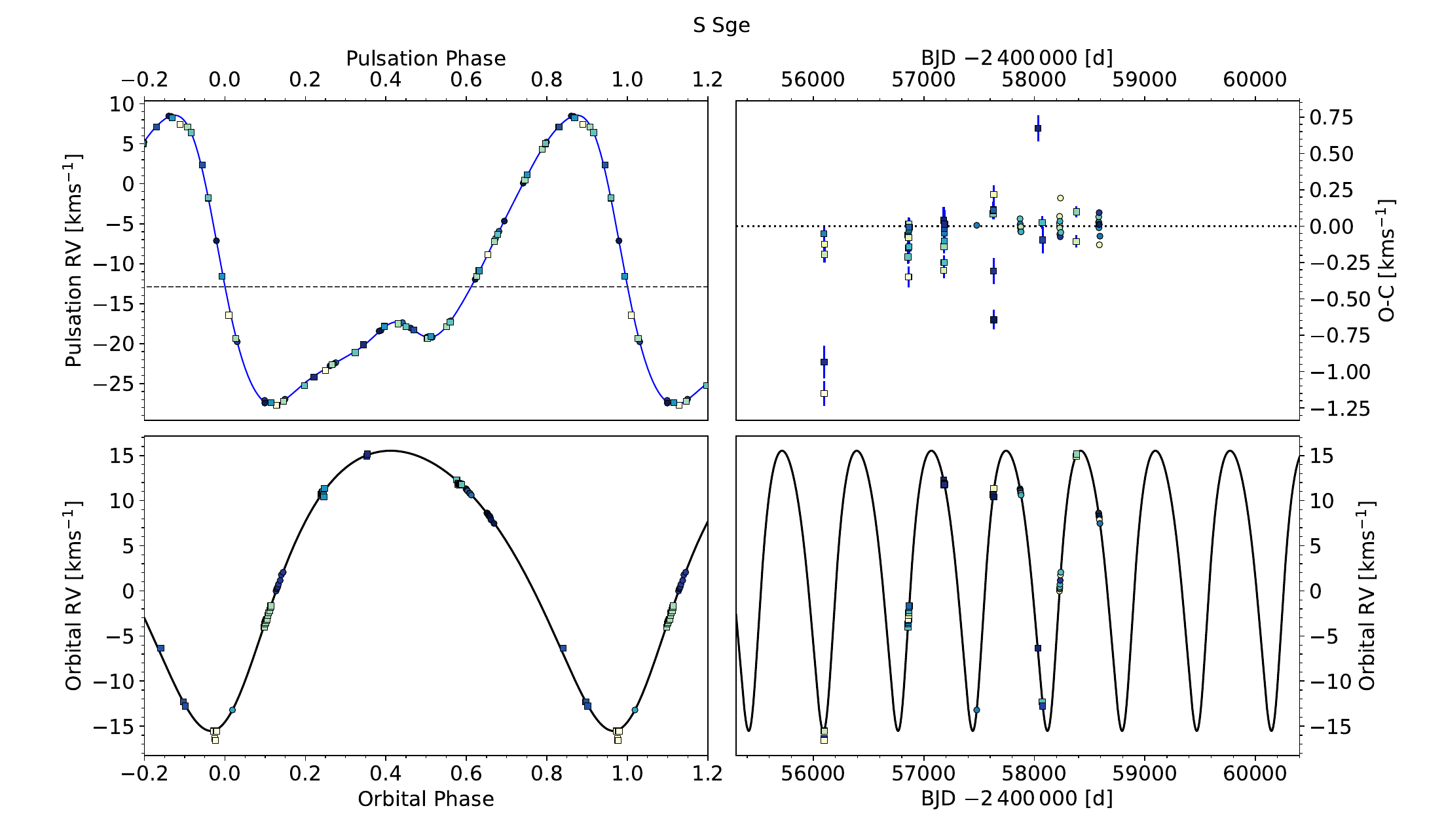}
    \caption{Pulsational and orbital fit of S Sge. Figure description is same as Figure~\ref{fig:Example_kepler1}.}
    \label{fig:SSgeOrbit}
\end{figure*}

\begin{figure*}
    \centering
    \includegraphics[scale=0.5]{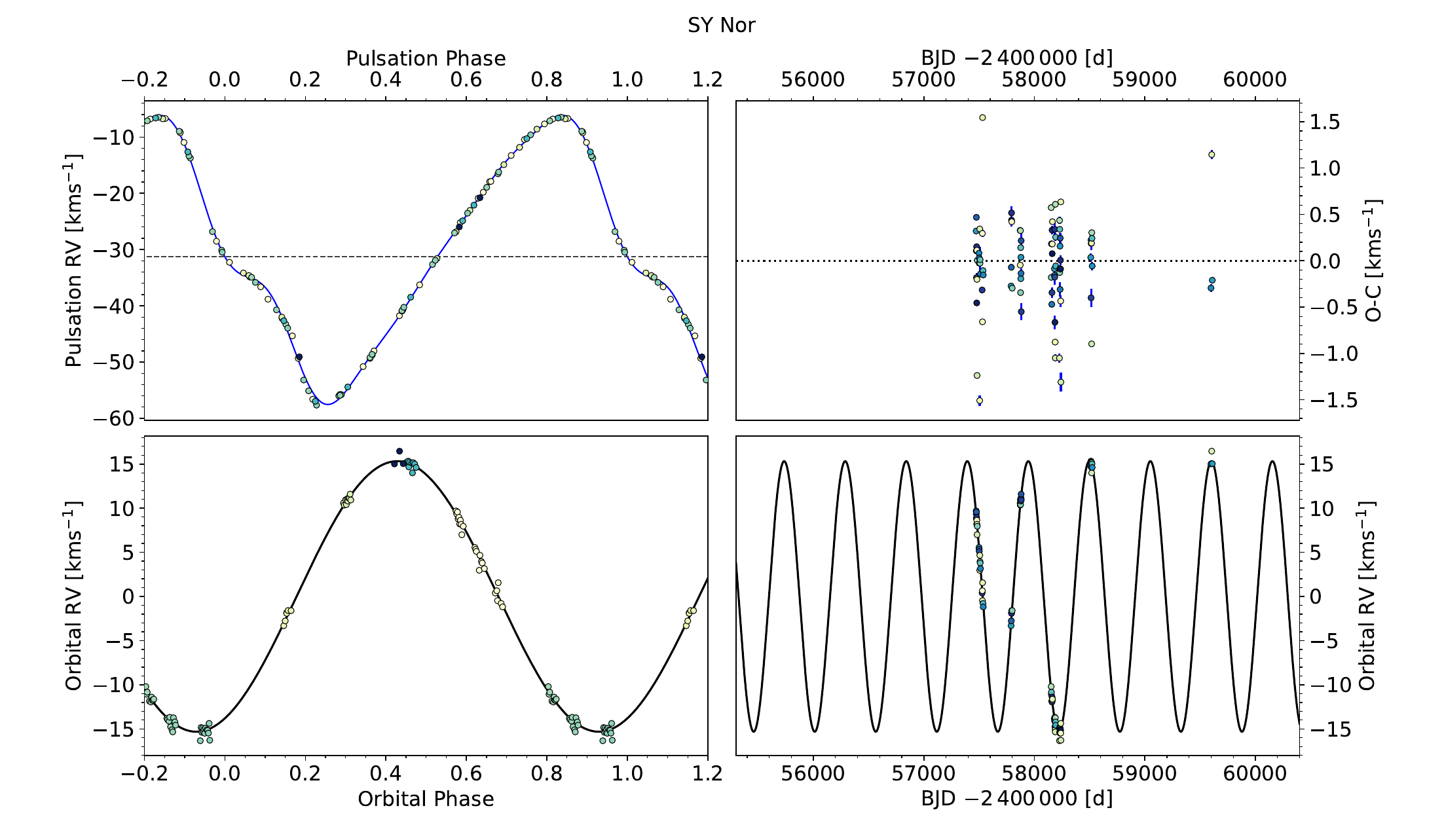}
    \caption{Pulsational and orbital fit of SY Nor. Figure description is same as Figure~\ref{fig:Example_kepler1}.}
    \label{fig:SYNorOrbit}
\end{figure*}

\begin{figure*}
    \centering
    \includegraphics[scale=0.5]{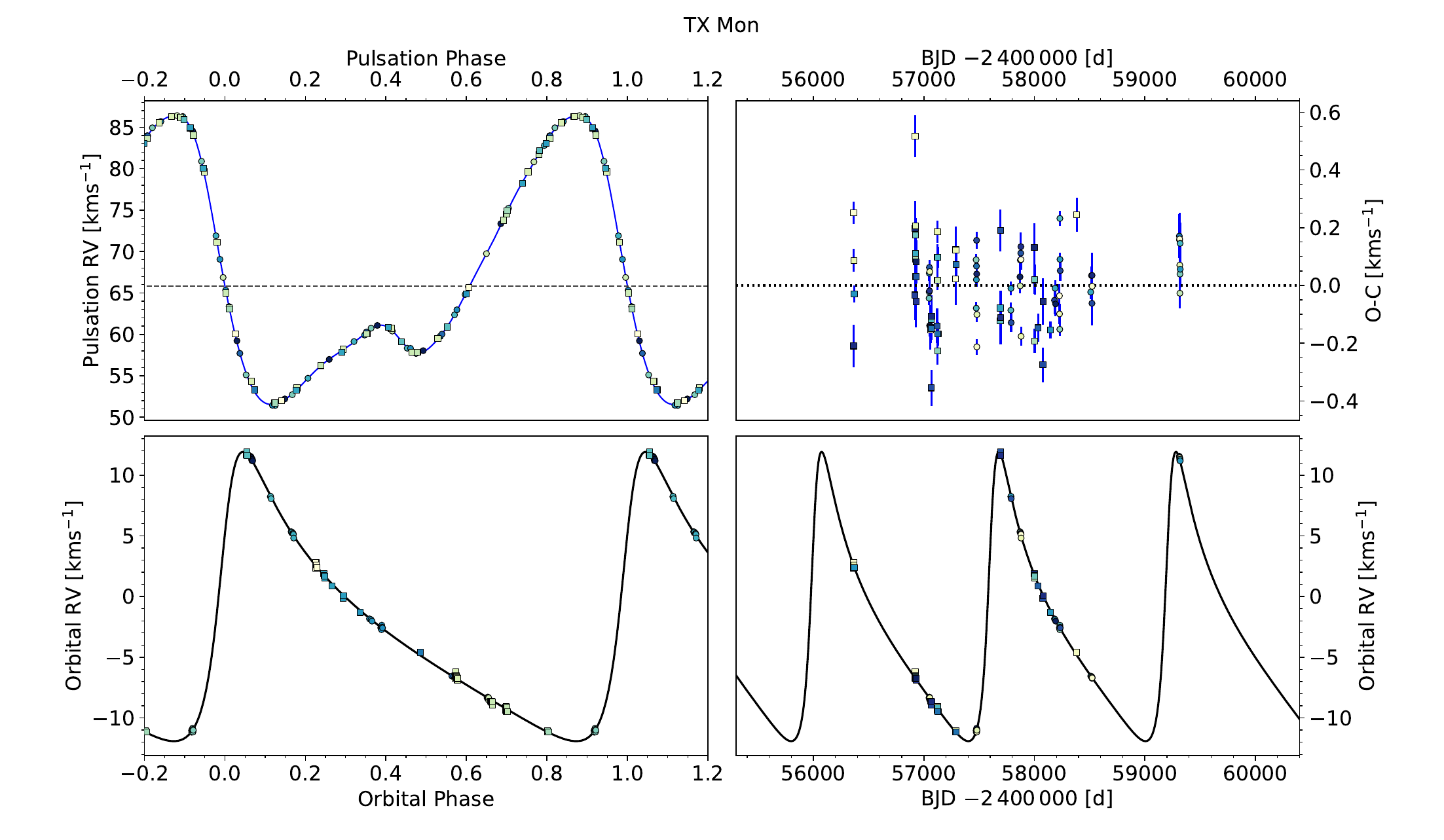}
    \caption{Pulsational and orbital fit of TX Mon. Figure description is same as Figure~\ref{fig:Example_kepler1}.}
    \label{fig:TXMonOrbit}
\end{figure*}

\begin{figure*}
    \centering
    \includegraphics[scale=0.5]{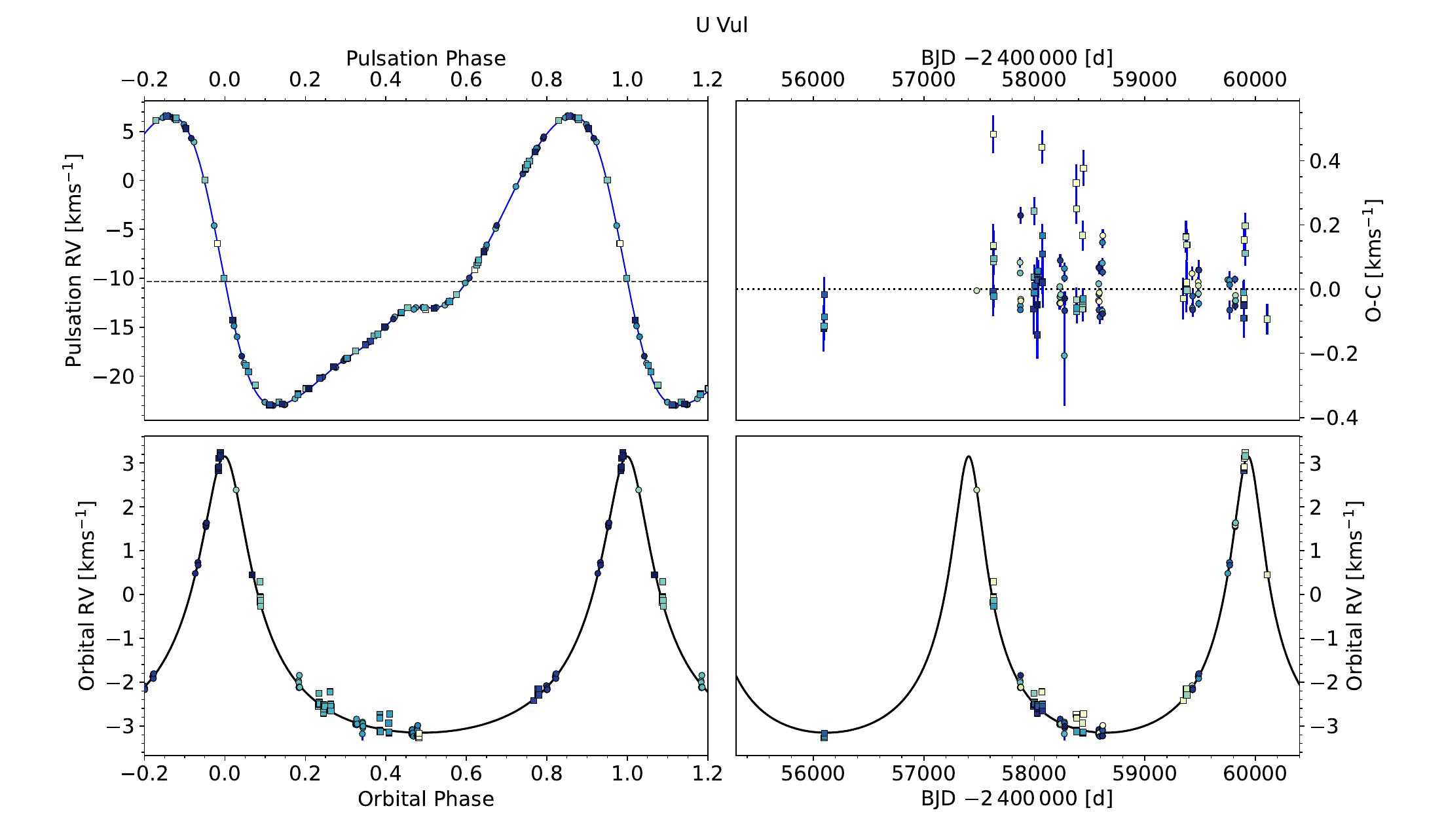}
    \caption{Pulsational and orbital fit of U Vul. Figure description is same as Figure~\ref{fig:Example_kepler1}.}
    \label{fig:UVulOrbit}
\end{figure*}

\begin{figure*}
    \centering
    \includegraphics[scale=0.5]{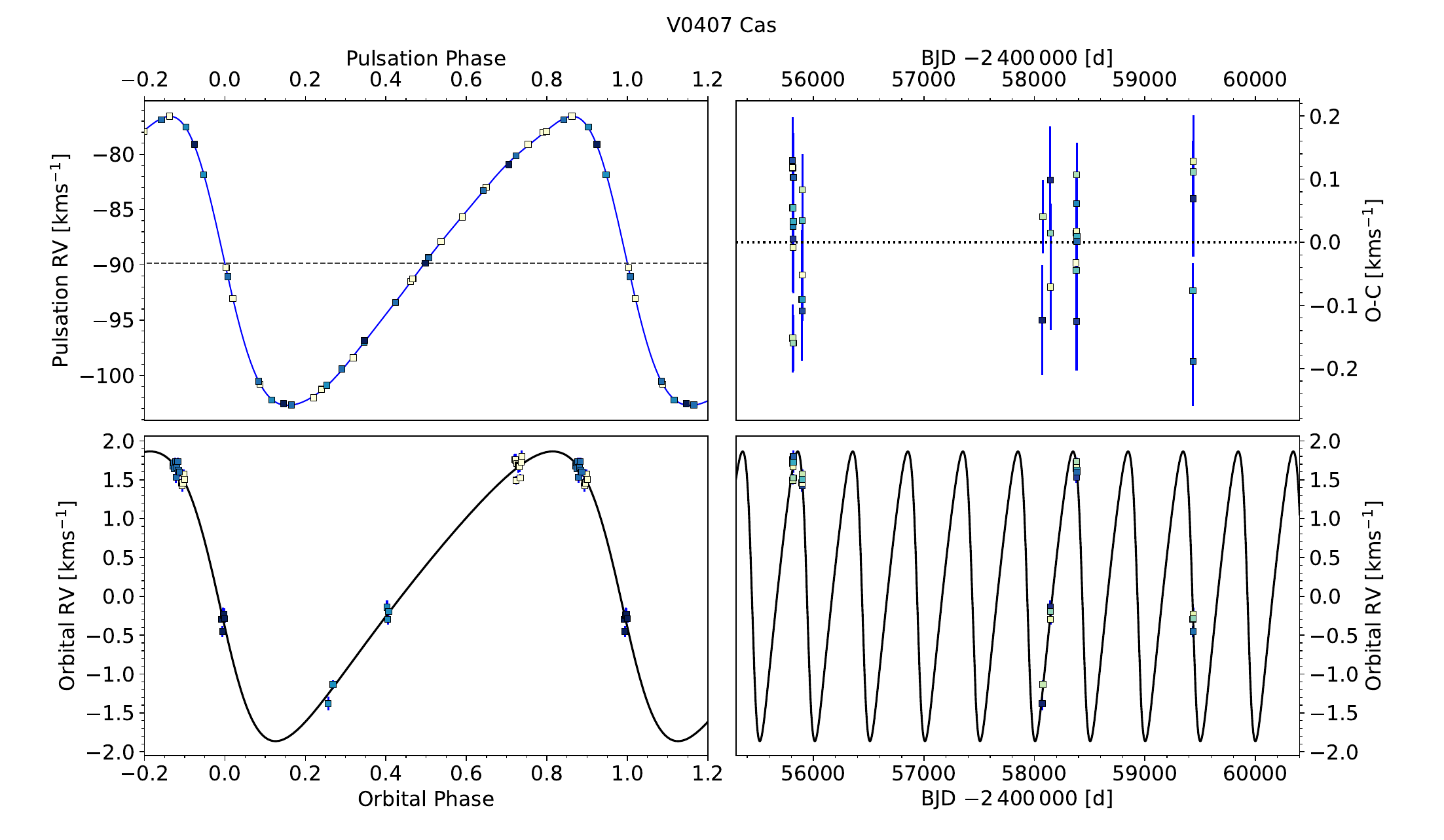}
    \caption{Pulsational and orbital fit of V0407 Cas. Figure description is same as Figure~\ref{fig:Example_kepler1}.}
    \label{fig:V0407CasOrbit}
\end{figure*}

\begin{figure*}
    \centering
    \includegraphics[scale=0.5]{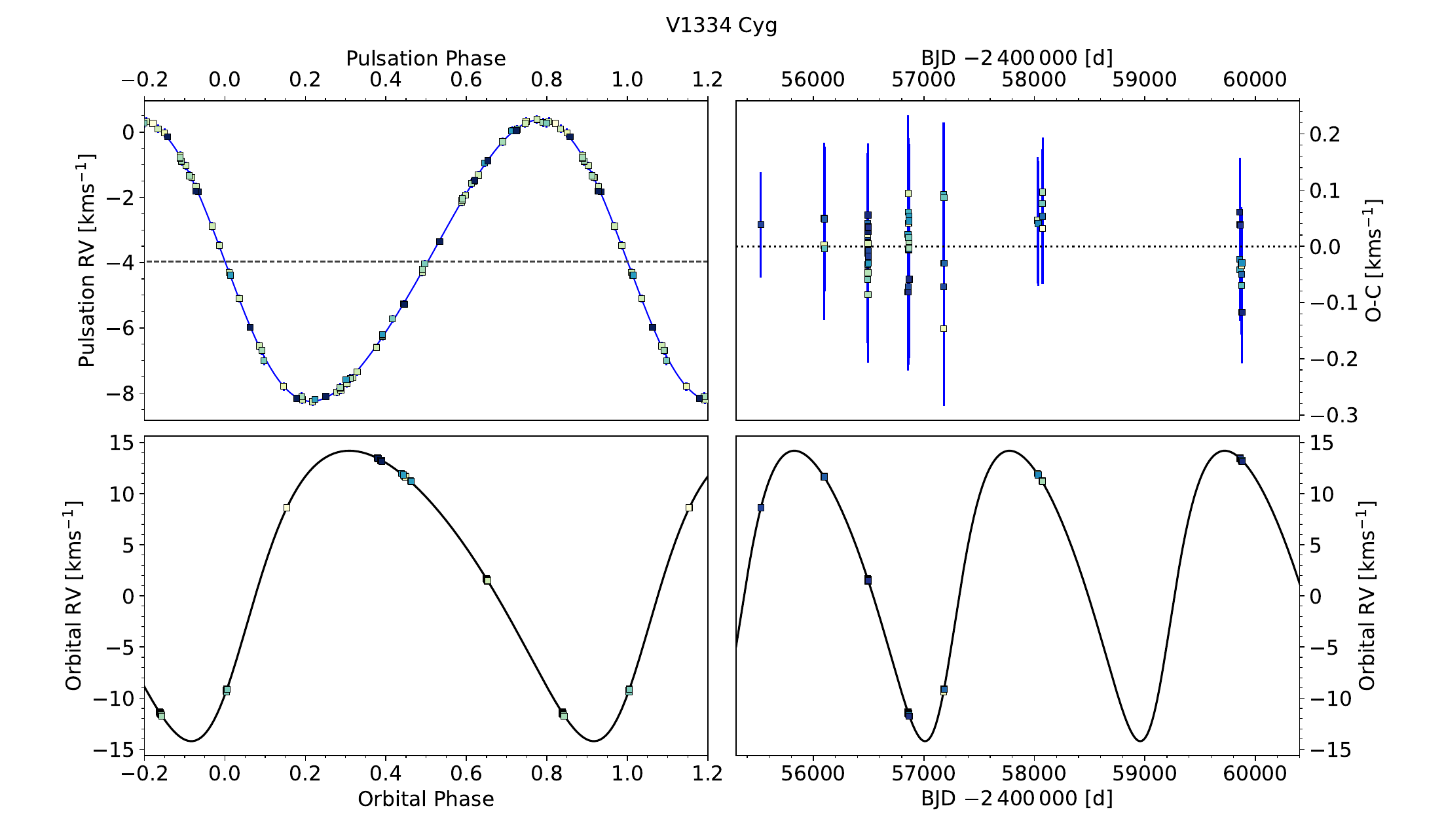}
    \caption{Pulsational and orbital fit of V1334 Cyg. Figure description is same as Figure~\ref{fig:Example_kepler1}.}
    \label{fig:V1334CygOrbit}
\end{figure*}

\begin{figure*}
    \centering
    \includegraphics[scale=0.5]{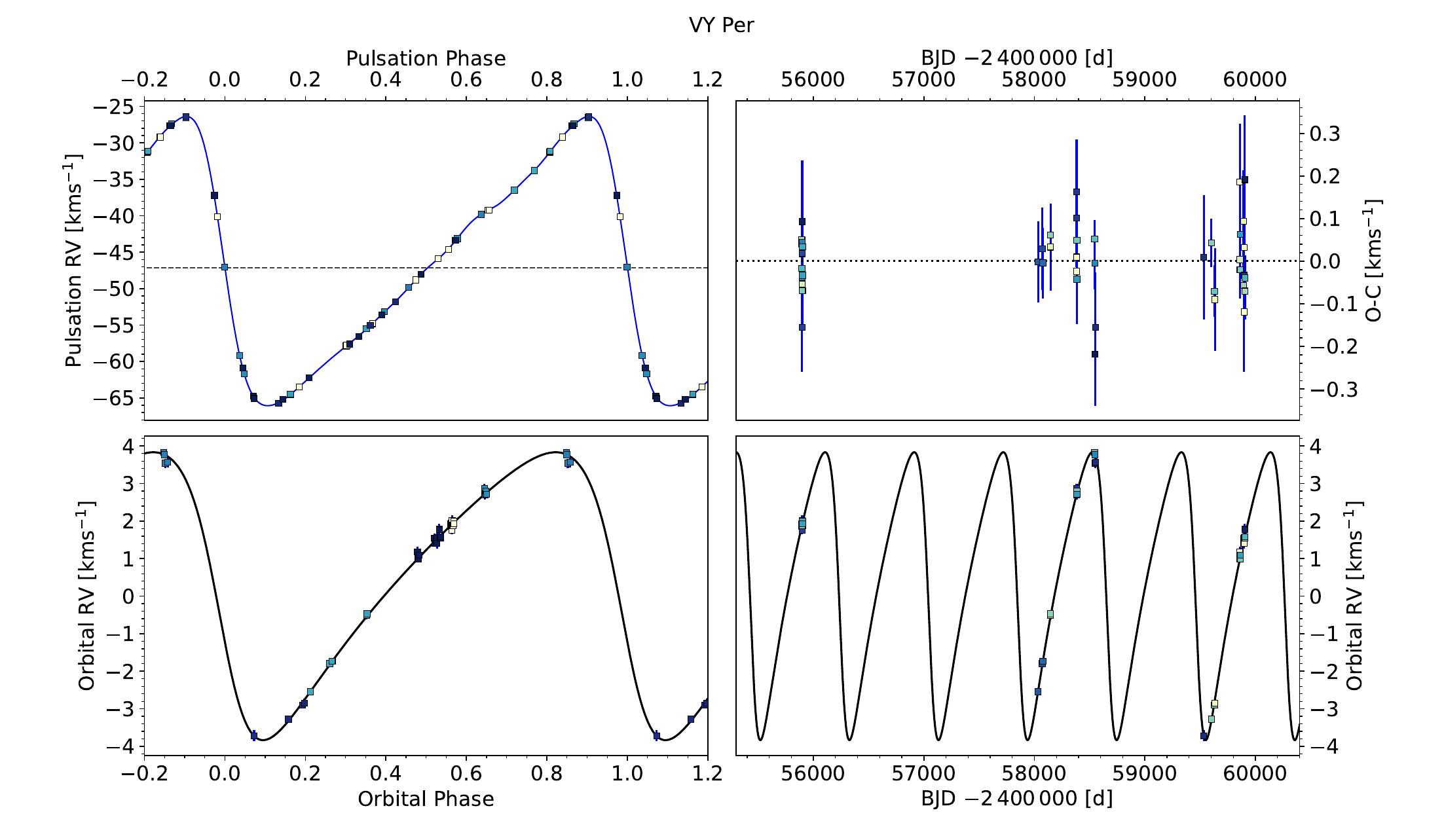}
    \caption{Pulsational and orbital fit of VY Per. Figure description is same as Figure~\ref{fig:Example_kepler1}.}
    \label{fig:VYPerOrbit}
\end{figure*}

\begin{figure*}
    \centering
    \includegraphics[scale=0.5]{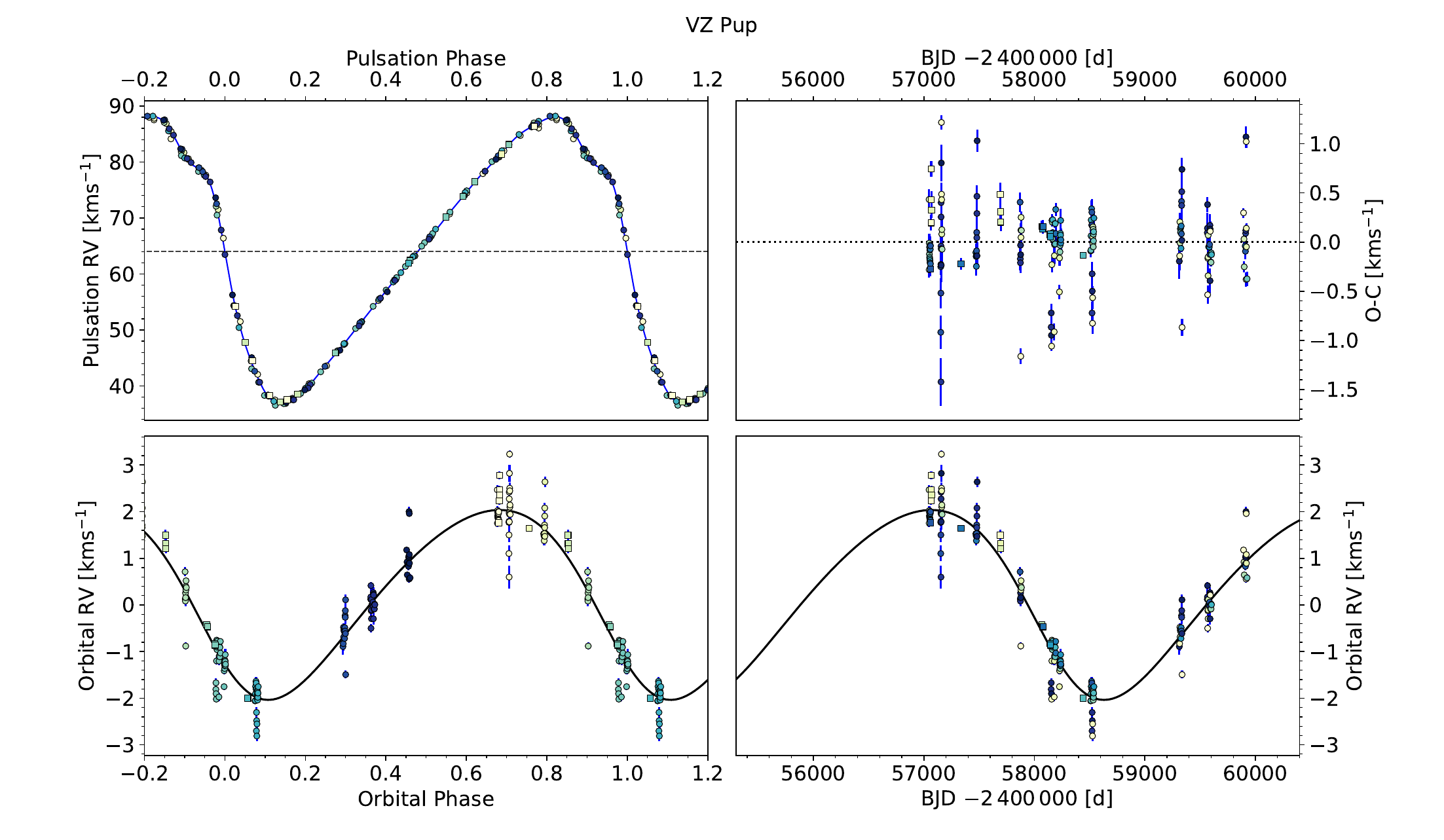}
    \caption{Pulsational and orbital fit of VZ Pup. Figure description is same as Figure~\ref{fig:Example_kepler1}.}
    \label{fig:VZPupOrbit}
\end{figure*}

\begin{figure*}
    \centering
    \includegraphics[scale=0.5]{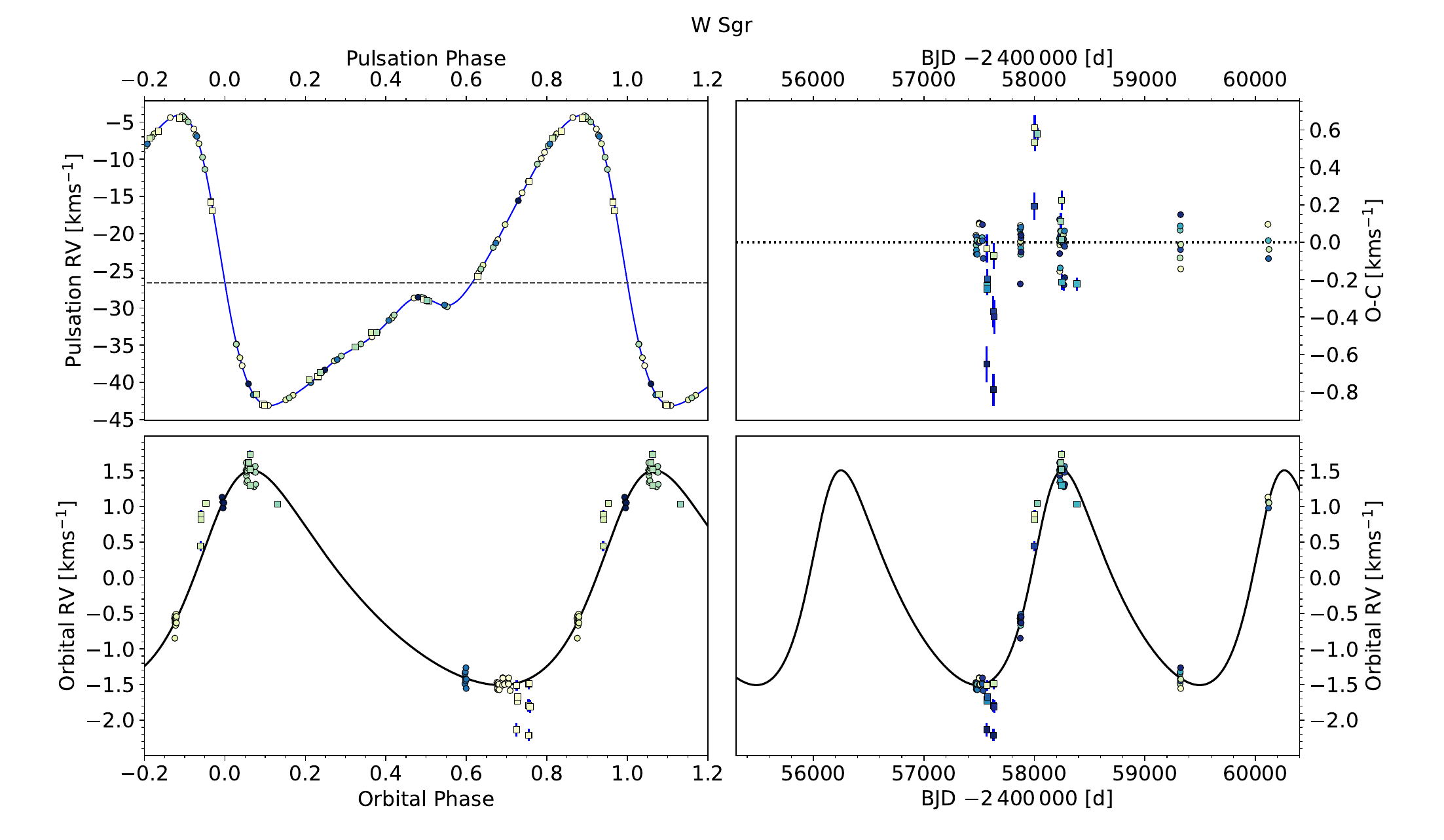}
    \caption{Pulsational and orbital fit of W Sgr. Figure description is same as Figure~\ref{fig:Example_kepler1}. The bottom right panel suggests that the orbit of W~Sgr could be more eccentric than the best fit model obtained here.}
    \label{fig:WSgrOrbit}
\end{figure*}

\begin{figure*}
    \centering
    \includegraphics[scale=0.5]{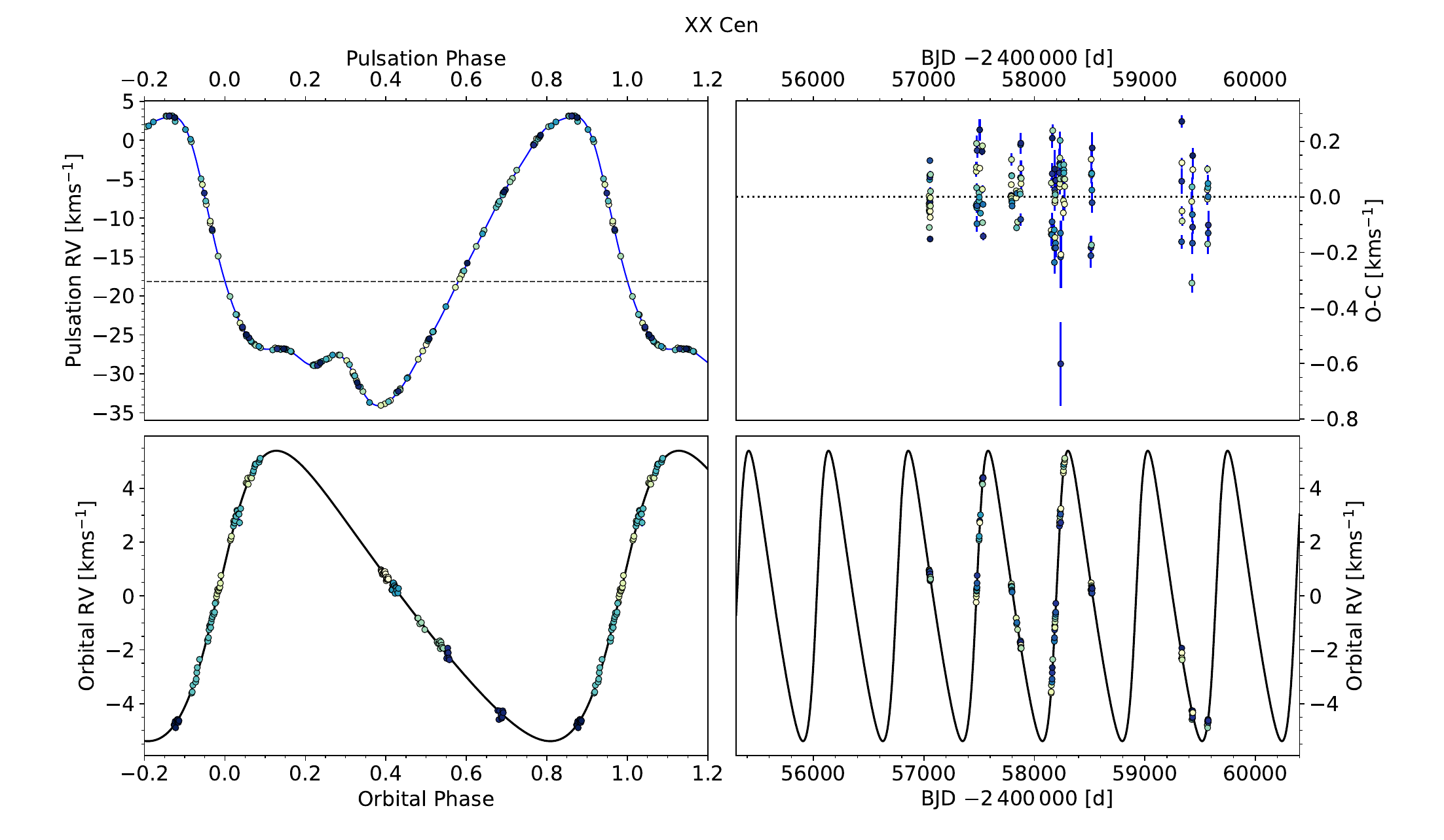}
    \caption{Pulsational and orbital fit of XX Cen. Figure description is same as Figure~\ref{fig:Example_kepler1}.}
    \label{fig:XXCenOrbit}
\end{figure*}

\begin{figure*}
    \centering
    \includegraphics[scale=0.5]{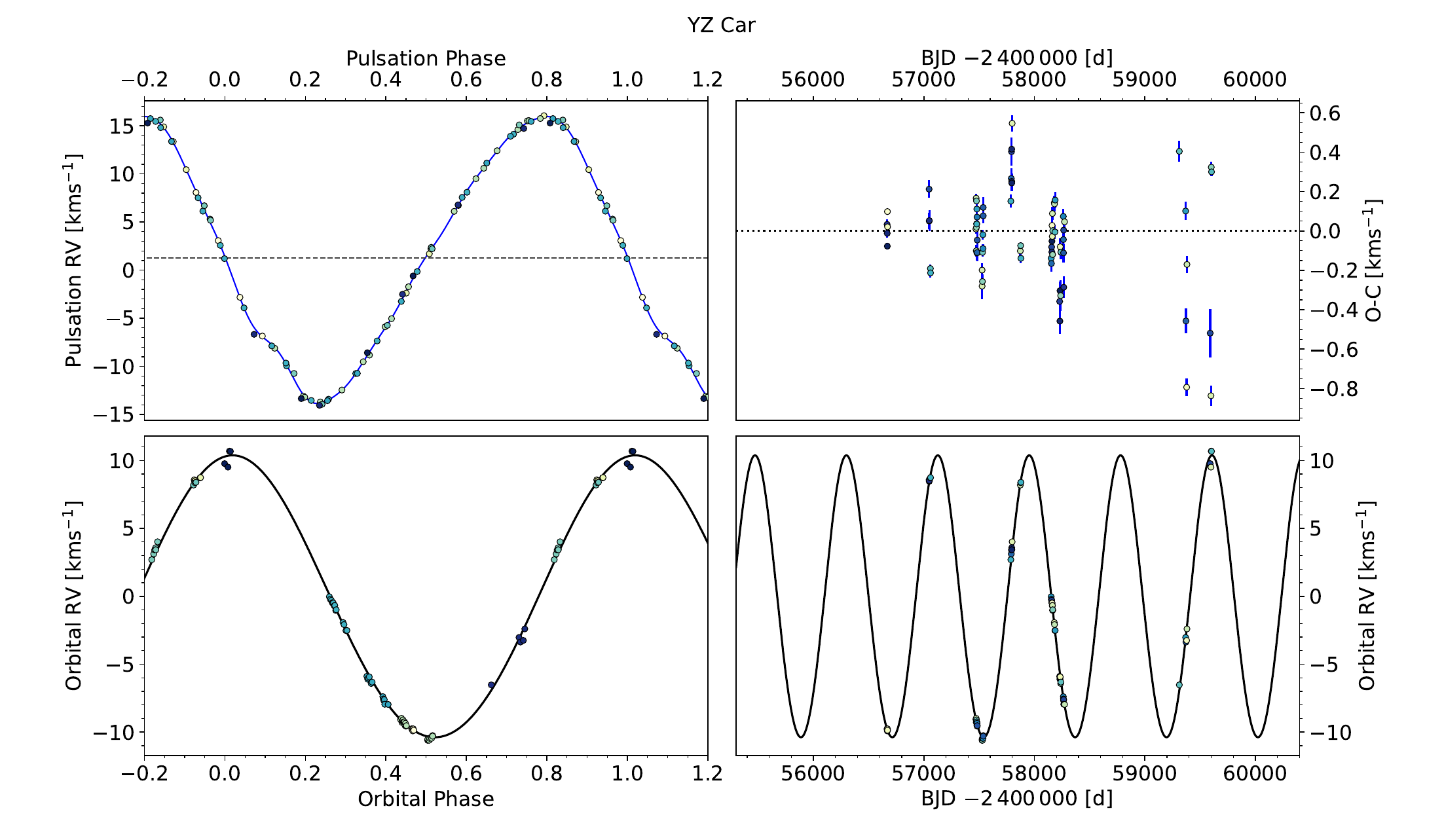}
    \caption{Pulsational and orbital fit of YZ Car. Figure description is same as Figure~\ref{fig:Example_kepler1}.}
    \label{fig:YZCarOrbit}
\end{figure*}

\begin{figure*}
    \centering
    \includegraphics[scale=0.5]{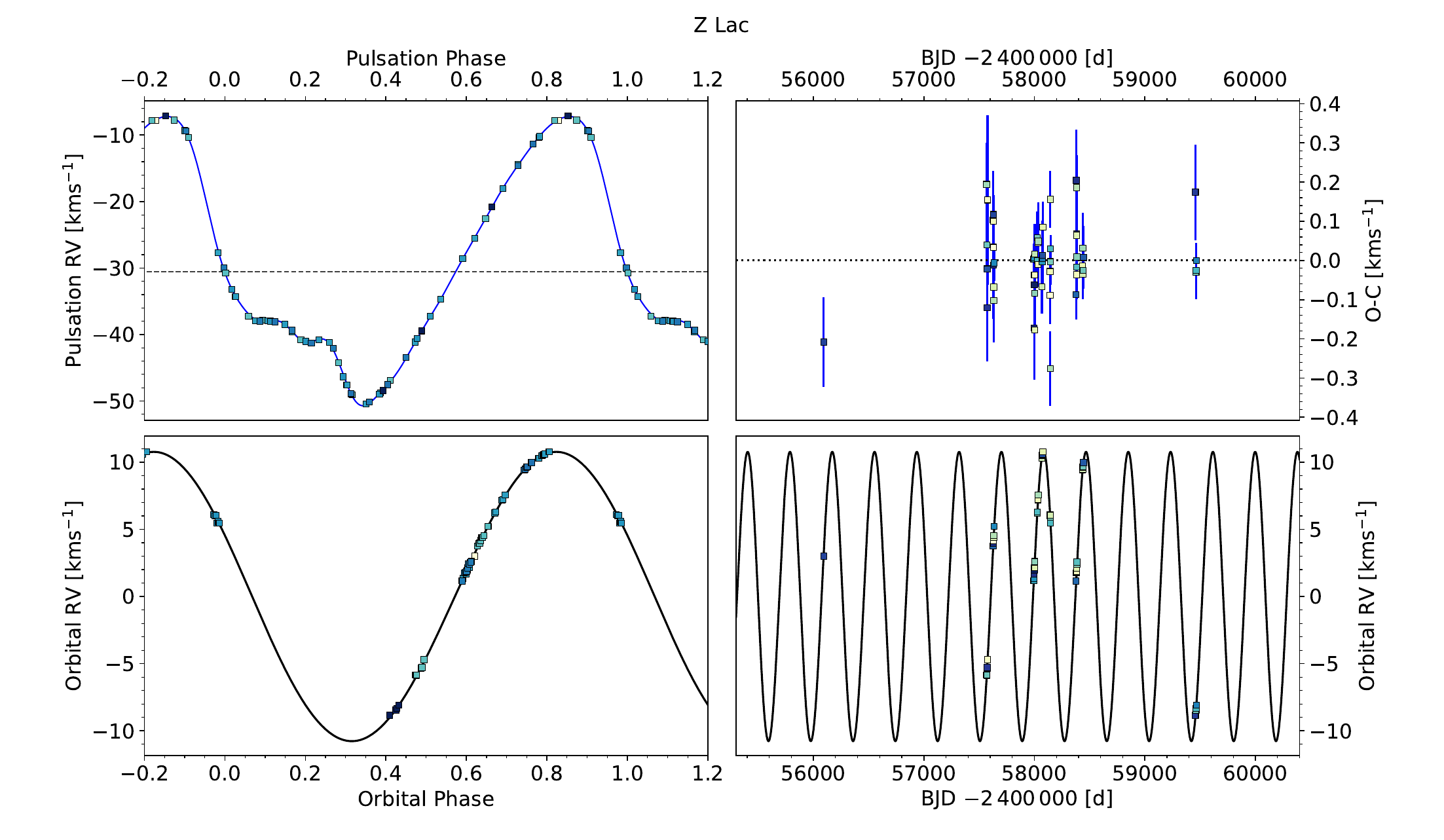}
    \caption{Pulsational and orbital fit of Z Lac. Figure description is same as Figure~\ref{fig:Example_kepler1}.}
    \label{fig:ZLacOrbit}
\end{figure*}

From Figure \ref{fig:ASAS0330Orbit} to \ref{fig:ZLacOrbit}, we present all the orbital fits obtained using \gc~data alone. The orbital fittings were derived using the method described in Section \ref{Sect:Kepfits}.

\section{V$_\gamma$ Orbits}\label{Appendix:Vgamma_Orbit_figures}
\begin{figure*}
    \centering
    \includegraphics[scale=0.55]{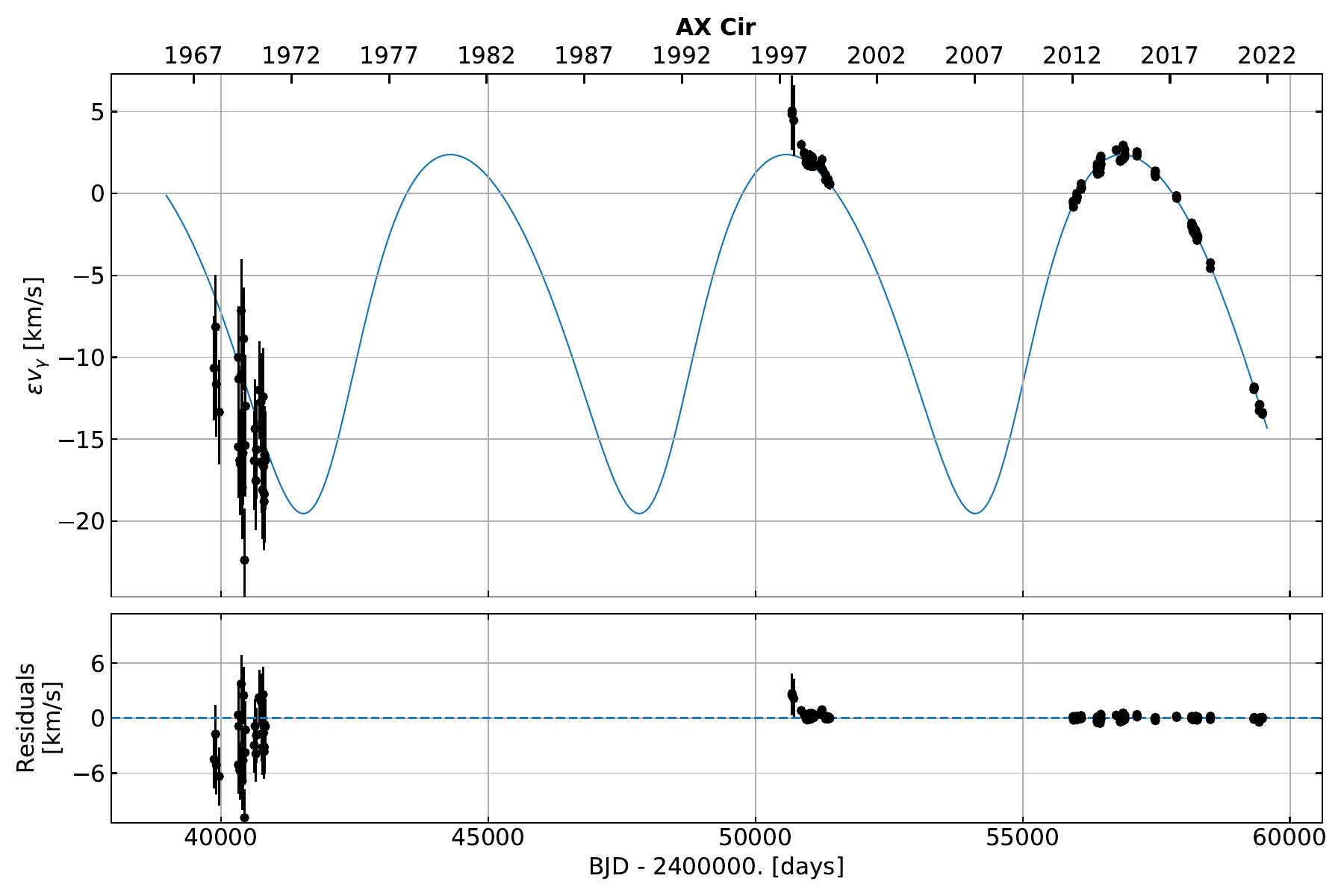}
    \caption{V$_\gamma$ residuals Orbit fitting for AX~Cir.}\label{fig:VgammaOrbitAXCir}
\end{figure*}

\begin{figure*}
\centering
     \includegraphics[scale=0.55]{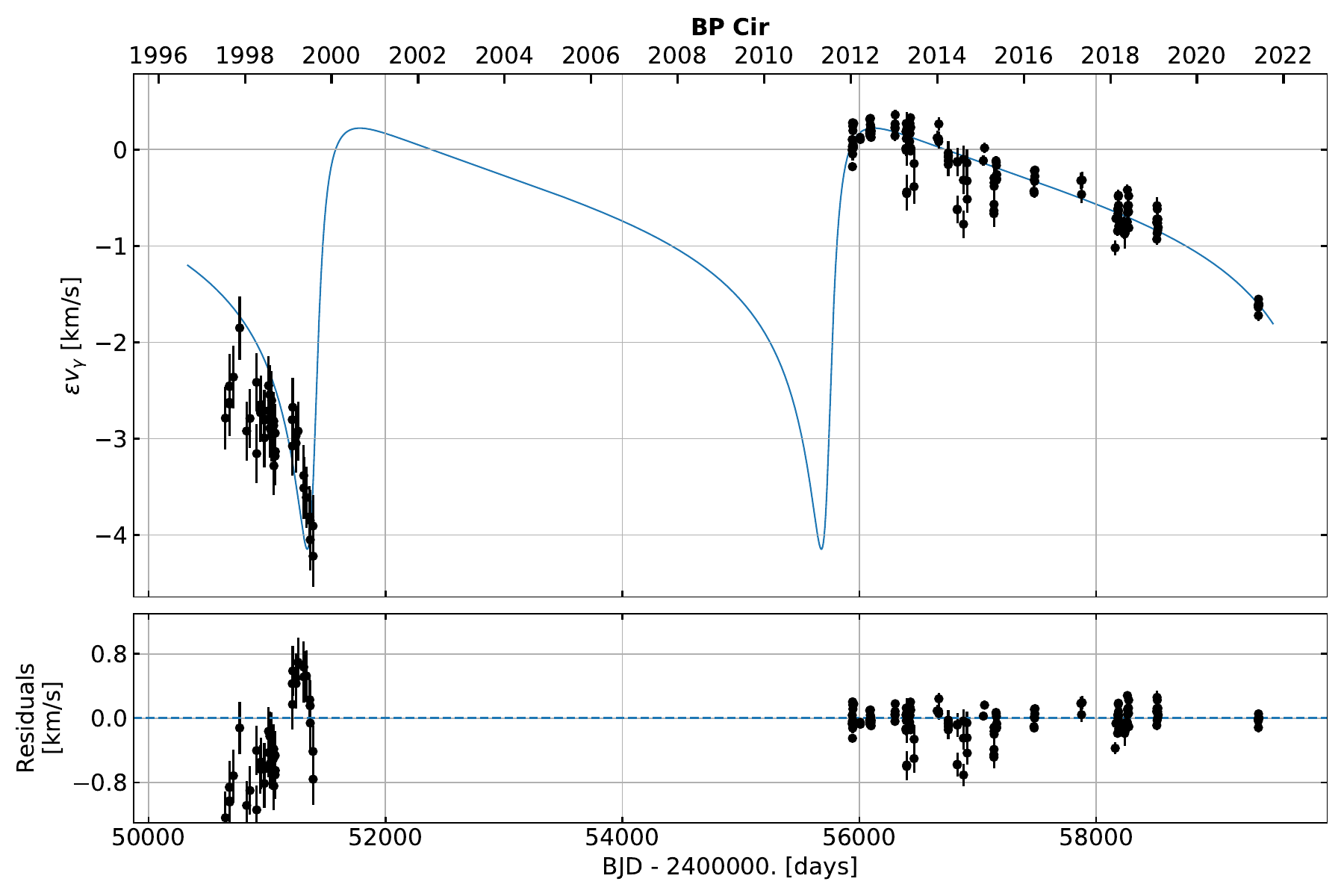}
    \caption{V$_\gamma$ residuals Orbit fitting for BP~Cir.}
    \label{fig:VgammaOrbitBPCir}
\end{figure*}

\begin{figure*}
    \centering
    \includegraphics[scale=0.55]{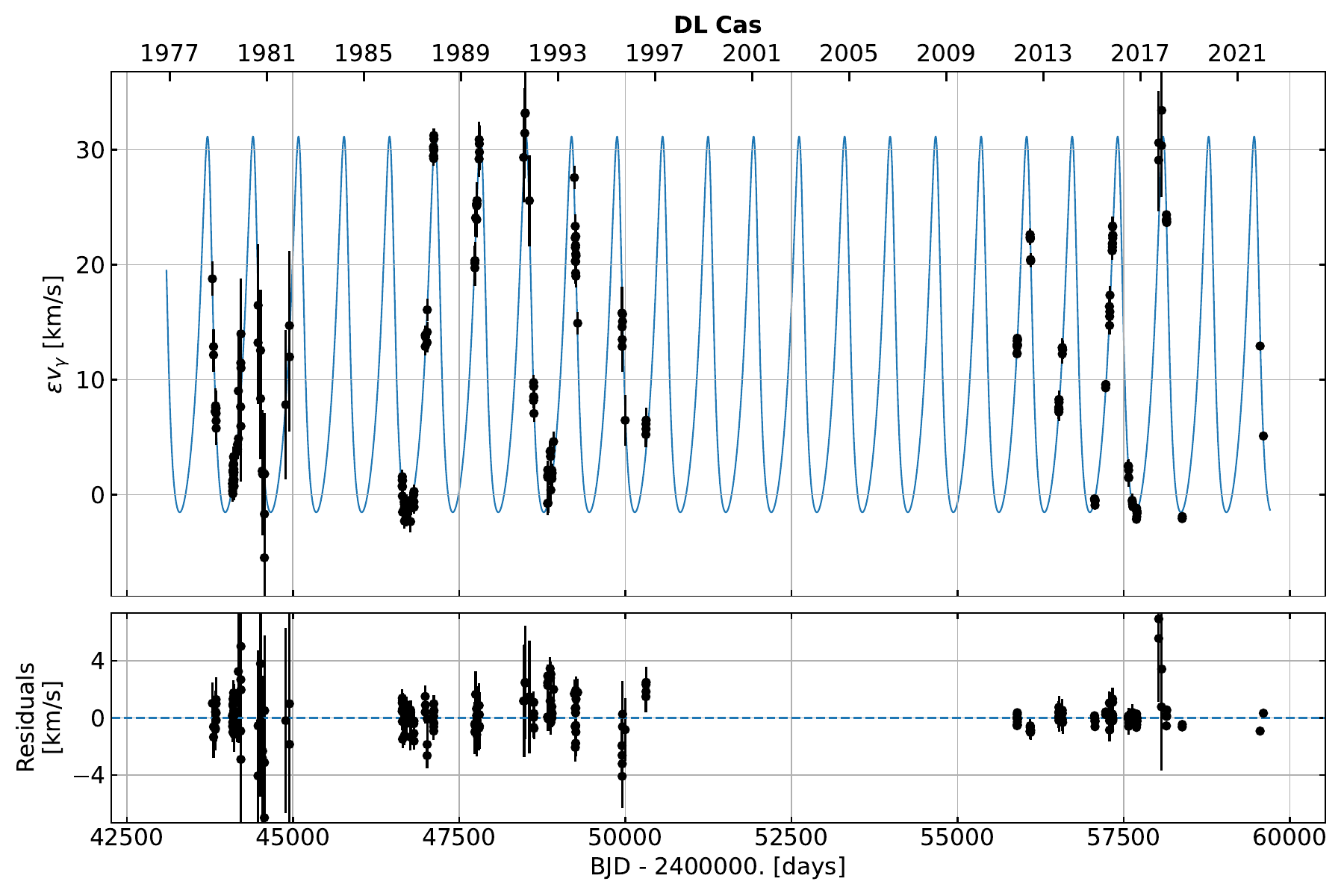}
    \caption{V$_\gamma$ residuals Orbit fitting for DL~Cas.}\label{fig:VgammaOrbitDLCas}
\end{figure*}

\begin{figure*}
\centering
    \includegraphics[scale=0.55]{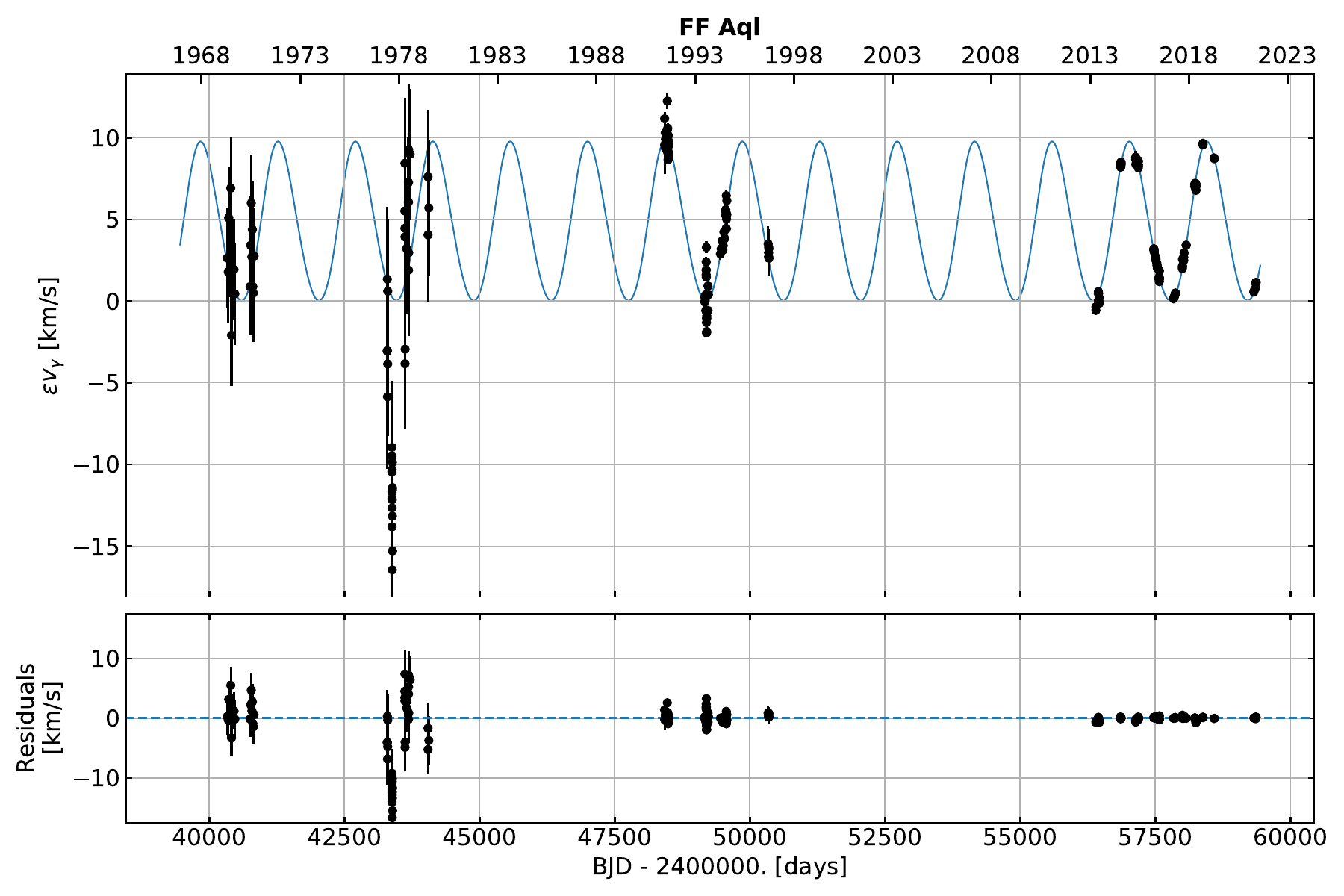}
    \caption{V$_\gamma$ residuals Orbit fitting for FF~Aql.}
    \label{fig:VgammaOrbitFFAql}
\end{figure*}

\begin{figure*}
    \centering
    \includegraphics[scale=0.55]{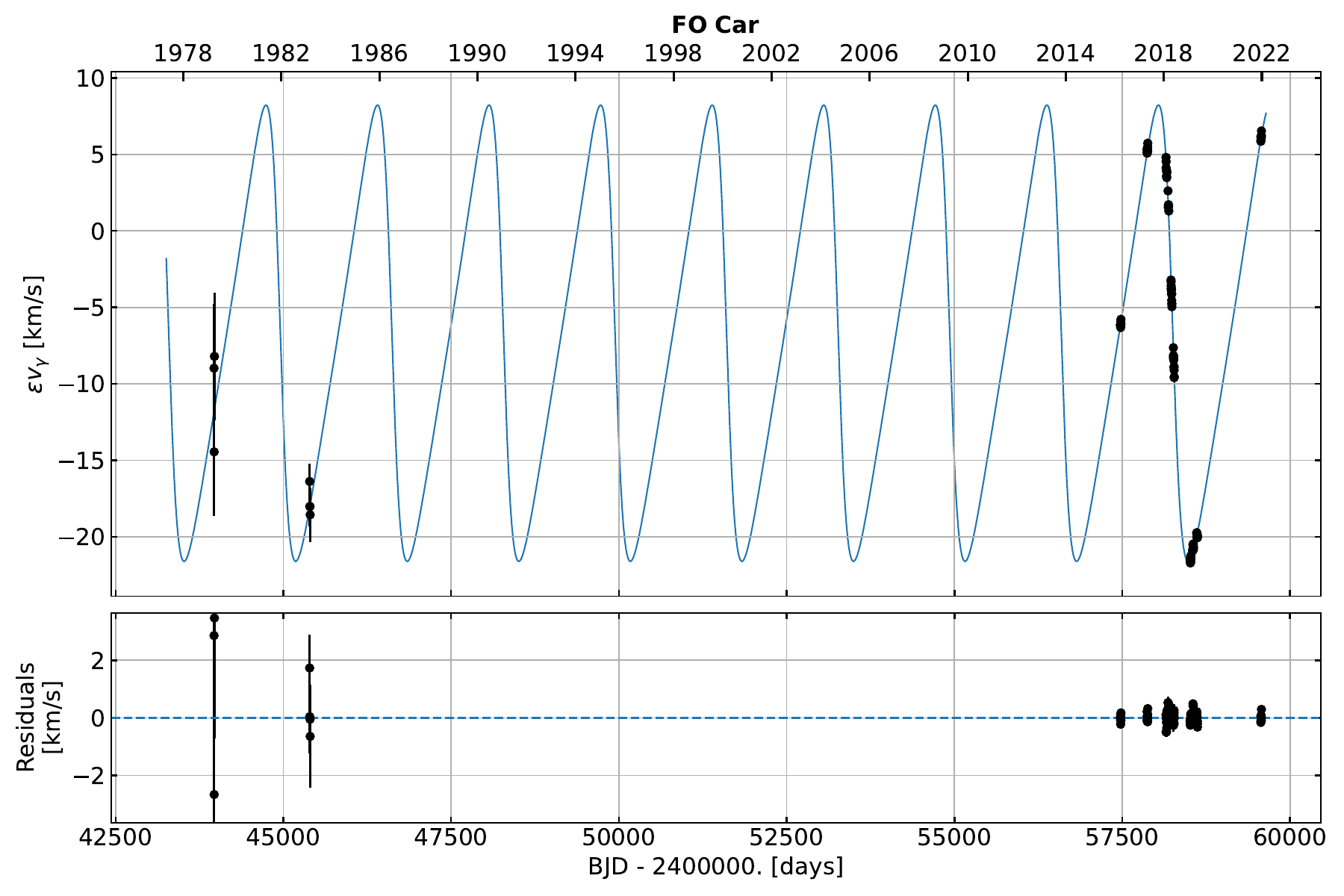}
    \caption{V$_\gamma$ residuals Orbit fitting for FO~Car.}
    \label{fig:VgammaOrbitFOCar}
\end{figure*}

\begin{figure*}
    \centering
    \includegraphics[scale=0.55]{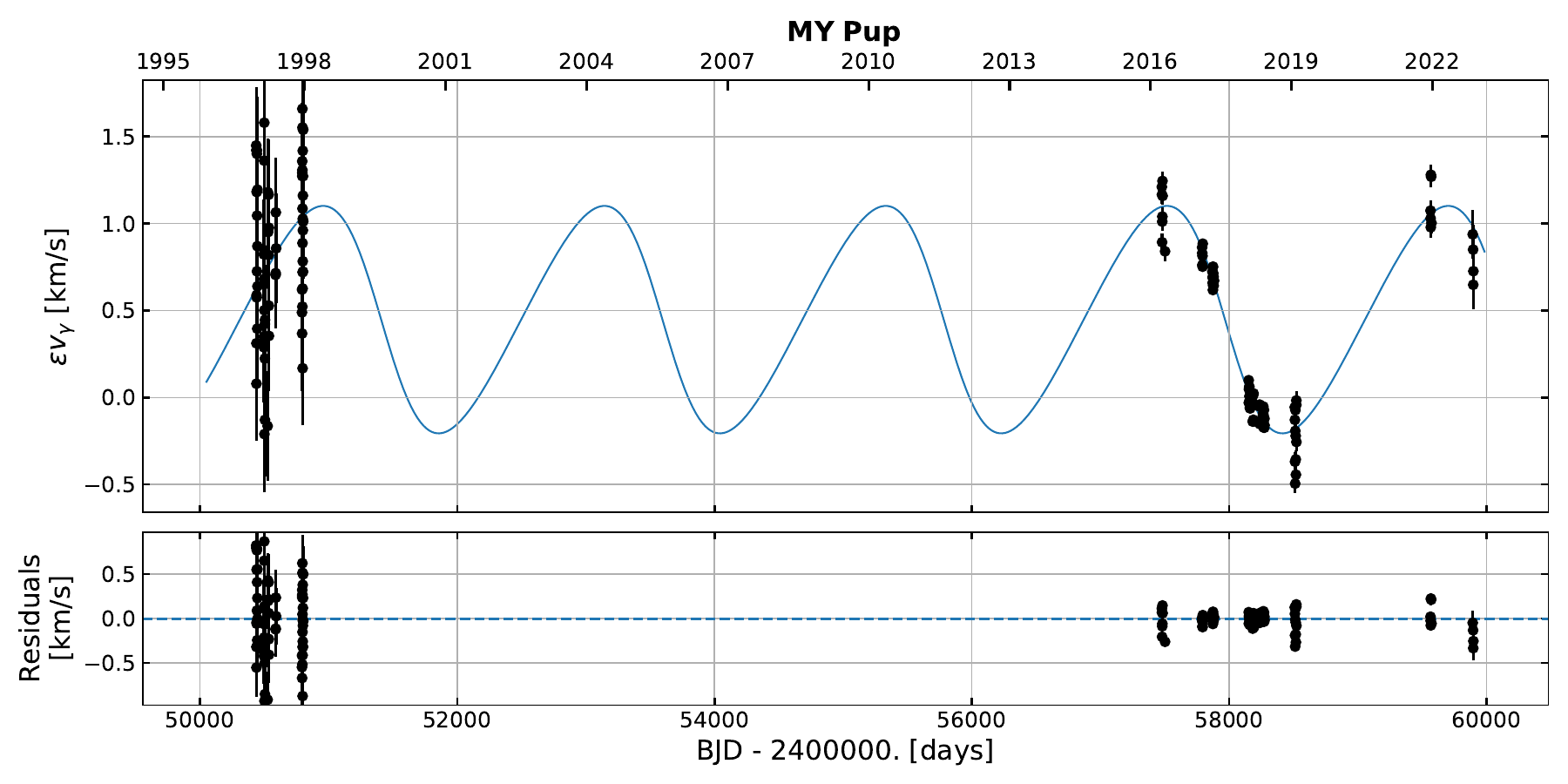}
\caption{V$_\gamma$ residuals Orbit fitting for MY Pup.}
    \label{fig:VgammaOrbitMYPup}
\end{figure*}

\begin{figure*}
\centering
    \includegraphics[scale=0.55]{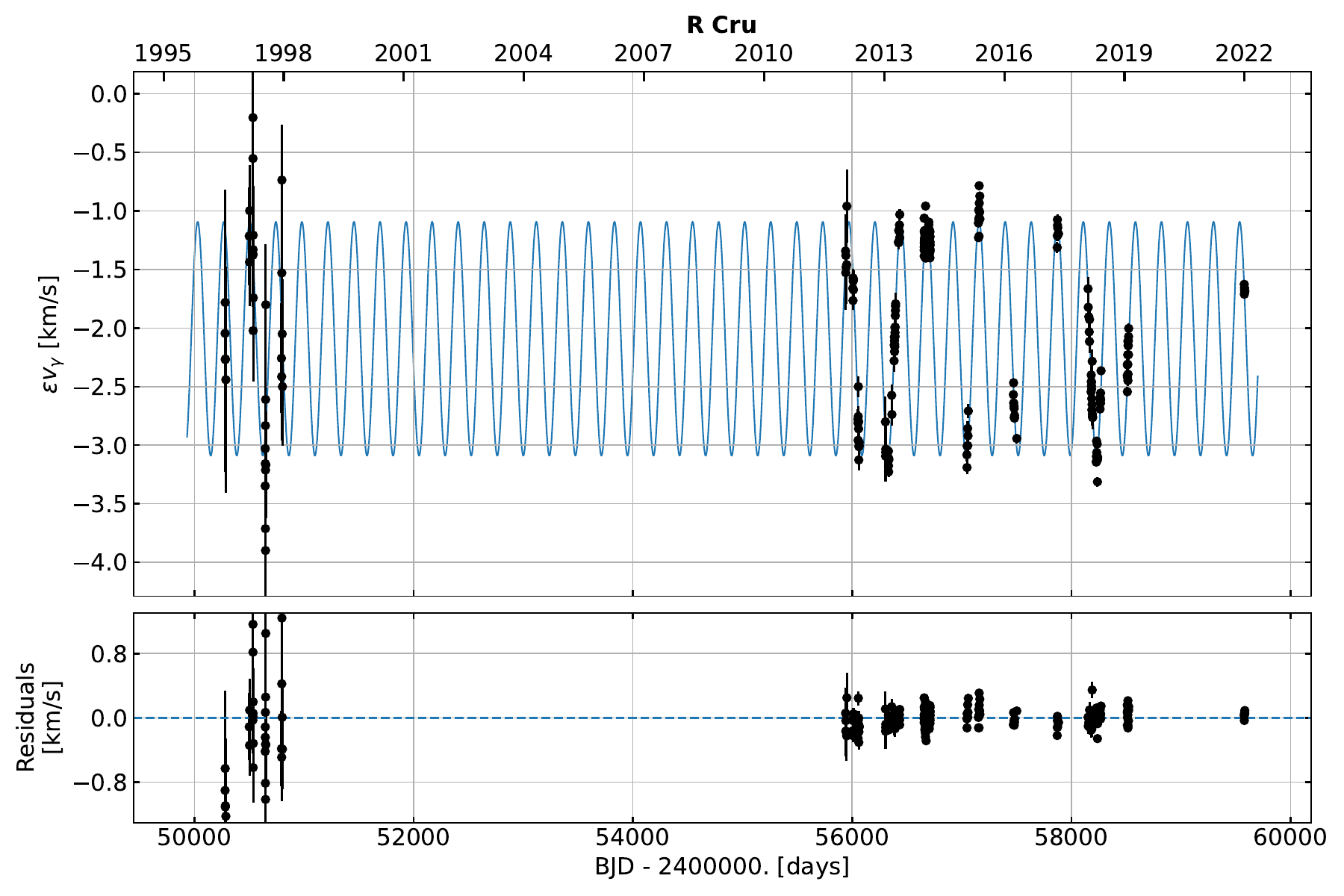}
    \caption{V$_\gamma$ residuals Orbit fitting for R~Cru.} 
     \label{fig:VgammaOrbitRCru}
\end{figure*}

\begin{figure*}
    \centering
    \includegraphics[scale=0.55]{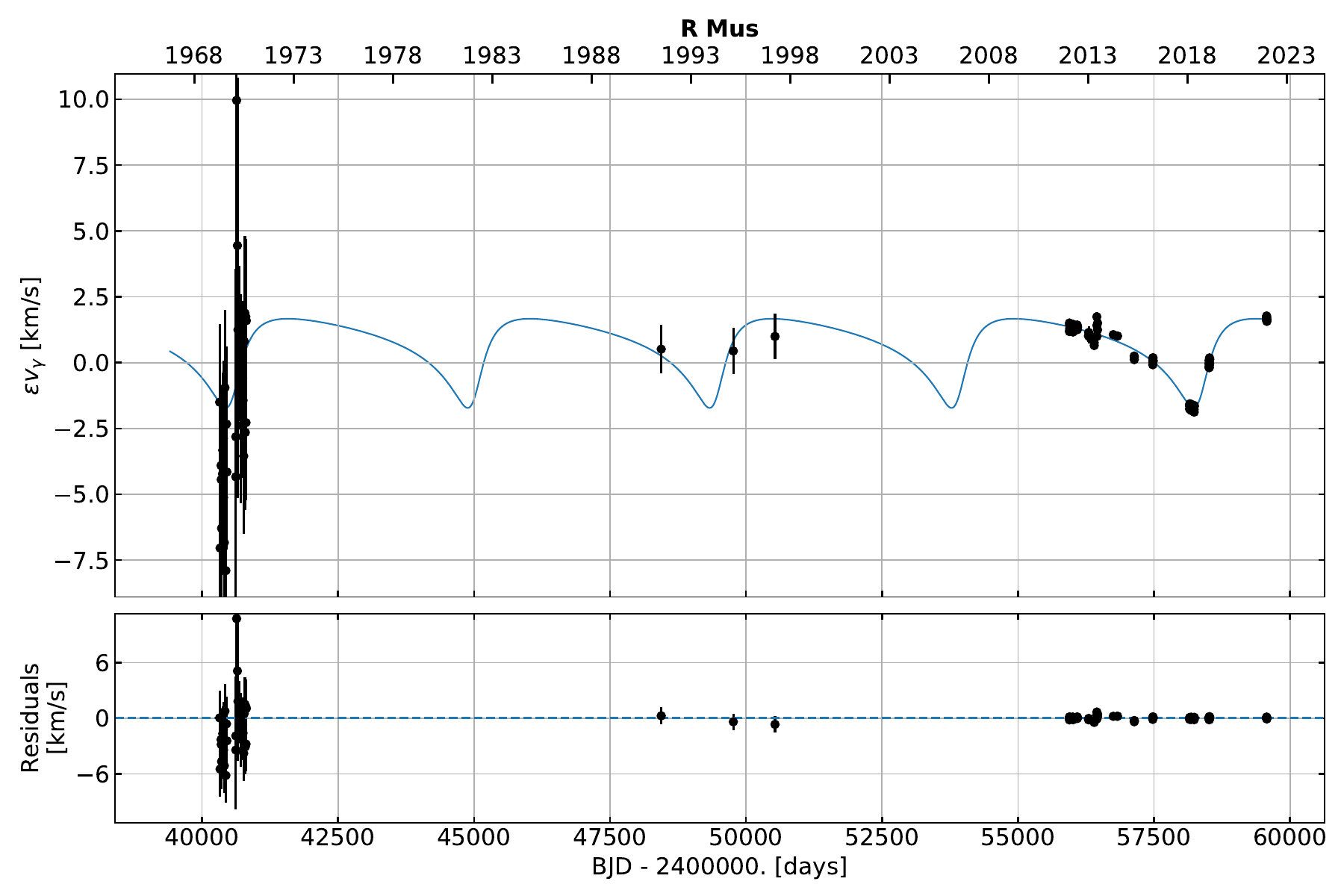}
\caption{V$_\gamma$ residuals Orbit fitting for R~Mus.} 
     \label{fig:VgammaOrbitRMus}
\end{figure*}

\begin{figure*}
\centering
     \includegraphics[scale=0.55]{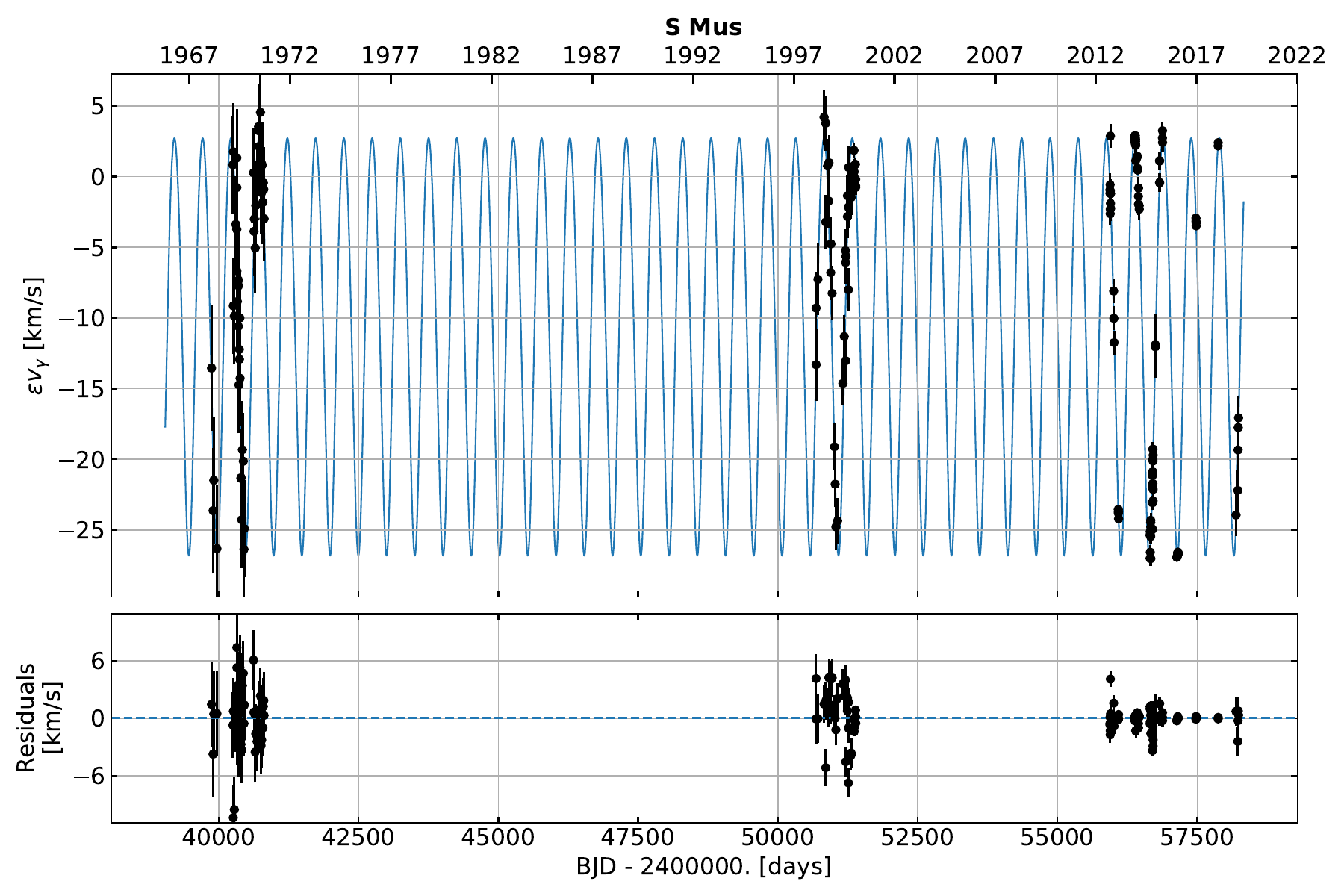}
    \caption{V$_\gamma$ residuals Orbit fitting for S~Mus.}
   \label{fig:VgammaOrbitSMus}
\end{figure*}

\begin{figure*}
    \centering
    \includegraphics[scale=0.55]{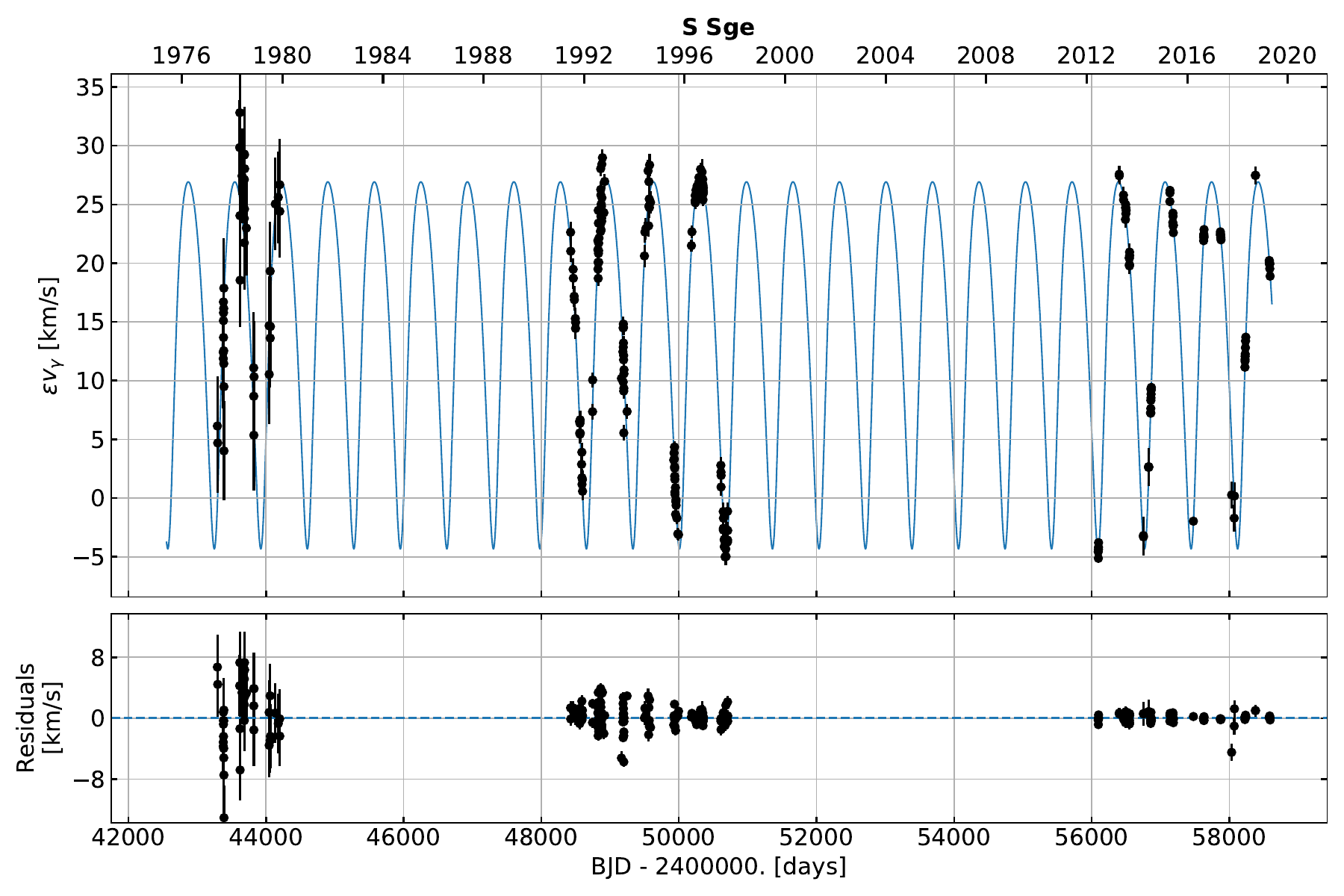}
\caption{V$_\gamma$ residuals Orbit fitting for S~Sge.}
   \label{fig:VgammaOrbitSSge}
\end{figure*}

\begin{figure*}
\centering
    \includegraphics[scale=0.55]{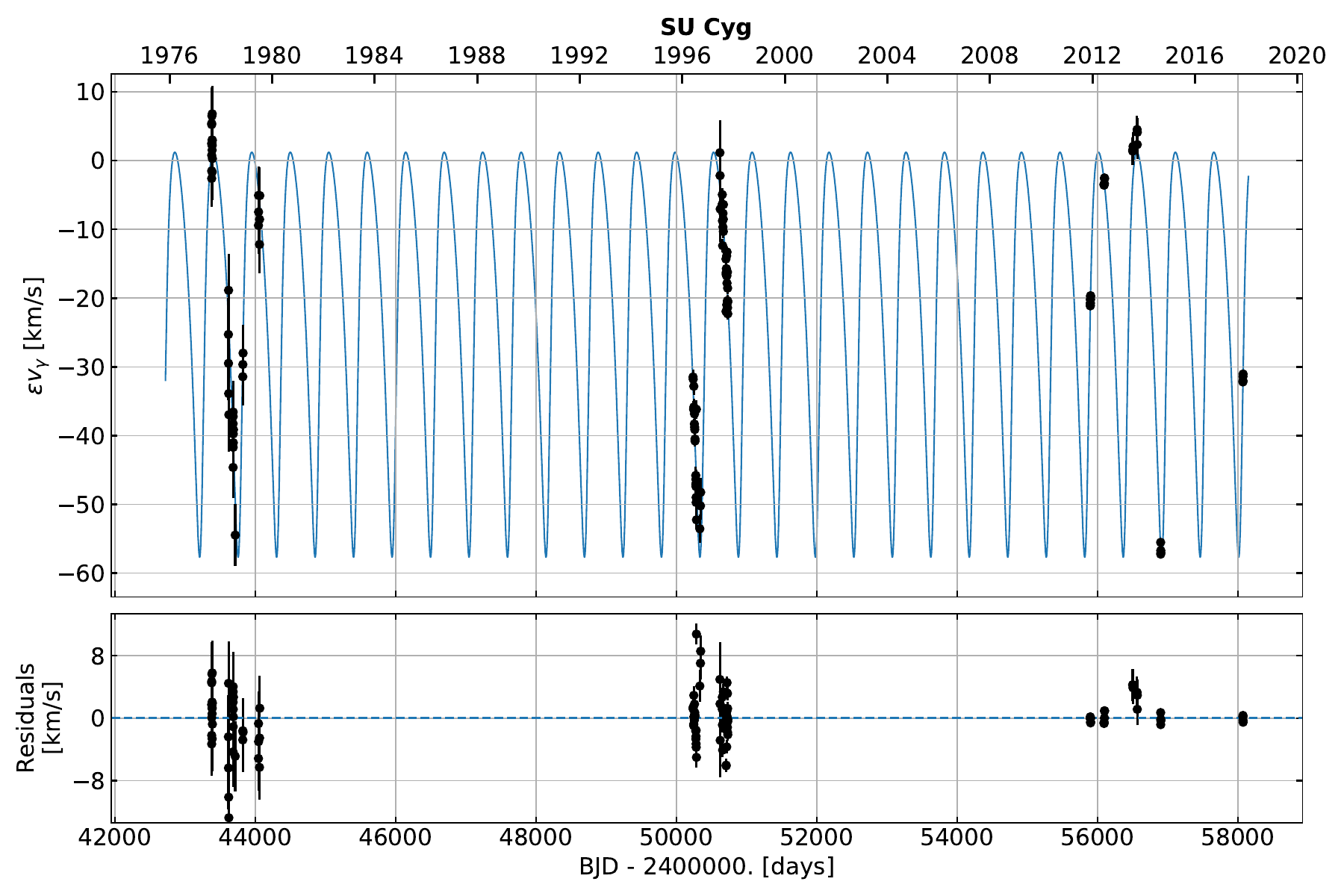}
    \caption{V$_\gamma$ residuals Orbit fitting for SU~Cyg.}
     \label{fig:VgammaOrbitSUCyg}
\end{figure*}

\begin{figure*}
    \centering
    \includegraphics[scale=0.55]{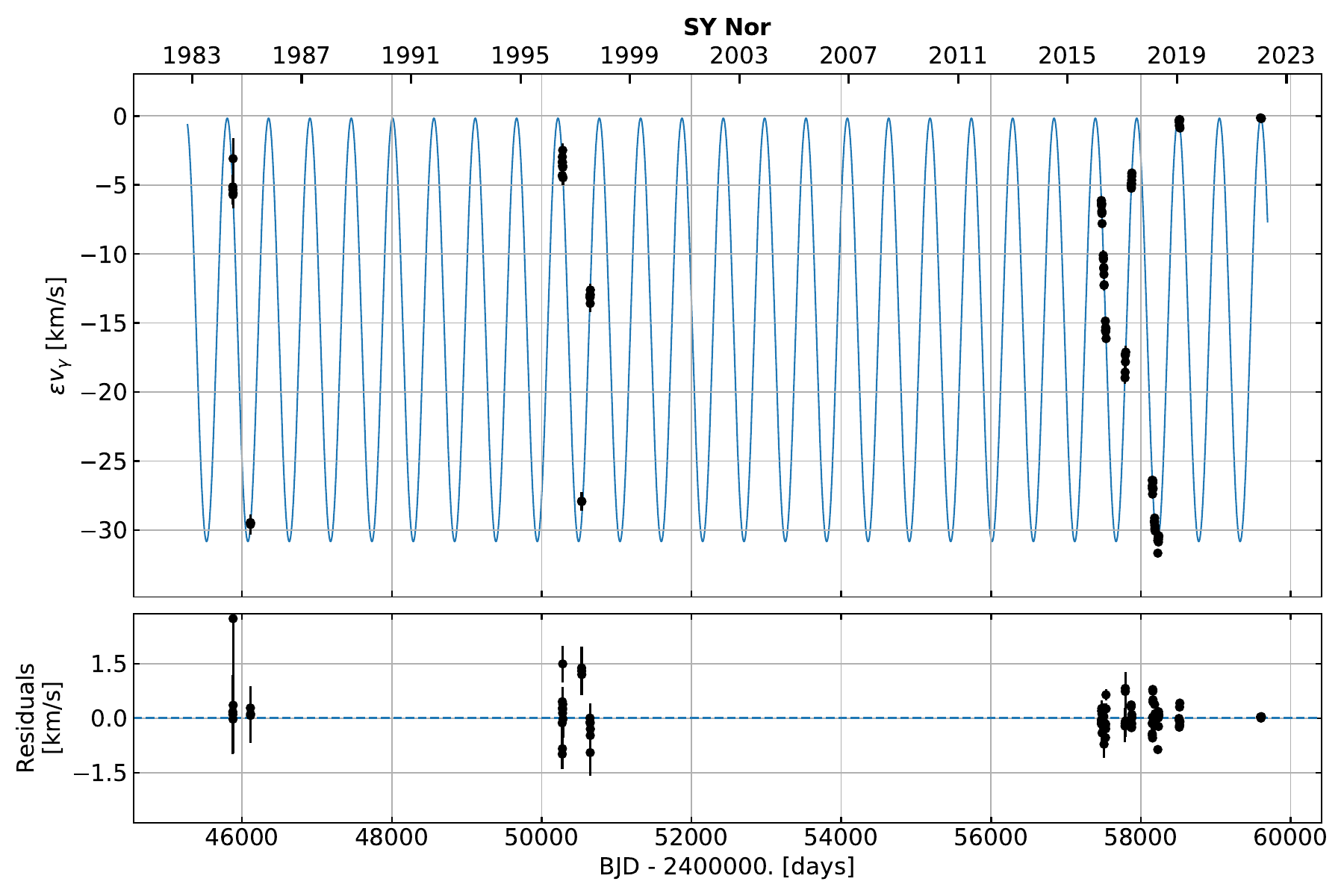}
\caption{V$_\gamma$ residuals Orbit fitting for SY Nor.}
     \label{fig:VgammaOrbitSYNor}
\end{figure*}

\begin{figure*}
    \centering
    \includegraphics[scale=0.55]{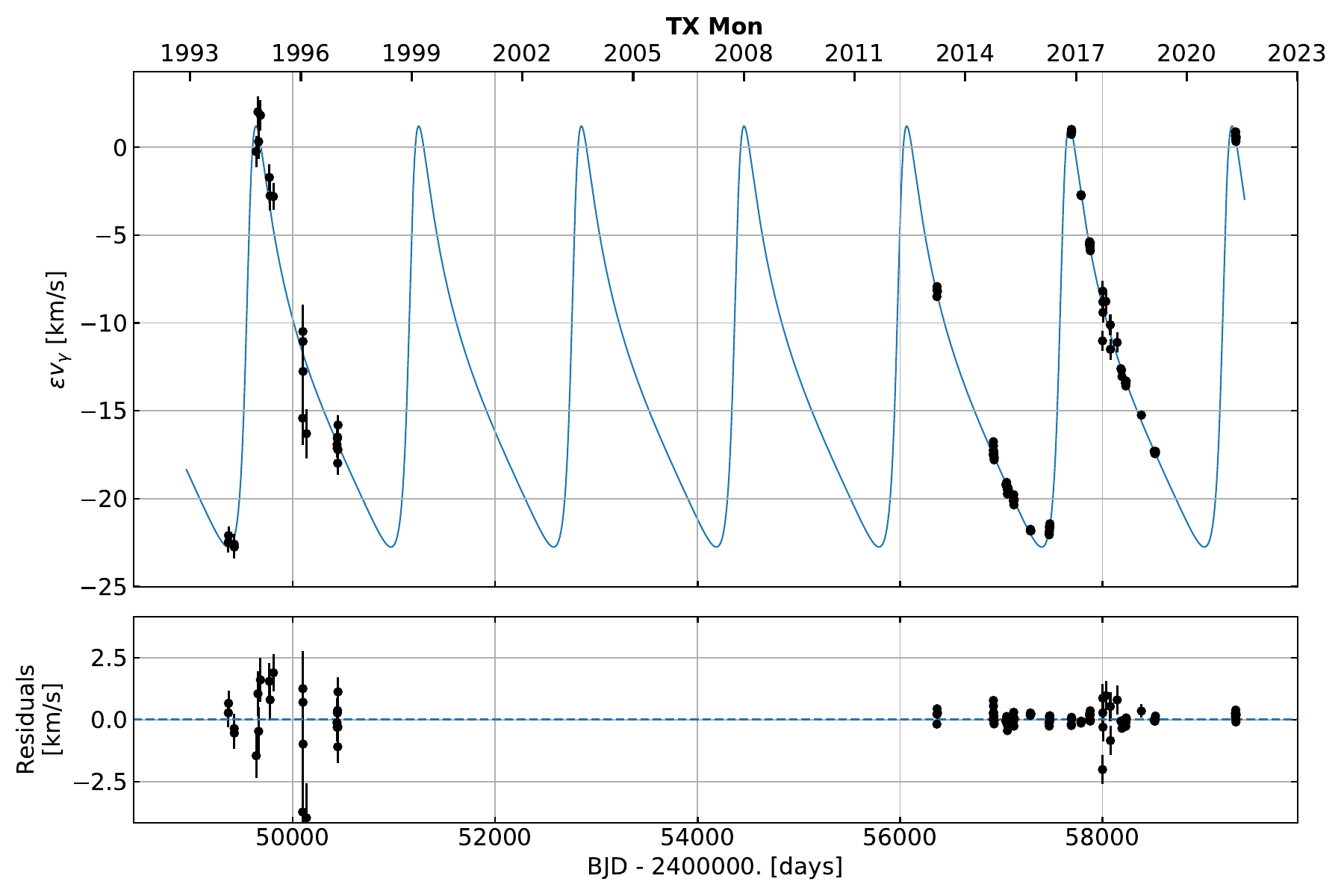}
    \caption{V$_\gamma$ residuals Orbit fitting for TX~Mon.}
     \label{fig:VgammaOrbitTXMon}
\end{figure*}

\begin{figure*}
    \centering
    \includegraphics[scale=0.55]{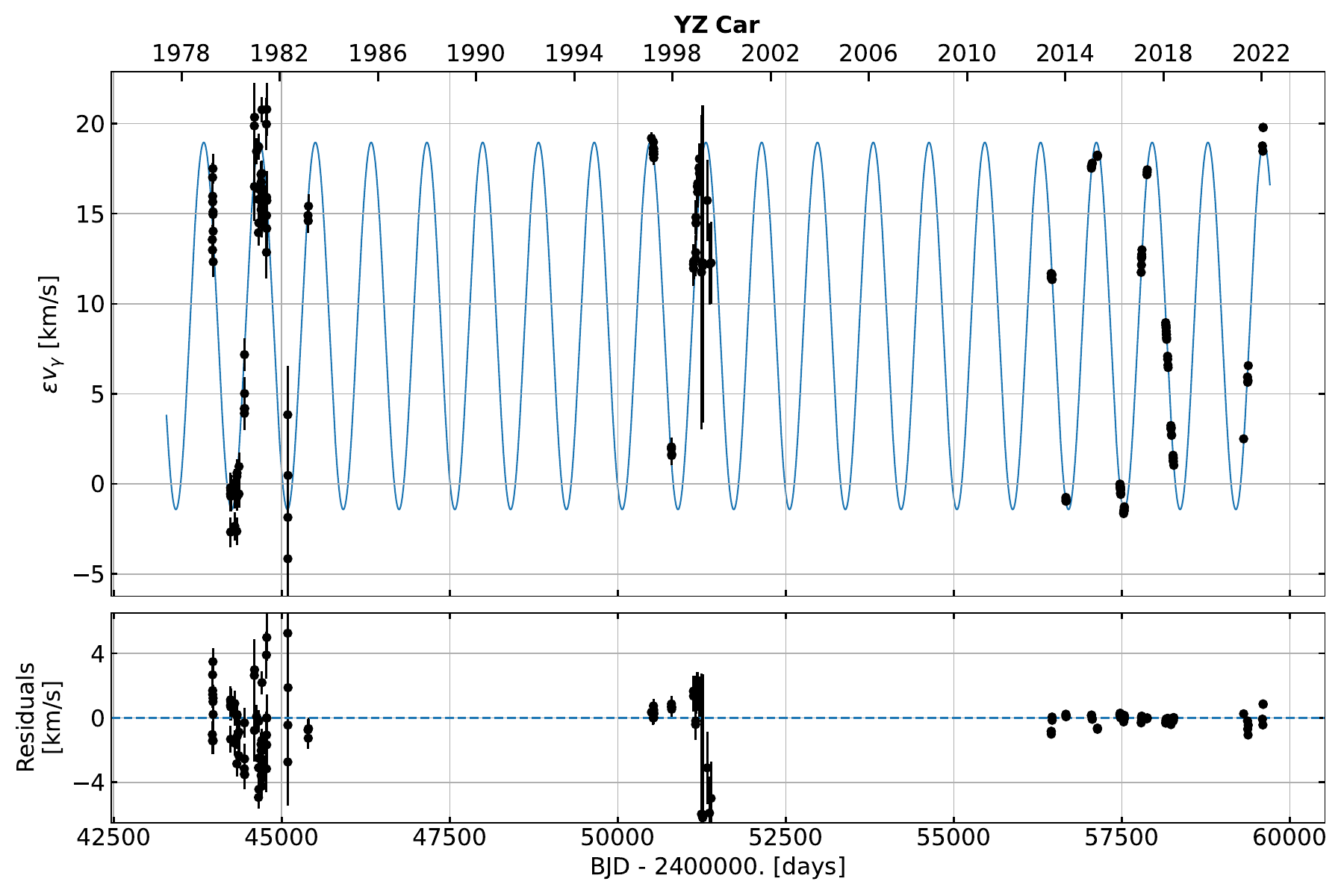}
    \caption{V$_\gamma$ residuals Orbit fitting for YZ~Car.}
     \label{fig:VgammaOrbitYZCar}
\end{figure*}

\begin{figure*}
    \centering
     \includegraphics[scale=0.55]{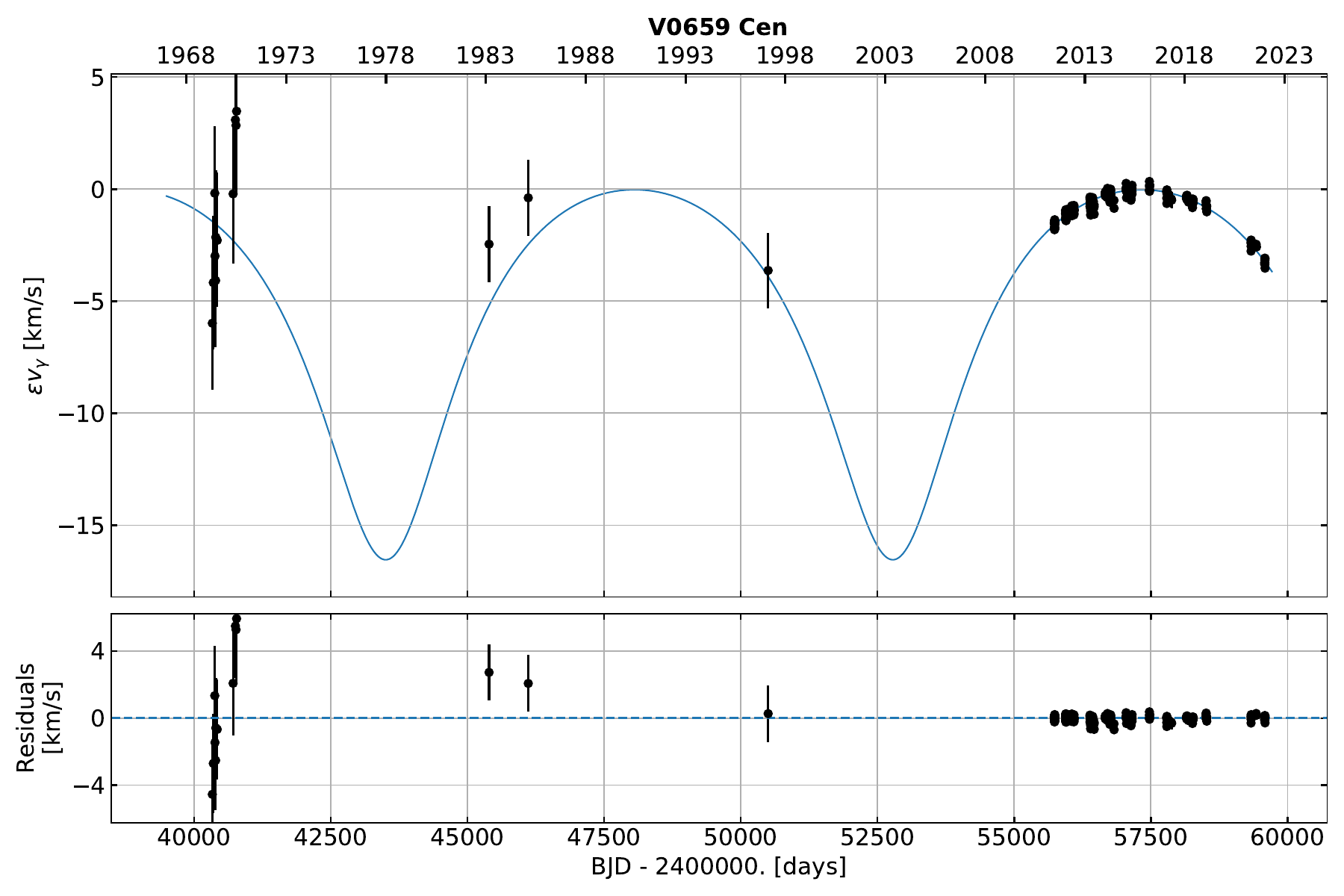}
\caption{V$_\gamma$ residuals Orbit fitting for V0659 Cen.}
     \label{fig:VgammaOrbitV0659}
\end{figure*}

\begin{figure*}
\centering
    \includegraphics[scale=0.55]{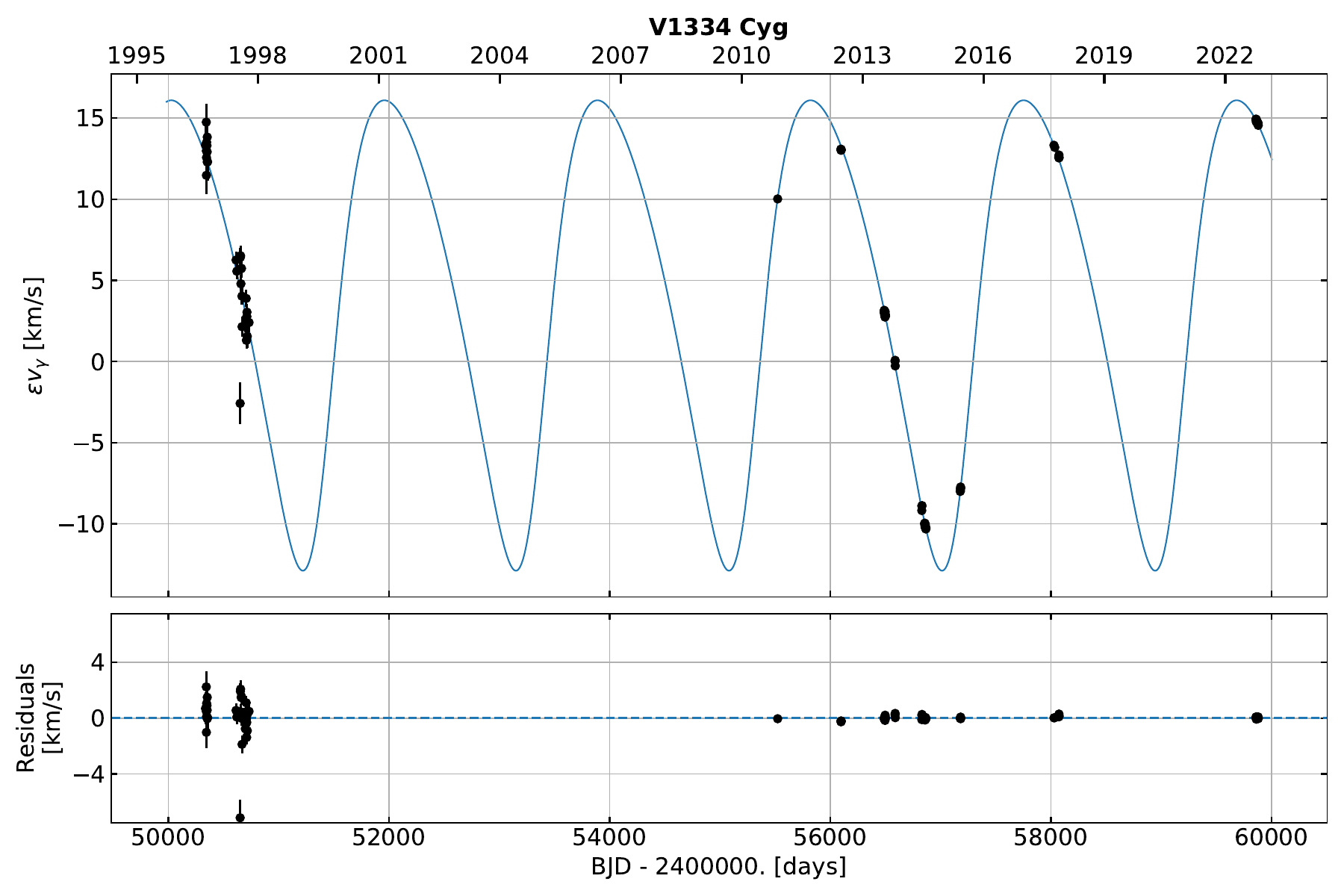}
    \caption{V$_\gamma$ residuals Orbit fitting for V1334~Cyg.}
    \label{fig:VgammaOrbitV1334}
\end{figure*}

\begin{figure*}
\centering
    \includegraphics[scale=0.55]{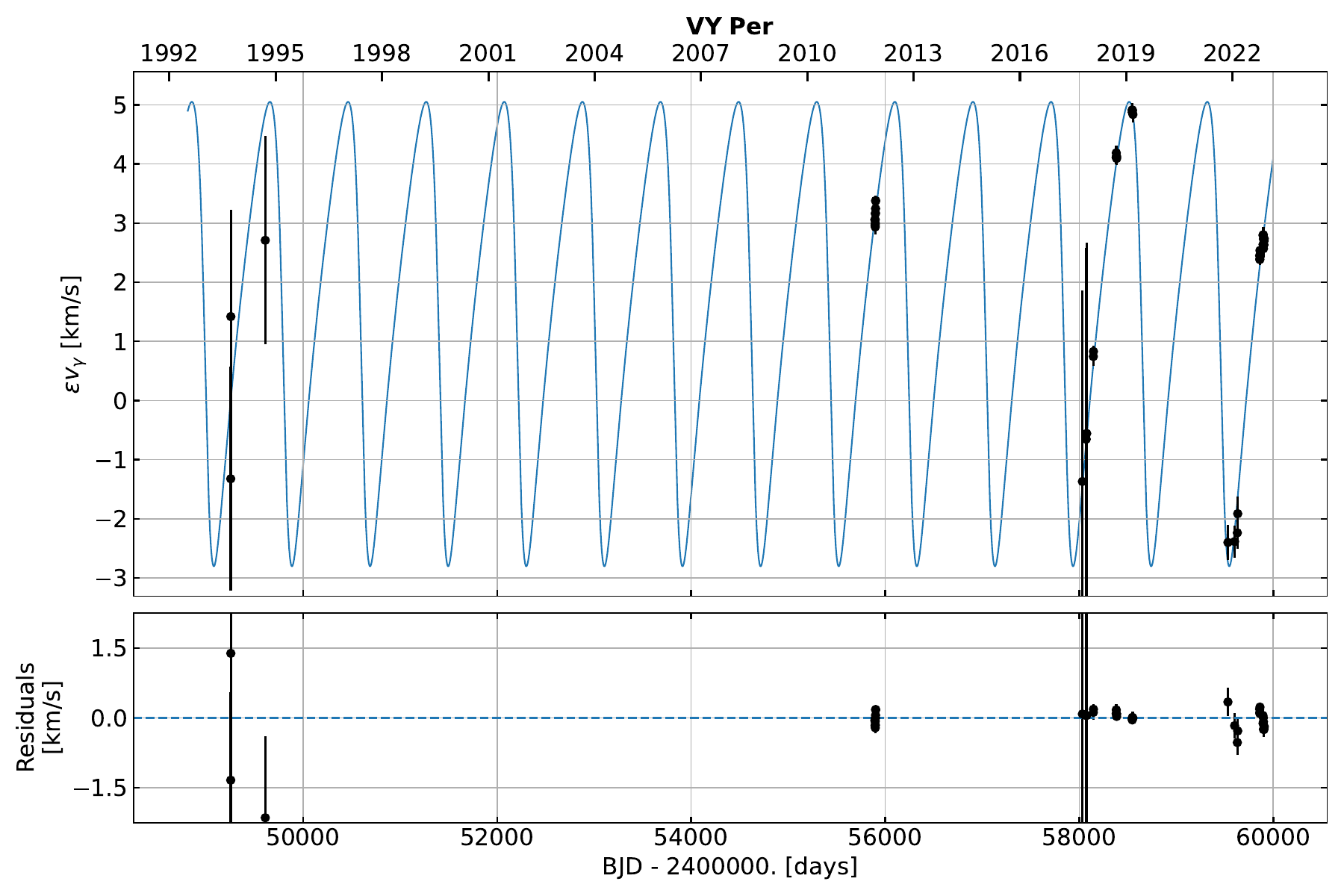}
\caption{V$_\gamma$ residuals Orbit fitting for VY Per.\label{fig:VgammaOrbitVYPer}}
\end{figure*}

\begin{figure*}
\centering
    \includegraphics[scale=0.55]{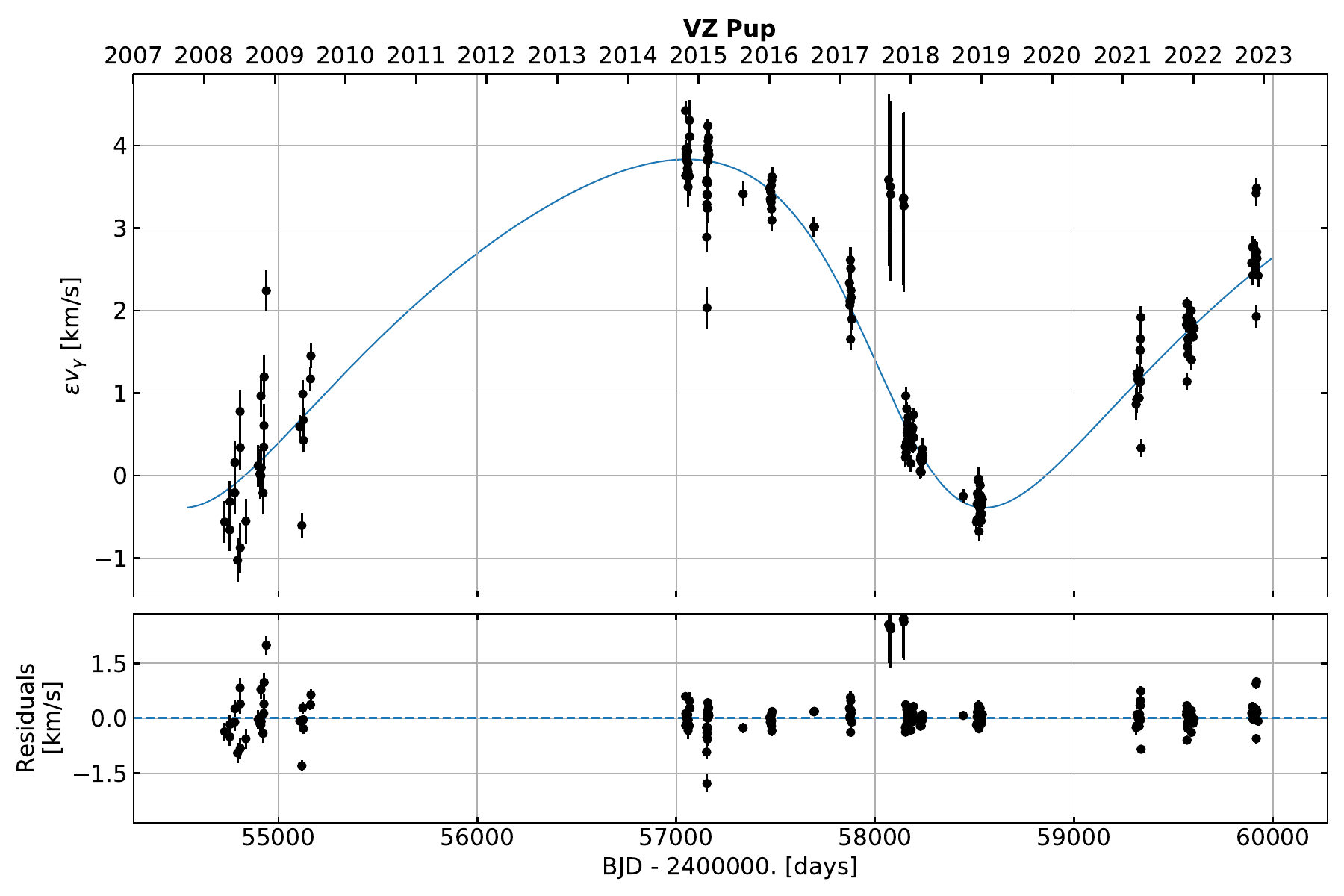}
\caption{V$_\gamma$ residuals Orbit fitting for VZ Pup.\label{fig:VgammaOrbitVZPup}}
\end{figure*}

In Figure \ref{fig:VgammaOrbitAXCir} to Figure \ref{fig:VgammaOrbitVZPup} we present the orbital fittings obtained using the combination of RV data from \gc~and literature datasets. These orbital fits were derived using the MCMC method described in Section~\ref{sec:mcmc}.

\section{Gaia astrometry\label{app:Gaiaastrometry}}
In Table~\ref{tab:GaiaAstrometry}, we present all the \gaia~RUWE and PMa flags used in this study for all our sample stars.

\begin{table*}
    \centering
    \caption{\textit{Gaia} astrometry and PMa tags for the confirmed SB1 Cepheids within \gc. Here, binary flag of `1' means that the star was flagged as binary through PMa, and binary flag of `0' means the PMa did not detect them as binaries. \label{tab:GaiaAstrometry}}
\sisetup{round-mode=places}
\setlength{\tabcolsep}{1pt}
    \begin{tabular}{lS[round-precision=4]c|lS[round-precision=4]c|lS[round-precision=4]c}
    \toprule
    \hline
    Star & {RUWE} & Bin flag & Star & {RUWE} & Bin flag & Star & {RUWE} & Bin flag \\
& & from PMa  & & & from PMa  && & from PMa  \\
    \hline
    \multicolumn{3}{c|}{SB1 with Orbits} & \multicolumn{3}{c|}{SB1 with Trends} & \multicolumn{3}{c}{SB1 with RVTF-Yes} \\
    \hline
   ASAS J064540+0330.4 & ~ & ~ & AD Pup & 1.36213 & 0 & $\beta$ Dor & 4.53476 & 0 \\ 
        ASAS J084951-4627.2 & 1.0863 & ~ & AH Vel & 0.8293 & 1 & $\eta$ Aql & 2.561 & 0 \\ 
        ASAS J100814-5856.6 & 1.1193 & ~ & AQ Pup & 1.1803056 & 0 & CD Cyg & 1.007 & 0 \\ 
        ASAS J174603-3528.1 & 0.6934 & ~ & AQ Car & 1.0672319 & 0 & RS Ori & 1.119 & 0 \\ 
        AX Cir & 7.80521 & ~ & ASAS J064553+1003.8 & 1.499896 & ~ &  SS CMa & 1.1134018 & 0 \\
        BP Cir & 1.0495 & ~ & ASAS J103158-5814.7 & 0.8688451 & ~ & SZ Aql & 0.94 & 0 \\ 
        $\delta$ Cep & 2.7130983 & ~ & ASAS J155847-5341.8 & 0.8680487 & ~ & SZ Cyg & 0.956 & 0 \\ 
        DL Cas & 1.882963 & 0 & ASAS J174108-2328.5 & 0.75773543 & ~ & V0340 Ara & 0.9316048 & 0 \\
        FF Aql & 1.0553534 & 1 & AW Per & 1.1555363 & 1 &  V0402 Cyg & 0.92 & 0 \\ 
        FN Vel & 1.6964414 & 0 & DR Vel & 1.0049819 & ~ & V0916 Aql & 0.9160 & \\
        FO Car & 0.9022557 & ~ & FR Car & 1.0235187 & 0 & VY Sgr & 0.8056516 & ~ \\ 
        GX Car & 1.0175964 & 1 & KN Cen & 1.0335958 & 0 &  X Cyg & 1.277353 & 1 \\
        IT Car & 1.0755398 & 1 & LR TrA & 0.9513159 & 1 & ~ & ~ & ~ \\ 
        MU Cep & 0.99277467 & ~ & OX Cam & 0.9843955 & ~ & ~ & ~ & ~ \\ 
        MY Pup & 1.0086893 & 0 & RY Vel & 1.075225 & 0 & ~ & ~ & ~ \\ 
        NT Pup & 0.98610914 & ~ & RX Aur & 0.9820967 & 1 & ~ & ~ & ~ \\ 
        R Cru & 1.1609027 & ~ & RV Sco & 0.804823 & 1 & ~ & ~ & ~ \\ 
        R Mus & 1.0741858 & 1 & RW Cam & 8.013022 & 1 & ~ & ~ & ~ \\ 
        S Mus & 4.494788 & 1 & RY Sco & 0.7325544 & 0 & ~ & ~ & ~ \\ 
        S Sge & 4.003331 & 1 & SX Vel & 1.022172 & 0 & ~ & ~ & ~ \\ 
        SU Cyg & 3.4429314 & 0 & T Mon & 1.7238975 & 1 & ~ & ~ & ~ \\ 
        SY Nor & 1.5524937 & 0 & UX Per & 1.1667587 & 1 & ~ & ~ & ~ \\ 
        TX Mon & 1.6885945 & 1 & UZ Sct & 0.9132 & ~ & ~ & ~ & ~ \\ 
        U Vul & 2.882503 & 1 & V0391 Nor & 0.832 & ~ & ~ & ~ & ~ \\ 
        V0407 Cas & 0.8948287 & ~ & V0492 Cyg & 0.99664867 & ~ & ~ & ~ & ~ \\ 
        V0659 Cen & 2.9895537 & 1 & V0827 Cas & 1.16 & ~ & ~ & ~ & ~ \\ 
        V1334 Cyg & 2.7684453 & 1 & V1162 Aql & 0.95236385 & 0 & ~ & ~ & ~ \\ 
        VY Per & 1.1543808 & ~ & V1803 Aql & 0.9414667 & ~ & ~ & ~ & ~ \\ 
        VZ Pup & 1.2365955 & 0 & V2475 Cyg & 0.99432325 & ~ & ~ & ~ & ~ \\ 
        W Sgr & 3.9523914 & 0 & XZ Car & 1.0491393 & 0 & ~ & ~ & ~ \\ 
        XX Cen & 1.2384809 & 0 & ~ & ~ & ~ & ~ & ~ & ~ \\ 
        YZ Car & 1.1657516 & 0 & ~ & ~ & ~ & ~ & ~ & ~ \\ 
        Z Lac & 1.0548457 & 0 & ~ & ~ & ~ & ~ & ~ & ~ \\ 
        \hline 
        \bottomrule
    \end{tabular}
\end{table*}

\section{RV Template fits after zero-point offset correction for sample stars with no signs of SB1} \label{app:RVTF-noFigs}
Figures~\ref{fig:VgammafornoSB1signs} and \ref{fig:VgammafornoSB1signs1} showcase the long-terms trends in \vgamma~for sample stars where we find no signs of binarity through the RVTF analysis of Section~\ref{sec:RVTF}.

\begin{figure*}
    \begin{subfigure}{.35\textwidth}
    \includegraphics[scale=0.25,trim={4cm 0 0 1.5cm}]{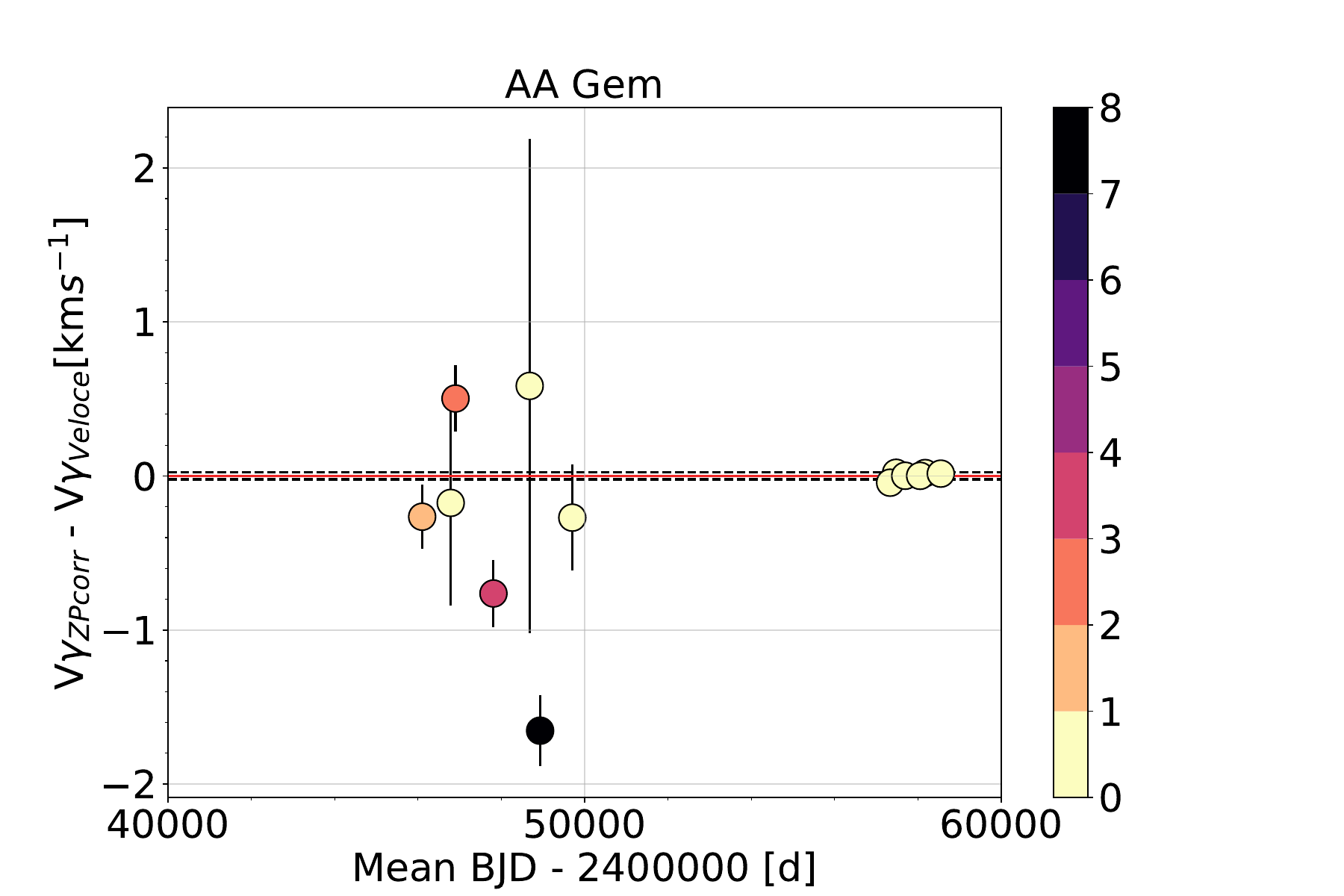}
    \end{subfigure}
    \begin{subfigure}{.35\textwidth}
    \includegraphics[scale=0.25,trim={4cm 0 0 1.5cm}]{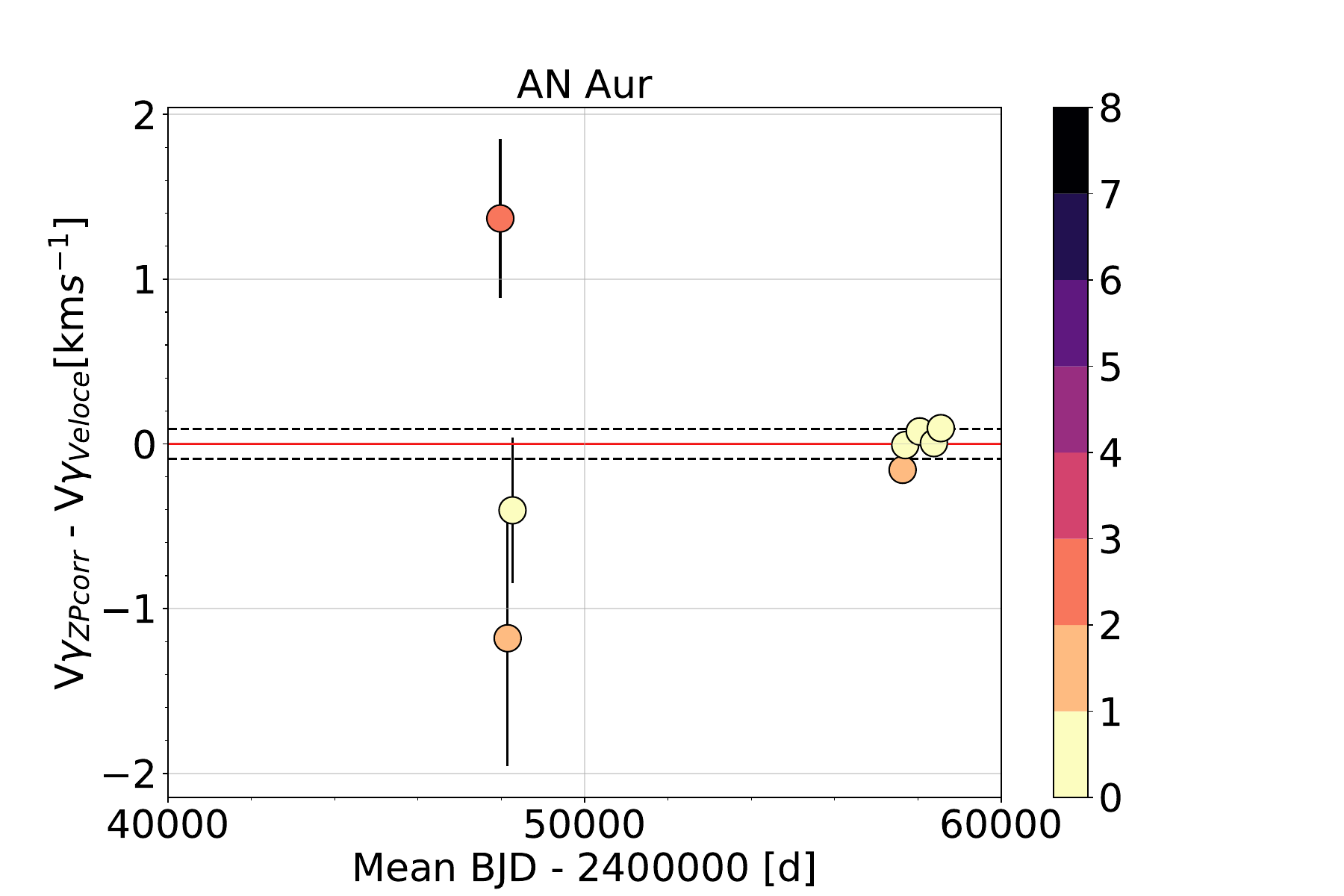}
    \end{subfigure}
    \begin{subfigure}{.35\textwidth}
    \includegraphics[scale=0.25,trim={4cm 0 0 1.5cm}]{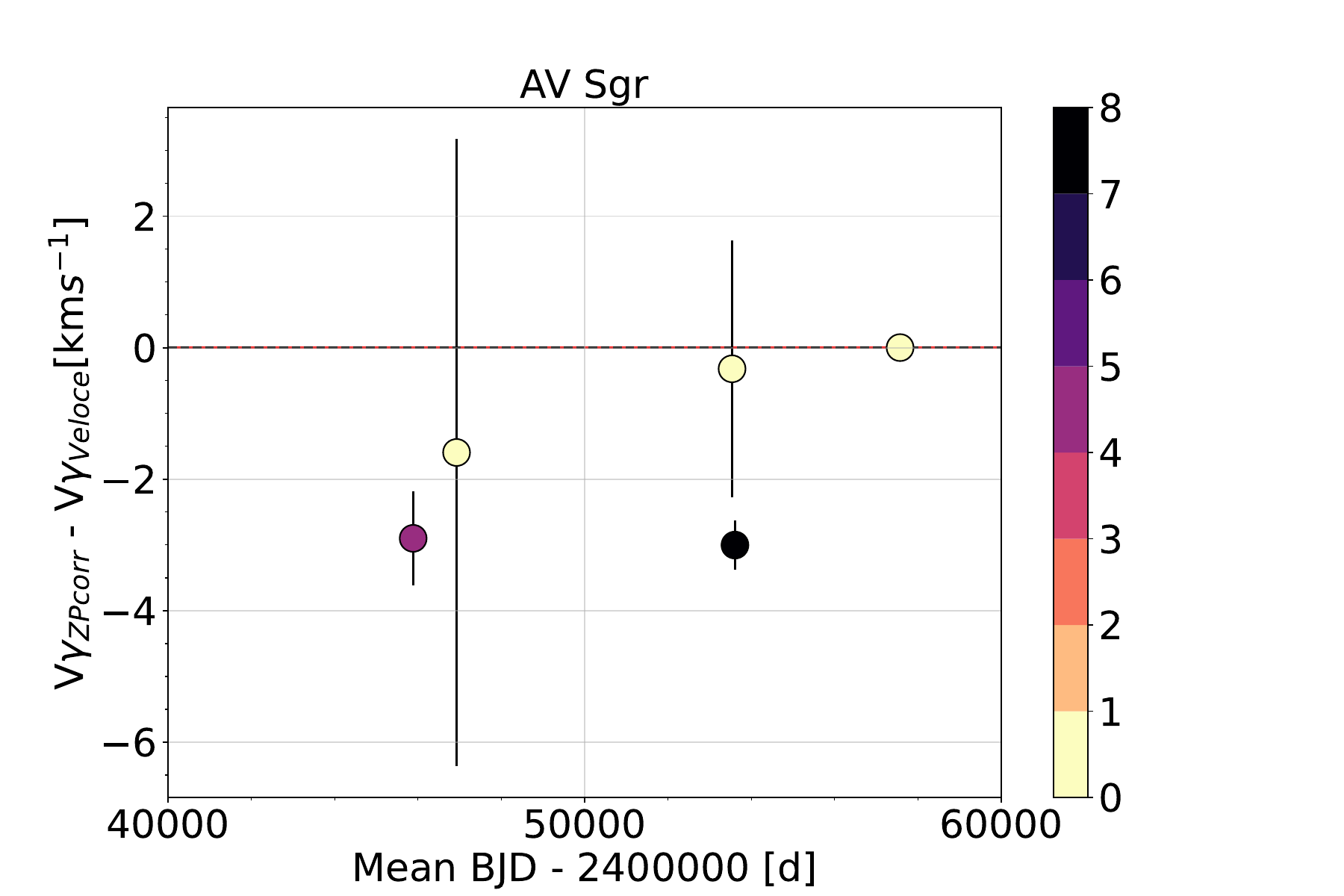}
    \end{subfigure}
    \newline
    
    \begin{subfigure}{.35\textwidth}
    \includegraphics[scale=0.25,trim={4cm 0 0 0cm}]{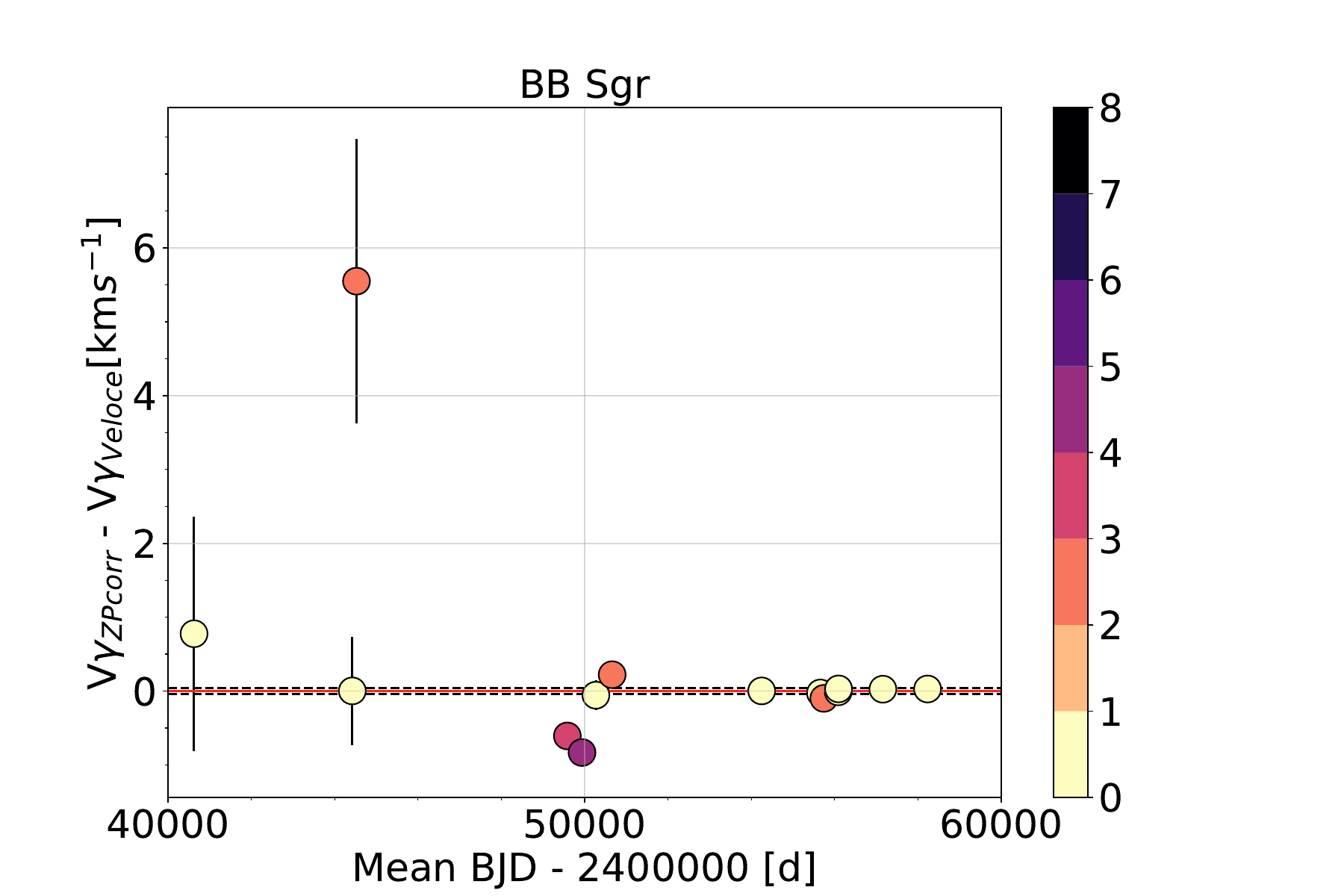}
    \end{subfigure}
    \begin{subfigure}{.35\textwidth}
    \includegraphics[scale=0.25,trim={4cm 0 0 0cm}]{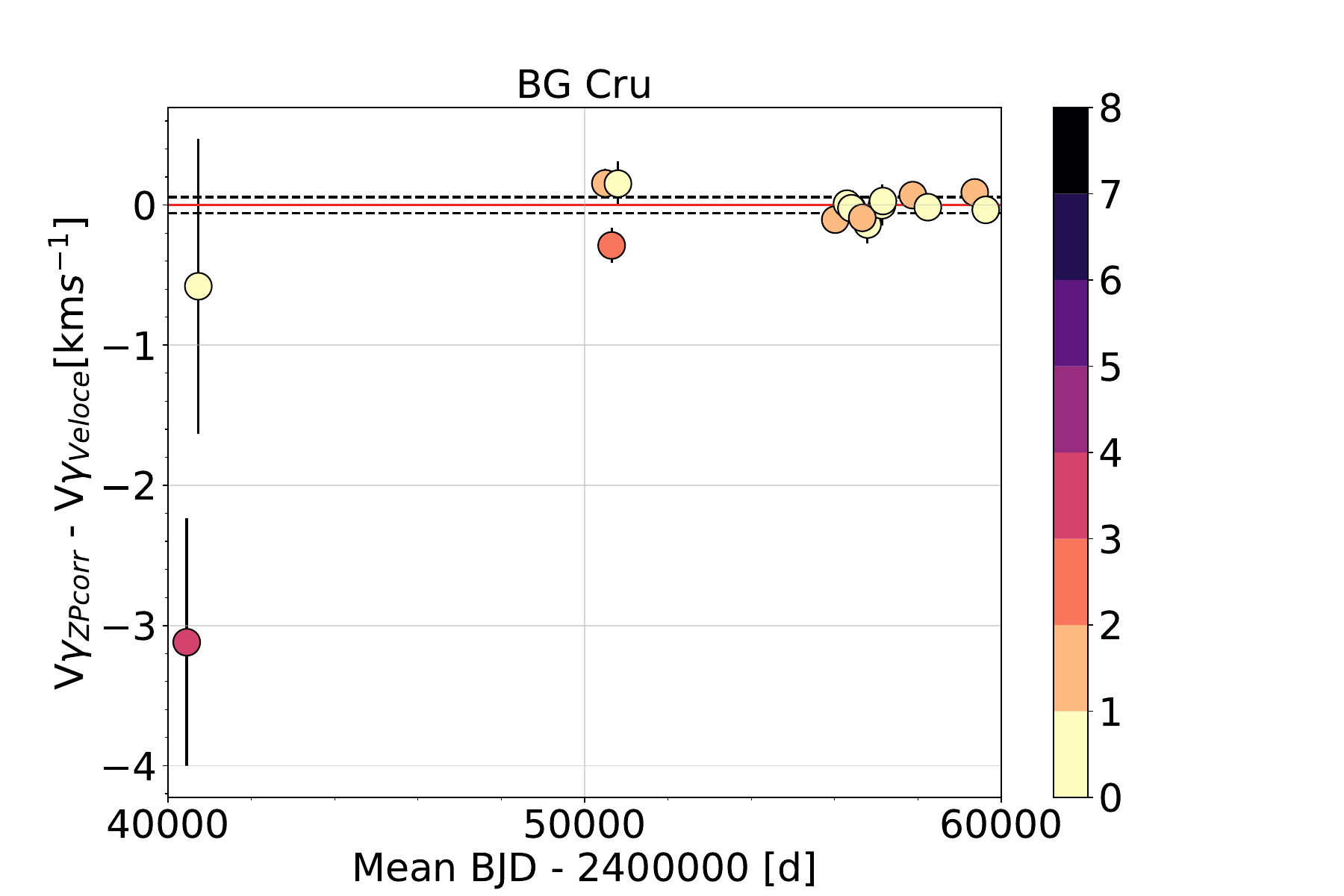}
    \end{subfigure}
    \begin{subfigure}{.35\textwidth}
    \includegraphics[scale=0.25,trim={4cm 0 0 0cm}]{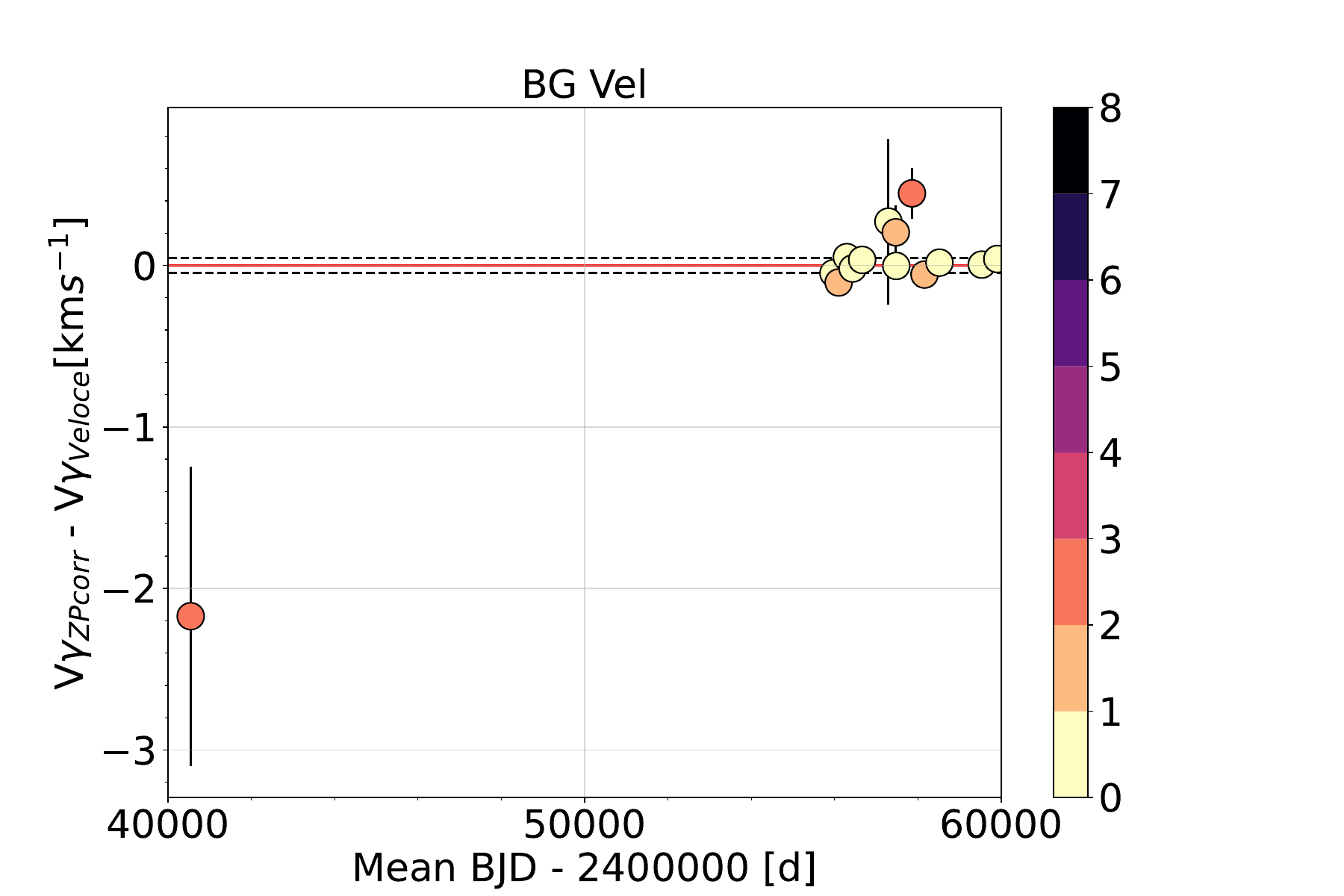}
    \end{subfigure}
    \newline
    
    \begin{subfigure}{.35\textwidth}
    \includegraphics[width=6cm,height=5cm,trim={4cm 0 3cm 0}]{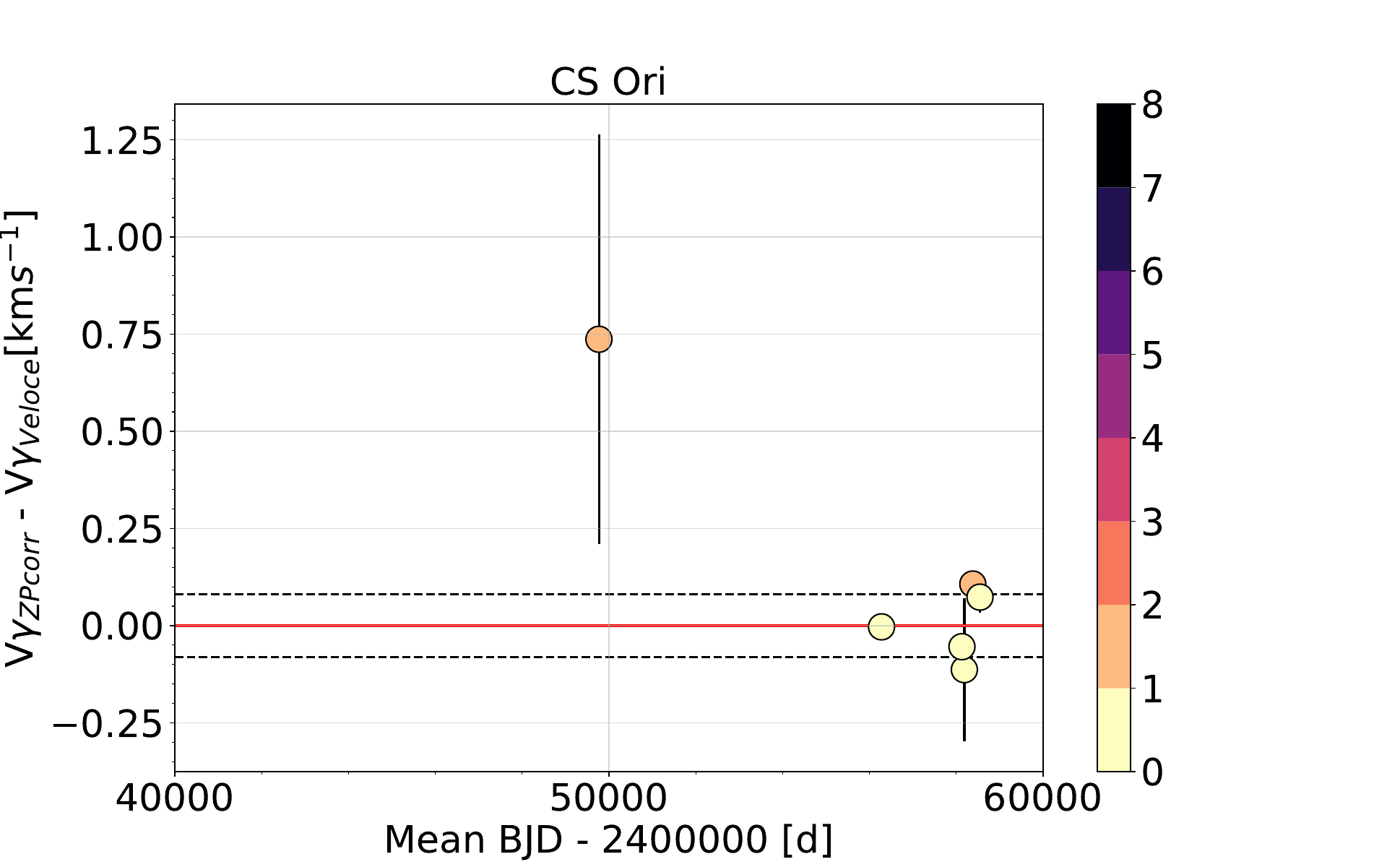}
    \end{subfigure}
    \begin{subfigure}{.35\textwidth}
    \includegraphics[scale=0.25,trim={4cm 0 0 0cm}]{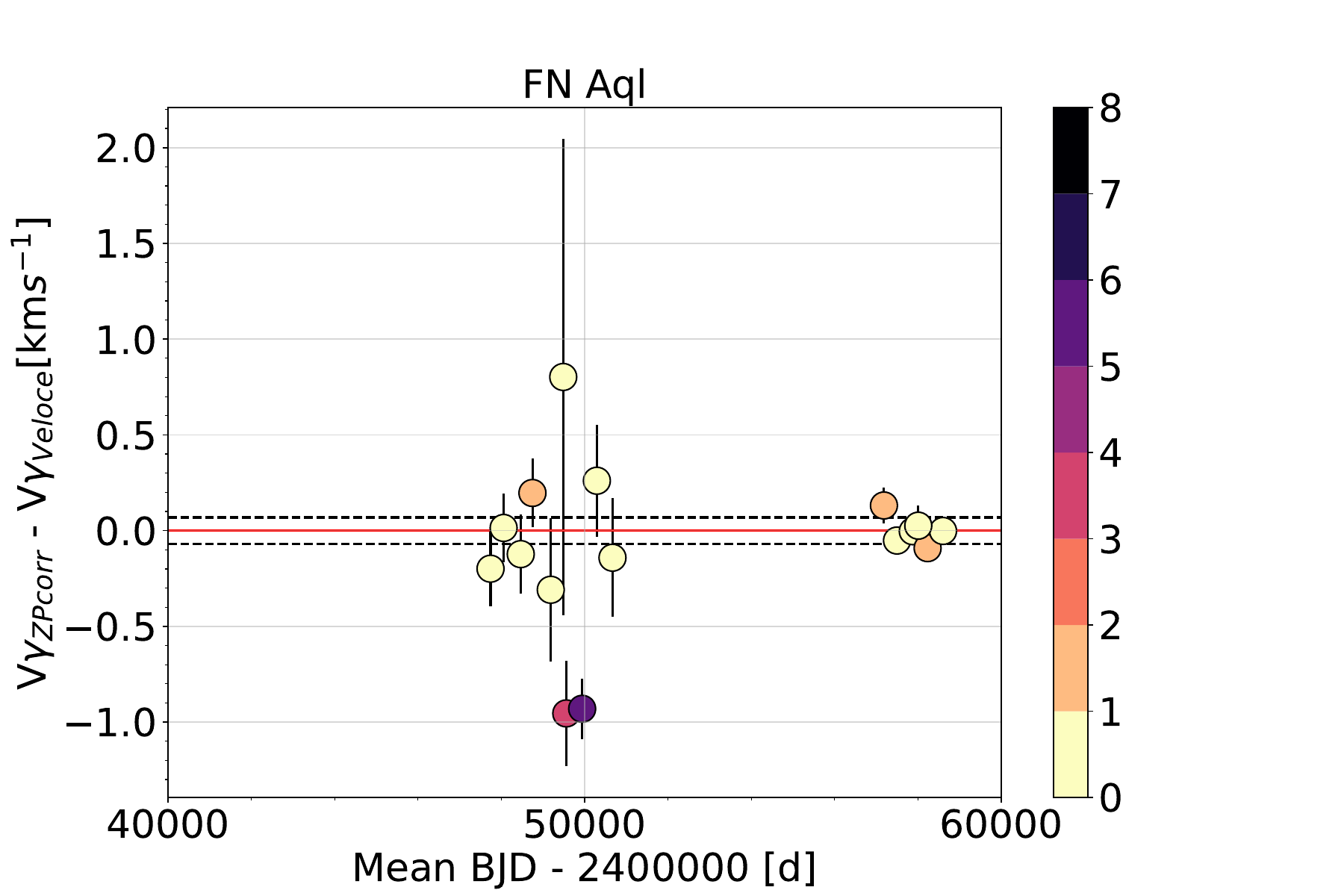}
    \end{subfigure}
    \begin{subfigure}{.35\textwidth}
    \includegraphics[scale=0.25,trim={4cm 0 0 0cm}]{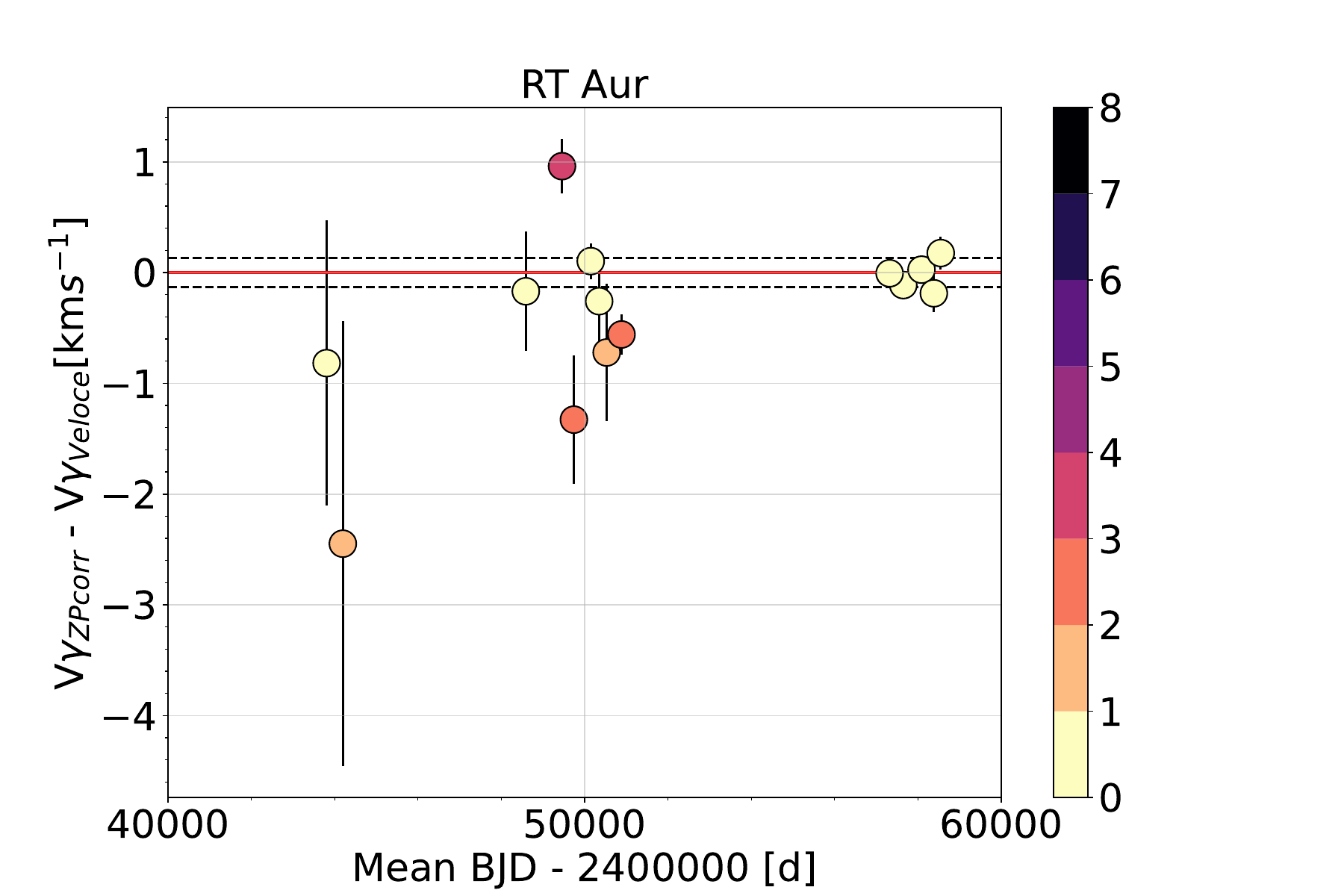}
    \end{subfigure}
    \newline
    
    \begin{subfigure}{.35\textwidth}
    \includegraphics[width=6cm,height=5cm,trim={4cm 0 3cm 0cm}]{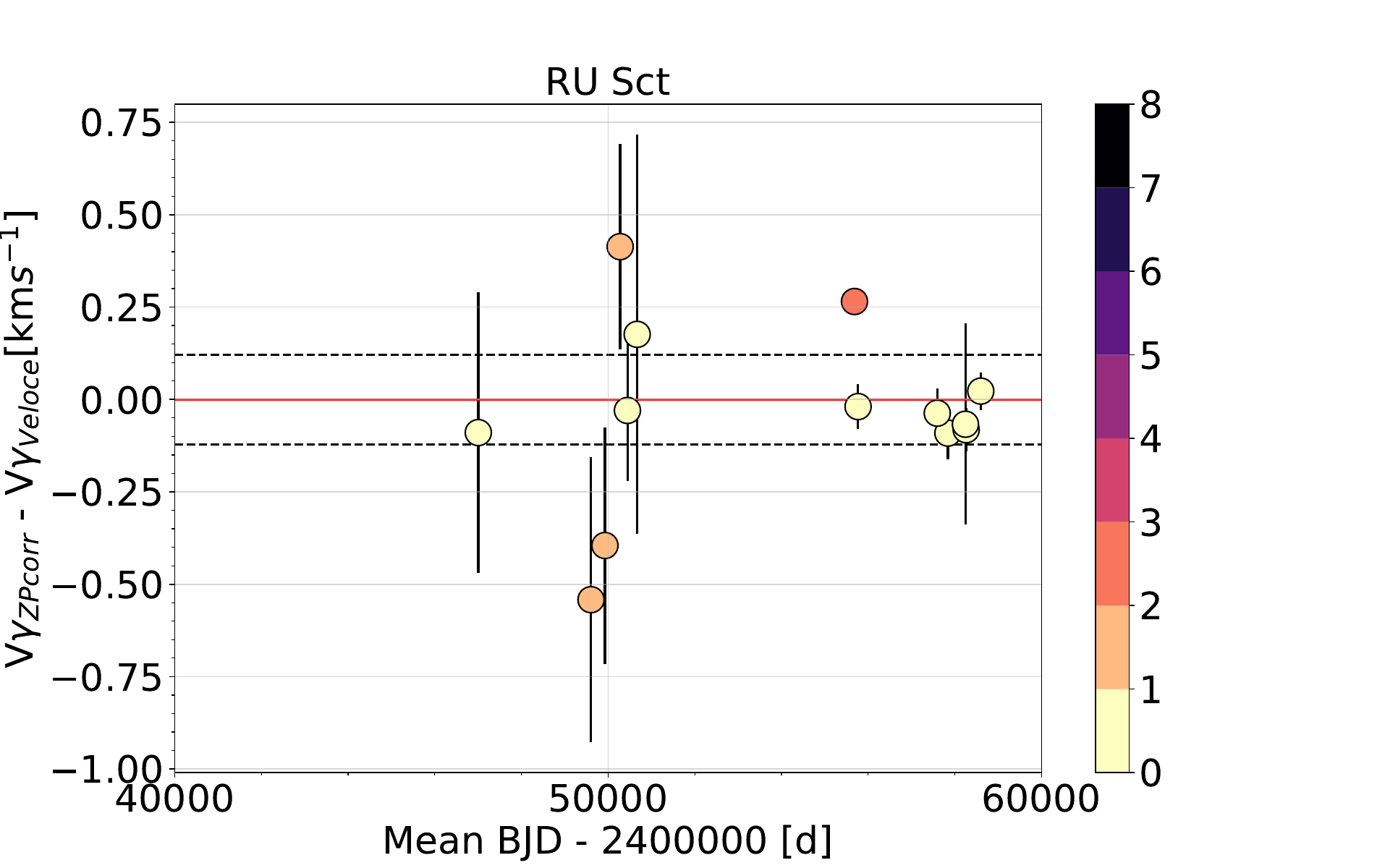}
    \end{subfigure}
    \begin{subfigure}{.35\textwidth}
    \includegraphics[scale=0.25,trim={4cm 0 0 0cm}]{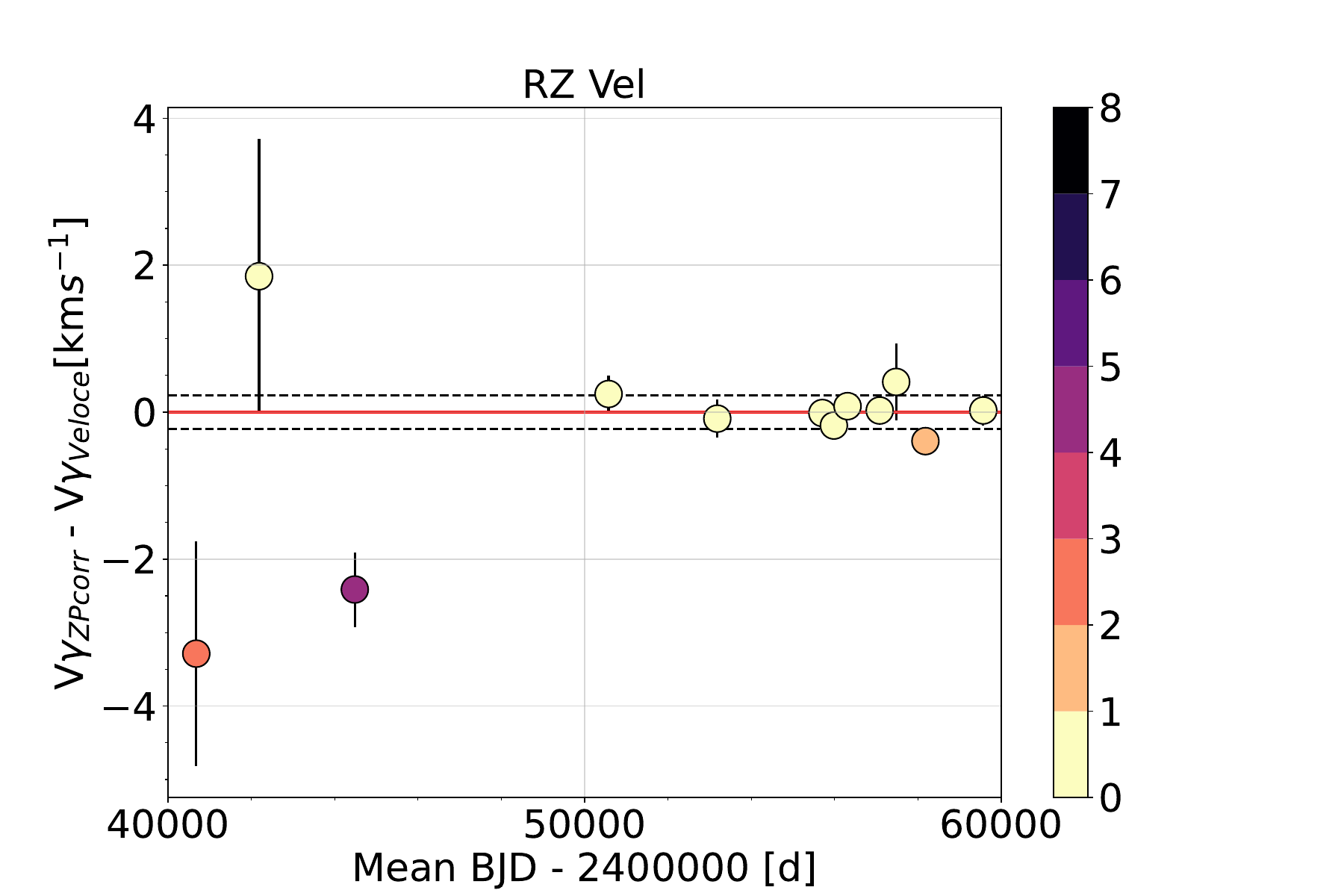}
    \end{subfigure}
    \begin{subfigure}{.35\textwidth}
    \includegraphics[scale=0.25,trim={4cm 0 0 0cm}]{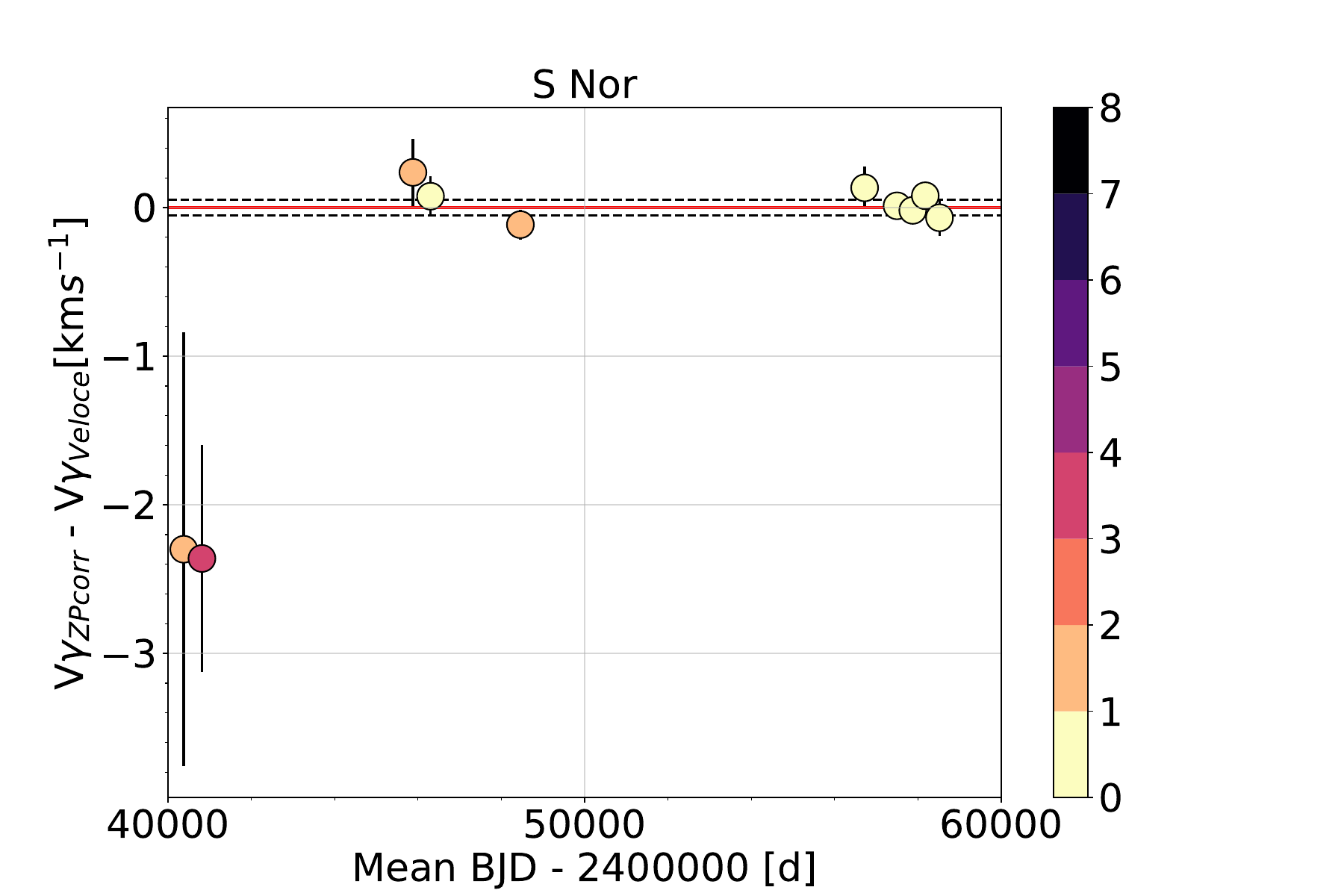}
    \end{subfigure}
    \newline
    
    \begin{subfigure}{.35\textwidth}
    \includegraphics[scale=0.25,trim={4cm 0 0 0cm}]{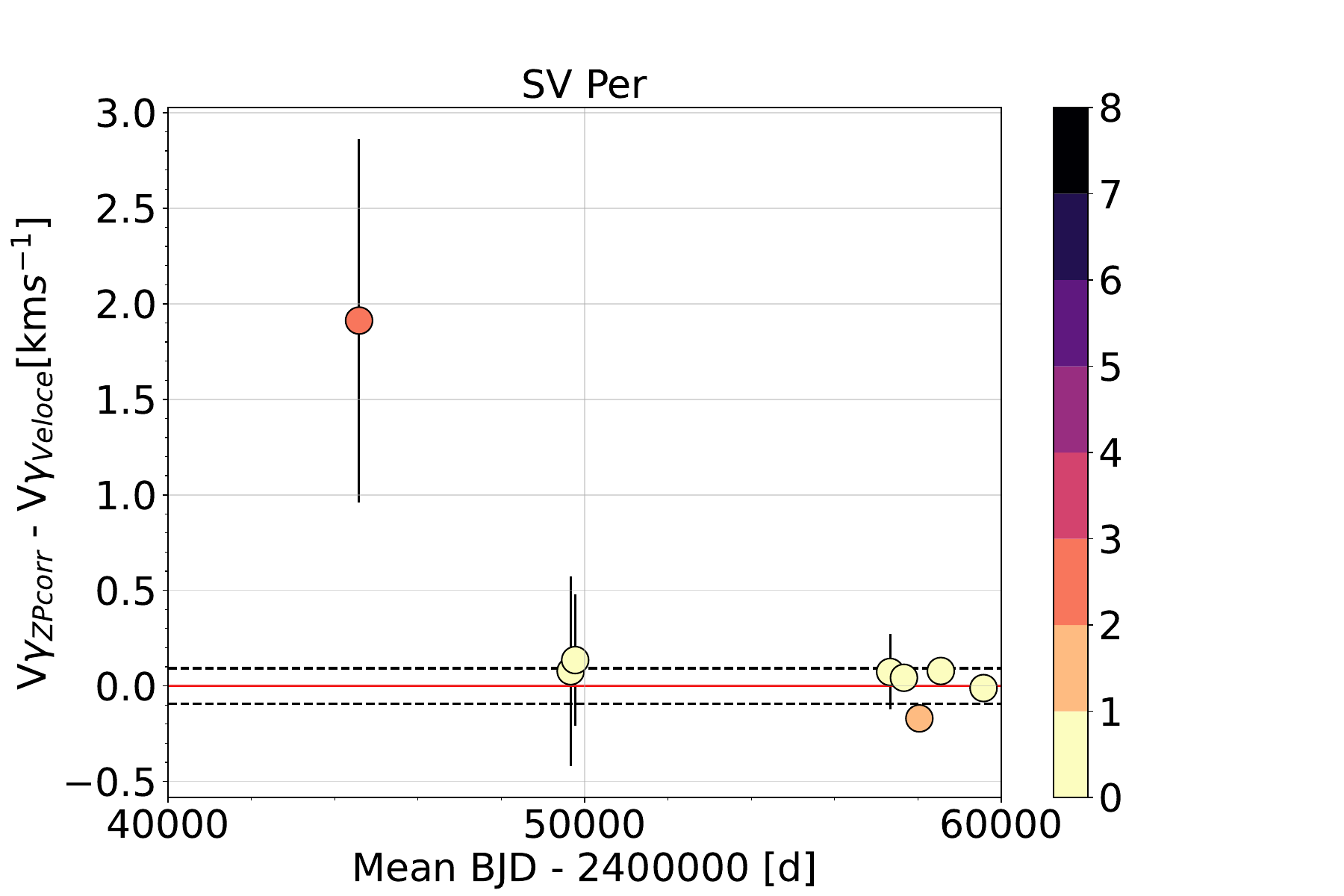}
    \end{subfigure}
    \begin{subfigure}{.35\textwidth}
    \includegraphics[scale=0.25,trim={4cm 0 0 0cm}]{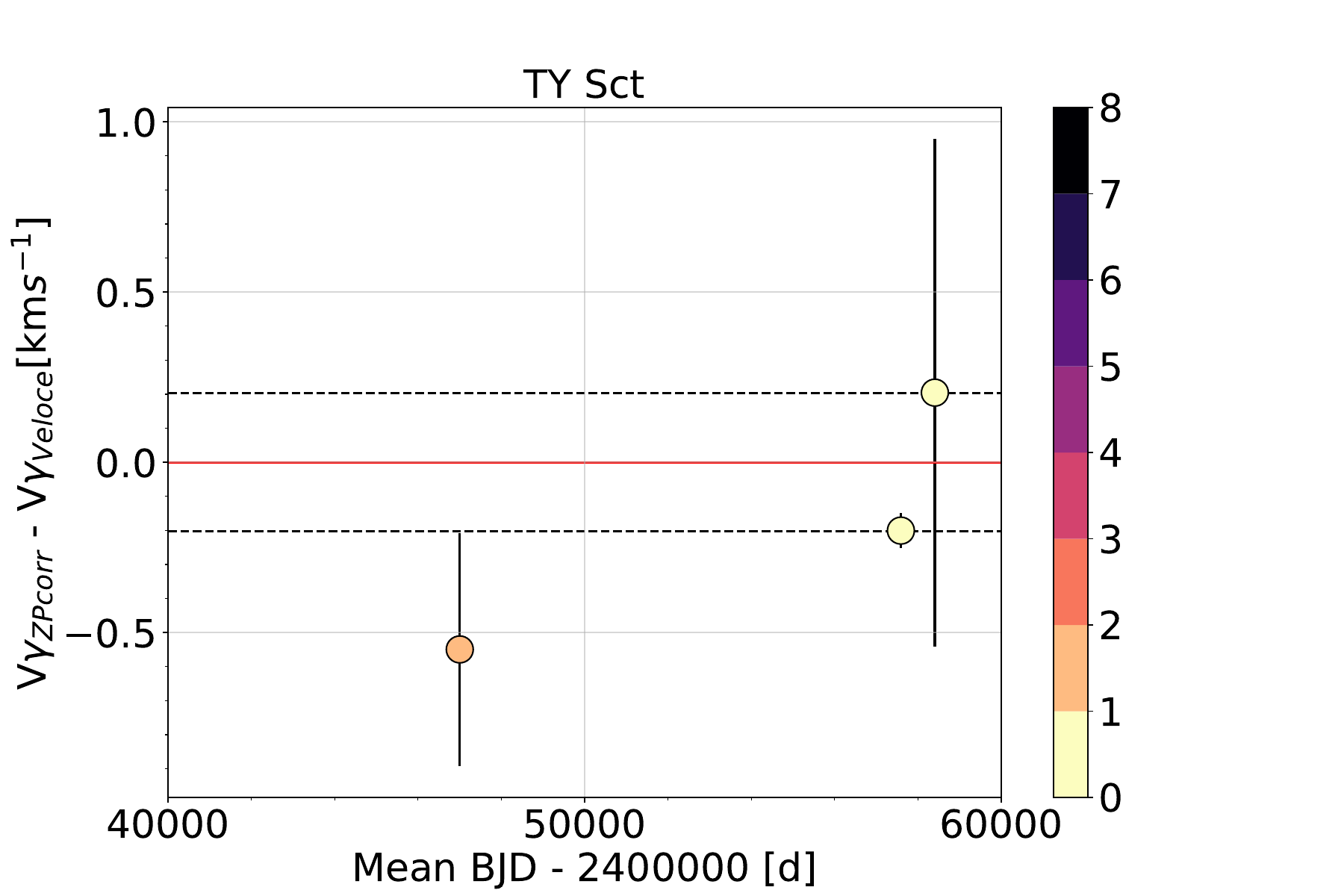}
    \end{subfigure}
    \begin{subfigure}{.35\textwidth}
    \includegraphics[scale=0.25,trim={4cm 0 0 0cm}]{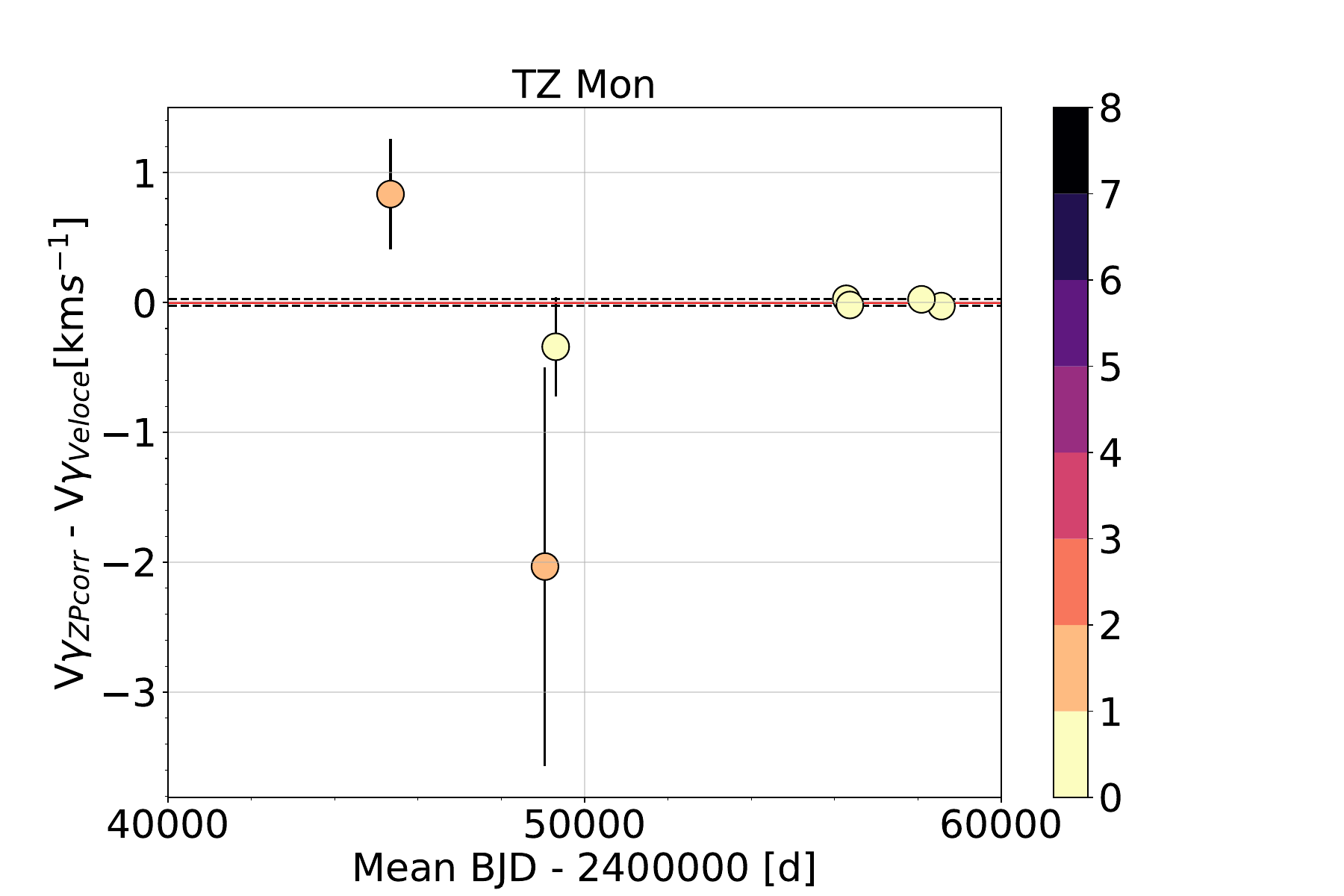}
    \end{subfigure}
    \newline

\caption{Same as Figure~\ref{fig:VgammaforSB1signs} for stars within \gc~with no signs of SB1 from their $v_{\gamma}$ RVTF analysis (contd. on the next page.)} 
\label{fig:VgammafornoSB1signs1}
\end{figure*}

\begin{figure*}\ContinuedFloat
\begin{subfigure}{.35\textwidth}
    \includegraphics[scale=0.25,trim={4cm 0 0 0cm}]{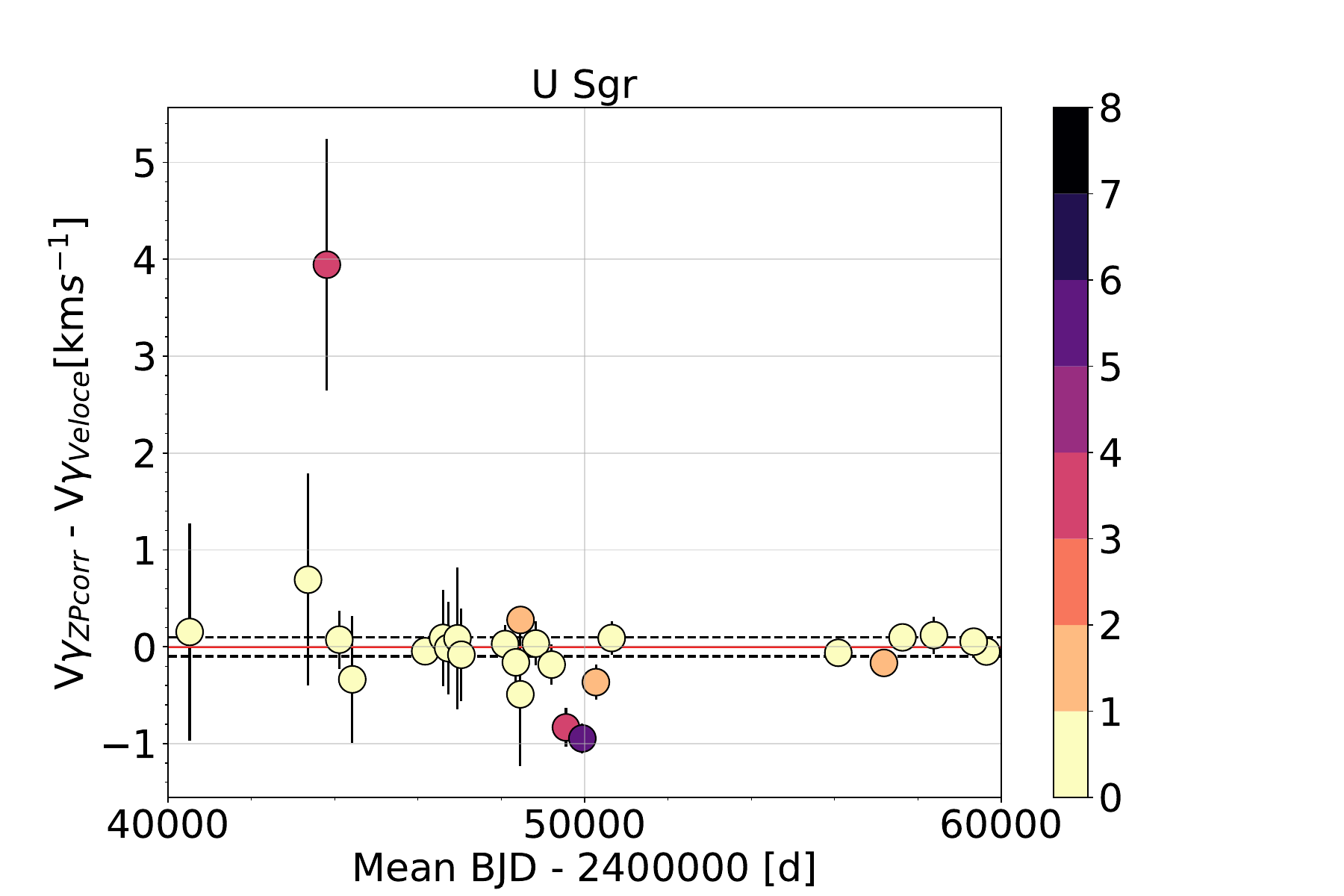}
    \end{subfigure}
\begin{subfigure}{.35\textwidth}
    \includegraphics[scale=0.25,trim={4cm 0 0 0cm}]{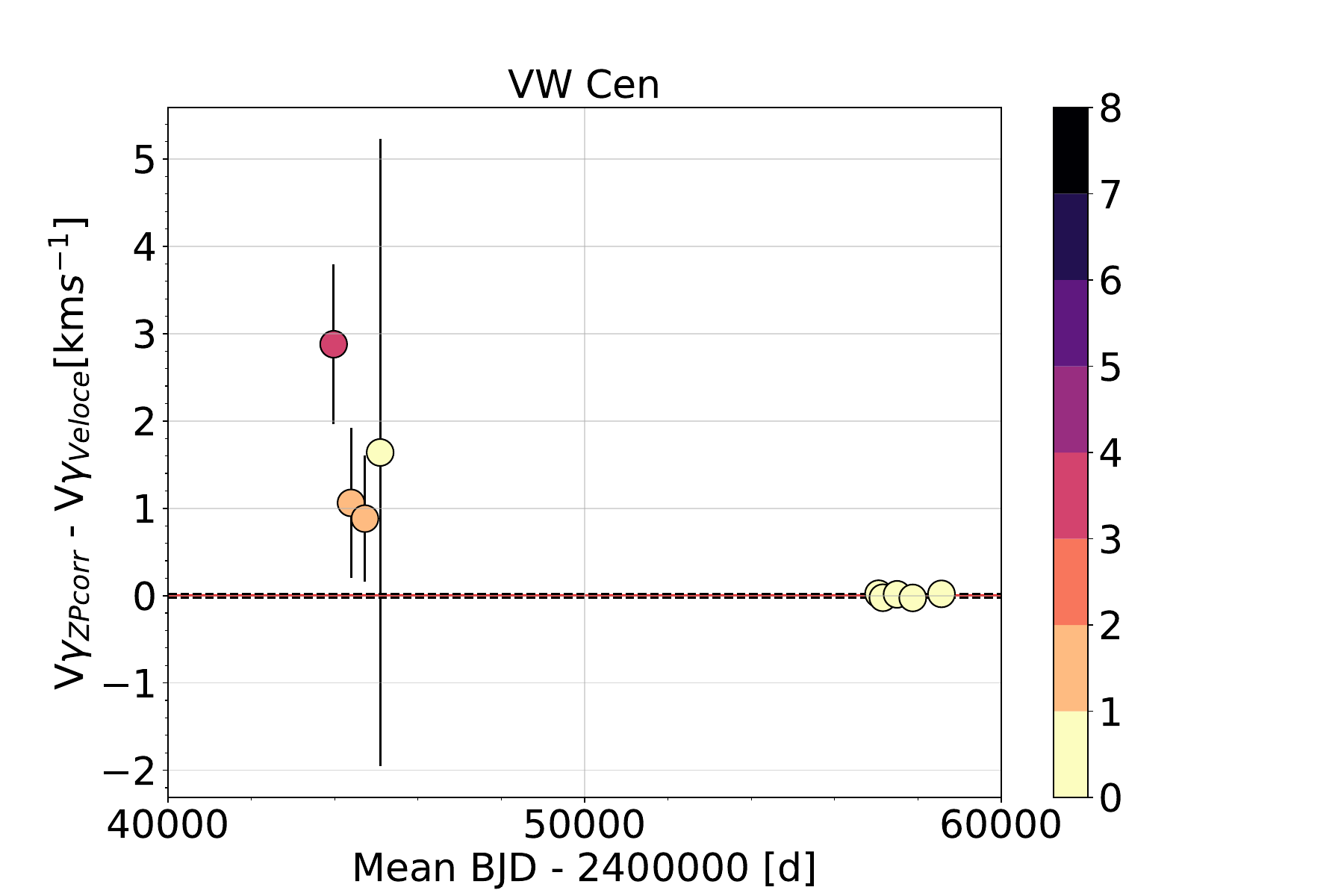}
    \end{subfigure}
    \begin{subfigure}{.35\textwidth}
    \includegraphics[scale=0.25,trim={4cm 0 0 0cm}]{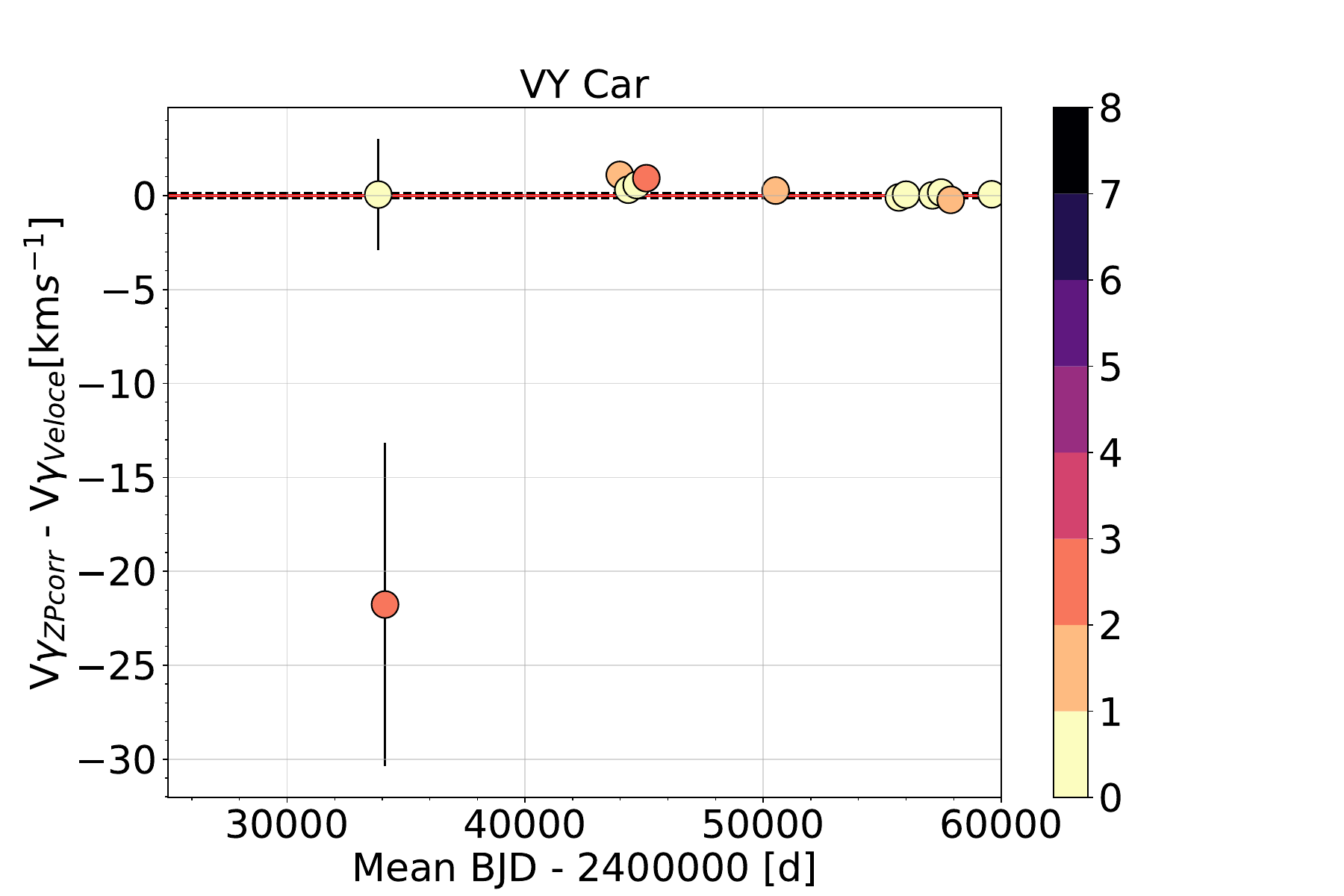}
    \end{subfigure}
\newline

    \begin{subfigure}{.35\textwidth}
    \includegraphics[scale=0.25,trim={4cm 0 0 0cm}]{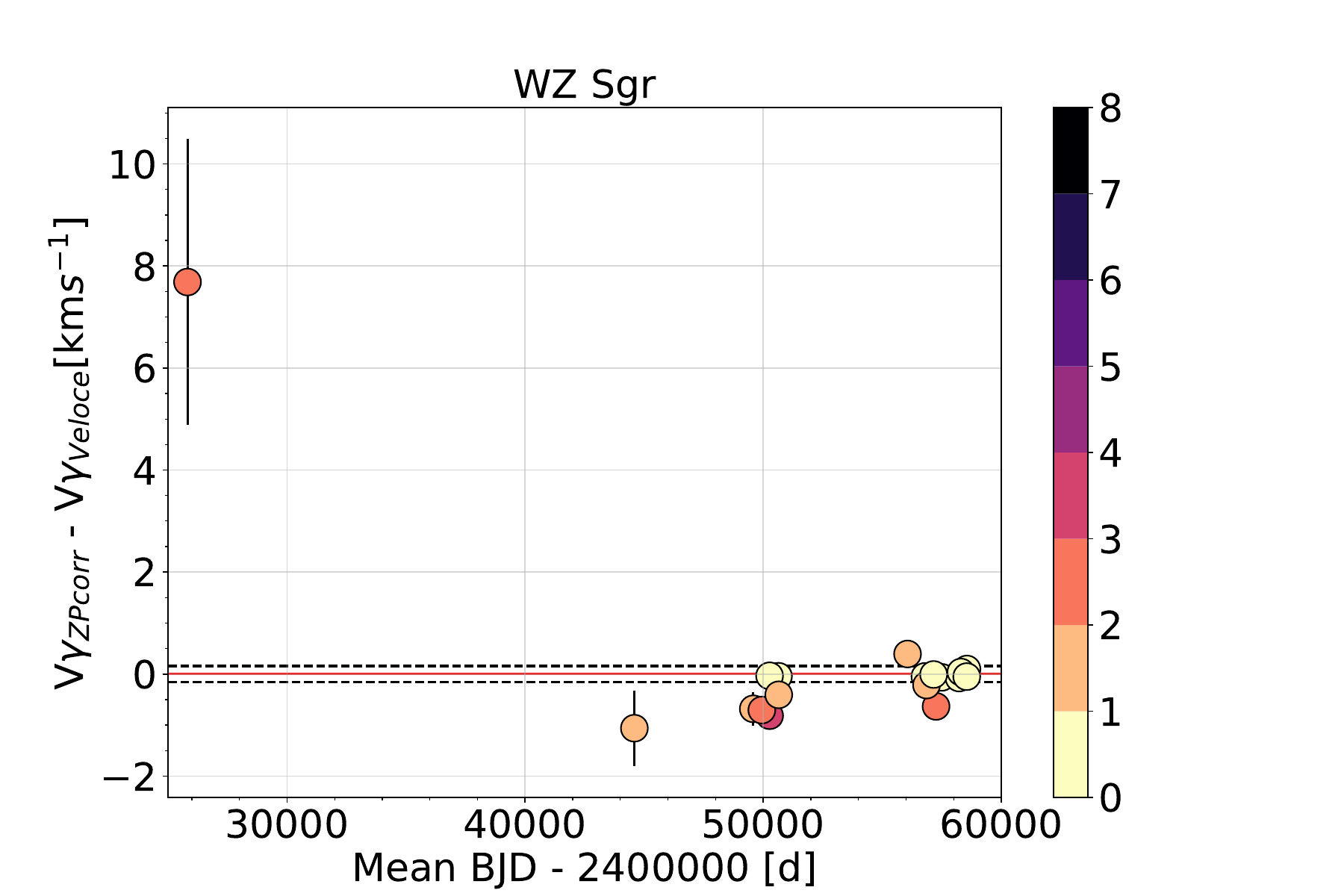}
    \end{subfigure}
    \begin{subfigure}{.35\textwidth}
    \includegraphics[scale=0.25,trim={4cm 0 0 0cm}]{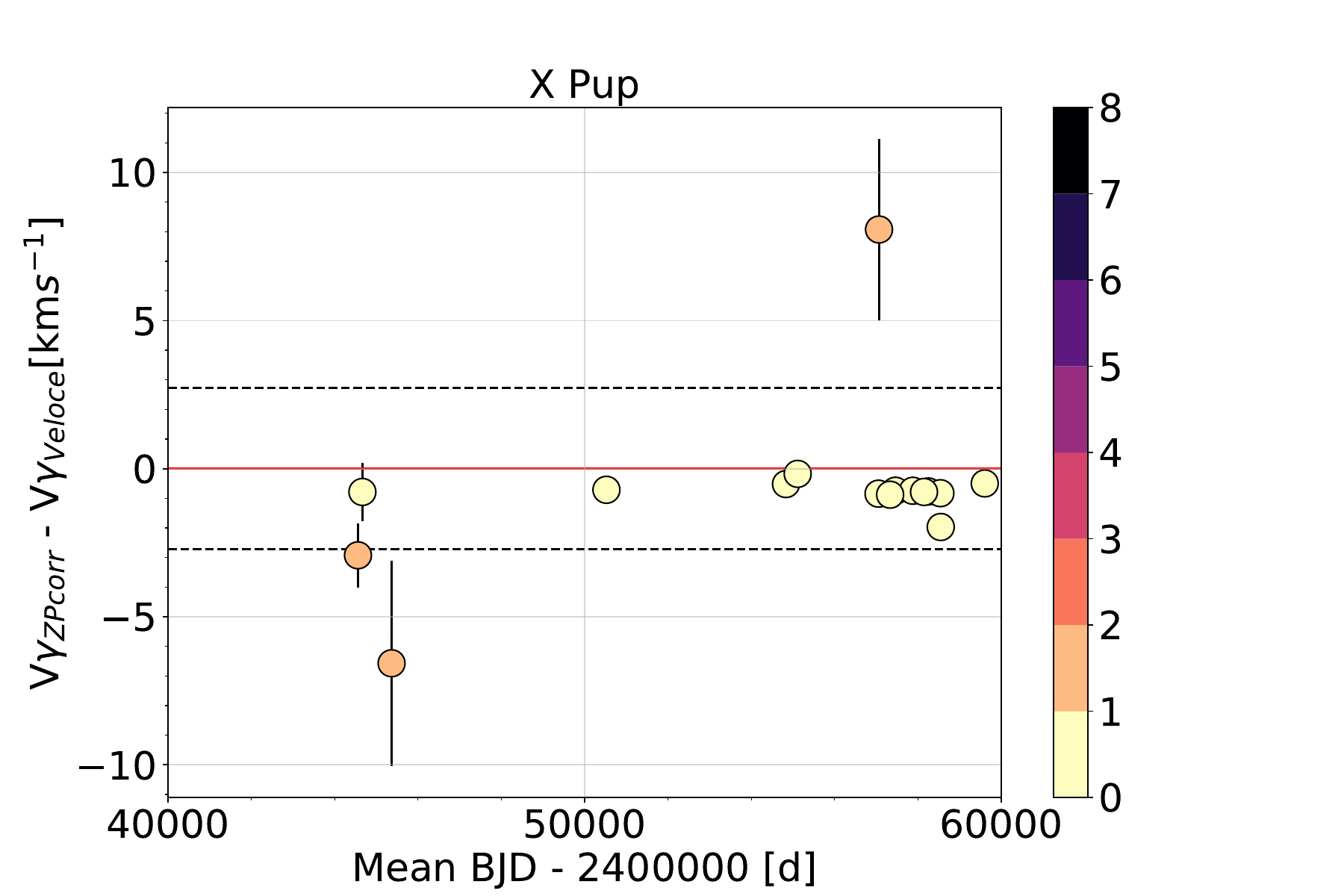}
    \end{subfigure}
    \begin{subfigure}{.35\textwidth}
    \includegraphics[scale=0.25,trim={4cm 0 0 0cm}]{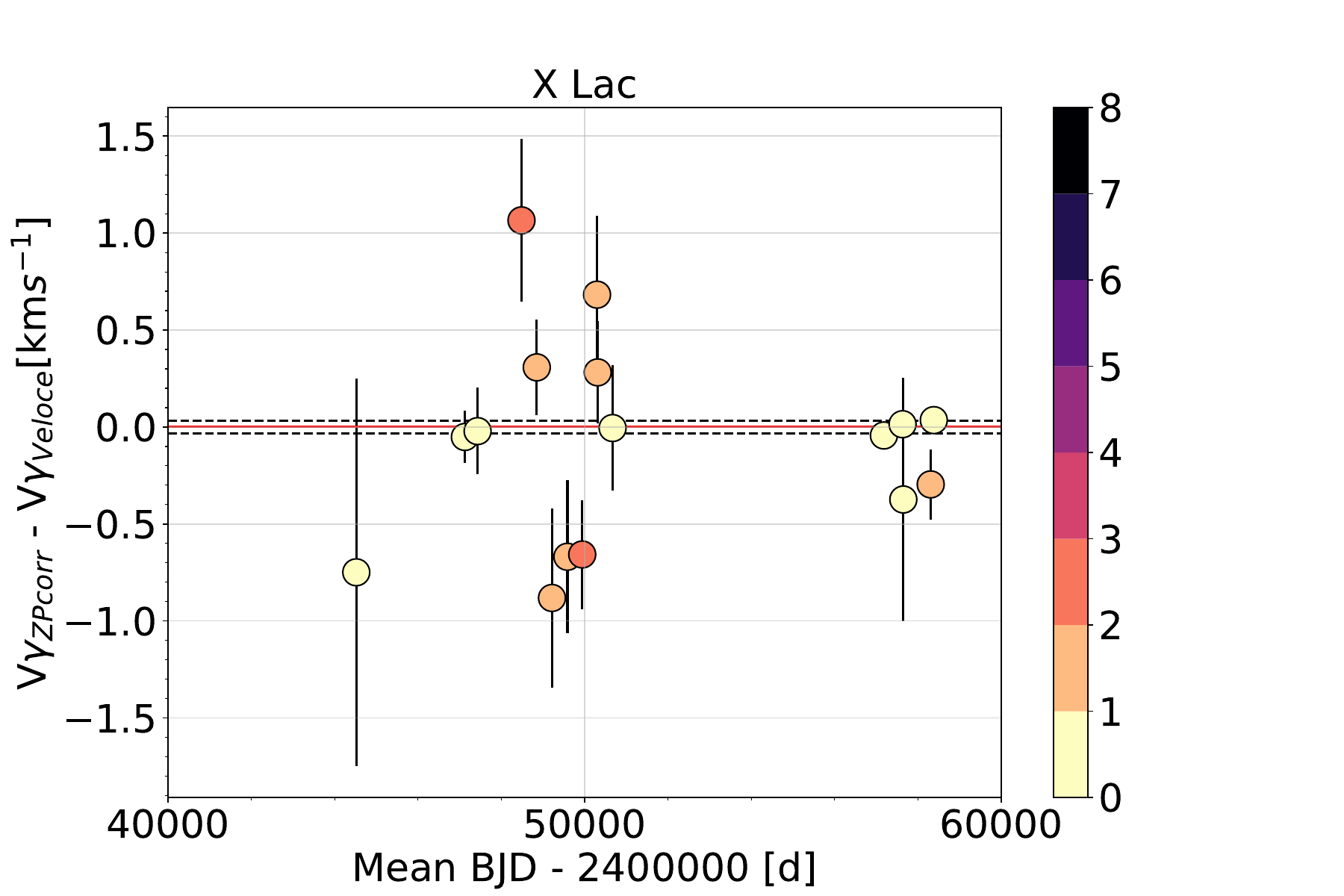}
    \end{subfigure}
    \newline
    
    \begin{subfigure}{.35\textwidth}
    \includegraphics[scale=0.25,trim={4cm 0 0 0cm}]{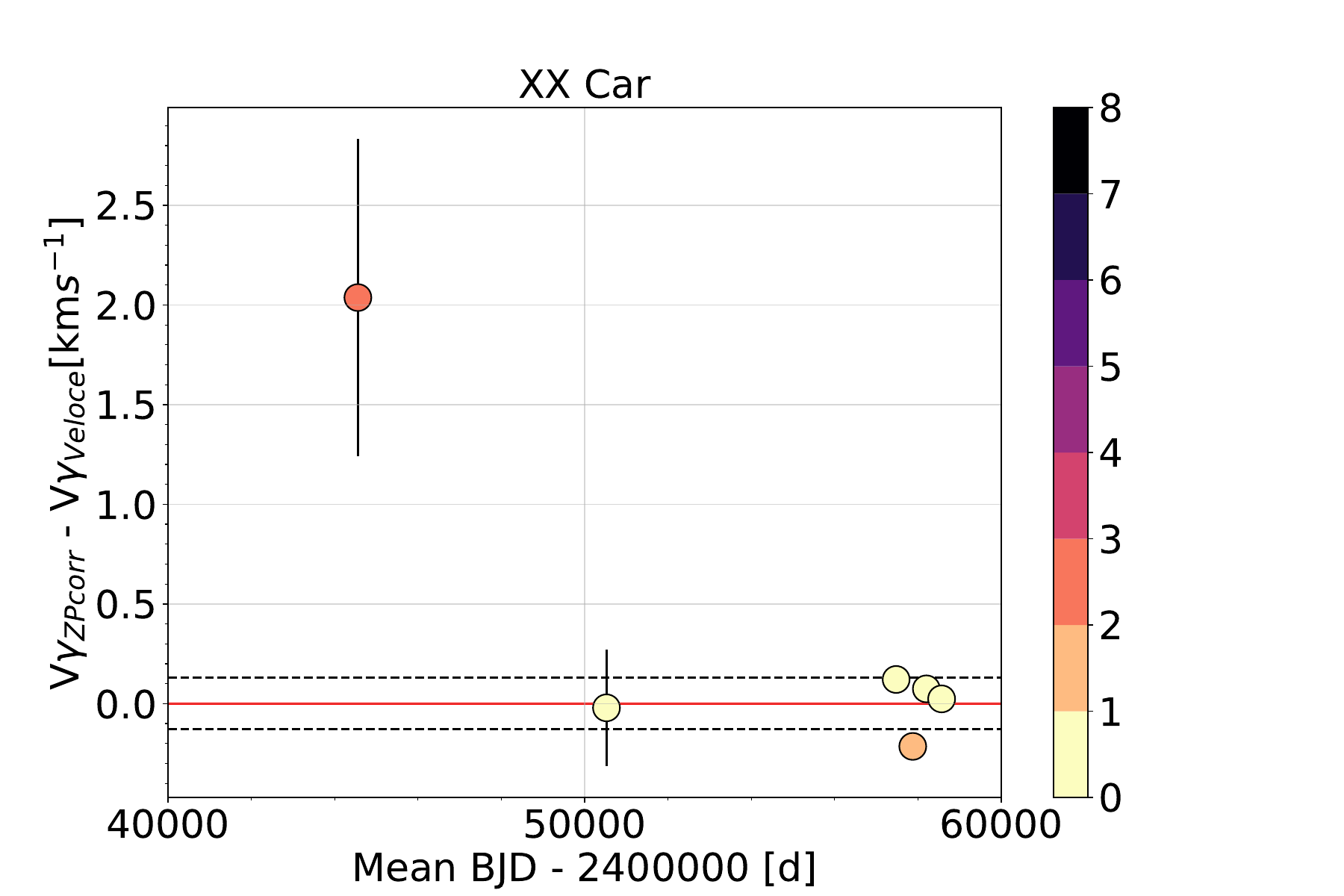}
    \end{subfigure}
    \begin{subfigure}{.35\textwidth}
    \includegraphics[scale=0.25,trim={4cm 0 0 0cm}]{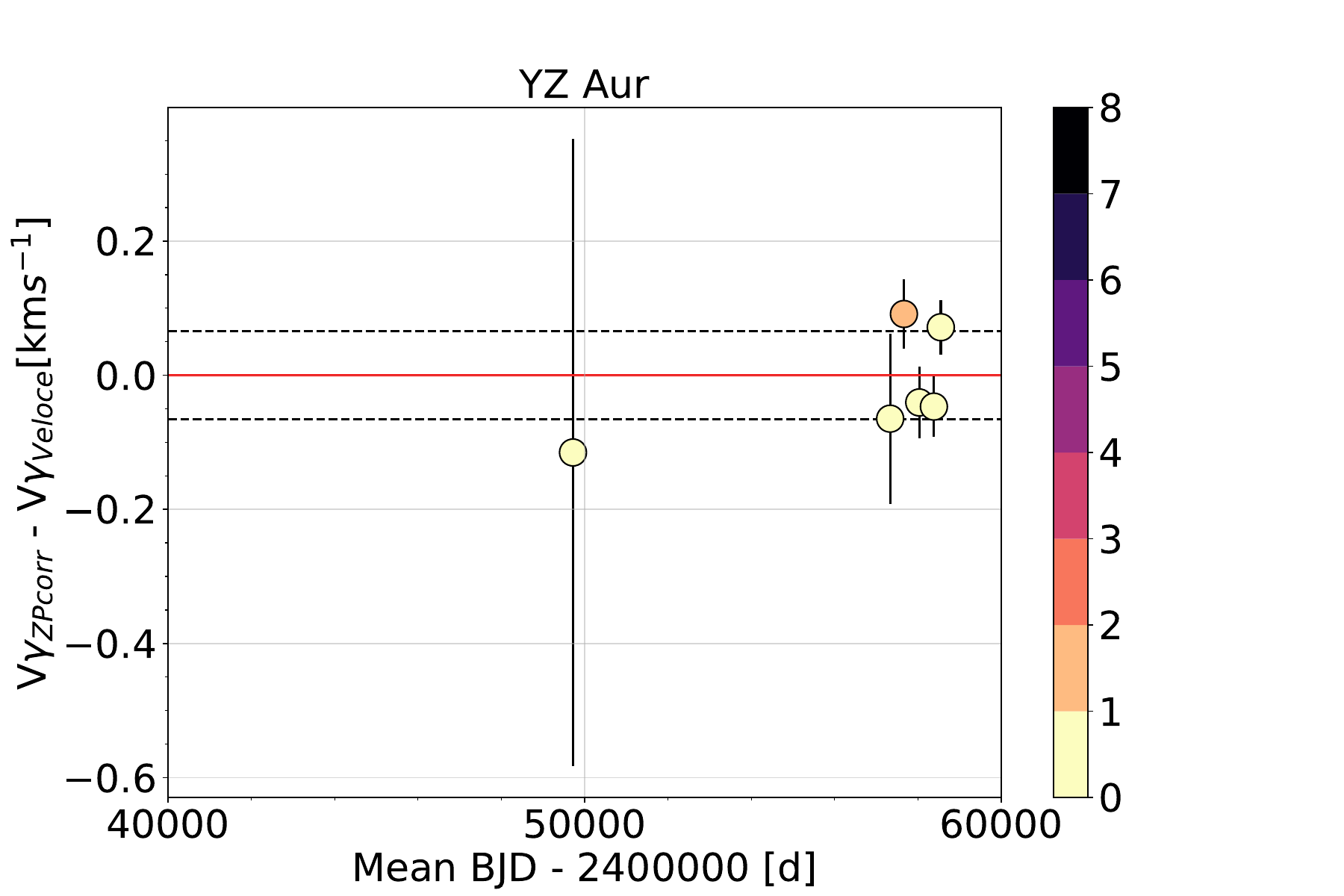}
    \end{subfigure}
    \begin{subfigure}{.35\textwidth}
    \includegraphics[scale=0.25,trim={4cm 0 0 0cm}]{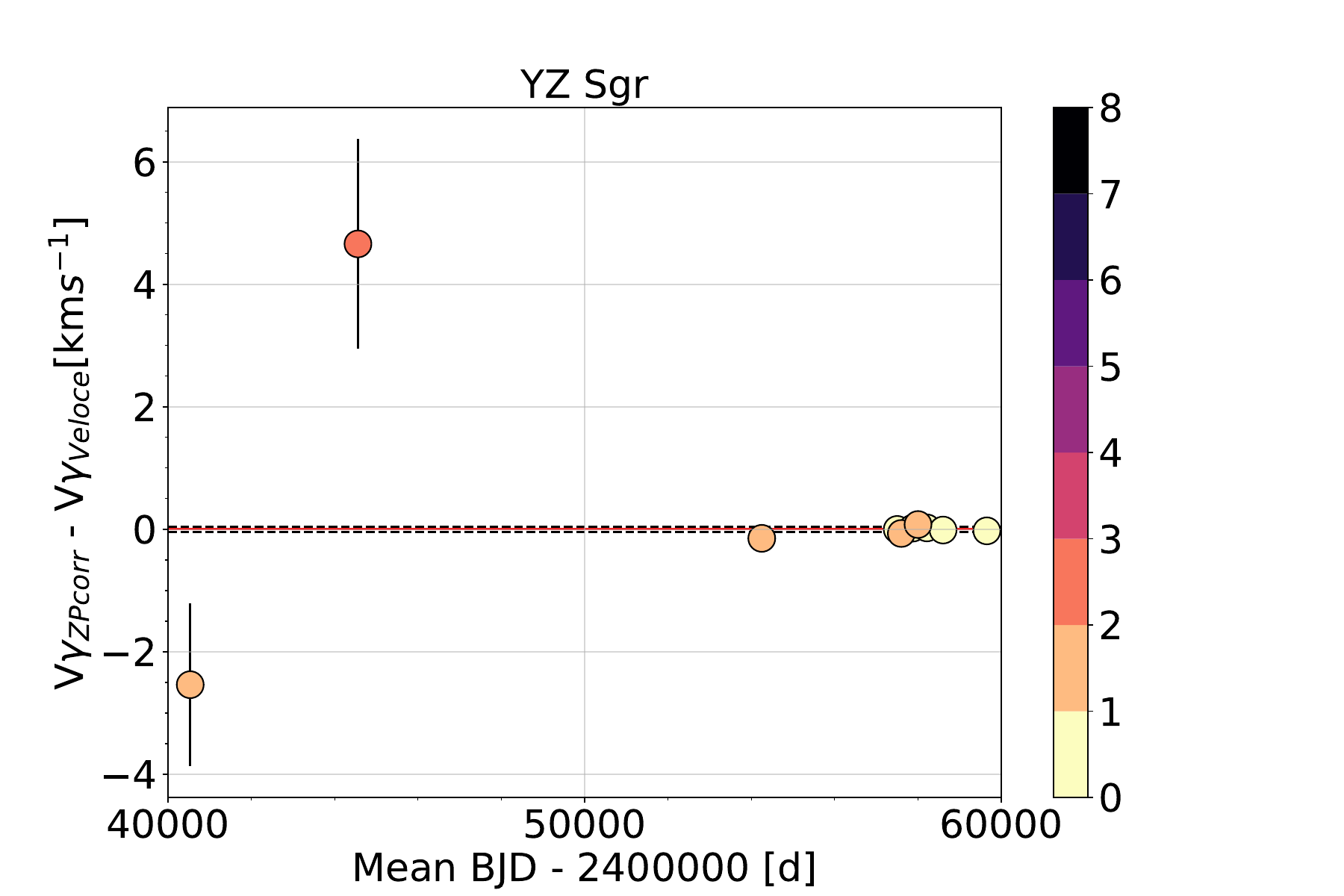}
    \end{subfigure}
\caption{Same as Figure~\ref{fig:VgammaforSB1signs} for stars within \gc~with no signs of SB1 from their $v_{\gamma}$ RVTF analysis.}
\label{fig:VgammafornoSB1signs}
\end{figure*}

\section{SU Cyg RVTF}\label{app:SUCygRVTF}
In Table~\ref{tab:SUCygVgamma}, we present the zero-point corrected \vgamma~residuals (\vgammares) of SU Cyg. The details of how these were obtained are provided in Section~\ref{sec:individual}.

\begin{table*}
\centering
\caption{\label{tab:SUCygVgamma}
 RV data used for the `V+L' orbit estimation of SU~Cyg. }
\sisetup{round-mode=places}
\setlength{\tabcolsep}{1pt}
    \begin{tabular}{lS[round-precision=2]S[round-precision=2]S[round-precision=2]|lS[round-precision=2]S[round-precision=2]S[round-precision=2]|lS[round-precision=2]S[round-precision=2]S[round-precision=2]}
    \hline
Reference & {BJD} & {\vgammares} & {$\sigma \epsilon v_\gamma$} & Reference & {BJD} & {\vgammares} & {$\sigma \epsilon v_\gamma$}& Reference & {BJD} & {\vgammares} & {$\sigma \epsilon v_\gamma$} \\
& {(d)} & {(\kms)} & {(\kms)} &  & {(d)} & {(\kms)} & {(\kms)}  & {(d)} & {(\kms)} & {(\kms)}\\
    \hline

        Ba87 & 43616.976 & -25.28023047 & 5.34293901 & Bo19 & 56566.333 & 4.11970908 & 2.05593092 & Go92 & 50705.305 & -14.31117146 & 0.74274243 \\ 
        Ba87 & 43617.976 & -29.48715733 & 5.34293901 & Bo19 & 56898.519 & -55.51837238 & 0.39148889 & Go92 & 50707.324 & -16.53896313 & 0.80874367 \\ 
        Ba87 & 43618.974 & -18.86346588 & 5.34293901 & Bo19 & 56899.57 & -57.23200214 & 0.39101861 & Go92 & 50708.346 & -21.88836071 & 0.88032172 \\ 
        Ba87 & 43620.987 & -33.88445608 & 5.34293901 & Bo19 & 56900.441 & -56.70302236 & 0.39101861 & Go92 & 50708.352 & -21.97346751 & 0.79433389 \\ 
        Ba87 & 43622.956 & -36.95755498 & 5.34293901 & Bo19 & 56901.565 & -57.03608269 & 0.39077302 & Go92 & 50709.287 & -16.2597389 & 0.76037249 \\ 
        Ba87 & 43683.857 & -44.61876332 & 4.49633488 & Go92 & 50236.429 & -31.46581342 & 1.11497536 & Go92 & 50711.246 & -15.71130328 & 0.77650906 \\ 
        Ba87 & 43683.859 & -38.25956644 & 4.49633488 & Go92 & 50238.403 & -31.7893406 & 1.1088598 & Go92 & 50713.282 & -16.35894371 & 0.75297166 \\ 
        Ba87 & 43684.842 & -41.69601284 & 4.49633488 & Go92 & 50245.479 & -36.12168694 & 1.14641618 & Go92 & 50714.38 & -16.70102517 & 0.76822283 \\ 
        Ba87 & 43684.843 & -36.56370212 & 4.49633488 & Go92 & 50246.511 & -36.21576377 & 1.20231862 & Go92 & 50715.252 & -13.86145184 & 0.81373603 \\ 
        Ba87 & 43684.845 & -37.21905639 & 4.49633488 & Go92 & 50247.497 & -32.81919954 & 1.22178969 & Go92 & 50716.29 & -20.97562062 & 0.87433765 \\ 
        Ba87 & 43685.771 & -39.74771192 & 4.49633488 & Go92 & 50248.467 & -35.83014869 & 1.15826165 & Go92 & 50718.269 & -16.76915087 & 0.77231232 \\ 
        Ba87 & 43686.806 & -41.02834678 & 4.49633488 & Go92 & 50255.418 & -38.29848857 & 0.7798799 & Go92 & 50719.242 & -13.28993823 & 0.80384471 \\ 
        Ba87 & 43688.835 & -39.12571853 & 4.49633488 & Go92 & 50257.462 & -36.84090296 & 0.78905809 & Go92 & 50719.247 & -13.30910086 & 0.74944401 \\ 
        Ba87 & 43713.943 & -54.4614314 & 4.49633488 & Go92 & 50259.409 & -38.76487488 & 0.76276645 & Go92 & 50720.34 & -17.8255758 & 0.80874367 \\ 
        Ba87 & 43821.584 & -31.42453168 & 4.13751386 & Go92 & 50261.414 & -39.1094383 & 0.80356248 & Go92 & 50726.327 & -20.37146428 & 0.80731126 \\ 
        Ba87 & 43822.573 & -29.6469279 & 4.13751386 & Go92 & 50264.479 & -40.48347103 & 0.78441868 & Go92 & 50727.287 & -16.21734745 & 0.87232533 \\ 
        Ba87 & 43825.665 & -27.99658678 & 4.13751386 & Go92 & 50265.406 & -40.77488312 & 0.85773694 & Go92 & 50728.332 & -21.39518947 & 0.79589665 \\ 
        Ba87 & 44044.873 & -9.42635861 & 4.15036544 & Go92 & 50275.458 & -45.7788587 & 1.28988889 & Go92 & 50729.313 & -18.54641752 & 0.84702507 \\ 
        Ba87 & 44045.888 & -5.06195871 & 4.15036544 & Go92 & 50276.432 & -47.11672825 & 1.2670096 & Go92 & 50730.301 & -20.42124266 & 0.82398512 \\ 
        Ba87 & 44046.878 & -7.47513445 & 4.15036544 & Go92 & 50277.405 & -46.35299683 & 1.39158663 & Go92 & 50731.251 & -22.28141884 & 0.78882918 \\ 
        Ba87 & 44059.867 & -12.19289049 & 4.15036544 & Go92 & 50278.42 & -47.39613221 & 1.29881228 & Go92 & 50732.233 & -20.55278154 & 0.83291745 \\ 
        Ba87 & 44060.893 & -8.54991985 & 4.15036544 & Go92 & 50279.435 & -46.92342507 & 1.28705607 & Hermes & 55896.31976 & -21.12224825 & 0.59814161 \\ 
        Ba87 & 44063.887 & -5.08920102 & 4.15036544 & Go92 & 50280.414 & -48.98266309 & 1.32563696 & Hermes & 55896.32064 & -21.05852109 & 0.60074403 \\ 
        W89 & 43378.674 & 5.41679782 & 4.11317732 & Go92 & 50281.384 & -49.71601883 & 1.30511047 & Hermes & 55897.31818 & -20.75098405 & 0.58684017 \\ 
        W89 & 43378.677 & -2.5983225 & 4.11317732 & Go92 & 50284.423 & -36.16898399 & 1.34852265 & Hermes & 55897.31988 & -20.71593417 & 0.58684017 \\ 
        W89 & 43378.762 & -1.49027621 & 4.11317732 & Go92 & 50285.368 & -52.24315992 & 1.35255807 & Hermes & 55898.32335 & -20.19881475 & 0.59169366 \\ 
        W89 & 43378.764 & 2.45681142 & 4.11317732 & Go92 & 50333.327 & -53.5419998 & 2.06624766 & Hermes & 55898.32483 & -20.21686332 & 0.59253809 \\ 
        W89 & 43378.823 & 5.23700572 & 4.11317732 & Go92 & 50342.264 & -50.20834422 & 2.05072655 & Hermes & 55900.35343 & -19.64654849 & 0.59788994 \\ 
        W89 & 43378.825 & 0.77661137 & 4.11317732 & Go92 & 50345.296 & -48.23638834 & 2.05715322 & Hermes & 55900.35566 & -19.76350546 & 0.59916391 \\ 
        W89 & 43381.855 & 2.54430822 & 4.11317732 & Go92 & 50619.493 & 1.15020366 & 4.75839269 & Hermes & 56091.50302 & -3.43372903 & 0.19627055 \\ 
        W89 & 43381.857 & 6.49301724 & 4.11317732 & Go92 & 50621.521 & -2.17907043 & 4.76149147 & Hermes & 56092.56907 & -3.57536873 & 0.24356134 \\ 
        W89 & 43384.665 & 3.01070792 & 4.11317732 & Go92 & 50623.49 & -7.04717834 & 4.76082986 & Hermes & 56093.48335 & -3.50443462 & 0.21692194 \\ 
        W89 & 43384.667 & 2.1753127 & 4.11317732 & Go92 & 50652.37 & -6.34042214 & 0.94654293 & Hermes & 56097.61345 & -3.35857349 & 0.21741694 \\ 
        W89 & 43384.824 & 2.26118771 & 4.11317732 & Go92 & 50653.489 & -4.94554154 & 0.93201047 & Hermes & 56098.60545 & -2.52270761 & 0.23736497 \\ 
        W89 & 43384.826 & 1.52557908 & 4.11317732 & Go92 & 50655.421 & -8.75187071 & 0.9273314 & Hermes & 56099.5956 & -2.63235567 & 0.2454692 \\ 
        W89 & 43385.652 & -1.67001949 & 4.11317732 & Go92 & 50658.47 & -12.37399697 & 0.9273314 & Hermes & 58070.37301 & -32.18905058 & 0.54018969 \\ 
        W89 & 43386.648 & 6.80409712 & 4.11317732 & Go92 & 50662.391 & -9.66269061 & 0.98318031 & Hermes & 58071.37081 & -32.01511877 & 0.56924854 \\ 
        W89 & 43387.664 & 2.95782103 & 4.11317732 & Go92 & 50663.444 & -7.67405698 & 0.97772364 & Hermes & 58071.3752 & -31.40363149 & 0.54832554 \\ 
        W89 & 43388.629 & 0.29105696 & 4.11317732 & Go92 & 50667.466 & -8.52608822 & 0.88540585 & Hermes & 58073.33571 & -31.03450755 & 0.54656738 \\ 
        Bo19 & 56500.515 & 1.37047747 & 2.05595718 & Go92 & 50668.475 & -10.31080233 & 0.9723392 & ~ & ~ & ~ & ~ \\ 
        Bo19 & 56501.573 & 1.51611887 & 2.05597688 & Go92 & 50669.46 & -6.39555779 & 0.9723392 & ~ & ~ & ~ & ~ \\ 
        Bo19 & 56504.602 & 1.60773439 & 2.05593773 & ~ & ~ & ~ & ~ & ~ & ~ & ~ & ~ \\ 
        Bo19 & 56505.509 & 2.05959348 & 2.05599244 & ~ & ~ & ~ & ~ & ~ & ~ & ~ & ~ \\ 
        Bo19 & 56563.322 & 4.51752246 & 2.05600096 & ~ & ~ & ~ & ~ & ~ & ~ & ~ & ~ \\ 
        Bo19 & 56565.431 & 2.32941593 & 2.05600096 & ~ & ~ & ~ & ~ & ~ & ~ & ~ & ~ \\ 
\hline
    \end{tabular}
\tablefoot{The column labeled `Reference' lists the sources of various RV datasets, with corresponding abbreviations as follows: , Ba87 for \cite{1987ApJS...65..307B}, W89 for \cite{1989ApJS...69..951W}, Bo19 for \cite{Borgniet2019}, Go92 for \cite{Gorynya1992}. In column 2 is the Barycentric Julian Date (BJD) of the observation, in column 3 is the residuals between the measurement and the \gc~template (\vgammares) after applying the zero-point offset (shifted according to the phase shift of the epoch) \vgamma~residuals.
Finally, in column 4 we present the uncertainty on the zero-point corrected \vgammares.}
\end{table*}

\end{appendix}
\end{document}